\RequirePackage{fix-cm}
\documentclass[letterpaper,11pt]{article}
\usepackage[T1]{fontenc}

\usepackage[english]{babel}
\usepackage{csquotes}
\usepackage{amsfonts,amssymb,amsmath,amsthm,mathtools}
\usepackage{fullpage}
\usepackage{relsize}
\usepackage{microtype,xspace,bm,floatrow}
\usepackage{hyperref}
\usepackage{booktabs}
\usepackage{multirow}
\usepackage{graphicx}
\usepackage{caption, subcaption}
\DeclareCaptionFont{figurecaption}{\fontsize{10.5}{13}\selectfont}
\usepackage{float}
\restylefloat{table}
\usepackage{siunitx}
\usepackage{bbm, bm}
\usepackage{xcolor}
\providecommand{\mathcolor}[2]{{\color{#1}#2}}
\definecolor{blueviolet}{rgb}{0.2,0.2,0.6}
\definecolor{webgreen}{rgb}{0,0.5,0}
\definecolor{webbrown}{rgb}{0.6,0,0}
\hypersetup{colorlinks=true,linkcolor=blueviolet,citecolor=webgreen,urlcolor=webbrown}
\usepackage{enumitem}
\usepackage{algorithm, algpseudocode}
\usepackage{dsfont}
\usepackage{authblk}
\usepackage{cancel}
\usepackage{braket}
\usepackage{appendix}
\usepackage[capitalise,nameinlink]{cleveref}
\usepackage{xurl}
\usepackage{macros}
\usepackage[numbers,sort&compress]{natbib}
\usepackage{physics}

\makeatletter
\newcounter{algorithmicH}
\let\oldalgorithmic\algorithmic
\renewcommand{\algorithmic}{%
\stepcounter{algorithmicH}\oldalgorithmic}
\def\theHALG@line{ALG\arabic{algorithmicH}.\arabic{ALG@line}}
\makeatother

\title{Classical algorithms for sparse random Hamiltonians \\via free probability}

\author[1,2]{Matthew Ding}
\author[1,3,4,5]{Robbie King}
\author[1,3]{Thomas Schuster}

\affil[1]{Google Quantum AI, Venice, CA 90291, USA}
\affil[2]{Stanford University, Stanford, CA 94305, USA}
\affil[3]{California Institute of Technology, Pasadena, CA 91125, USA}
\affil[4]{Simons Institute for the Theory of Computing, Berkeley, CA 94720, USA}
\affil[5]{Oratomic, Pasadena, CA 91105, USA}

\date{November 24, 2026}
\begin{document}
\maketitle

\begin{abstract}\normalsize
The search for quantum advantage in average-case problems is a pressing open question in physics and computer science.
The widest conjectured example of such an advantage is in simulation of quantum many-body systems, where sign problems and entanglement growth cause the breakdown of many classical methods.
Nonetheless, theoretical evidence for or against such an advantage remains scarce.
Here, we investigate this in sparse random non-local Hamiltonians.
We design a classical algorithm for such Hamiltonians that enables accurate computation of spectral fluctuations, thermal  values, and time-ordered and out-of-time-order correlation functions.
For any subexponential precision $\varepsilon$, our algorithm runs in time $(1/\varepsilon)^{2+o(1)}$, nearly identical to a quantum simulation. 
Crucially, this applies even at late evolution times, low temperatures, and near critical points, where existing classical methods require exponential time.
Rather than directly simulating a quantum state or operator, our algorithm first computes the model's mean behavior and then a short hierarchy of instance-dependent corrections.
These corrections predict long-time behavior based on short-time data using free probability theory.
\end{abstract}

\addtocontents{toc}{\protect\setcounter{tocdepth}{-2}}
\section{Introduction}
Quantum computers are advancing at a remarkable pace~\cite{bluvstein2024logical,google2025belowthreshold,paetznick2026logical}.
However, our understanding of the landscape of problems that admit useful quantum speedups is surprisingly uncharted.
Seminal works identify highly-structured algebraic tasks where quantum algorithms provide large advantages over classical ones~\cite{shor1997polynomial,kammerlander2026compendium}.
Nonetheless, hopes for quantum computing lie also in broader unstructured settings, such as simulation of physical systems~\cite{feynman1982simulating,daley2022practical}, optimization~\cite{farhi2014qaoa,jordan2025dqi}, and machine learning~\cite{biamonte2017qml}.
The simulation of quantum many-body systems is a premier example.
Compared to usual computer science tasks, quantum many-body systems of physical interest are highly unstructured: aside from locality and a small number of conservation laws, their qualitative behavior is famously independent of precise system details~\cite{wilson1974renormalization,deutsch1991thermalization}.
One thus expects that the computational difficulty of predicting their quantitative properties should be similarly independent.
Meanwhile, despite some great successes~\cite{white1992dmrg,vidal2004tebd,sandvik2019sse,carleo2017neural,evenbly2026loops,georges1996dmft,bartlett2007coupledcluster},
classical algorithms for simulating many-body systems encounter substantial  repeated barriers.
In thermal states, these arise from sign problems~\cite{loh1990sign} and strong correlations due to critical effects~\cite{qin2022hubbard},
and in dynamics, from the rapid spread of quantum information and entanglement in time~\cite{paeckel2019time,calabrese2005entanglement,sekino2008scramblers}.
These barriers are naturally surmounted by quantum computers, raising hope for large and ubiquitous quantum advantages in many-body simulation.

Incredibly, however, theoretical evidence  for or against any such advantage is largely absent.
Evidence for advantage is restricted to unnatural `worst-case' constructions~\cite{childs2013universal,zhou2026universal,aharonov2007adiabatic}, empirical observations~\cite{haghshenas2025digital,google2025observation}, or sampling tasks without physical applications~\cite{aaronson2011linearoptics,bremner2016averagecase,bouland2019complexity,movassagh2023hardness,arute2019supremacy}.
Meanwhile, evidence against advantage is mostly limited to regimes where classical easiness is intuitive, such as those with weak interactions~\cite{bravyi2008weak,chen2025weakfermions}, high temperatures~\cite{kliesch2014locality,harrow2020partition,bakshi2024gibbs,zk2026syk}, short evolution times~\cite{wild2023shorttime}, or large noise~\cite{gao2018noisy,aharonov2023noisy,schuster2025noisy}.
This leaves a fundamental question still open: 
Do quantum computers yield generic, scalable advantages for simulating quantum many-body systems in traditionally complex regimes~\cite{anschuetz2025strongly,anschuetz2026glassiness,zlokapa2025glassiness,ding2026sparse,naghdi2026gibbs,hastings2022optimizing,putterman2026thermal,chen2024sparse,zlokapa2026rigorous,schmidhuber2026meanfield,swingle2026syk,liu2026otocs,lewis2026otocs,schuster2026reactivity,google2025observation}?

In this work, we answer this question in a model Hamiltonian introduced by Chen et al.~\cite{chen2024sparse}.
The Hamiltonian consists of $m = \poly(n)$ independently drawn random Pauli operators on $n$ qubits,
\begin{equation} \label{eq:1}
    H = \frac{1}{\sqrt{m}} \sum_{i=1}^m J_i P_i, \quad \quad P_i \sim \{ \mathbb{I}, X, Y, Z \}^{\otimes n},
\end{equation}
with random coefficients $J_i = \pm 1$.
Importantly, the Pauli operators are  fully non-local: this differs from natural systems, but opens an entry point to theoretical analysis~\cite{chen2024sparse}.
The Hamiltonian can be viewed as a sparsification of a Gaussian (GUE) random matrix, a prototypical model of quantum chaos~\cite{cotler2017blackholes,altland2021chaos}, and displays all features commonly associated with classical simulation hardness.
The randomness of the model also allows computation of its properties to be viewed as a well-defined computational problem, whose average-case complexity can be studied.

\begin{figure}[!t]
    \centering
    \includegraphics[width=\textwidth]{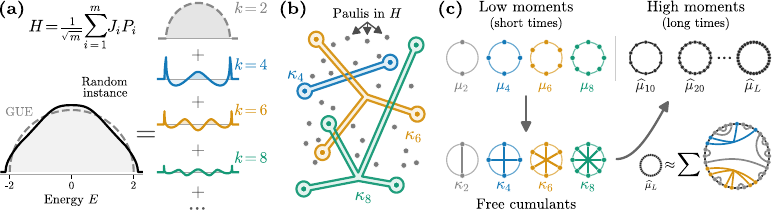}
    \caption{Illustration of our classical algorithm. 
    \textbf{(a)} The spectrum of the sparse Pauli Hamiltonian features instance-to-instance fluctuations about its limiting semicircle behavior.
    Our classical algorithm decomposes these fluctuations into a hierarchy of orders, $k = 2,4,6,8,\ldots$, corresponding to the free cumulants of the model.
    A similar decomposition applies to thermal values and correlation functions.
    \textbf{(b)} The free cumulants $\kappa_k$ measure averaged  properties of commutation relations between subsets of $k/2$ Pauli operators in $H$ [Eq.~(\ref{eq:connected-cumulant-expansion})].
    Their values require an exponentially increasing time to compute, $\Os(m^{k/2})$, due to the high number of subsets, yet also feature an exponentially decreasing magnitude, $\Os(m^{-k/4})$, due to random cancellations between subsets.
    \textbf{(c)} Our algorithm first computes the lowest $\ell$ free cumulants from the low moments of the Hamiltonian.
    It then uses these to approximate the high moments of the Hamiltonian,
    using a diagrammatic resummation from free probability.
    Here, we show $\ell = 8$; the highest moment $L = \Os(t), \Os(\beta)$ is set by the evolution time or inverse temperature of interest.}
    \label{fig:1}
\end{figure}

Our main result is a rigorous and practical classical algorithmic framework for computing a broad range of properties of the sparse random Pauli model, in a time nearly identical to a quantum simulation. 
These properties include nearly all those of physical interest: spectral fluctuations, partition functions, thermal expectation values, and time-ordered and out-of-time-order correlation functions.
We also investigate a variant of the model that undergoes a spectral phase transition, and design a more general algorithm that simulates the critical point of the transition.
In every case, our algorithms compute a  property to accuracy $\varepsilon$ in time $\mathcal{O}(n/\varepsilon^{2+o(1)})$,
for any accuracy, $\varepsilon = 1/2^{o(\frac{n}{\log n})}$, down to near-exponentially small in $n$.\footnote{Here, $g(x) = o(f(x))$ means that $g(x)$ is negligible relative to $f(x)$ as $x$ increases, i.e.~$\lim_{x \rightarrow \infty} g(x)/f(x) = 0$.}
This nearly matches the runtime of a quantum simulation, $\Omega(n/\varepsilon^2)$, for a fundamental reason that we identify.
Notably, our algorithms succeed up to evolution times and inverse temperatures that scale nearly linearly in the system size, $\Os(n/\log n)$.
This exponentially improves upon conventional  methods, which would be limited to $\Os(\log n)$ values.

Our algorithm is built on a general approach, in which we first compute the mean behavior of the model, and then a short hierarchy of instance-dependent corrections around it.
Our results hence illustrate a common mechanism by which seemingly complex average-case problems can fail to yield quantum advantage~\cite{schuster2026reactivity,vonkeyserlingk2022backflow}, despite featuring large instance-to-instance fluctuations~\cite{google2025observation,mele2026otocs,angrisani2025structure}.
In the sparse Pauli model, we find that this hierarchy is most cleanly viewed through the lens of free probability~\cite{Voiculescu1991matrices,speicher1994multiplicative,nica2006lectures,pappalardi2022eigenstate,fava2025designs,vardhan2026free}.
Our algorithm first computes a small handful of \emph{free cumulants} of the model---an analog of the standard cumulants in probability theory---from short-time data.
It then processes these cumulants to make provably accurate predictions about long-time behavior.
The success of this approach follows from an exponential concentration of the free cumulants, which, intuitively, arises due to a free-probabilistic analogue of the central limit theorem.
Our algorithm can be equivalently viewed as a diagrammatic resummation, and bears a similar conceptual spirit to cluster expansions~\cite{kubo1962cumulants,bartlett2007coupledcluster}, the memory matrix~\cite{mori1965transport,zwanzig1961memory,forster2018hydrodynamic}, and concurrent works on other average-case quantum many-body problems~\cite{liu2026otocs,lewis2026otocs,schmidhuber2026meanfield,swingle2026syk}.

Intriguingly, our results leave open the possibility that quantum computers may yield polynomial advantages for computing properties of the sparse Pauli model at late time evolutions and extremely small precisions which vanish exponentially in the system size $n$.
At such values, our algorithm's expansion fundamentally breaks down, and the only remaining known classical approach is state-vector simulation.

\section{Main results}
We now review the sparse Pauli model [\cref{eq:1}], and summarize our main results in  greater detail. We describe the workings of our classical algorithms in the next section, and provide full details and proofs in the Appendix.

\subsection{Review of the sparse Pauli model}

The sparse Pauli model was introduced by Chen et al.~\cite{chen2024sparse}, as an elegant potential candidate for average-case quantum advantage.
This proposal was motivated by the following findings.
In the limit of many terms, $m \gg 4^n$, the model approaches a Gaussian (GUE) random matrix.
In this limit, all properties of the model concentrate to GUE predictions; for example, its spectrum approaches a semicircle distribution [\cref{fig:1}(a)].
Meanwhile, for any physically realizable number of terms, $m = \poly(n)$, the model features significant $\Omega(1/m)$ fluctuations about GUE behavior~\cite{chen2024sparse}.
These fluctuations depend on the random instance of the model (in particular, the choice of terms $P_i$), and are large enough that they can be efficiently resolved on a quantum computer.
At the same time, they are small enough that the model inherits many characteristics of a GUE matrix.
These include standard signatures of quantum chaos and scrambling, as well as a compact spectrum with a semicircle-like edge~\cite{chen2024sparse}.
The latter guarantees that properties of the model, such as its spectral amplitude $Z(t) \equiv \nTr(e^{-iHt})$, where $\nTr(\cdot) \equiv \frac{1}{2^n} \Tr(\cdot)$, display non-trivial signals up to polynomially late times.
It can also be used to prove that low-energy states of the model can be efficiently prepared using quantum phase estimation~\cite{chen2024sparse};
these states were further proven to have at least a polynomial quantum circuit complexity, suggesting they may be difficult to represent classically~\cite{chen2024sparse}.

Together, these findings raise a natural hope: that the long-time and low-energy properties of the sparse Pauli model are easy to compute quantumly, yet difficult classically.
Similar arguments have been put forth to support other average-case quantum advantages, such as simulation of random-circuit OTOCs~\cite{google2025observation,liu2026otocs,lewis2026otocs}.
This hope was buoyed by the fact that the sparse Pauli model displays many other features that stymie classical methods, such as the rapid growth of entanglement, non-local correlations, and sign problems, as the evolution time or inverse temperature are increased.

\subsection{Classical algorithm for sparse random Hamiltonians}
Let us now turn to our results.
We prove that, despite the accuracy of all findings above, the sparse Pauli model nonetheless fails to yield any average-case quantum speedup over classical computation.

\begin{figure}[!t]
    \centering
    \includegraphics[width=\textwidth]{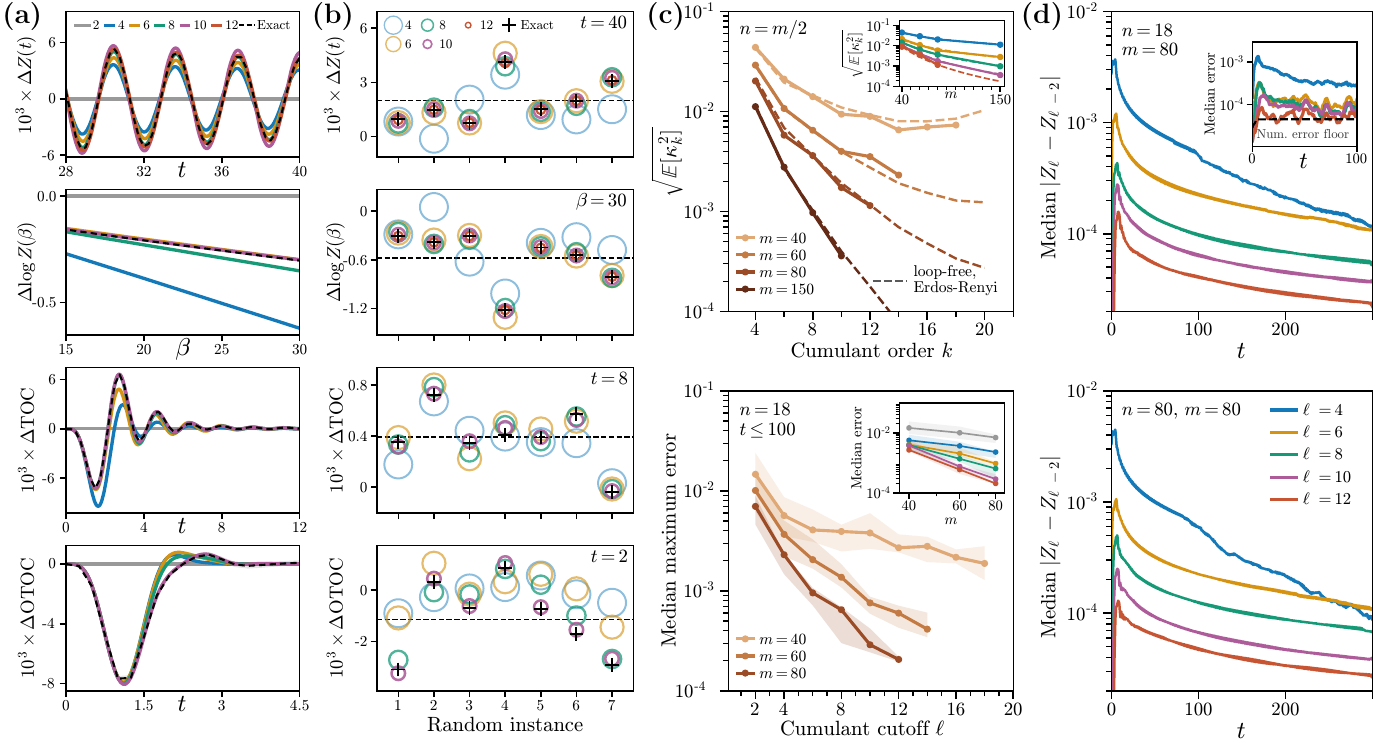}
    \caption{Numerical demonstration of our classical algorithms.
    \textbf{(a)} Convergence of our algorithms (solid) to the exact result (dashed) as $\ell$ is increased from $2$ to $12$ for the spectral amplitude, partition function, TOC, and OTOC, on a single random instance with $m = 80, n = 18$.
    In each case, we subtract the semicircle contribution to isolate instance-dependent fluctuations.
    We note that the sparse Pauli model and our algorithms display similar behavior at the same $m$ regardless of $n$.
    \textbf{(b)} The same comparison, across different random instances at a fixed time or temperature.
    Algorithm predictions correspond to the center of each circle.
    The empirical mean (black dashed) differs from zero at finite $m$.
    \textbf{(c)} Decay of the root-mean-square free cumulants with the order $k$ (top), and corresponding decay of our algorithms' error (bottom).
    Top: Solid lines denote mean over 60 instances; dashed lines denote theoretical predictions.
    Bottom: Solid lines denote median over 10 instances of the maximum error in the spectral amplitude over $0 \leq t \leq 100$; shaded area denotes 25th and 75th percentiles.
    \textbf{(d)} Absolute value change in our algorithms' prediction when increasing $\ell$ by two, smoothed over time windows of width $\Delta t = 5$.
    The fast decay with $\ell$ shows convergence to a precise prediction even at very late times.
    At small $n$, this prediction agrees with the exact result (inset) up to a small numerical error floor (see Appendix).
    At large $n$, similar convergence is observed, though the exact result cannot be checked.}
    \label{fig:2}
\end{figure}

We consider the following broad set of observables to which our algorithms apply.
The qualitative behavior of each observable in the sparse Pauli model mimics that in a GUE Hamiltonian, but with quantitative $\mathcal{O}(1/m)$ corrections that depend on the specific random Hamiltonian instance.
\begin{enumerate}
    \item The spectral amplitude, $Z(t) \equiv \nTr(e^{-iHt})$.
    This is the Fourier transform of the spectrum's density of states, $D(E)$. 
    In the GUE limit, the amplitude displays long-time oscillations with an envelope $\sim t^{-3/2}$, owing to the square-root edge in $D(E)$~\cite{cotler2017blackholes}.
    \item The partition function, $\Zs(\beta) \equiv \text{Tr}(e^{-\beta H})$. This displays exponential growth $\sim e^{2\beta}$ asymptotically, due to the compact support of the semicircle spectrum, $\lVert H \rVert \approx 2$~\cite{chen2024sparse}.
    Our algorithm will only aim to achieve a relative error in computing $\Zs(\beta)$, owing to this exponential factor.
    \item\label{item:thermal-expectation} Thermal expectation values, $\langle A\rangle_\beta \equiv \Tr(A\rho_\beta)$, where $\rho_\beta \equiv e^{-\beta H}/\Zs(\beta)$ is the thermal state.
    We assume $A$ is a Pauli operator in $H$, as otherwise the value is zero with high probability.
    \item Time-ordered correlation functions (TOCs), $G_\beta^A(t) \equiv \Tr(A(t)A(0)\rho_\beta)$.
    \item Out-of-time-order correlation functions (OTOCs), $F_\beta^{A,B}(t) \equiv \Tr(A(t)B(0)A(t)B(0)\rho_\beta)$.
\end{enumerate}
For TOCs and OTOCs, we assume that the operators $A$, $B$ are fixed and independent of the operators in $H$, although our results can be straightforwardly generalized to other scenarios.
Examples of the behavior of each observable can be seen in Fig.~\ref{fig:2}(a,b).
The normalization of the sparse Pauli model is chosen so that all observables begin to fluctuate at order-one evolution time and temperature scales; this is parametrically earlier than the near-linear scales up to which our algorithm will apply.

Our main theorem shows that all of these observables can be classically computed in nearly the same runtime as a quantum simulation with high probability over the random Hamiltonian instance:
\begin{maintheorem}[Classical algorithm for the sparse Pauli model]\label{thm:main-classical-algorithm}
    Fix $m = \Omega(n^3)$.
    For any evolution time $t = o\big(\frac{n}{\log n}\big)$, inverse temperature $\beta = o\big(\frac{n}{\log n}\big)$, accuracy $\varepsilon = 1/2^{o(\frac{n}{\log n})}$,
    and failure probability $\delta = \frac{1}{\poly(n)}$, all observables 1.--5. of Eq.~(\ref{eq:1}) can be classically computed in time  $(1/\varepsilon)^{2+o(1)}(n/\delta) \, \text{\emph{poly}}(t,\beta)$.
\end{maintheorem}
\noindent The exact polynomial degrees in $t, \beta$ are mild and summarized in Table~\ref{tab:complete-runtimes}.
The theorem is proven in the Appendix, and follows from our classical algorithms described in the next section.

The scaling of the runtimes with $n$ and $\varepsilon$ are nearly identical to that of a quantum simulation.
In standard approaches, the quantum gate count for evolving an $n$-qubit Hamiltonian with $m$ non-commuting terms to time $t$ for $\Os(1/\varepsilon^2)$ repetitions is $\tilde{\mathcal{O}}(nm^{3/2}t/\varepsilon^2)$ using qubitization~\cite{Low_2019};
the $\varepsilon$-dependence can be further improved to $\tilde{\mathcal{O}}(1/\varepsilon)$ using amplitude estimation.
Our algorithm shares the trivial linear scaling in $n$ (which arises from the space needed to represent a single Hamiltonian term), mildly improves the joint dependence on $m$ and $t$, and, remarkably, \emph{precisely matches} the original quadratic scaling in $1/\varepsilon$.
The scaling with the accuracy $\varepsilon$ is an especially important attribute of algorithms built on expansions about average-case behavior, as the desired accuracy determines the expansion order.
Our results leave only a standard quadratic speedup for quantum simulations of the sparse Pauli model in this regard, arising from amplitude estimation.
This applies even at extremely small precisions diminishing almost exponentially in $n$.

We remark that the asymptotic scaling of our algorithm in $\varepsilon$ can be further improved to $(1/\varepsilon)^{2\omega/3+o(1)}$ whenever $1/\varepsilon$ is sub-exponential in $m$, where $2\leq\omega \leq 2.25$, using a standard application~\cite{NesetrilPoljak1985} of fast matrix multiplication algorithms~\cite{strassen1969gaussian,openai2026matrix}.
The assumption $m = \Omega(n^3)$ can also be improved to $m = \Omega(n)$ under a plausible combinatorial conjecture.
We refer to the Appendix for the details of these improvements, focusing on the spectral amplitude for simplicity.

\begin{table}[t]
    \centering
    \renewcommand{\arraystretch}{1.2}
    \begin{tabular*}{\linewidth}{@{\extracolsep{\fill}}ll@{}}
        \hline\hline
        Observable & Classical runtime \\
        \midrule
        Spectral amplitude & $(n/\delta)(1/\varepsilon)^{2+\Os(1/\log m)}$ \\
        Partition function & $(n/\delta) (1/\varepsilon)^{2+\Os(1/\log m)}(1+\beta)^2$ \\
        Thermal expectation & $(n/\delta) (1/\varepsilon)^{2+\Os(1/\log m)}(1+\beta)^5$ \\
        TOC and OTOC & $(n/\delta)(1/\varepsilon)^{2+\Os(1/\log m)}(1+\beta)^3(1+\beta+t)^2 $ \\
        Spectral amplitude (spiked model) & $(n/\delta) (1/\varepsilon)^{2+\Os(1/\log m)}(1+t)^2$ \\
        \hline\hline
    \end{tabular*}
    \caption{The runtimes of our classical algorithms in Theorems~\ref{thm:main-classical-algorithm} and~\ref{thm:spiked-classical-algorithm}.
    We refer to Theorems~\ref{thm:scalar-runtime}, \ref{thm:spiked-runtime}, \ref{thm:thermal-partition}, \ref{thm:thermal-correlator-algorithm}, and~\ref{thm:thermal-expectation} in the Appendices for detailed statements and proofs.
    }
    \label{tab:complete-runtimes}
\end{table}

\subsection{Classical algorithm for spectral phase transitions}
A common heuristic argument posits that quantum advantages may be larger near phase transitions in system behavior~\cite{haghshenas2025digital,google2025observation}, as classical algorithm approximations lead to larger  errors, due to non-local and long-time correlations.
Sparked by this argument, we investigate whether a phase transition in the spectrum of the sparse Pauli model leads to any increase in its classical complexity.

\begin{figure}[!t]
    \centering
    \includegraphics[width=\linewidth]{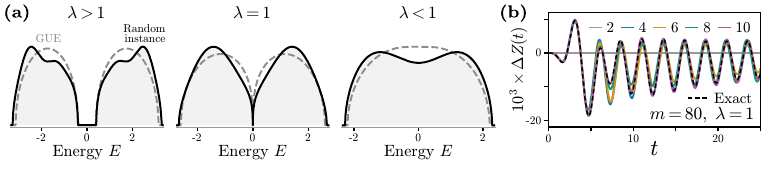}
    \caption{Spectral phase transition in the spiked Pauli model [Eq.~(\ref{eq:2})].
    \textbf{(a)} For $\lambda > 1$, the spiked term dominates, and the spectrum  consists of two disjoint semicircle-like supports at energies $E = \pm \lambda$.
    At $\lambda = 1$, the two supports meet, and display a sharp cusp at $E=0$ in which the density of states vanishes as $D(E) \sim |E|^{1/3}$~\cite{bleher2007externalSourceCritical}.
    This is parametrically sharper than the square-root scaling, $D(E) \sim |E-E_{\text{edge}}|^{1/2}$, associated with a semicircle-like edge.
    For $\lambda < 1$, the random Pauli terms dominate, and the spectrum consists of a single semicircle-like support at $E = 0$.
    \textbf{(b)} Numerical demonstration of our  algorithm for the spectral amplitude at $\lambda = 1$; see Fig.~\ref{fig:2} for further details.}
    \label{fig:3}
\end{figure}

We do so by adding a ``spiked'' term, $\lambda Q$, to the sparse Pauli Hamiltonian,
\begin{equation} \label{eq:2}
    H_{\text{spiked}} = \lambda Q + \frac{1}{\sqrt{m}} \sum_{i=1}^m J_i P_i, \quad \quad P_i \sim \{\mathbb{I},X,Y,Z\}^{\otimes n},
\end{equation}
where $Q$ is a fixed non-identity Pauli operator independent of $H$ and $\lambda\geq 0$ is a tunable constant.
In the GUE limit, the spectrum of the spiked model undergoes a sharp phase transition as $\lambda$ is tuned from above to below one, in which it displays a cube-root cusp at energy zero [see Fig.~\ref{fig:3}(a)]~\cite{bleher2007externalSourceCritical}.
This transition is mimicked when $m = \poly(n)$, where it also becomes sensitive to instance-to-instance fluctuations.
In the GUE limit, the  cusp produces a $t^{-4/3}$ tail in the spectral amplitude at late times, which decays parametrically slower than the $t^{-3/2}$ tail away from the transition; these powers follow by Fourier transforming the cube-root cusp and square-root spectral edges~\cite{bleher2007externalSourceCritical,cotler2017blackholes}.

Unfortunately, we find that this phase transition does not lead to any increase in the complexity of classically simulating the model.
We develop a more sophisticated variant of our classical algorithm, which simulates the spiked Pauli model in the same time as the original model for any $\lambda$.
\begin{maintheorem}[Classical algorithm for the spiked Pauli model]\label{thm:spiked-classical-algorithm}
    Fix $m = \Omega(n^3)$.
    For any spike  $\lambda = \Os(1)$, evolution time $t=o\big(\frac{n}{\log n}\big)$, accuracy $\varepsilon = 1/2^{o(\frac{n}{\log n})}$, and failure probability $\delta = \frac{1}{\poly(n)}$,
    the spectral amplitude of Eq.~(\ref{eq:2}) can be classically computed in time $(1/\varepsilon)^{2+o(1)}(n/\delta)(1+t)^2$.
\end{maintheorem}
\noindent We focus on the spectral amplitude solely for brevity, as it is the simplest observable that is sensitive to the transition. 
The relative factor of $(1+t)^2$ compared to the spectral amplitude of the unspiked model (Table~\ref{tab:complete-runtimes}) is a proof artifact and, we believe, not fundamental.

Our algorithm is described in the next section, and proven in the Appendix.
A surprising finding is that our algorithm is not particularly sensitive to the phase transition, $\lambda = 1$, compared to other values of $\lambda$.
Intuitively, this arises because the polynomial difference in the long-time decay at the transition is overwhelmed by the exponential convergence of our algorithm in its expansion order.

\section{Our algorithms}
We now turn to our classical algorithms.
As aforementioned, the key idea behind our algorithm is that the long-time and low-temperature behaviors of the sparse Pauli model do \emph{not} depend on its randomness in an exponentially complex manner, but rather in a simple manner captured by a small number of classically computable quantities.
We describe this in several steps as follows.

\subsection{Loop-free approximation}
We focus our exposition on classically computing the spectral amplitude, $\nTr(e^{-iHt})$, in the sparse Pauli model, and turn to correlation functions and the spiked model afterwards.
To compute $\nTr(e^{-iHt})$, we provide an algorithm to  compute the Hamiltonian moments, $\nTr(H^k)$, from which $\nTr(e^{-iHt})$ follows by its Taylor expansion.
A similar approach gives the partition function, $\text{Tr}(e^{-\beta H})$.

To compute the moments, we first perform a rigorous diagrammatic decomposition of their values.
By definition, the $k$-th Hamiltonian moment can be written as a sum over $m^k$ Pauli terms, 
\begin{equation}
    \nTr(H^k) = \frac{1}{m^{k/2}} \sum_{i_1,\ldots,i_k} J_{i_1} J_{i_2} \ldots J_{i_k}  \nTr( P_{i_1} P_{i_2} \ldots P_{i_k} ),
\end{equation}
where each index runs from $1$ to $m$, and corresponds to a copy of $H$ appearing in $H^k$.
To proceed, we observe that, with high probability over the Pauli terms appearing in $H$, the summation over terms can be restricted to \emph{pairings}---or more generally, \emph{partitions into even-sized blocks}---of these $k$ copies.
This enforces that all odd moments are zero; hence, from hereon, we assume $k$ is even.
As the simplest example, for the second moment we have
\vspace{1mm}
\begin{equation}
    \vcenter{\hbox{\includegraphics[scale=1.0]{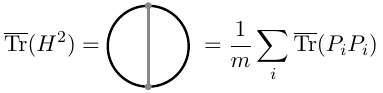}}}
    \label{eq:moment-two-diagrams}
\vspace{1mm}
\end{equation}
where the equality of the two indices is enforced by the trace whenever all Paulis appearing in $H$ are distinct, which occurs with probability exponentially close to one when the terms are random.
Here, we depict the trace as a circle, with each copy of $H$ corresponding to a dot on its perimeter, and the pairing of the two copies denoted by a gray line.
Meanwhile, for the fourth moment, we have
\vspace{1mm}
\begin{equation}
    \vcenter{\hbox{\includegraphics[scale=0.95]{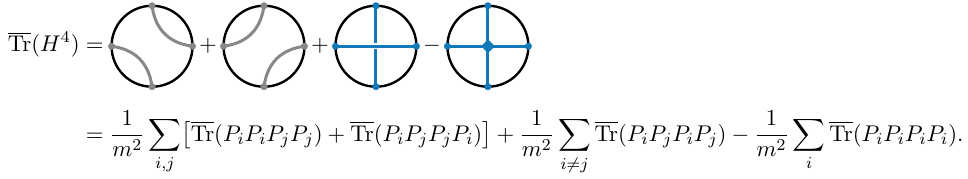}}}
    \label{eq:moment-four-diagrams}
\end{equation}
where the pairing or quadrupling of the indices is enforced whenever there is no ``loop'' of four Pauli terms whose product is proportional to the identity, i.e. $|\nTr(P_{i} P_{j} P_{k} P_{l})| = 1$ for $i,j,k,l$ all distinct.
Similar to before, this is true with probability exponentially close to one when each Pauli term is random.
In the above, the first two terms correspond to pairings that do not ``cross'' each other in the trace, while the third term corresponds to a pairing that crosses.
The final term corresponds to all four indices being equal, which contributes with a negative sign, $+1-2=-1$, to cancel overcounting in the first two terms.
Our choice of coloring will become clear in the following section.

This decomposition can be systematically extended to higher moments, which we refer to as the \emph{loop-free approximation}.
For each moment, we first show that, with high probability, there are no length-$k$ loops of terms such that $|\nTr(P_{i_1} \ldots P_{i_k})| = 1$ while having an odd number of any particular term.
This enforces that non-zero contributions to $\nTr(H^k)$ must arise from partitions of the $k$ copies of $H$ into even-sized ``blocks''.
From this, one finds that each moment can be decomposed as
\begin{equation}\label{eq:loop-free-moment-expansion}
    \nTr(H^{k}) = \frac{1}{m^{k/2}} \sum_{\pi} 
    \,\,\,\,
    \chi(\pi) 
    \!\!\!\!\!\!\!\!\!\!\!\!\!\!\!
    \sum_{\substack{i_1,\ldots,i_{|\pi|} \\ \text{distinct w/in components}}}
    \!\!\!\!\! \!\!\!\!\!\!\!\!\!\!
    \nTr\big(P_{i_{\pi(1)}}\cdots P_{i_{\pi(k)}}\big),
\end{equation}
where $\pi$ are the partitions, $i_{\pi(a)}$ is the index of the block in $\pi$ containing copy $a$, and the factor $\chi(\pi)$ accounts for overcounting.
Similar to the third term in Eq.~(\ref{eq:moment-four-diagrams}), we find it convenient to restrict the sums over $i_1,\ldots,i_{|\pi|}$ to be pairwise distinct within each ``component'' of $\pi$, i.e.~each subset of blocks that cross one another in a connected fashion.
Each partition $\pi$ can be represented diagrammatically, similar to above.
For example, for $k = 6$, we have
\vspace{1mm}
\begin{equation}
    \vcenter{\hbox{\includegraphics[scale=0.96]{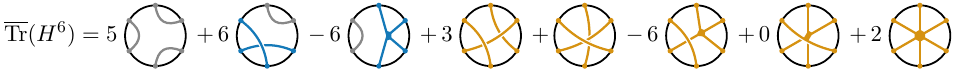}}}
    \label{eq:moment-six-diagrams}
\vspace{1mm}
\end{equation}
In the Appendix, we use techniques from coding theory to prove that, for $m=\poly(n)$, the loop-free approximation is exact for all moments of order $k\leq\gamma n/\log n$ with probability at least $1-2^{-\alpha n}$, for constants $\gamma,\alpha>0$.
This corresponds to times, $t = \mathcal{O}\big( \frac{n}{\log n} \big)$, since the  norm of $H$ is constant.

We remark that the loop-free approximation resembles diagrammatic expansions typically obtained by \emph{averaging} over the random couplings $J_i$ in $H$~\cite{berkooz2025cordial,lin2022bulk,tao2023topics}.
Our results show that, in the sparse Pauli model, such expansions hold even for a single random instance with high probability.

\subsection{Free cumulant expansion}
While the loop-free approximation provides an important simplification, it does not by itself suggest a route to classically computing the moments of $H$.
A typical diagram in the $k$-th moment will involve $\Omega(k)$ many indices, and hence $m^{\Omega(k)}$ many Pauli terms.
The diagram's contribution can depend on these terms in a non-trivial manner, and hence seems to require time $m^{\Omega(k)}$ to compute.
This, as well as the total number of diagrams, quickly explode as $k$ increases.

To alleviate this, the key idea of our algorithm is to prioritize the important (i.e.~largest) diagrams in the loop-free decompositions.
We can motivate this within the fourth moment [Eq.~(\ref{eq:moment-four-diagrams})].
The first two diagrams, $m^{-2} \sum_{i, j} \nTr(P_i P_i P_j P_j)$ and $m^{-2} \sum_{i, j} \nTr(P_i P_j P_j P_i)$, each factorize into a product of pairings and so give value one.
Meanwhile, the third diagram, $m^{-2} \sum_{i \neq j} \nTr(P_i P_j P_i P_j)$, 
is an average of $\mathcal{O}(m^2)$ terms with mean zero and unit variance, since, for each fixed $i$ and $j$, the random Pauli terms $P_i$ and $P_j$ are equally likely to commute or anti-commute. 
The diagram itself thus has mean zero\footnote{This explains our choice to restrict $i$, $j$ to be distinct, which ensures that the mean is zero. A similar motivation underlies the distinctness condition imposed on $i_1,\ldots,i_{|\pi|}$ in our more general formula, Eq.~(\ref{eq:loop-free-moment-expansion}).} and standard deviation $\mathcal{O}(m^{-1})$.
Finally, the fourth diagram, $m^{-2} \sum_{i} \nTr(P_i P_i P_i P_i)$, has value $1/m$, which is small due to the low probability that two out of $m$ indices collide.

To build further intuition, we continue to the sixth moment [Eq.~(\ref{eq:moment-six-diagrams})].
The first diagram is a product of three pairings and has value one.
Meanwhile, the second and third diagrams have mean zero and standard deviation $\mathcal{O}(m^{-1})$, and value $m^{-1}$, respectively, as they involve the product of a two-copy pairing (gray) and an intrinsically four-copy interaction (blue).
The higher diagrams arise from intrinsically six-copy interactions (yellow) and are even further suppressed:
the fourth and fifth diagrams are sums over $\mathcal{O}(m^3)$ random variables with mean zero and unit variance, and thus have mean zero and standard deviation $\mathcal{O}(m^{-3/2})$, while the remaining diagrams are even smaller, $\mathcal{O}(m^{-2})$.
We thus notice a pattern: Diagrams with fewer-copy interactions have larger magnitude, while those with many-copy interactions have smaller magnitude.
This arises because the many-copy interactions either involve many collisions, each of which incurs a cost $\mathcal{O}(m^{-1})$, or are averages of many random variables, $\mathcal{O}(m^j)$, which concentrate to zero with standard deviation $\mathcal{O}(m^{-j/2})$.

This hierarchy is elegantly captured by the notion of \emph{free cumulants} in free probability theory.
Free probability theory extends usual probability theory for scalar random variables to non-commuting matrix random variables. 
A central object in usual probability theory is the normal distribution, which is the limiting distribution of sums of many independent random scalars. 
The analogous central object in free probability theory is the semicircle distribution, which is the limiting \emph{spectrum} of sums of many \emph{freely independent} random matrices. Here, two matrices $A$ and $B$ are freely independent if $\nTr\big(\prod_i [A^{p_i} - \nTr(A^{p_i})]\,[B^{q_i} - \nTr(B^{q_i})]\big) = 0$ for all $p_i, q_i$.
For example, this holds with exponentially small error if $A$ is rotated with respect to $B$ by a Haar-random unitary, $A = U A_0 U^\dagger$, where $A_0$ and $B$ are arbitrary.
This immediately motivates application to our problem: While the Pauli terms in $H$ are not freely independent, they are random with respect to one another via Clifford rotations, and converge to a semicircle distribution in the limit of many terms.
Hence, one might expect that tools from free probability theory are applicable.

In usual probability theory, the \emph{cumulants} of a probability distribution characterize its deviation from a normal distribution. 
Specifically, the $k$-th cumulant isolates the component of the deviation between the $k$-th moment of the distribution and that of the normal distribution that is not explainable by deviations in lower moments.
In free probability theory, the deviation of a random matrix spectrum from the semicircle distribution is similarly captured by its \emph{free cumulants}.
For any spectra whose moments can be expanded as diagrams, such as ours, the $k$-th free cumulant is equal to the sum of all intrinsically $k$-copy diagrams in the $k$-th moment~\cite{Lehner_2002},
\begin{equation}\label{eq:connected-cumulant-expansion}
    \kappa_k(H)=\frac{1}{m^{k/2}}
    \sum_{\substack{\pi\\\text{connected}}} \chi(\pi)
    \sum_{\substack{i_1,\ldots,i_{|\pi|}\\\text{distinct}}}
    \nTr\big(P_{i_{\pi(1)}}\cdots P_{i_{\pi(k)}}\big).
\end{equation}
Here, $\pi$ is summed over all \emph{connected} partitions, i.e.~those whose blocks cross each other in a fully connected fashion.
For example, at $k=10$, the free cumulant includes these diagrams,
\begin{equation}\label{eq:cumulant-ten-diagrams}
    \vcenter{\hbox{\includegraphics[scale=1.0]{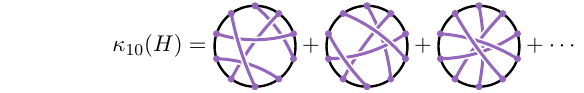}}}
\vspace{1mm}
\end{equation}
but excludes disconnected diagrams that factorize into fewer-copy components,
\begin{equation}\label{eq:excluded-ten-diagrams}
    \vcenter{\hbox{\includegraphics[scale=1.0]{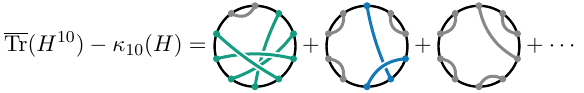}}}
\end{equation}
This is intuitive, as the disconnected diagrams factorize and so are specified by data that is contained in lower moments.
From the free cumulants through order $k$, the $k$-th moment can be obtained as
\begin{equation}\label{eq:moment-cumulant-main-text}
    \nTr(H^k)=\sum_{\pi\in NC(k)}\prod_{V\in\pi}\kappa_{|V|}(H),
\end{equation}
where $NC(k)$ denotes the \emph{non-crossing} partitions of $k$ sites, and $V$ runs over the blocks of $\pi$.
The right side can be efficiently computed by recursion in practice.
The equation can also be inverted to efficiently compute the $k$-th free cumulants from the moments through order $k$ instead.

With this framework in hand, the key observation of our classical algorithm is that the free cumulants of the sparse Pauli model decay exponentially in their order $k$.
This can be seen already in our case-by-case analysis of the low moments.
The fourth free cumulant is the sum of the final two terms (blue) in Eq.~(\ref{eq:moment-four-diagrams}), and so is a random variable with root-mean-square value $\mathcal{O}(m^{-1})$. 
Meanwhile, the sixth free cumulant is the sum of the final five terms (yellow) in Eq.~(\ref{eq:moment-six-diagrams}), and so is $\mathcal{O}(m^{-3/2})$.
In the Appendix, we rigorously prove that the $k$-th free cumulant of the sparse Pauli model has an absolute value 
\begin{equation}\label{eq:cumulant-decay-main-text}
    | \kappa_k |  \leq \mathcal{O}\bigg( \frac{1}{\sqrt{\delta}} \, \bigg[\frac{Ck}{m}\bigg]^{k/4} \, \bigg),
\end{equation}
with probability $1-\delta$. Here, $C$ is a constant, and our bound applies for any $k\leq \gamma n /\log n$, $k = \mathcal{O}(m^{1/3})$.
We also provide evidence that this bound is tight, i.e.~that $|\kappa_k|$ is precisely this order.

The decay of the free cumulants is similar to the decay of the cumulants in usual probability theory, in which the average of $m$ scalar random variables has cumulants $\mathcal{O}(m^{1-k})$~\cite{nica2006lectures}. 
In our setting, the free cumulant decay arises because the connected $k$-copy diagrams are either averages of a high number of random variables, or contain a large number of collisions between indices.
The former provide the leading-order contribution\footnote{This follows because diagrams with collisions between indices are suppressed by a factor of $m^{-1}$ per collision, whereas adding an additional pairing to a connected component leads only to a decrease of its typical magnitude by a factor of $\mathcal{O}(m^{-1/2})$ due to central limit arguments. For example, a consequence of this is that the diagram where all $k$ copies collide has value $m^{1-k/2}$, which is much less than that of diagrams with $k/2$ connected pairings, $\mathcal{O}(m^{-k/4})$.}.
A diagram with $k/2$ pairings crossing in a connected manner is an average over $\mathcal{O}(m^{k/2})$ random variables, and hence has typical magnitude $\mathcal{O}(m^{-k/4})$.
Our proof of Eq.~(\ref{eq:cumulant-decay-main-text}) combines this intuition with a rigorous combinatorial bound on the number of diagrams contributing at each order, which yields the remaining factor of $(Ck)^{k/4}$ in Eq.~(\ref{eq:cumulant-decay-main-text}).

Given the rapid decay of the free cumulants, it is natural to consider approximating the sparse Pauli model's spectrum using only its low free cumulants, approximating higher cumulants as zero.
This leads to our classical algorithm, which we specify in detail in the following subsection.
When keeping only the second free cumulant ($k=2$), our algorithm sums all diagrams of solely non-crossing pairings and yields a semicircle spectrum.
This follows from a well-known counting of the number of non-crossing pairings by the Catalan numbers, $C_{k/2}$, which are equal to the semicircle distribution's moments.
When keeping higher cumulants ($k \leq \ell$, for a chosen threshold $\ell$), our algorithm sums all diagrams that factorize into products of $(k\leq \ell)$-body interactions.
These summations are performed automatically  using the moment-cumulant relation, Eq.~(\ref{eq:moment-cumulant-main-text}).

\subsection{Classical algorithm}

Building on the free cumulant expansion, our classical algorithm is extraordinarily simple:
\begin{enumerate}
    \item First, we compute the moments of $H$ through order $\ell \sim \log(1/\varepsilon)/\log m$. 
    \item Second, we compute the free cumulants of $H$ through order $\ell$ from the moments. 
    \item Third, we compute approximations to the higher moments of $H$ through order $L \sim \log(1/\varepsilon) + t$, using Eq.~(\ref{eq:moment-cumulant-main-text}) and approximating all free cumulants beyond $\ell$ as zero.
    \item Finally, we sum the moments to compute a truncated Taylor expansion of $\nTr(e^{-iHt})$.
\end{enumerate}
The computation in Step 1, described below, is the dominant step and requires time $\mathcal{O}(n(m/C'\ell)^{\ell/2})$, where $C'$ is a constant.
Meanwhile, from the decay of the free cumulants, we rigorously upper bound the error of our algorithm with probability $1-\delta$ by
\begin{equation}
    \varepsilon \leq \mathcal{O}\!\left(\frac{1}{\sqrt{\delta}}\left[\frac{C(\ell+2)}{m}\right]^{(\ell+2)/4}+\left[\frac{2et}{L}\right]^L\right).
\end{equation}
Combining the two yields the runtime\footnote{For example, to achieve an accuracy $\varepsilon < \mathcal{O}(m^{-1})$, one sets $\ell = 4$, which gives a runtime $\mathcal{O}(m^2)$. To achieve $\varepsilon < \mathcal{O}(m^{-3/2})$, one sets $\ell = 6$, which gives a runtime $\mathcal{O}(m^3)$. And so on.} $(n/\delta)(1/\varepsilon)^{2+\mathcal{O}(1/\log m)}$ in Table~\ref{tab:complete-runtimes}.

To compute the moments in Step 1, for each even $k \leq \ell$, we first compute the coefficients of $H^{k/2}$ in the Pauli basis, $H^{k/2} = \sum_R c_R R$, from which the moments are $\nTr(H^k) = \sum_R |c_R|^2$.
This can be done in time $\Os(m^{\ell/2}/(\ell/2)!)$, since there are at most this many Pauli operators appearing in $H^{k/2}$. 
This follows because each operator is the product of $k/2$ out of $m$ terms in $H$, and the product is independent of the order up to a sign.
Steps 2 and 3 then follow easily from the moment-cumulant relation, Eq.~(\ref{eq:moment-cumulant-main-text}).
Step 4 is also immediate, where truncating the Taylor expansion of $\nTr(e^{-iHt})$ at order $L$ produces error at most $\mathcal{O}((2e t / L)^L)$ from the Taylor remainder theorem.

Beyond the free cumulant bound, the key technical innovation in our proof is a method to translate our bounds on the higher free cumulants to bounds on the approximation error in $\nTr(e^{-iHt})$.
To do so, we leverage a known re-interpretation of the moment-cumulant formula [Eq.~(\ref{eq:moment-cumulant-main-text})] in terms of dynamics on a small effective Fock space.
The Fock space contains states, $\{ \ket{0}, \ket{1}, \ket{2}, \ldots \}$, and a non-Hermitian Hamiltonian, $H_{\text{eff}} = a^\dagger + a + \sum_{k > 2} \kappa_k a^{k-1}$, where $a^\dagger \ket{n} = \ket{n+1}$ and $a \ket{n} = (1-\delta_{n,0}) \ket{n-1}$.
Remarkably, the moments of $H_{\text{eff}}$ in the zero state are exactly equal to the moments of $H$, $\bra{0} H_{\text{eff}}^k \ket{0} = \nTr(H^k)$.
Hence, the spectral amplitude corresponds to the return amplitude of a quantum walk in the Fock space, $\nTr(e^{-iHt}) = \bra{0} e^{-iH_{\text{eff}}t} \ket{0}$.
Our algorithm truncates terms in $H_{\text{eff}}$ with $k > \ell$.
Since $a$ has spectral norm one, the sum of these terms has norm at most $\sum_{k>\ell} |\kappa_k|$.
This produces an error at most $t \sum_{k>\ell} |\kappa_k| \approx t | \kappa_{\ell+2} |$, since $H_{\text{eff}}$ is Hermitian to leading order in $m$.
The factor of $t$ can be further eliminated by analyzing the specific dynamics of $H_{\text{eff}}$.
We refer to the Appendix for details and a pedagogical derivation of the Fock space mapping.

\subsection{Correlation functions and the spiked model}
The free cumulants contain data regarding the moments of $H$, but not its interactions with other operators.
To extend our classical algorithm to thermal values, TOCs, OTOCs, and the spiked model, we extend the free cumulant expansion to more general moment quantities, such as $\nTr(HQHQH^2QH^4Q)$. These appear when such quantities are Taylor expanded in powers of $H$.
We term this extension the \emph{color cumulant expansion}, for reasons that will become clear.

Let us introduce our extension in the context of the spectral amplitude of the spiked model, $H_{\text{spiked}} = H + \lambda Q$.
The Taylor expansion of the spectral amplitude in $t$ and $\lambda$ consists of moment quantities as above, where powers of $H$ are interspersed with single applications of the spike $Q$ (recalling $Q^2 = 1$).
With high probability, these quantities are loop-free, similar to the moments of $H$.
Hence, they can be expanded as sums of diagrams, such as
\begin{equation}
    \vcenter{\hbox{\includegraphics[scale=1.0]{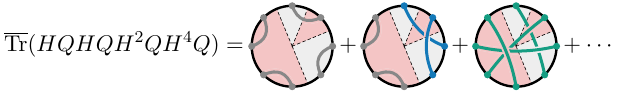}}}
    \label{eq:color-moment-eight-diagrams}
    \vspace{1mm}
\end{equation}
The four applications of $Q$ correspond to the four locations where the background color switches between gray and red.
This depiction is motivated by the fact that any pairing that begins and ends in the same color region will cross $Q$ an even number of times, while any pairing that begins and ends in opposite-color regions will cross $Q$ an odd number of times.
These lead to either no sign, or a sign  $\nTr(P_i Q P_i Q) = \pm 1$, in the diagram's contribution, where $P_i$ is the Pauli term corresponding to the pairing.
An analogous statement can be made for larger even-size blocks.

From these diagrams, our color cumulant expansion is straightforward.
For every moment order $k \leq L$, and every coloring within that moment order, we perform an analogous free cumulant expansion as in the original sparse Pauli model.
That is, we keep all diagrams that factorize into products of interactions between $k'$ copies of $H$, for any $k' \leq \ell$.
While easily phrased, turning this statement into an efficient algorithm encounters two subtleties, associated with the explosion of the number of colorings with the order $k$.
First, our algorithm must compute more low-order quantities than before.
Specifically, there are $\mathcal{O}(2^{\ell})$ more ``color cumulants'' than usual free cumulants, corresponding to the number of colorings of $k' \leq \ell$ vertices.
This increases the runtime by $\mathcal{O}(2^{\ell})$, which is negligible asymptotically. 
Second, translating the low-order color free cumulants into approximations of the high moments of $H_{\text{spiked}}$ is challenging, since there are $\sim 2^L$ possible placements of $Q$ at order $L$, which cannot be brute-force summed.
We solve this by deriving a generalized moment-cumulant relation using ideas from operator-valued free probability theory~\cite{voiculescu1995operations,speicherCombinatorialTheoryFree1998}.

We emphasize that this expansion differs from a naive expansion in terms of the \emph{mixed free cumulants}~\cite{nica2006lectures} of $H$ and $Q$, which is essential to the efficiency of our algorithm.
The mixed free cumulant expansion cannot be easily truncated, since the spectrum of $Q$ is very far from a semicircle and so its free cumulants do not decay as the order increases.
Our algorithm instead builds an expansion around the simple defining property of $Q$, that it has zero trace and squares to one.

Our algorithm immediately extends to computing the TOCs and OTOCs of the sparse Pauli model, which involve subsets of $\binom{k}{2}$ and $\binom{k}{4}$ possible colorings, respectively.
We can also apply it to thermal expectation values, $\Tr(A \rho_\beta)$, for $A$ in $H$, by decomposing $H = \frac{1}{\sqrt{m}} J_1 A + \frac{1}{\sqrt{m}} \sum_{i=2}^m J_i P_i$, assuming $A$ is the first term in $H$ without loss of generality.
We refer to the Appendix for details.

\section{Discussion}
Our work shows a stark mechanism by which average-case quantum many-body problems can be classically easy despite displaying trademark features of hardness.
This arises due to a fundamental competition between complexity and concentration when expanding instance-to-instance fluctuations about average-case behavior.
To our knowledge, our results also represent the first rigorous classical algorithm for computing long-time and low-temperature properties of a  generic class of strongly-interacting quantum Hamiltonians.

Perhaps the most striking feature of our results is how \emph{precisely} the classical runtime collapses to the quantum value, $\mathcal{O}(n/\varepsilon^2)$.
In retrospect, however, this collapse is quite understandable.
The leading-order contributions to the $\ell$-th free cumulant are averages of $\mathcal{O}(m^{\ell/2})$ random variables.
The classical time to compute the average is $\mathcal{O}(m^{\ell/2})$ since there are this many terms.
Meanwhile, the quantum time to resolve the average is also $\mathcal{O}(m^{\ell/2})$, since the average of $\mathcal{O}(m^{\ell/2})$ many terms has typical magnitude $\mathcal{O}(m^{-\ell/4})$.
The classical time to resolve the average using Monte Carlo sampling of terms is also $\mathcal{O}(m^{\ell/2})$,  for the same reason.
Hence, the lack of any quantum speedup arises for the simplest reason possible: 
a signal receiving random contributions from many sources requires the same number of random shots to resolve on a quantum computer as random samples on a classical computer.
Our analysis using free probability theory translates this fact into predictions for observables in the sparse Pauli model.
However, the fundamental cause of the collapse of the classical runtime to the quantum runtime appears plausibly general.

To this end, although the sparse Pauli model has fully non-local interactions, we hope that our results may provide a starting point for understanding complexity in systems with few-body interactions as well.
An analogy can be made to the history of  spin glass physics, in which the study of highly non-local random models enabled early progress, before extending to more realistic settings~\cite{mezard2009information}.
In this direction, a natural next target would be the Sachdev--Ye--Kitaev (SYK) model~\cite{SachdevYe1993,Kitaev2015,MaldacenaStanford2016}.
Compared to recent concurrent algorithms for the SYK model~\cite{zlokapa2026rigorous,swingle2026syk,schmidhuber2026meanfield}, our results extend to longer times and lower temperatures, and display an exact agreement between the classical and quantum runtimes rather than a polynomial separation.
It would be interesting to understand whether these features can extend to the SYK model;
a potential route may involve the double-scaled SYK model~\cite{berkooz2025cordial,lin2022bulk} and $q$-Gaussian algebras~\cite{Speicher1992clt}, which continuously interpolate from the free probability regime of our work to the standard SYK limit.

More broadly, our work joins a small handful of other recent classical algorithms based on expansions about mean or limiting quantum behavior.
These include algorithms for random circuit expectation values~\cite{angrisani2025structure} and random circuit OTOCs~\cite{liu2026otocs,lewis2026otocs}, as well as the SYK model~\cite{swingle2026syk,schmidhuber2026meanfield}.
From a physical perspective, an elegant explanation~\cite{liu2026otocs} of these results is that quantum chaos, by dispersing microscopic information into complex unobservable features, can lead to simpler classical algorithms rather than more complex ones.
How broadly these approaches extend---and what physical phenomena might evade them---is a pressing question.

In light of our results, one specifically interesting question is whether our quadratic dependence on the accuracy $\varepsilon$, which matches the quantum runtime without amplitude estimation, might extend broadly to this class of algorithms above.
Currently, these algorithms obtain significantly worse dependencies on $\varepsilon$, which leave open a possibility of large polynomial or quasi-polynomial quantum advantages at small accuracies.
Obtaining an optimal dependence on $\varepsilon$ can be challenging for such algorithms, as the dependence is determined by a delicate competition between an exponential increase in classical runtime, and an exponential decrease in the algorithm error, as the expansion order increases.
Small modifications of the details of the expansion can alter the ratio of these exponents, and lead to polynomial changes in the runtime.
The simplicity of the sparse Pauli model allowed us to identify an `optimal' basis for this expansion, i.e.~the free cumulants, in our work.

Several more specific questions regarding the sparse Pauli model remain open as well.
While our numerical results (Fig.~\ref{fig:2}) suggest that our algorithm might extend beyond linear times, a rigorous handle on this regime remains lacking.
The model's behavior below inverse linear temperatures also remains unexplored.
Perhaps most intriguing, however, is that our algorithm appears fundamentally unable to compute properties at very high inverse accuracies relative to the system size: specifically, in the regime where we have $1/\varepsilon$ scaling exponentially in $m$, late-time evolutions $t=\Omega(n)$, and $m = \Theta(n)$ is of moderate size.
To reach these accuracies, the free cumulant expansion would require order $\ell = \Omega(m)$; however, at this order, the expansion begins to diverge [see e.g.~Fig.~\ref{fig:2}(c)] due to a combinatorial explosion in the number of diagrams.
We perform a  heuristic analysis of this regime in the Appendix, which suggests that a polynomial quantum advantage over known classical algorithms could remain here.
A common theme in all of these questions is that their answers require understanding the behavior of higher moments, $k = \Omega(n)$, where our techniques break down.

\subsection*{Acknowledgments}
We are grateful to Ryan Babbush, Fernando Brand\~{a}o, Chi-Fang Chen, Jeongwan Haah, and Shreya Vardhan for valuable discussions and insights.
MD was supported in part by the National Science Foundation Graduate Research Fellowship Program under Grant No.\ DGE-2146755.
TS acknowledges support early in the project from the Walter Burke Institute for Theoretical Physics, the Institute for Quantum Information and Matter, and the U.S. Department of Energy, Office of Science, National Quantum Information Science Research Centers, Quantum Systems Accelerator.

\subsection*{AI disclosure}
Our main classical algorithm and its proof of efficiency were developed solely by the authors, one to two years ago, without the use of AI. 
This includes all main ideas described in the main text and Part~\ref{part:sfa-algorithm} of the Appendix.

We then more recently used OpenAI's ChatGPT 5.6 and 6 Pro collaboratively to optimize and extend our results.
These include improving the maximum evolution time from cube-root to near-linear in $n$ via a tighter combinatorial analysis, improving the failure probability from fixed to arbitrary inverse polynomial, and improving the asymptotic runtime using fast matrix multiplication.
The algorithms and proofs for correlation functions and the spiked model were also developed by the authors, with the idea to use operator-valued free probability to compute moments from the color cumulants coming from conversations with ChatGPT 5.6 Pro.
Minor technical assistance from AI was also used to apply the color framework, primarily completing combinatorial and complex-analytic proofs.
We acknowledge the use of OpenAI's Codex in numerical simulations.

Our main text and Parts~\ref{part:sfa-algorithm} to~\ref{part:numerical-results} of our appendix are written entirely by the authors.
Our final appendix, Part~\ref{part:technical-proofs}, contains select technical proofs generated by AI and verified by the authors; as mentioned, these improve the parameters of our results but are not necessary to their understanding.

\newpage
\appendix
\noindent\textbf{\LARGE Appendices}\par
\vspace{2em}

\tableofcontents
\addtocontents{toc}{\protect\setcounter{tocdepth}{2}}

\newpage

\part{Computing spectral amplitudes}\label{part:sfa-algorithm}

We begin by considering the simplest task for our classical algorithms, which is computing the spectral amplitude of the sparse Pauli model. This gives an introduction to our techniques which we will generalize to derive all of our other classical algorithms. 

\cref{sec:preliminaries} gives preliminaries and notation which we use in \cref{part:sfa-algorithm} as well as the remainder of the paper. We formalize the idea of loop-freeness in \cref{sec:loop-free} and give an exact expression of the free cumulants under loop-freeness in \cref{sec:commutation-graphs}. Using this expression, the concentration of free cumulants is proved in \cref{sec:free-cumulant-concentration}.

In \cref{sec:fock-space}, we give an introduction to the Fock space interpretation and formalize the connections with our algorithm. Finally in \cref{sec:classical-sfa-algorithm}, we utilize this Fock space interpretation along with the concentration of free cumulants to prove the correctness of our classical algorithm for the spectral amplitude.

\section{Preliminaries}\label{sec:preliminaries}
We begin with some general notation and definitions we use in the paper. Then within the two subsections we give broad and extensive background on a few relevant subfields of mathematics which we utilize in the construction and analysis of our algorithm. \textbf{Only Subsections~\ref{subsubsec:nc-partitions} and \ref{subsubsec:free-cumulants} are essential for understanding the rest of the paper}, while the remainder of the preliminaries may initially be skipped during a first reading and referred to when referenced in later passages.

In \cref{subsec:prelims-combinatorics}, we provide definitions from the combinatorics of partitions, partially-ordered sets, and lattices that form the combinatorial basis of free probability theory~\cite{nica2006lectures}. Specifically, in \cref{subsubsec:nc-partitions} we introduce non-crossing partitions and their relation to free cumulants; the main definitions in this subsection will be essential for the entire paper. In \cref{subsubsec:nc-multiplicative} we define $NC$-multiplicative functions, a key property of free cumulants. In \cref{subsubsec:partition-factorization} we give the relation between non-crossing partitions and connected partitions and in \cref{subsubsec:incidence-algebras} we introduce incidence algebras, which allow us to bound free cumulants using combinatorial methods. The dependency on \cref{subsubsec:nc-multiplicative,subsubsec:partition-factorization,subsubsec:incidence-algebras} is fairly localized and is primarily used to bound the free cumulants of the sparse Pauli Hamiltonian in \cref{sec:commutation-graphs,sec:free-cumulant-concentration}. 

Then in \cref{subsec:prelims-free-probability} we begin with an expository introduction to free probability. We formally introduce the concept of free cumulants in \cref{subsubsec:free-cumulants} and the $R$-transform in \cref{subsubsec:R-transform}. Only \cref{subsubsec:free-cumulants} is essential for the entire paper, while \cref{subsubsec:R-transform} is used primarily to derive the algorithm in \cref{sec:classical-sfa-algorithm}.

\paragraph{General notation and conventions.} 
\begin{itemize}
    \item Pauli operators: For $n$-qubit Pauli operators $P_i$ and $P_j$, we define
    \begin{equation}
        \langle P_i,P_j\rangle =
        \begin{cases}
            +1 & \text{if $P_i$, $P_j$ commute} \\
            -1 & \text{if $P_i$, $P_j$ anticommute} \\
        \end{cases}
    \end{equation}
    to denote their commutation relation. When the sparse Pauli Hamiltonian instance 
    \[
        H = \frac{1}{\sqrt{m}} \sum_{i=1}^m J_i P_i
    \]
    is clear from context, we define its \emph{Pauli support} as the sequence $\Pm\equiv (P_1, \ldots, P_m)$ and its coefficient vector as the vector $\va{J}\equiv (J_1, \ldots, J_m)\in \{\pm 1\}^m$.
    
    For an $n$-qubit Pauli operator $P$, we use $[P]$ to denote its equivalence class modulo phase.

    \item Set and function notation: We denote the set $[n]\equiv \{1,2,\ldots, n\}$. Let $(m)_{r} \equiv m(m-1)\dots(m-r+1)$, with $(m)_0 \equiv 1$. 
    
    For disjoint finite sets $A,B \subseteq [N]$, $N\in\N$ let
    \begin{equation}
        \operatorname{inv}(A,B)\equiv|\{(a,b)\in A\times B:\ b<a\}|
    \end{equation}
    denote the number of \emph{inversions}, and let
    \begin{equation}
        \operatorname{cr}(A,B)\equiv |\{(a_1, a_2, b_1, b_2)\in A^2\times B^2: a_1<b_1<a_2<b_2\ \text{or}\ b_1<a_1<b_2<a_2\}|
    \end{equation}
    denote the number of \emph{crossings} (alternating quadruples).
    
    A function $f$ with domain $A$ and range $B$ is denoted $f:A\rightarrow B$. We may also use $f:A\hookrightarrow B$ if $f$ is additionally an injective function.
    
    For non-commutative random variables $(x_i)_{i\in V}$ over a set of indices $V$, we assume any product $\prod_{i\in V} x_i$ is taken in increasing order 
    \[
        \prod_{i\in V} x_i = x_{i_1}x_{i_2}\cdots x_{i_k},\qquad V=\{i_1, \ldots, i_{k}\},\quad i_1<\cdots<i_k,
    \]
    unless noted otherwise.

    \item Runtimes: All running times count arithmetic operations on exact real numbers, each at unit cost, together with $\Os(n)$-time operations on $n$-qubit Pauli strings; we do not track the bit complexity of the arithmetic.
    
    We say $x = \poly(n)$ if there exists a constant $c>0$ such that $x=\Os(n^c)$. For the sparse Pauli Hamiltonian, we assume the number of terms satisfies $m\geq 2n$ and $m=\poly(n)$ throughout. Constants may implicitly depend on the polynomial degree of $m=\poly(n)$ and $1/\delta = \poly(n)$.
\end{itemize}

\subsection{Combinatorics}\label{subsec:prelims-combinatorics}
A partially-ordered set (``poset'') is a pair $(P,\leq)$ where $P$ is a set and $\leq$ defines a partial-order over it \cite{Stanley_vol1}.
\begin{definition}[Partial order]
    A partial order on set $P$ is a binary relation $R\subseteq P\times P$, where we use the notation $x_1\leq x_2 \Longleftrightarrow (x_1,x_2)\in R$, satisfying the following axioms:
    \begin{enumerate}
        \item Reflexivity: for all $x\in P$, $x\leq x$.
        \item Antisymmetry: for all $x,y\in P$, if $x\leq y$ and $y\leq x$, then $x = y$.
        \item Transitivity: for all $x,y,z\in P$, if $x\leq y$ and $y\leq z$, then $x\leq z$.
    \end{enumerate}
\end{definition}
A lattice is a type of poset with special structure.
\begin{definition}[Lattice]
    A partially-ordered set $P$ is called a \emph{lattice} if for any elements $\pi, \sigma \in P$, there exists a \emph{meet} and \emph{join} of $\pi$ and $\sigma$:
    \begin{itemize}
        \item Meet ($\pi \wedge \sigma $): if the set of $\{\rho\in P\mid (\rho \leq \pi) \wedge (\rho\leq \sigma)\}$ is non-empty and has a maximum element, $\pi \wedge \sigma $ is defined as this maximum element.

        \item Join ($\pi \vee \sigma $): if the set of $\{\tau\in P\mid (\tau \geq \pi) \wedge (\tau\geq \sigma)\}$ is non-empty and has a minimum element, $\pi \vee \sigma $ is defined as this minimum element.
    \end{itemize}
    A \emph{sublattice} is a nonempty subset of a lattice that is closed under its meet and join.
\end{definition}

\begin{definition}[Intervals]
Given a partially-ordered set $P$ and elements $\pi, \sigma \in P$ satisfying $\pi\leq \sigma$, the interval $[\pi, \sigma] \subseteq P$ is defined as $\{\tau\mid \pi \leq \tau \leq \sigma\}$. If $P$ is a lattice, the interval $[\pi, \sigma]$ is also a sublattice of $P$. 
\end{definition}
We can also extend this notion by defining the set of length $k-1$ \emph{multi-chains} in $P$: $P^{(k)}=\{(\pi_1, \pi_2,\dots,\pi_k): \pi_1 \leq \pi_2\leq \cdots \leq \pi_k\}$, e.g., $P^{(1)}=P$, and $P^{(2)} = \{(\pi, \sigma):\pi,\sigma\in P, \pi\leq\sigma\}$ is the set of all intervals on $P$. 

One can also take the direct product of partially-ordered sets $P_1\times P_2\times\dots\times P_n$, which has an induced partial-order defined as $(\pi_1, \pi_2,\dots,\pi_n) \leq (\sigma_1,\sigma_2,\dots,\sigma_n) \Longleftrightarrow \forall i: \pi_i\leq\sigma_i$. 

\subsubsection{Non-crossing partitions}\label{subsubsec:nc-partitions}
Partitions and non-crossing partitions are key combinatorial objects appearing in (free) probability theory \cite{nica2006lectures}.
\begin{definition}[Partitions and non-crossing partition]
    A set $\pi = \{V_1, V_2,\dots,V_r\}$ is a partition of finite set $S$ if all sets $V_i \subseteq S$ are nonempty and pairwise disjoint and $V_1\sqcup V_2\sqcup \cdots\sqcup V_r = S$. A partition is \emph{non-crossing} if for all $a<b<c<d$, $a\sim_\pi c$ and $b \sim_\pi d$ together imply $b \sim_\pi c$. We let $\Ps(n)$ denote the set of all partitions of $[n]$ and let $NC(n)$ denote the set of all non-crossing partitions of $[n]$. Thus $NC(n)\subseteq \Ps(n)$.
\end{definition}

The set of non-crossing partitions forms a lattice with partial order defined as $\forall \pi, \sigma \in NC(n)$, $\pi \leq \sigma$ if $\pi$ is a refinement of the blocks of $\sigma$, called ``reverse-refinement order'' \cite{kreweras1972partitions}. Thus the maximum element is the partition of a single block (denoted $1_n$) and the minimum element is the partition of singletons (denoted $0_n$). For a partition $\pi\in \Ps(n)$, we use $|\pi|$ to denote the number of blocks in $\pi$ and $\pi(i)$ to denote the index of the block containing element $i\in[n]$ for some predefined block labeling. The number of non-crossing partitions of $n$ elements is given by the Catalan numbers:
\begin{equation}
    |NC(n)|=C_n \equiv \frac{1}{n+1} \binom{2n}{n}.
\end{equation}
There exists a canonical factorization of intervals in $NC(n)$: for any $\pi\leq \sigma \in NC(n)$, there exist integers $k_1,\ldots k_n \geq 0$ such that
\begin{equation}
    [\pi, \sigma] \cong NC(1)^{k_1}\times NC(2)^{k_2}\times\dots\times NC(n)^{k_n}.
\end{equation}

\subsubsection{\texorpdfstring{$NC$}{NC}-multiplicative functions}\label{subsubsec:nc-multiplicative}
\begin{definition}[Multiplicative functions on $NC$ \cite{speicher1994multiplicative}]\label{def:nc-multiplicative}
    Let $(\alpha_i)_{i\in\N}$ be a sequence of complex numbers. Consider elements $\pi, \sigma \in NC(n)$, with $\pi\leq\sigma$ and $[\pi, \sigma]$ has canonical factorization $NC(1)^{k_1}\times NC(2)^{k_2}\times\dots\times NC(n)^{k_n}$. A (bivariate) family of functions $(F_n)_{n\in \N}$ is called a \emph{multiplicative family of functions on $NC(n)$} (or $NC$-multiplicative) if
    \begin{equation}
        F_n(\pi, \sigma) = \alpha_1^{k_1}\alpha_2^{k_2}\dots\alpha_n^{k_n}.
    \end{equation}
    for any choice of $\pi, \sigma \in NC(n)$. Note that given a multiplicative family $F_n$, $\alpha_i$ is fixed for all $i\in\N$ since $\alpha_i = F_i(0_i, 1_i)$.

    A (univariate) family of functions $f_n: NC(n) \rightarrow \C$ is NC-multiplicative if for all $\pi = \{V_1, \dots, V_r\}$
    \begin{equation}
        f_n(\pi) = \alpha_{|V_1|}\alpha_{|V_2|}\dots \alpha_{|V_r|}.
    \end{equation}
\end{definition}

Given a multiplicative family of functions $(f_n)_{n\in\N}$ with associated scalar sequence $(\alpha_n)_{n\in\N}$, the \emph{multiplicative extension} $(\alpha_\pi)_{n\in \N, \pi\in NC(n)}$ is defined as $\alpha_{\pi} \equiv f_n(\pi)$ for all $n\in\N$ and $\pi$ in $NC(n)$. This is an extension because $\alpha_{1_n} = \alpha_n$, and the extension map $\pi \mapsto \alpha_\pi$ is a multiplicative function on $NC(n)$. This definition of multiplicative functions on a poset can be extended to multilinear functionals on an algebra.
\begin{definition}[Multiplicative functionals on an algebra]\label{def:multiplicative-function}
    Let $\As$ be an algebra and $(\rho_k)_{k\in \N}$ be a sequence of multilinear functions on $\As$ denoted $\rho_k: \As^k \rightarrow \C$, where $\rho_k$ acts as the map $(a_1, \ldots, a_k)\mapsto \rho_k(a_1,\ldots,a_k)$. We can extend this to a family of multiplicative functionals on $NC$ denoted as $\rho_\pi: \As^k \rightarrow \C$, where $\rho_\pi$ acts as the map $(a_1,\ldots, a_k)\mapsto\rho_\pi[a_1,\ldots,a_k]$, defined as follows: if $\pi = \{V_1, \ldots, V_r\}\in NC(k)$, then
    \begin{equation}
        \rho_\pi[a_1,\ldots,a_k]\equiv \rho(V_1)[a_1,\ldots,a_k]\cdots \rho(V_r)[a_1,\ldots,a_k]
    \end{equation}
    where 
    \begin{equation}
        \rho(V)[a_1,\ldots,a_k] \equiv \rho_{k'}(a_{i_1},\ldots,a_{i_{k'}}),\qquad V=\{i_1, \ldots, i_{k'}\},\quad i_1<\cdots<i_{k'}.
    \end{equation}
\end{definition}

\subsubsection{Unique factorization of partitions}\label{subsubsec:partition-factorization}
A general theme in combinatorics is that many objects admit unique decompositions into irreducible components which allow them to be more easily understood or enumerated, e.g., prime factorizations of natural numbers or connected components in graphs. We specifically use the general framework derived by Beissinger \cite{beissinger1985enumeration} applied to partitions in a manner which will be important in the analysis of free cumulants. 

In our setting, the irreducible components of partitions are called \emph{connected partitions}:
\begin{definition}[Connected partitions \cite{Lehner_2002, Arizmendi_2015}]
    A partition $\pi\in\Ps(n)$ is called \emph{connected} if there does not exist any interval $I=\{k, k+1, \ldots, k+r-1, k+r\}\neq [n]$ such that $\pi=\pi_1\sqcup\pi_2$, where $\pi_1\in \Ps(I)$ and $\pi_2\in\Ps([n]\setminus I)$. We denote the poset of all connected partitions of $[n]$ as $\tilde{\Ps}(n)$.
\end{definition}
\begin{remark}
    Connected partitions are also called \emph{indecomposable} partitions by Nica--Speicher \cite{nica2006lectures}. Older works from enumerative combinatorics originally called them \emph{irreducible} partitions \cite{kleitman1970proportions,hsieh1973proportions, beissinger1985enumeration}, however we avoid this usage since it conflicts with a distinct definition in Arizmendi et al.~\cite{Arizmendi_2015}. The decomposition of $\Ps(n)$ into connected (partition) components is also analogous to the decomposition of graphs into connected (graph) components.
\end{remark}
The general factorization framework of \cite{beissinger1985enumeration} also relies on an instance-dependent closure operator acting on the irreducible components. The closure operator in our specific setting is the \emph{non-crossing closure}:
\begin{definition}[Non-crossing closure \cite{Lehner_2002,Arizmendi_2015}]
    Let $\cl_{NC}: \Ps(n) \rightarrow \Ps(n)$ be the closure operator which given a partition $\pi$ outputs the smallest non-crossing partition $\overline{\pi}$ such that $\pi \leq \overline{\pi}$. $\overline{\pi}$ is called the \emph{non-crossing closure} of $\pi$. 
\end{definition}
Ref.~\cite[Theorem 6.1]{beissinger1985enumeration} proves (as a special case) that any partition in $\Ps$ can be uniquely decomposed into connected components, which are disjoint, connected subpartitions in $\tilde{\Ps}$ that are mutually non-crossing.\footnote{Formally, $\Ps$ is the $NC$-complex of $\tilde{\Ps}$.} Combinatorially, the non-crossing closure of a partition is obtained by taking each connected component and merging together every block within the component into a single block \cite{Lehner_2002}. The resulting partition is non-crossing. Connected partitions can equivalently be defined as the subset of all partitions in $\Ps(n)$ which have a non-crossing closure of $1_n$.

Motivated by the sparse Pauli Hamiltonian, we consider the set of all \emph{even partitions}:
\begin{definition}[Even partitions]
    We say a partition $\pi$ of $\{1,\dots,k\}$ is \emph{even} if all blocks of $\pi$ have even size. Let $\peven(k)$ denote the set of all even partitions of $\{1,\dots,k\}$.
\end{definition}
The definition of connectivity naturally extends to even partitions, as the subpartitions of an even partition induced by unions of blocks are also even. \cite[Theorem 6.1]{beissinger1985enumeration} also extends to show that $\peven$ uniquely decomposes into connected, even partitions. We will denote the set of connected even partitions of $[n]$ as $\pevenconn(n)$. 

An important subset of even partitions is the set of \emph{chord diagrams} $\Ps_2(n)\subseteq \peven(n)$, i.e. the set of partitions where every block has size 2. More generally, we can define $\Ps_q(n)\subseteq \peven(n)$ for any even $q$, which denotes the set of all partitions where every block has size $q$. Let $\tilde{\Ps}_q(n)\subseteq \pevenconn(n)$ denote the set of all connected partitions where every block has size $q$.

\subsubsection{Incidence algebras}\label{subsubsec:incidence-algebras}
Incidence algebras over partially-ordered sets were introduced in the seminal work of Rota~\cite{rota1964foundations}. 

\begin{definition}[Algebra]
    An algebra over field $K$ is a vector space $\As$ over $K$ equipped with a bilinear product $\As\times\As\rightarrow\As$. We will assume all algebras are over $\C$ and are unital (have an identity element $\mathbf{1}_{\As}$) and associative ($\forall a,b,c\in\As: (ab)c = a(bc)$) throughout unless otherwise stated.
\end{definition}

For any (finite) poset $P$, we can define an incidence algebra whose elements are functions $P^{(2)}\rightarrow \C$. The binary operation of the algebra is \emph{convolution}:

\begin{definition}[Convolution on posets]
    For any two functions $F,G: P^{(2)}\rightarrow \C$, the convolution of $F$ and $G$ is defined as
    \begin{equation}
        (F*G)(\pi,\sigma) = \sum_{\substack{\tau \in P\\\pi\leq\tau\leq\sigma}} F(\pi, \tau)G(\tau,\sigma).
    \end{equation}
\end{definition}
Special elements of the incidence algebra of $P$ include the identity and zeta functions:
\begin{equation}
    \delta(\pi,\sigma) = \begin{cases} 1 \text{ if } \pi=\sigma
   \\ 0 \text { else,}\end{cases}\quad \zeta(\pi,\sigma) = \begin{cases} 1 \text{ if } \pi \leq \sigma
   \\ 0 \text { else.}
\end{cases}
\end{equation}
The inverse element of the zeta function with respect to convolution in the incidence algebra is the M\"{o}bius function $\mu$ defined by $\mu*\zeta = \zeta*\mu = \delta$:
\begin{definition}[M\"{o}bius function on a poset] \label{def:mobius}
    Let $P$ be a finite poset. The M\"{o}bius function $\mu : P \times P \rightarrow \Z$ on $P$ is defined by
    \begin{equation}\label{eq:mobius}
        \sum_{\substack{\tau \in P\\\pi \leq \tau \leq \sigma}} \mu(\pi, \tau) = \sum_{\substack{\tau \in P\\\pi \leq \tau \leq \sigma}} \mu(\tau, \sigma)= \delta(\pi, \sigma)
    \end{equation}
\end{definition}
Additionally, the M\"{o}bius function of any interval $[x,y]$ of poset $P$ is equal to $\mu_P$ restricted to the interval $[x,y]$ \cite[Proposition 4]{rota1964foundations}. The M\"{o}bius function of $NC$ is also a multiplicative family in the sense of \cref{def:nc-multiplicative}. The M\"{o}bius function of the entire $NC$ interval has value equal to the signed Catalan number $\mu_{NC}(0_n, 1_n) = (-1)^{n-1}C_{n-1}$ and has the uniform bound $|\mu_{NC}(\pi, \sigma)|\leq 4^n$ for all $\pi,\sigma\in NC(n)$. 

This function allows us to take inverses within the incidence algebra, using the process of \emph{M\"{o}bius inversion}: 
\begin{lemma}[M\"{o}bius inversion] \label{lem:mobius_inversion}
    Let $P$ be a finite poset. For any $f,g: P\rightarrow \C$,
    \begin{equation}
        \forall \pi\in P: f(\pi) = \sum_{\substack{\sigma \in P\\\sigma \leq \pi}} g(\sigma) \Longleftrightarrow \forall \pi\in P:  g(\pi) = \sum_{\substack{\sigma \in P\\\sigma \leq \pi}} f(\sigma) \mu(\sigma,\pi).
    \end{equation}
\end{lemma}

\subsection{Free probability}\label{subsec:prelims-free-probability}
We briefly introduce the theory of free probability, which was initially formalized by Voiculescu \cite{voiculescu1985symmetries,voiculescu1986addition}. The combinatorial perspective and relation to non-crossing partitions was due to Speicher~\cite{speicher1994multiplicative}. For a comprehensive treatment, we refer the reader to the textbook of Nica and Speicher~\cite{nica2006lectures}, where all definitions and results from this subsection can be found.

\paragraph{Non-commutative probability.} Free probability theory is an example of \emph{non-commutative probability}: in non-commutative probability, both the sample space and event space from classical probability/measure theory are abstracted away, while random variables (observables) and expectations (traces) are treated as the foundational objects. This formalism encapsulates classical probability theory but also non-commutative random variables, e.g., quantum observables, random matrices, etc.~\cite{tao2023topics}.

\begin{definition}[Non-commutative probability spaces]
    A pair $(\As, \tau)$ is called a \emph{non-commutative probability space} where $\As$ is a (potentially non-commutative) algebra over the field $\C$ equipped with the unital linear functional $\tau: \As\rightarrow \C$ (satisfies $\tau(\mathds{1}_\As)=1$). Elements $a\in\As$ are called non-commutative random variables.
\end{definition}

\begin{example}
    Some examples of non-commutative probability spaces include:
    \begin{itemize}
        \item Classical random variables over complex numbers: $(\As, \tau)=(L^{\infty-}, \E[\cdot])$, where $L^{\infty-}$ denotes the space of random variables over $\C$ with all moments finite. The product of $\As$ is pointwise multiplication of random variables. 
        
        \item Random matrices: $(\As, \tau) = (L^{\infty-}\otimes M_n(\C), \E\frac{1}{n}\mathrm{tr})$, where the bilinear operator corresponds to matrix multiplication and the distribution corresponds to its expected eigenvalue spectrum.
    \end{itemize}
\end{example}

In this paper we will primarily consider the finite-dimensional algebra $M_{2^n}(\C)$ of $2^n\times2^n$ matrices, which is spanned by the $n$-qubit Pauli group, and a linear functional given by the normalized trace $\nTr(\cdot)\equiv 2^{-n}\,\Tr(\cdot)$.

\paragraph{Free probability.} The field of free probability is the major sub-branch of non-commutative probability dealing with the concept of \emph{free independence}. Independence of random variables $X,Y$ in classical probability theory is equivalent to ``tensor independence'', which implies $X$ and $Y$ commute and their joint moments can be factorized (e.g., $\E[X^kY^j]=\E[X^k]\E[Y^j]$). Free independence is an alternative definition of independence for non-commutative random variables:
\begin{definition}[Free independence]
    Let $(\As,\tau)$ be a non-commutative probability space. Subalgebras $(\As_i)_{i\in [s]}$ are called \emph{freely independent} (or free) if
    \begin{equation}
        \tau(a_1)=\tau(a_2)=\dots=\tau(a_k)=0\implies \tau(a_1a_2\dots a_k)=0
    \end{equation}
    for all $a_j\in \As_{i_j}$ when neighboring elements are from different subalgebras ($i_j \neq i_{j+1}$). Non-commutative random variables $a_1,\ldots,a_s\in\As$ are also called \emph{freely independent} if subalgebras $\big(\mathrm{alg}(\mathbb{I},a_i)\big)_{i\in [s]}$ are freely independent.
\end{definition}

Many concepts in classical probability theory have analogous versions in free probability, including convolution, cumulants, the central limit theorem, and more~\cite{tao2023topics}. 
Additionally, the semicircular distribution plays the central role in free probability theory that the Gaussian distribution does in classical probability. The free central limit theorem states that if non-commutative random variables $a_1,\ldots, a_N$ are identically distributed, freely independent, self-adjoint, centered, and variance 1, $N^{-1/2}\sum_{j=1}^N a_j$ converges in distribution to the standard semicircular distribution as $N\rightarrow \infty$. Voiculescu~\cite{Voiculescu1991matrices} showed that several classes of $N\times N$ random matrices are asymptotically free in the limit $N\rightarrow \infty$, showing a key connection between the Wigner semicircle law in random matrix theory and free probability.
This central importance of the semicircular distribution in free probability makes it a natural framework to analyze the sparse Pauli Hamiltonian, which can be seen as a sparse version of the GUE matrix ensemble~\cite{chen2024sparse}.

\subsubsection{Free cumulants}\label{subsubsec:free-cumulants}
Given a noncommutative probability space $(\As,\tau)$, define the moment functional
\begin{equation}
    \tau_k[a_1,\ldots,a_k]\equiv\tau(a_1a_2\cdots a_k).
\end{equation}
The multiplicative extension of the moment functional is given by the collection of
\begin{equation}
    \tau_\pi[a_1,\ldots,a_k] \equiv \prod_{V\in\pi} \tau_{k'}[a_{i_1},\ldots,a_{i_{k'}}], \qquad V=\{i_1, \ldots, i_{k'}\},\quad i_1<\cdots<i_{k'}.
\end{equation}
for all $\pi\in NC(k)$. The free cumulants are the inverse of the moment functionals over the $NC$ incidence algebra:
\begin{definition}[Free cumulants]
    Let $(\As,\tau)$ be a non-commutative probability space. The \emph{free cumulant functionals} are $(\kappa_k)_{k\in\N}$ where
    \begin{equation}
        \kappa_k[a_1,\ldots,a_k] \equiv \sum_{\sigma\in NC(k)}
        \tau_{\sigma}[a_1,\ldots,a_k]\, \mu_{NC(k)}(\sigma,1_k),
    \end{equation}
    for all $a_1,\ldots, a_k\in \As$.
\end{definition}
We can also define the multiplicative extension of free cumulants: for all $\pi\in NC(k)$
\begin{equation}
    \kappa_\pi[a_1,\ldots,a_k] \equiv \prod_{V\in\pi} \kappa_{k'}[a_{i_1},\ldots,a_{i_{k'}}],\qquad V=\{i_1, \ldots, i_{k'}\},\quad i_1<\cdots<i_{k'}.
\end{equation}

The semicircular element has second-order free cumulant $\kappa_2=1$ and all other free cumulants equal to 0 (this is analogous to the standard Gaussian which has second-order cumulant equal to 1 and all other cumulants equal to 0).

\begin{proposition}[Moment--cumulant formulas]\label{prop:moment-cumulant}
    For $a_1,\ldots,a_k\in\As$,
    \begin{equation}
        \begin{aligned}
            \kappa_{\pi}[a_1,\ldots,a_k] &=\sum_{\substack{\sigma\in NC(k)\\\sigma\leq\pi}} \tau_\sigma[a_1,\ldots,a_k]\, \mu_{NC(k)}(\sigma,\pi),\\
        \tau_{\pi}[a_1,\ldots, a_k] &=\sum_{\substack{\sigma\in NC(k)\\\sigma\leq\pi}} \kappa_\sigma[a_1,\ldots,a_k].
        \end{aligned}
    \end{equation}
\end{proposition}
Using the moment--cumulant relation and decomposing a non-crossing partition according to the block containing element $1$ (also from \cite[Equation~16.23]{nica2006lectures}), we get
\begin{equation}\label{eqn:joint-R-recurrence}
    \tau(a_1\cdots a_k) = \sum_{k'=1}^k \sum_{1=v_1<\cdots<v_{k'}\leq k} \kappa_{k'}[a_{v_1},\ldots,a_{v_{k'}}] \prod_{h=1}^{k'} \tau(a_{v_h+1}\cdots a_{v_{h+1}-1}), \qquad v_{k'+1}\equiv k+1,
\end{equation}
where the moment of an empty word is $1$.

\subsubsection{The \texorpdfstring{$R$}{R}-transform}\label{subsubsec:R-transform}
The $R$-transform of a non-commutative random variable plays the analogous role in free probability as the (logarithm of the) characteristic function of classical random variables. It plays an important role in computing the spectrum of sums of freely independent variables~\cite{tao2023topics}.

\begin{definition}[$R$-transform]
    The (single-variable) $R$-transform of a bounded, self-adjoint element $a$ of a non-commutative probability space is given by
    \begin{equation}
        R_a(z)\equiv\sum_{k=1}^\infty
        \kappa_k(a,\ldots,a)z^k,
    \end{equation}
    where $z$ is an indeterminate.
\end{definition}

The $R$-transform of a bounded, self-adjoint noncommutative random variable also determines its moments. 
\begin{definition}[Moment series]
    Let $a\in\As$ be a bounded, self-adjoint element. Its moment series is the formal power series given by
    \begin{equation}
        M_{a}(z) \equiv \sum_{k=0}^\infty 
        \tau(a^k) z^k,
    \end{equation}
    where $z$ is an indeterminate. 
\end{definition}
The moment--cumulant relation is equivalent to the formal-series identity
\begin{equation}\label{eqn:scalar-R-moment-relation}
    M_a(z)=1 + R_a\bigl(zM_a(z)\bigr).
\end{equation}
A key property of the $R$-transform is that it is additive over free variables: for freely independent variables $a_1,a_2\in\As$, the $R$-transform of $a_1+a_2$ is
\begin{equation}
    R_{a_1+a_2}(z) = R_{a_1}(z) + R_{a_2}(z).
\end{equation}

\section{The loop-free regime of the sparse Pauli model}
\label{sec:loop-free}

In this section we will discuss how the randomness in the Paulis $P_i$ and the Rademacher coefficients $J_i$ each contribute to the spectrum of sparse Pauli Hamiltonians $H$. In \cref{subsec:loop-freeness}, we show that all moments and free cumulants of $H$ through order $\Omega(n/\log n)$ are completely independent of the values of the Rademacher coefficients with high probability over the random Paulis. We call this concept \emph{loop-freeness}.

In \cref{subsec:commutation-graph}, we show that under loop-freeness, the moments and free cumulants of the sparse Pauli Hamiltonian can be described by a single random graph called the \emph{commutation graph}. Under loop-freeness to order $k$, the $k$-th order free cumulant is a polynomial in the commutation signs of $H$, with each monomial depending on at most $\binom{k/2}{2}$ distinct edges in the commutation graph. We can therefore show concentration of free cumulants of the sparse Pauli Hamiltonian (\cref{sec:commutation-graphs,sec:free-cumulant-concentration}) by computing expectations of these polynomials, which is much simpler than a direct computation of the free cumulants.

\subsection{Loop-freeness of Pauli Hamiltonians}\label{subsec:loop-freeness}
We say that a set of Paulis $\{P_{1},\dots,P_{m}\}$ forms a loop if there exist distinct indices $i_1,\ldots, i_k\in[m]$ where $[P_{i_1} \dots P_{i_k}] = [\mathbb{I}]$. We will see that there is a \emph{loop-free} regime where the Paulis of $H$ do not form any small loops with high probability; this implies that the low-order moments and free cumulants of $H$ are independent of the Rademacher coefficients. As an example, consider expanding $\nTr(H^4)$:
\begin{equation}
    \nTr(H^4) = \frac{1}{m^2} \sum_{i_1,i_2,i_3,i_4=1}^m J_{i_1}J_{i_2}J_{i_3}J_{i_4} \nTr(P_{i_1}P_{i_2}P_{i_3}P_{i_4})
\end{equation}
The trace $\nTr(P_{i_1}P_{i_2}P_{i_3}P_{i_4})$ is nonzero if and only if the product of Paulis $P_{i_1}P_{i_2}P_{i_3}P_{i_4}$ is proportional to $\mathbb{I}$. This can happen in two conceptually distinct ways. Firstly, the Paulis could pair up and cancel each other, for example in $P_{1}P_{1}P_{2}P_{2} = \mathbb{I}$; or alternatively, the Paulis could form a loop and cancel out, for example $(Z \otimes \mathbb{I} \otimes \mathbb{I})(\mathbb{I} \otimes Z \otimes \mathbb{I})(\mathbb{I} \otimes \mathbb{I} \otimes Z)(Z \otimes Z \otimes Z) = \mathbb{I}$. Paired-up terms do not depend on the Rademacher coefficients, since the coefficients all square to 1; for example, $\nTr(P_1P_1P_2P_2)$ is accompanied by the product of coefficients $J_1^2 J_2^2 = 1$. On the other hand, a loop is accompanied by a monomial in $\{J_i\}$ in which at least one coefficient has an odd power, so it vanishes after averaging over the coefficients.

This motivates the following approximation: if we average over the Rademacher coefficients $J_i$, we eliminate the contributions from the loops and are left only with the contributions from the paired-up terms. Formally, we capture this approximation with the following definition:
\begin{definition}[Loop-freeness]\label{def:loop-free}
    Given a sequence of $m$ Pauli operators $\Pm = (P_1,\ldots, P_m)$, we say that $\Pm$ is \emph{loop-free through order $L$} if 
    \begin{equation}
        [P_{i_1}P_{i_2}\cdots P_{i_k}] \neq [\mathbb{I}]
    \end{equation}
    for any $1\leq k \leq L$ and any sequence of distinct indices $i_1, i_2,\ldots, i_{k}\in [m]$. We say that a Pauli Hamiltonian instance $H$ is loop-free through order $L$ if its support on Pauli strings is loop-free through order $L$.
\end{definition}

\paragraph{Loop-freeness via random linear codes.} We use an argument based on error-correcting codes to help us understand the presence of loops in the sparse Pauli Hamiltonian. Each Pauli $P_i$ can be associated to a $2n$-bit vector $\va{x}_i$ via its binary symplectic representation: 
\begin{equation}
    [P_i] = [X_1^{\va{x}_i(1)} Z_1^{\va{x}_i(n+1)}] \otimes \cdots \otimes [X_n^{\va{x}_i(n)} Z_n^{\va{x}_i(2n)}]
\end{equation}
We place these vectors $\{\va{x}_i\}_{i\in[m]}$ into a $2n \times m$ matrix
\begin{equation}
    \hat{P} \equiv \begin{bmatrix}
        \uparrow & & \uparrow \\
        \va{x}_1 & \dots & \va{x}_m \\
        \downarrow & & \downarrow
    \end{bmatrix}.
\end{equation}
Since the Paulis $P_i$ are chosen randomly, $\hat{P}$ is a uniformly random $2n\times m$ binary matrix that we can view as the parity-check matrix of a random binary linear code of block length $m$. A set of Paulis $P_{i_1},\dots,P_{i_k}$ forms a loop if and only if the vector $\va{y} \equiv \va{e}_{i_1} + \dots + \va{e}_{i_k}$ is in the kernel of $\hat{P}$, and thus $\va{y}$ forms a codeword of weight $k$. 

We use this argument to formally prove the following statement:
\begin{lemma}[Loop-freeness of the sparse Pauli Hamiltonian]\label{lem:loop-freeness}
    There exist constants $\gamma,\alpha>0$ such that a Hamiltonian $H$ drawn from the sparse Pauli model with $m = \poly(n)$ is loop-free through order $\gamma n/\log n$ with probability at least $1-2^{-\alpha n}$.
\end{lemma}
\begin{proof}
    The proof follows the probabilistic method for the Gilbert--Varshamov bound for random binary linear codes \cite{varshamov1957estimate}. Since $\hat{P}\va{u}$ is uniformly random in $\F^{2n}_2$ for non-zero vector $\va{u}\in \F_2^m$, the probability that $\va{u}\in\ker\hat{P}$ is $4^{-n}$. Thus the probability that the code has a nonzero codeword of weight at most $L$ is
    \begin{equation}
        \Pr(\exists \va{u} \in \ker\hat{P} \setminus \{\va{0}\}: \mathrm{wt}(\va{u})\leq L) \leq 4^{-n}\sum_{j=1}^{L}\binom{m}{j} \leq 4^{-n}\cdot \left(\frac{em}{L}\right)^{L}
    \end{equation}
    when $L\leq m/2$. Setting $L=\gamma n/\log n$ for some small enough constant $\gamma>0$ proves the lemma.
\end{proof}
\begin{remark}\label{rem:relative-loop-free}
    The union bound in the proof of \cref{lem:loop-freeness} holds even if we fix some of the columns to be arbitrary linearly independent vectors and then sample the remaining columns randomly. We will use this fact in \cref{part:operator-valued}.
\end{remark}

\paragraph{Free cumulant expression in the loop-free regime.} \cref{lem:loop-freeness} showed that $H$ is loop-free through order $\Omega(n/\log n)$ with high probability; we now show that this rigorously implies the intuition that free cumulants $\kappa_k$ are well-approximated by their expectation over the Rademacher variables $\E_{\va{J}} \kappa_k$ through order $L$.

\begin{proposition}\label{prop:loop-free-average}
    Let $H$ be a sparse Pauli Hamiltonian instance with $n$ qubits, $m$ terms, and free cumulants $(\kappa_k)_{k\in\N}$.  If $H$ is loop-free through order $L$, then
    \begin{equation}
        \kappa_k[H,\ldots,H] = \kappa_k[H^{(\Pm, \va{J})},\ldots, H^{(\Pm, \va{J})}]
    \end{equation}
    for all $k\leq L$ and $\va{J}=(J_1,\ldots, J_m)\in \{\pm 1\}^m$, where $H^{(\Pm, \va{J})}\equiv m^{-1/2}\sum_{i=1}^m J_iP_i$. Specifically, $\E_{\va{J}} \kappa_k[H^{(\Pm, \va{J})},\ldots, H^{(\Pm, \va{J})}] = \kappa_k[H,\ldots,H]$.
\end{proposition}
\begin{proof}
    Recall $\Pm=(P_1,\ldots,P_m)$ and
    $H=m^{-1/2}\sum_{i=1}^m J_iP_i$. For each $k\leq L$, we have
    \[
        \nTr(H^k)
        = m^{-k/2}\sum_{i_1,\ldots,i_k=1}^m
        J_{i_1}\cdots J_{i_k}\,
        \nTr(P_{i_1}\cdots P_{i_k}).
    \]
    Since Pauli operators commute or anticommute and square to the identity, each word $P_{i_1}\cdots P_{i_k}$ equals, up to a sign, the product of the Pauli terms occurring an odd number of times.
    If this product is nonempty, it contains at most $k\leq L$ distinct terms. Loop-freeness (\cref{def:loop-free}) implies that it is not proportional to the identity and therefore has zero trace.
    Thus only Pauli words in which every term occurs an even number of times contribute to the sum. For each such word, $J_{i_1}\cdots J_{i_k}=1$, so $\nTr(H^k)$ is independent of $\va{J}$.

    By the moment--cumulant relation, $\kappa_k[H,\ldots H]$ is determined by the moments of order at most $k$. Therefore, the free cumulants of order $k\leq L$ are also independent of $\va{J}$, proving the desired result.
\end{proof}
For convenience, we will generally write $\E_{\va{J}} \kappa_k$ for a specified Hamiltonian $H^{(\Pm, \va{J})}$ to denote $\E_{\va{J}}\kappa_k[H^{(\Pm, \va{J})},\ldots, H^{(\Pm, \va{J})}]$, e.g.,
\[
    \kappa_k[H,\ldots, H] = \E_{\va{J}}\kappa_k.
\]

\subsection{Commutation graphs of sparse Pauli Hamiltonians}\label{subsec:commutation-graph}
We have shown that the moments of a loop-free Pauli Hamiltonian $H$ through order $L$ depend only on the labels of the random Paulis $\Pm$ and not the Rademacher variable coefficients $\va{J}$. Nevertheless, these moments can still fluctuate in non-trivial ways even within this loop-free regime. Let's return to our example of $\nTr(H^4)$:
\begin{equation*}
    \begin{aligned}
       \E_{\va{J}}{\nTr(H^4)} &= \frac{1}{m^2} \left(\sum_i \nTr(P_iP_iP_iP_i) + \sum_{i \neq j} \left(\nTr(P_iP_iP_jP_j) + \nTr(P_iP_jP_iP_j) + \nTr(P_iP_jP_jP_i)\right)\right) 
       \\&= 2 - \frac{1}{m} + \frac{1}{m^2} \sum_{i \neq j} \langle P_i,P_j\rangle
    \end{aligned}
\end{equation*}
We can see that $\nTr(H^4)$ depends on the number of pairs of Paulis $P_i, P_j$ that anticommute with each other. Higher moments $\nTr(H^k)$ will depend on higher order properties of the commutation relations. We can capture the commutation relations of the Paulis in their \emph{commutation graph}, defined as follows.
\begin{definition}[Commutation graph]
    The \emph{commutation graph} of a sequence of $m$ Pauli operators $\Pm \equiv \{P_a\}_{a\in[m]}$ has vertices identified with each Pauli operator and an edge between two vertices $P_a$ and $P_b$ if they anticommute.
    
    Given a Pauli Hamiltonian instance $H$, we also define the commutation graph of $H$, denoted $G^{(H)}$, to be the commutation graph of $\Pm$.
\end{definition}

Lastly, we show that on any small set of pairs the commutation graph behaves like an Erd\H{o}s--R\'enyi graph with edge probability $1/2$: every nonempty product of distinct commutation signs has expectation at most $4^{-n}$ in absolute value, so the edges are approximately unbiased and approximately $k$-wise independent. The statement also covers commutation signs with fixed Pauli insertions, which are used in \cref{part:operator-valued}. We use the following definition for representing commutation signs:
\begin{definition}[Commutation sign matrices]\label{def:commutation-sign-matrices}
    Given a sparse Pauli Hamiltonian $m^{-1/2}\sum_{i=1}^m J_i P_i$, we denote $W\in\{\pm1\}^{m\times m}$ as the \emph{commutation sign matrix} of a Pauli Hamiltonian, defined as
    \[
        W_{ij}\equiv\langle P_i,P_j\rangle, \qquad \forall i,j\in [m]
    \]
    and $V\in\{\pm1\}^{s\times m}$ as the \emph{commutation spike matrix} of a Pauli Hamiltonian for fixed Paulis $(Q_1, Q_2,\ldots, Q_s)$, defined as
    \[
        V_{ai}=\langle Q_a,P_i\rangle\qquad\forall a\in[s], i\in[m].
    \]
\end{definition}

\begin{lemma}\label{lemma:k-wise-independence}
    Let $\Pm \equiv (P_1,\ldots, P_m)$ be a sequence of $m$ i.i.d.\ sampled $n$-qubit Paulis. Let $Q_1,\ldots,Q_s$ be fixed independently of $\Pm$ and independent modulo phase. Every nonempty square-free monomial $M$ of entries of $W_{ij}$ with $i<j$ and $V_{ia}$ satisfies
    \[
        |\E_{\Pm} M|\leq 4^{-n}.
    \]
\end{lemma}
\begin{proof}
    If no $W$ factor occurs, group the factors by the sampled Pauli term $P_i$. The factors involving $P_i$ combine to $\langle Q,P_i\rangle$ with $Q=\prod_{a\in S_i}Q_a$ for a nonempty set $S_i\subseteq[s]$. Since the generators are independent modulo phase, $[Q]\neq[\mathbb{I}]$, so the expectation over $P_i$ is zero.
    
    Otherwise choose a Pauli term $P_i$ incident to a $W$ factor. Conditioning on all other sampled Paulis, all factors involving $P_i$ combine to $\langle P_i,R_i\rangle$. Averaging over $P_i$ gives the indicator that $R_i$ is identity modulo phase. The other factors have absolute value one. The product $R_i$ contains at least one independent uniform neighboring Pauli, hence is uniform modulo phase. Therefore $|\E_{\Pm} M|\leq\Pr([R_i]=[\mathbb{I}])=4^{-n}$.
\end{proof}

\section{Commutation graph statistics and free cumulants in the loop-free regime} \label{sec:commutation-graphs}
In this section, our main result (\cref{thm:cumulants-from-graph-statistics}) gives an expression for the free cumulants of sparse Pauli Hamiltonian instances $H$ up to their loop-free order in terms of its commutation graph. 

For an even order $k$, if the set of Paulis $\{P_{1},\dots,P_{m}\}$ is loop-free through order $k$, the trace of a product of terms $\nTr(P_{i_1}\dots P_{i_{k}})$ is zero unless each distinct $P_j$ appears an even number of times. Thus we can write
\begin{equation}\label{eq:moment_distinct_indices}
    \overline{\Tr}(H^{k}) = \sum_{\pi\in \peven(k)} \frac{1}{m^{k/2}} \sum_{\substack{j_1,\ldots, j_{|\pi|}\in[m]\\j_1,\ldots, j_{|\pi|} \text{ distinct}}} \overline{\Tr}(P_{j_{\pi(1)}}P_{j_{\pi(2)}}\cdots P_{j_{\pi(k)}})
\end{equation}
where the blocks of $\pi$ group together the positions in a Pauli word of equivalent terms. Then, the moments will be functions defined completely by certain statistics of the commutation graph $G^{(H)}$. Each tracial product $\nTr(P_{j_{\pi(1)}} \dots P_{j_{\pi(k)}})$ is equal to either $+1$ or $-1$ depending on the total sign picked up when we commute the $P_j$ through each other to cancel them out. The value of $\nTr(P_{j_{\pi(1)}} \dots P_{j_{\pi(k)}})$ will involve an XOR (addition $\bmod\ 2$) over at most $\binom{k/2}{2}$ edges of $G^{(H)}$. 

However, if we were to directly compute an expression for the free cumulants of $H$ using the moment-cumulant relation, the M\"{o}bius inversion creates non-trivial dependencies between different partitions. Our main goal in this section is to put this moment expression into a form which factorizes over the unique decomposition of even partitions, which allows us to compute a free cumulant expression for $H$ which features a distinct XOR of edges from the commutation graph (similar methods have been used in \cite{Lehner_2002,Arizmendi_2015}).

\subsection{Commutation graph statistics}
We begin by introducing the odd-intersection graph $g(\pi)$ which arises from an even partition $\pi \in \peven(k)$.

\begin{definition}[Odd-intersection graphs] 
\label{def:odd-intersection-graph}
    Given a partition $\pi =\{V_1, V_2, \ldots, V_r\}\in \peven(k)$, we define its \emph{odd-intersection graph} $g(\pi)$ as the simple graph where the blocks in $\pi$ are identified with the vertices of $g(\pi)$, and there is an edge between two blocks $V_i,V_j$ if and only if $\operatorname{cr}(V_i,V_j)$ is odd.
\end{definition}

Note that some vertices in the odd-intersection graph may be disconnected from all others if a block crosses all others an even number of times each. This motivates the following definition:
\begin{definition}[Edge-induced subgraphs]
    Given graph $g$, we denote $\overline{g}$ as the induced subgraph of all vertices which have at least one adjacent edge (the non-isolated vertices); we call this the \emph{edge-induced subgraph}. For a partition $\pi\in\peven(k)$, let $\overline{g}(\pi)$ denote the edge-induced subgraph of $g(\pi)$.
\end{definition}
For a graph $g$, let $V(g)$ denote its vertex set, $E(g)$ denote its edge set, and $|g|\equiv |V(g)|$ denote the number of vertices in graph $g$. Thus $V(\overline{g})$ is the set of non-isolated vertices in $g$, and $E(\overline{g}) = E(g)$. Let $\hat{\Ps}_{\mathrm{even}}(k)$ denote the set of connected even partitions on $\{1,\dots,k\}$ where the blocks are \emph{labeled}, and for a block-labeled partition $\hat{\pi} \in \hat{\Ps}_{\mathrm{even}}(k)$ let $\hat{g}(\hat{\pi})$ denote the vertex-labeled odd-intersection graph associated with $\pi$, where the labeling of $\hat{g}(\hat{\pi})$ is inherited from the labeling of $\hat{\pi}$. Now
\begin{equation}\label{eq:labeled-automorphism-factor}
    \big|\{\hat{\pi}\in\hat{\Ps}_{\mathrm{even}}(k):\hat{g}(\hat{\pi})=\hat{g}\}\big| = \big|\Aut(g)\big| \big|\{\pi\in\pevenconn(k):g(\pi)\cong g\}\big|.
\end{equation}
Additionally if we considered only graphs with a fixed number of vertices $r$ (partitions with exactly $r$ blocks), we have
\begin{equation}\label{eq:labeled-factorial-factor}
    \begin{aligned}
    \sum_{|\hat{g}|=r} \big|\{\hat{\pi}\in\hat{\Ps}_{\mathrm{even}}(k):\hat{g}(\hat{\pi})=\hat{g}\}\big|^2 &= \sum_{|\hat{g}|=r} \big|\{\pi\in\pevenconn(k):g(\pi)\cong g\}\big|^2 \big|\Aut(g)\big|^2 \\
    &= r! \sum_{|g|=r} \big|\{\pi\in\pevenconn(k):g(\pi)\cong g\}\big|^2 \big|\Aut(g)\big|.
\end{aligned}
\end{equation}
The same holds for other partition families invariant under labelings, e.g., $ \Ps_2(k), \tilde{\Ps}_2(k)$, etc.

For a partition $\pi=\{V_1,\ldots,V_r\}\in\peven(k)$ and a map $\phi:\{V_1,\ldots,V_r\}\to[m]$, write
\begin{equation}
    P_{\phi\circ\pi}\equiv P_{\phi(\pi(1))}P_{\phi(\pi(2))}\cdots P_{\phi(\pi(k))}
\end{equation}
for the Pauli word that carries the label $\phi(V)$ at every position of the block $V$. For a set of positions $U=\{U_1<\cdots<U_{|U|}\}\subseteq[k]$, write $P_{\phi\circ\pi}[U]\equiv P_{\phi(\pi(U_1))}\cdots P_{\phi(\pi(U_{|U|}))}$ for the subword at the positions of $U$. When $\phi=\iota$ is injective, we call it a \emph{labeling} of $\pi$, written $\iota:\pi\hookrightarrow[m]$; its labels $j_V\equiv\iota(V)$ are the distinct indices of \eqref{eq:moment_distinct_indices}.

For a loop-free Hamiltonian, a partition will give a contribution to the free cumulants based on its odd-intersection graph combined with the Hamiltonian's commutation graph. We will call these contributions \emph{commutation graph statistics}, which are parameterized by the odd-intersection graphs $g(\pi)$.

\begin{definition}[Commutation graph statistics] \label{def:graph-statistics}
    Fix a sparse Pauli Hamiltonian $H$ with support on $P_1, P_2, \ldots, P_m$. Given a simple graph $g$ with $0\leq r\leq m$ vertices, define
    \begin{equation}
        C[g] \equiv \frac{1}{(m)_{r}} \sum_{\iota:V(g)\hookrightarrow[m]} \left(\prod_{(u,v) \in E(g)} \langle P_{\iota(u)},P_{\iota(v)}\rangle\right).
    \end{equation}
\end{definition}
\begin{remark}[Normalization]
    The conventional $1/(m)_r$ factor ensures that $C[g] \in [-1,+1]$ and $C[\overline{g}]=C[g]$.
\end{remark}

\begin{lemma}[Trace of an even Pauli word]\label{lem:even-word-trace}
    For every even $k$, every $\pi\in\peven(k)$ and every map $\phi$ from the blocks of $\pi$ to $[m]$,
    \begin{equation}\label{eq:even-word-trace}
        \nTr\big(P_{\phi\circ\pi}\big)=\prod_{(u,v)\in E(g(\pi))}\big\langle P_{\phi(V_u)},P_{\phi(V_v)}\big\rangle,
    \end{equation}
    with the convention $\langle P,P\rangle=1$ for coinciding labels. In particular, for distinct labels $j_1,\ldots,j_r$,
    $\nTr(P_{j_{\pi(1)}}\cdots P_{j_{\pi(k)}})=\prod_{(u,v)\in E(g(\pi))}\langle P_{j_u},P_{j_v}\rangle$. Moreover, for disjoint finite sets $A,B\subset\N$ of even size,
    \begin{equation}\label{eq:inv-cr-parity}
        \operatorname{inv}(A,B)\equiv\operatorname{cr}(A,B)\pmod 2.
    \end{equation}
\end{lemma}
\begin{proof}
    Sort the word by adjacent transpositions so that equal positions of $\pi$ become consecutive, say in the order $V_1,V_2,\ldots,V_r$. A transposition of two letters with labels $j\neq j'$ contributes the sign $\langle P_j,P_{j'}\rangle$, and a transposition of two equal letters contributes nothing; the number of transpositions between a letter of $V_u$ and a letter of $V_v$ with $u<v$ is $\operatorname{inv}(V_u,V_v)$. After sorting, each block contributes $P^{|V|}=\mathbb{I}$ because $|V|$ is even. Hence
    \begin{equation}
        \nTr\big(P_{\phi\circ\pi}\big)=\prod_{u<v}\big\langle P_{\phi(V_u)},P_{\phi(V_v)}\big\rangle^{\operatorname{inv}(V_u,V_v)},
    \end{equation}
    and since $\langle P,P'\rangle^2=1$ it remains to prove \eqref{eq:inv-cr-parity}, which is what makes the exponent the indicator of an edge of $g(\pi)$.

    For $a<a'$ in $A$ let $n(<a)$, $n(a,a')$ and $n(>a')$ denote the number of elements of $B$ below $a$, strictly between $a$ and $a'$, and above $a'$. The alternating quadruples whose two $A$-elements are $a,a'$ are counted by $n(a,a')\,n(>a')+n(<a)\,n(a,a')$. Since $|B|$ is even, $n(>a')\equiv n(<a)+n(a,a')\pmod 2$, so this count is congruent to $n(a,a')^2+2n(<a)n(a,a')\equiv n(a,a')=n(<a')-n(<a)$. Summing over the pairs $a<a'$, and writing $\rho(a)$ for the rank of $a$ in $A$,
    \begin{align}
        \operatorname{cr}(A,B)&\equiv\sum_{a\in A}n(<a)\big[(\rho(a)-1)-(|A|-\rho(a))\big]=\sum_{a\in A}n(<a)\,(2\rho(a)-1-|A|)\nonumber\\
        &\equiv\sum_{a\in A}n(<a)=\operatorname{inv}(A,B)\pmod 2,
    \end{align}
    using that $|A|$ is even in the last step.
\end{proof}

Note that in \eqref{eq:moment_distinct_indices}, each term is a product of commutation signs over a set of edges from the commutation graph, therefore the value of the moment $\overline{\Tr}(H^{k})$ is completely determined by the values of $C[g]$ for graphs $g$ on at most $k/2$ vertices. Consequently, for all even $k\leq2m$,
\begin{equation}\label{eq:moments-graph-statistics}
    \nTr(H^{k})=\frac{1}{m^{k/2}}\sum_{\pi\in\peven(k)}(m)_{|\pi|}\,C[g(\pi)]
    \qquad\text{if $H$ loop-free through order $k$}.
\end{equation}

To be able to bound concentration of free cumulants in \cref{sec:free-cumulant-concentration}, we begin by deriving upper bounds for mixed second moments of the commutation graph statistic $C[g]$ over the randomness of the commutation graph using \cref{lemma:k-wise-independence}.
\begin{lemma}[Mixed second moments of commutation graph statistic]\label{lem:graph-properties-covariance}
    Consider any pair of graphs $g$ and $g'$ with at most $m$ vertices. The commutation graph statistics satisfy the following:
    \begin{equation}\label{eq:graph-statistic-covariance}
        \E\left[C[g]\,C[g']\right] = \begin{cases}
        \frac{|\Aut(\overline{g})|}{(m)_{|\overline{g}|}} \pm \mathcal O(4^{-n}) & \text{if } \overline{g}\cong \overline{g}',\\
         \pm\mathcal{O}(4^{-n}) & \text{if } \overline{g}\ncong \overline{g}'.
   \end{cases}
    \end{equation}
\end{lemma}
\begin{proof}  
    Given an injection $\iota:V(g)\hookrightarrow[m]$, define
    \begin{equation}
        X_{g,\iota} \equiv \prod_{\{u,v\}\in E(g)}
        \langle P_{\iota(u)},P_{\iota(v)}\rangle.
    \end{equation}
    
    For two injections $\iota:V(g)\hookrightarrow[m]$ and $\iota':V(g')\hookrightarrow[m]$, denote the symmetric difference of their labeled edge sets by
    \begin{equation}
        S_{\iota,\iota'} \equiv \iota(E(g))\mathbin{\triangle}\iota'(E(g')).
    \end{equation}
    Since every repeated commutation sign squares to one, \cref{lemma:k-wise-independence} gives
    \begin{equation}
        \left|\E\left[X_{g,\iota}\,X_{g',\iota'}\right]\right|
        \leq 4^{-n}, \quad \text{if } S_{\iota,\iota'}\neq\varnothing.
    \end{equation}
    When $\overline{g}\ncong \overline{g}'$, $S_{\iota,\iota'}$ is nonempty for every pair of injections. Therefore linearity of expectation and triangle inequality give
    \begin{equation}
        \left|\E\left[C[g]\,C[g']\right]\right| \leq \frac{1}{(m)_{|g|}(m)_{|g'|}} \sum_{\iota, \iota'} \left|\E\left[X_{g,\iota}\,X_{g',\iota'}\right]\right| \leq \mathcal O(4^{-n}),
    \end{equation}
    since there are a total of $(m)_{|g|}$ unique injections $\iota:V(g)\hookrightarrow[m]$. This completes Case 2. 
    
    For Case 1, we take $\overline{g}\cong\overline{g}'$, and thus we can simply consider $\E\left[C[\overline{g}]\, C[\overline{g}]\right]$ due to the equivalence of commutation graph statistics with their edge-induced subgraph. The symmetric difference
    $S_{\iota,\iota'}$ is then empty if and only if the two injections are equivalent up to an automorphism. There are $(m)_{|\overline{g}|}|\Aut(\overline{g})|$ such ordered pairs. For each of them, $\E\left[X_{\overline{g},\iota}\, X_{\overline{g},\iota'}\right]=1$. Every other pair with nonempty $S_{\iota,\iota'}$ contributes $\E\left[X_{g,\iota}\, X_{g,\iota'}\right] = \Os(4^{-n})$ as previously shown. Summing the two contributions gives
    \begin{equation}
        \E\left[C[g]\,C[g']\right] = \frac{1}{(m)_{|\overline{g}|}^2} \sum_{\iota, \iota'} \E\left[X_{\overline{g},\iota}\, X_{\overline{g},\iota'}\right] = \frac{1}{(m)_{|\overline{g}|}}|\Aut(\overline{g})| \pm \mathcal O(4^{-n}),
    \end{equation}
    as claimed.
\end{proof}

\subsection{The \texorpdfstring{$\chi$}{χ}-function}
We define the following coefficient arising from M\"{o}bius transforms in our free cumulant expansions. Recall that $\overline{\tau}\equiv\cl_{NC}(\tau)$ denotes the smallest non-crossing partition coarser than $\tau$.
\begin{definition}\label{def:chi-function}
    For $\tau\in\Ps(k)$, define
    \begin{equation}\label{eq:chi-NC}
        \chi(\tau) \equiv \sum_{\substack{\rho\in NC(k)\\\rho\leq\overline{\tau}}} \mathbf{1}_{\peven}(\tau\wedge\rho)\cdot \mu_{NC}(\rho, \overline{\tau}),
    \end{equation}
    where $\mathbf{1}_{\peven}: \Ps(k)\rightarrow \Z$ is the indicator function for whether the partition is even.
\end{definition}
\begin{remark}
    For a connected even partition $\tau\in\peven(k)$, the coefficient has the simplified form
    \begin{equation}
        \chi(\tau) = \sum_{\substack{\rho\in NC(k)\\\tau\wedge\rho \text{ even}}} \mu_{NC}(\rho, 1_{k}).
    \end{equation}
\end{remark}
\begin{remark}
    Let $\epsilon_1,\epsilon_2,\ldots, \epsilon_k$ be independent commuting uniform signs. For a partition $\tau$ and $\rho\in NC(k)$, 
    \[
        \E_\rho[\epsilon_{\tau(1)},\ldots,\epsilon_{\tau(k)}]
        =\mathbf{1}_{\peven}(\tau\wedge\rho).
    \]
    M\"obius inversion gives
    \[
    \chi(\tau)=\kappa_{\overline\tau}[\epsilon_{\tau(1)},\ldots,\epsilon_{\tau(k)}].
    \]
    When $\tau$ is connected, this reduces to a single free cumulant $\kappa_{k}[\epsilon_{\tau(1)},\ldots,\epsilon_{\tau(k)}]$.
\end{remark}

To state two equivalent combinatorial descriptions, let $\operatorname{icr}_{\pi}(D)$ denote the number of crossings between two chords of $D$ that are contained in the same block of $\pi$. Also, for even partitions $\tau$ and $\pi$, write $\tau\preceq\pi$ when $\tau\leq\pi$ and no block of $\pi$ contains two distinct blocks belonging to the same connected component of $\tau$; write $\tau\prec\pi$ when additionally $\tau\neq\pi$.

\begin{proposition}[Equivalent characterizations of $\chi$]\label{prop:chi-characterizations}
    Let $\pi\in\peven(k)$.
    \begin{enumerate}
        \item 
        \begin{enumerate}[label=\textnormal{(\alph*)}]
            \item   If $\pi_1,\ldots,\pi_s$ are the connected components of $\pi$, then
            \begin{equation}\label{eq:chi-multiplicative}
                \chi(\pi)=\prod_{i=1}^s\chi(\pi_i).
            \end{equation}
            \item If $\pi$ is connected, then
            \begin{equation}\label{eq:chi-signed-pairings}
                \chi(\pi)=\sum_{\substack{D\in\Ps_2(k)\\D\leq\pi,\ D\text{ connected}}}(-1)^{\operatorname{icr}_{\pi}(D)}.
            \end{equation}
        \end{enumerate}
      
        \item The coefficient $\chi$ satisfies the recursive identity
        \begin{equation}\label{eq:chi-recursion}
            \sum_{\substack{\tau\in\peven(k)\\\tau\preceq\pi}}\chi(\tau)=1.
        \end{equation}
    \end{enumerate}
\end{proposition}
\begin{proof}
    The interval of non-crossing partitions less than $\overline{\pi}$ factorizes over the blocks of $\overline{\pi}$, whose blocks are the connected components of $\pi$. Both the indicator function $\mathbf{1}_{\peven}$ and the M\"{o}bius function $\mu_{NC}$ in \eqref{eq:chi-NC} also factorize over the connected components of $\pi$, proving part~\textnormal{(1a)}.

    For any finite ordered set $A$ of even cardinality, we have the signed chord-diagram identity
    \begin{equation}\label{eq:signed-pairing-identity}
        \sum_{D\in\Ps_2(A)}(-1)^{\operatorname{cr}(D)}=1,
    \end{equation}
    where $\operatorname{cr}(D)$ is the number of crossings in $D$. This can be proved, for example, by expanding the Pfaffian of the skew-symmetric matrix whose entries above the diagonal are all $1$ and below the diagonal are all $-1$. We can then apply \eqref{eq:signed-pairing-identity} inside every block of $\pi\wedge\rho$; chords contained in distinct blocks of the non-crossing partition $\rho$ cannot cross, so the product of all signs is $(-1)^{\operatorname{icr}_{\pi}(D)}$. Consequently, for every $\rho\in NC(k)$,
    \begin{equation}
        \mathbf{1}_{\peven}(\pi\wedge\rho)=\sum_{\substack{D\in\Ps_2(k)\\D\leq\pi\wedge\rho}}
        (-1)^{\operatorname{icr}_{\pi}(D)}.
    \end{equation}
    If $\pi$ is connected, substituting this identity into \eqref{eq:chi-NC} and interchanging the sums gives
    \begin{equation}
        \chi(\pi)=\sum_{\substack{D\in\Ps_2(k)\\D\leq\pi}}(-1)^{\operatorname{icr}_{\pi}(D)}
        \sum_{\substack{\rho\in NC(k)\\D\leq\rho}}\mu_{NC}(\rho,1_{k})=\sum_{\substack{D\in\Ps_2(k)\\D\leq\pi}}
        (-1)^{\operatorname{icr}_{\pi}(D)}
        \mathbf{1}\{\overline{D}=1_{k}\},
    \end{equation}
    where the last equality is M\"{o}bius inversion on the interval $[\overline{D},1_{k}]$. This proves part~\textnormal{(1b)}.

    Finally, apply \eqref{eq:signed-pairing-identity} independently inside every block of $\pi$ to obtain
    \begin{equation}
        1=\sum_{\substack{D\in\Ps_2(k)\\D\leq\pi}}(-1)^{\operatorname{icr}_{\pi}(D)}.
    \end{equation}
    Decompose each $D$ into its connected components and split every block of $\pi$ according to those components. The resulting even partition is a unique $\tau\preceq\pi$. Conversely, choosing such a $\tau$ and a connected chord diagram below each connected component of $\tau$ reconstructs $D$. The signs factor over components, so parts~\textnormal{(1a)} and~\textnormal{(1b)} identify the contribution of each $\tau$ as $\chi(\tau)$. This proves part~\textnormal{(2)}.
\end{proof}

In the remainder of the subsection we derive some properties of the $\chi$-function which will be useful for free cumulant derivations. We first show a uniform bound on the absolute value of $\chi$:
\begin{lemma}[$\chi$-function uniform upper bound]\label{lem:growth_chi}
    $|\chi(\sigma)| \leq 6^{k}$ for all even $k$ and $\sigma\in\Ps(k)$.
\end{lemma}
\begin{proof}
    Let $NC_{\mathrm{even}}(k)$ denote the set of non-crossing even partitions on $[k]$. Whenever $\sigma\wedge\rho$ is even, $\rho$ is itself even, so $|\rho|\leq k/2$. If no even $\rho$ lies below $\overline{\sigma}$, then $\chi(\sigma)=0$; otherwise, by the triangle inequality, the sum has at most $|NC_{\mathrm{even}}(k)|$ nonzero terms.
    The number of even non-crossing partitions is the Fuss--Catalan number~\cite{edelman1980chain}
    \begin{equation}
        |NC_{\mathrm{even}}(k)|
        =\frac{1}{k+1}\binom{3k/2}{k/2}
        \leq 8^{k/2}.
    \end{equation}
    The standard factorization of intervals in the non-crossing partition lattice gives
    \[
        |\mu_{NC}(\rho,\overline{\sigma})|
        \leq 4^{|\rho|-|\overline{\sigma}|}
        \leq 4^{|\rho|}
        \leq 4^{k/2}.
    \]
    Putting both bounds together,
    \[
        |\chi(\sigma)|\leq 32^{k/2}\leq 6^{k},
    \]
    as claimed.
\end{proof}

\begin{proposition}\label{prop:vanishing-NC-fiber}
    For any fixed $\pi\in\peven(k), \sigma\in NC(k)$,
    \begin{equation}
       \sum_{\substack{\rho\in NC(k)\\\rho\leq\sigma\\\pi\wedge\rho\in\peven(k)}} \mu_{NC}(\rho,\sigma) = \begin{cases} 
      \chi(\pi\wedge\sigma) & \text{if }\overline{\pi\wedge\sigma}=\sigma, \\
      0 & \text{otherwise}.
   \end{cases}
    \end{equation}
\end{proposition}
\begin{proof}
    Let $\tau\equiv \pi\wedge\sigma$. For every $\rho\leq\sigma$, we have $\pi\wedge\rho=\tau\wedge\rho$. Since $\tau\leq\sigma$ and $\sigma$ is non-crossing, $\overline{\tau}\leq \sigma$. The case $\overline{\tau}=\sigma$ follows directly from \eqref{eq:chi-NC}. It remains to show that the expression vanishes when $\overline{\tau}<\sigma$.

    Let $\widehat{\mathbf{1}}_{\peven}:\Ps(k)\rightarrow\Z$ be the M\"{o}bius transform of $\mathbf{1}_{\peven}$ on $\Ps(k)$, such that
    \begin{equation}
        \mathbf{1}_{\peven}(\tau\wedge\rho)
        = \sum_{\substack{\eta\in \Ps(k)\\\eta\leq\tau\wedge\rho}}
        \widehat{\mathbf{1}}_{\peven}(\eta).
    \end{equation}
    Because $\eta\leq\rho$ is equivalent to $\overline{\eta}\leq\rho$ for non-crossing $\rho$, interchanging the sums gives
    \begin{align*}
         \sum_{\substack{\rho\in NC(k)\\\rho\leq\sigma}}
        \mathbf{1}_{\peven}(\tau\wedge\rho)\mu_{NC}(\rho,\sigma)&=
        \sum_{\substack{\eta\in\Ps(k)\\\eta\leq\tau}}
        \widehat{\mathbf{1}}_{\peven}(\eta)
        \sum_{\substack{\rho\in NC(k)\\\overline{\eta}\leq\rho\leq\sigma}}
        \mu_{NC}(\rho,\sigma)\\
        &=\sum_{\substack{\eta\in\Ps(k)\\\eta\leq\tau}}
        \widehat{\mathbf{1}}_{\peven}(\eta)
        \delta(\overline{\eta},\sigma)=0.
    \end{align*}
    The last equality follows because $\eta\leq\tau$ implies $\overline{\eta}\leq\overline{\tau}<\sigma$.
\end{proof}

We lastly prove that connected partitions $\pi\in\peven(k)$, $k\geq4$, with a chord block that crosses every other block with even parity satisfy $\chi(\pi) = 0$.
\begin{lemma}[Vanishing condition of $\chi$]\label{lemma:chi-vanish}
    Let $\pi\in \peven(k)$ be a connected even partition of $[k]$, where $k\geq4$. If $g(\pi)$ contains an isolated vertex corresponding to a chord block $V_j\in\pi$, then $\chi(\pi)=0$.
\end{lemma}
\begin{proof}
    Write $V_j=\{a,b\}$ with $a<b$, and let $I=\{a+1,\ldots,b-1\}$. For every block $V_i\neq V_j$,
    \begin{equation}\label{eq:isolated-pair-parity}
        \operatorname{cr}(V_j,V_i)=|V_i\cap I|\,\big(|V_i|-|V_i\cap I|\big)\equiv|V_i\cap I|\pmod 2,
    \end{equation}
    since $|V_i|$ is even. Since $V_j$ is an isolated vertex of $g(\pi)$, every $|V_i\cap I|$ must also be even.

    Since $\pi$ is connected, \eqref{eq:chi-signed-pairings} expresses $\chi(\pi)$ as a signed sum over the connected chord diagrams $D\leq\pi$. We show that its terms cancel in pairs, using an involution $D\mapsto D'$ which has no fixed points and reverses the sign $(-1)^{\operatorname{icr}_\pi(D)}$.

    Every $D\leq\pi$ contains the chord $V_j$, and a chord of $D$ crosses $V_j$ if and only if exactly one of its endpoints lies in $I$. Every other chord has zero or two endpoints in $I$, so the number of chords crossing $V_j$ in a block $V_i$ has the parity of $|V_i\cap I|$, which is even. If $D$ is connected, then some chord crosses $V_j$, since $D$ has $k/2\geq2$ chords. Let $x_1$ be the smallest site of $I$ lying on a chord that crosses $V_j$, let $V_i$ be its block, and let $x_2$ be the smallest site of $V_i\cap I$ other than $x_1$ lying on a chord that crosses $V_j$, which exists by the parity count. Write $\{x_1,y_1\}$ and $\{x_2,y_2\}$ for these two chords, so that $y_1,y_2\notin I$, and let $D'$ be obtained from $D$ by replacing them with $\{x_1,y_2\}$ and $\{x_2,y_1\}$.

    The new chords lie in $V_i$ and still cross $V_j$, and all other chords are unchanged. Hence $D'\leq\pi$, the construction applied to $D'$ selects the same sites $x_1,x_2$ and returns $D$, and $D'\neq D$ since $y_1\neq y_2$. To see that the sign is reversed, recall that only crossings between chords in the same block count towards $\operatorname{icr}_\pi$. Since $I$ is an interval, $x_1$ and $x_2$ are adjacent in the order of the four sites $x_1,x_2,y_1,y_2$, so the chord diagram on these sites pairing $x_1$ with $x_2$ is non-crossing, and exactly one of the other two is crossing. Moreover, a chord $\{u,v\}$ with $u<v$ crosses a chord $\{x,y\}$ if and only if exactly one of $x,y$ lies strictly between $u$ and $v$. Hence for any other chord $\{u,v\}$, the number of its crossings with the two exchanged chords has the parity of the number of sites among $x_1,x_2,y_1,y_2$ strictly between $u$ and $v$, both before and after the exchange. Therefore $\operatorname{icr}_\pi(D')-\operatorname{icr}_\pi(D)$ is odd.

    It remains to show that $D'$ is connected. In both $D$ and $D'$, the chords crossing $V_j$ are adjacent to $V_j$, and the other chords, together with the crossings among them, are unchanged. Since $D$ is connected, every connected component of these other chords is joined to a chord crossing $V_j$, so it suffices to show that each of them crosses as many new chords as exchanged chords. If such a chord $\{u,v\}$ lies inside $I$, then no site outside $I$ lies strictly between $u$ and $v$, and if it lies outside $I$, then either all of $I$ or none of it does. Hence whether $\{u,v\}$ crosses a chord with one endpoint $x\in I$ and the other $y\notin I$ depends only on $x$ in the first case and only on $y$ in the second, so the exchange preserves the number of crossings.
\end{proof}

\subsection{Deriving the free cumulant expression in the loop-free regime}
Now we introduce our main theorem for the section, which gives an expression for sign-averaged free cumulants $\E_{\va{J}}\kappa_{k}$ of a Hamiltonian instance $H$ explicitly in terms of the edges of its commutation graph $G^{(H)}$:

\begin{definition}[Loop-free expression of free cumulants]\label{def:loop-free-expression}
    For any (not necessarily loop-free) sparse Pauli Hamiltonian instance $H$ and even integer $2\leq k\leq2m$, we define the \emph{loop-free expression} $\tilde{\kappa}_{k}$ of its order-$k$ free cumulants as
    \begin{equation}\label{eq:cumulants_graph_properties}
        \tilde{\kappa}_{k}(\Pm) \equiv \frac{1}{m^{k/2}} \sum_{\pi \in \pevenconn(k)} (m)_{|\pi|}\chi(\pi)\, C[g(\pi)].
    \end{equation}
\end{definition}
\begin{theorem}[Free cumulants from commutation graph statistics] \label{thm:cumulants-from-graph-statistics}
    Let $H$ be a sparse Pauli Hamiltonian instance with $n$ qubits, $m$ terms, and free cumulants $(\kappa_k)_{k\in\N}$ which is loop-free through order $k\leq 2m$. The sign-averaged free cumulant satisfies $\E_{\va{J}}\kappa_k=0$ when $k$ is odd and equals its loop-free expression $\E_{\va{J}}\kappa_k=\tilde{\kappa}_{k}(\Pm)$ when $k$ is even.
\end{theorem}

The proof of \cref{thm:cumulants-from-graph-statistics} relies on (a direct corollary of) our main lemma which gives an exact expression for (the multiplicative extension of) sign-averaged free cumulants as sums over traces of Pauli monomials:
\begin{lemma}\label{lemma:loop-free-expression-partition}
    Let $H$ be a sparse Pauli Hamiltonian instance with $n$ qubits, $m$ terms, and free cumulants $(\kappa_k)_{k\in\N}$. For every even integer $k\leq2m$ such that $H$ is loop-free through order $k$ and every $\sigma\in NC(k)$, the sign-averaged free cumulant $\E_{\va{J}}\kappa_{\sigma}$ is given by
    \begin{equation}
        \sum_{\substack{\pi\in\peven(k)\\\overline{\pi\wedge\sigma}=\sigma}} \frac{\chi(\pi\wedge\sigma)}{m^{k/2}} \sum_{\iota:\pi\hookrightarrow[m]}\ \prod_{V\in\sigma} \overline{\Tr}\big(P_{\iota\circ\pi}[V]\big).
    \end{equation}
\end{lemma}
As a direct corollary by setting $\sigma=1_{k}$, we get:
\begin{corollary}[Free cumulants from commutation graph in the loop-free regime] \label{corollary:loop-free-expression}
    Let $H$ be a sparse Pauli Hamiltonian instance with $n$ qubits, $m$ terms, and free cumulants $(\kappa_k)_{k\in\N}$. For every even integer $2\leq k\leq2m$ such that $H$ is loop-free through order $k$, the sign-averaged free cumulant $\E_{\va{J}}\kappa_{k}$ is given by
    \begin{equation}
        \sum_{\substack{\pi\in\peven(k)\\\pi\text{ connected}}} \frac{\chi(\pi)}{m^{k/2}} \sum_{\iota:\pi\hookrightarrow[m]} \overline{\Tr}\big(P_{\iota\circ\pi}\big).
    \end{equation}
\end{corollary}
\begin{proof}[Proof of \cref{lemma:loop-free-expression-partition}]
    By definition, for all $\sigma\in NC(k)$,
    \begin{equation}
        \tau_{\sigma}[H,H,\ldots,H] \equiv \prod_{V\in\sigma}\tau_{|V|}(H,\ldots,H)=\frac{1}{m^{k/2}} \sum_{i_1,\ldots, i_{k}\in[m]} J_{i_1}\cdots J_{i_{k}} \prod_{V\in\sigma} \overline{\Tr}(P_{i_{V_1}}P_{i_{V_2}}\cdots P_{i_{V_{|V|}}}),
    \end{equation}
    where $V_1<\cdots<V_{|V|}$ are the elements of $V$. Averaging over the signs keeps exactly the index tuples in which every index occurs an even number of times. Each such tuple is of the form $i_t=\iota(\pi(t))$ for a unique even partition $\pi\in\peven(k)$, the partition of the positions with equal indices, and a unique labeling $\iota:\pi\hookrightarrow[m]$ of its blocks. Therefore
    \begin{equation}
       \E_{\va{J}} \tau_{\sigma}[H,H,\ldots,H] = \frac{1}{m^{k/2}} \sum_{\pi\in\peven(k)}\ \sum_{\iota:\pi\hookrightarrow[m]}\ \prod_{V\in\sigma} \overline{\Tr}\big(P_{\iota\circ\pi}[V]\big).
    \end{equation}
    For a block $V\in\sigma$, the labels occurring in the subword $P_{\iota\circ\pi}[V]$ are the labels of the blocks $W\in\pi$ meeting $V$, each with multiplicity $|V\cap W|$. By the loop-free assumption, the trace of the subword vanishes unless all of these multiplicities are even, so only partitions with $\pi\wedge\sigma$ even contribute:
    \begin{equation}
       \E_{\va{J}}\tau_{\sigma}[H,H,\ldots,H] = \frac{1}{m^{k/2}} \sum_{\substack{\pi\in\peven(k)\\\pi\wedge\sigma\in\peven(k)}}\ \sum_{\iota:\pi\hookrightarrow[m]}\ \prod_{V\in\sigma} \overline{\Tr}\big(P_{\iota\circ\pi}[V]\big).
    \end{equation}
 
    If $\rho\leq\sigma$ and $\pi\wedge\rho$ is even, then every label occurs an even number of times in each subword $P_{\iota\circ\pi}[U]$, $U\in\rho$, so each such subword is a scalar multiple of the identity. Since the restriction of $\rho$ to every block of $\sigma$ is non-crossing, recursively removing its interval blocks gives
    \begin{equation}\label{eq:trace-factorization}
        \prod_{U\in\rho}\nTr\big(P_{\iota\circ\pi}[U]\big)
        =\prod_{V\in\sigma}\nTr\big(P_{\iota\circ\pi}[V]\big).
    \end{equation}
 
Now applying the standard moment-cumulant formula
\begin{align}
     \E_{\va{J}} &\kappa_{\sigma}[H, \ldots, H] = \sum_{\substack{\rho\in NC(k)\\\rho\leq\sigma}} \mu_{NC}(\rho, \sigma)\cdot \E_{\va{J}} \tau_{\rho}[H,\ldots,H]\\
     &=\frac{1}{m^{k/2}}\sum_{\substack{\rho\in NC(k)\\\rho\leq\sigma}} \mu_{NC}(\rho, \sigma) \sum_{\substack{\pi\in\peven(k)\\\pi\wedge\rho\in\peven(k)}}\ \sum_{\iota:\pi\hookrightarrow[m]}\ \prod_{U\in\rho} \nTr\big(P_{\iota\circ\pi}[U]\big)\\
     &= \frac{1}{m^{k/2}}\sum_{\pi\in\peven(k)}\ \sum_{\iota:\pi\hookrightarrow[m]}\ \prod_{V\in\sigma} \nTr\big(P_{\iota\circ\pi}[V]\big) \sum_{\substack{\rho\in NC(k)\\\rho\leq\sigma\\\pi\wedge\rho\in\peven(k)}} \mu_{NC}(\rho,\sigma),
\end{align}
where the last line uses \eqref{eq:trace-factorization}.
To complete the proof we use the expression from \cref{prop:vanishing-NC-fiber}, and plugging this in gives us
\begin{equation}
     \E_{\va{J}} \kappa_{\sigma}[H, \ldots, H] = \sum_{\substack{\pi\in\peven(k)\\\overline{\pi\wedge\sigma}=\sigma}} \frac{\chi(\pi\wedge\sigma)}{m^{k/2}} \sum_{\iota:\pi\hookrightarrow[m]}\ \prod_{V\in\sigma} \overline{\Tr}\big(P_{\iota\circ\pi}[V]\big),
\end{equation}
as desired.
\end{proof}

We now complete the final proof of \cref{thm:cumulants-from-graph-statistics}:
\begin{proof}[Proof of \cref{thm:cumulants-from-graph-statistics}]
    Sign-averaging causes all odd moments and all odd free cumulants to vanish by symmetry. For even $k$, every surviving Pauli index occurs with even multiplicity. By \cref{lem:even-word-trace},
    \begin{equation}
        \nTr(P_{j_{\pi(1)}} \dots P_{j_{\pi(k)}}) = \prod_{(u,v) \in E(g(\pi))} \langle P_{j_u},P_{j_v}\rangle.
    \end{equation}
    \cref{corollary:loop-free-expression} and \cref{def:graph-statistics} complete the proof.
\end{proof}

\section{Concentration of free cumulants of the sparse Pauli model} \label{sec:free-cumulant-concentration}
In this section, we prove upper bounds on second moments of the loop-free regime expression of the even free cumulants $\tilde{\kappa}_{k}$, as defined in \cref{sec:commutation-graphs}. This will rest on two main premises: (a) the loop-free expressions of \cref{def:loop-free-expression} are expressed in terms of commutation graph statistics of $H$, and (b) \cref{lemma:k-wise-independence} shows that the edges of the commutation graph $G^{(H)}$ are almost $k$-wise independent.

Our main result proves tight asymptotic scaling bounds on the second moments of the loop-free expressions of free cumulants for the sparse Pauli model through order $\sim\!\min\{m^{1/3},n/\log n\}$. As a corollary, we prove a high probability bound on the absolute series of the true free cumulants $\kappa_k$ in \cref{thm:cumulant-concentration}, which will be the main result we utilize in analyzing our classical algorithm. In addition, we also believe that the condition on the order can be improved up to $m$ instead of $m^{1/3}$, and we provide a purely combinatorial conjecture (\cref{conjecture:combinatorial-linear}) which is sufficient to prove this claim.

\begin{theorem}[Second moment scaling of the loop-free expression of free cumulants]\label{thm:cumulant-scaling}
    Let $H$ be a sparse Pauli Hamiltonian instance with $n$ qubits, $m$ terms, and free cumulants $(\kappa_k)_{k\in\N}$. There exists a constant $\gamma>0$ such that for all even integers $k$ and integers $m$ satisfying
    \begin{equation}
        4\leq k \leq 2\gamma n/\log n,\quad (k/2)^3 \leq m \leq \poly(n),
    \end{equation}
    the loop-free expression of the free cumulants $\tilde{\kappa}_{k}$ satisfies
    \begin{equation}\label{eq:conjecture-variance-bound}
        \E_{\Pm}\left[\left(\tilde{\kappa}_{k}\right)^2\right]= \left(\frac{\Theta(k/2)}{m}\right)^{k/2}.
    \end{equation}
\end{theorem}

\subsection{Bounding the dominating second moment contribution}
The only remaining randomness which affects $\tilde{\kappa}_{k}$ is the distribution over commutation graphs $G^{(H)}$, which itself depends only on the random Pauli operators $\Pm$. We prove that the main contribution to the second moment is governed by pairs of graphs with isomorphic edge-induced subgraphs.

\begin{proposition}[Dominating second moment contribution]\label{prop:dominating-variance-contribution}
     Let $H$ be a sparse Pauli Hamiltonian instance with $n$ qubits, $m$ terms, and free cumulants $(\kappa_k)_{k\in\N}$. There exists a constant $\gamma>0$ such that for all even integers $4\leq k\leq2m$ satisfying $k\leq2\gamma n/\log n$, the loop-free expression $\tilde{\kappa}_{k}(\Pm)$ satisfies
    \begin{equation}
        \left|\mu_2^*(k) - \E_{\Pm}\left[\tilde{\kappa}_{k}^2\right]\right| \leq e^{-\Omega(n)},
    \end{equation}
    where we define
    \begin{equation}\label{eq:core-variance-contribution}
        \mu_2^*(k) \equiv \frac{1}{m^{k}}\sum_{\substack{\pi, \sigma \in \pevenconn(k)\\\Bar{g}(\pi)\cong \Bar{g}(\sigma)}} (m)_{|\pi|}(m)_{|\sigma|}\chi(\pi)\chi(\sigma)\cdot  \frac{|\Aut(\Bar{g})|}{(m)_{|\Bar{g}|}}.
    \end{equation}
    Note we use $\Bar{g}$ to denote the graph $\Bar{g}(\pi)$ and $\Bar{g}(\sigma)$ when $\Bar{g}(\pi)\cong \Bar{g}(\sigma)$.
\end{proposition}
\begin{proof}
    Expanding the second moment expression from \cref{def:loop-free-expression} gives
    \begin{equation}
        \E_{\Pm}[\tilde{\kappa}_{k}^{\,2}]
        =\frac1{m^{k}}
        \sum_{\pi,\sigma\in\pevenconn(k)}
        (m)_{|\pi|}(m)_{|\sigma|}\chi(\pi)\chi(\sigma)
        \E_{\Pm}[C[g(\pi)]C[g(\sigma)]].
    \end{equation}
    Apply \cref{lem:graph-properties-covariance} to every pair $(\pi, \sigma)$. Subtracting $\mu_2^*$ gives a uniform error bounded by $C4^{-n}$ for some constant $C>0$, for both isomorphic and non-isomorphic edge-induced subgraphs of $g(\pi)$ and $g(\sigma)$. Therefore,
    \begin{equation}
        \left|\E_{\Pm}[\tilde{\kappa}_{k}^{\,2}]-\mu_2^*(k)\right|
        \leq \frac{C}{4^n} \left(\frac1{m^{k/2}}\sum_{\pi\in\pevenconn(k)}(m)_{|\pi|}|\chi(\pi)|\right)^2.
    \end{equation}
     Applying the $\chi$ upper bound from \cref{lem:growth_chi} and using $|\pevenconn(k)|\leq (k/2)^{k}$, we get
     \begin{equation}
         \left|\E_{\Pm}[\tilde{\kappa}_{k}^{\,2}]-\mu_2^*(k)\right| \leq \frac{C}{4^n}
            \left(6^{k}|\pevenconn(k)|\right)^2
        \leq \frac{C(3k)^{2k}}{4^n}.
     \end{equation}
    For sufficiently small $\gamma>0$ and $k\leq2\gamma n/\log n$, this final expression is upper bounded by $e^{-\Omega(n)}$.
\end{proof}

From this, we also prove a lower bound for $\mu_2^*(k)$, which will also directly give a lower bound for $\E_{\mathbf{P}}[\tilde{\kappa}_{k}^2]$. We utilize the asymptotic expression on the total number of connected chord diagrams:
\begin{proposition}[{\cite[Theorem 4.6]{stein1978b}}]\label{lem:indecomp-chord-lower} For even $k$,
    \begin{equation}
        \big|\tilde{\Ps}_2(k)\big| = (k-1)!!(e^{-1} - o(1)).
    \end{equation}
\end{proposition}
    
\begin{lemma}[Dominating contribution lower bound]\label{lemma:second-moment-lower}
    Let $H$ be an instance drawn from the sparse Pauli model on $n$ qubits with $m$ terms and free cumulants $(\kappa_k)_{k\in\N}$. There exists a constant $\gamma>0$ such that for all even integers $4\leq k \leq 2\gamma n/\log n$,
    \begin{equation}
        \mu_{2}^*(k) \geq \Omega\left(\left(\frac{e^{-2}k}{m}\right)^{k/2}\right).
    \end{equation}
\end{lemma}
\begin{proof}
    Grouping \eqref{eq:core-variance-contribution} by isomorphism classes $\Bar{g}$ of edge-induced subgraphs, 
    \begin{equation}
        \mu_2^*(k)=\frac1{m^{k}}
        \sum_{\Bar g}
        \frac{|\Aut(\Bar{g})|}{(m)_{|\Bar{g}|}}
        \left(
            \sum_{\substack{\pi\in\pevenconn(k)\\
                \Bar{g}(\pi)\cong\Bar{g}}}
            (m)_{|\pi|}\chi(\pi)
        \right)^2.
    \end{equation}
    We lower bound the contribution of the subset of even connected partitions which are connected chord diagrams ($|\Bar{g}|=k/2$). Since each term in this summation is nonnegative, this is also a lower bound of the entire sum.
    
    For any connected chord diagram $\pi\in \tilde{\Ps}_2$, $\chi(\pi) = 1$. We get
    \begin{equation}
        \mu_2^* \geq \frac{1}{m^{k}}\sum_{\substack{\pi, \sigma \in \tilde{\Ps}_2(k)\\\Bar{g}(\pi)\cong \Bar{g}(\sigma)}} (m)_{k/2}^2 \frac{|\Aut(\Bar{g})|}{(m)_{k/2}} \geq \frac{(m)_{k/2}}{m^{k}} \big|\tilde{\Ps}_2(k)\big| \geq \frac{(m)_{k/2} (k-1)!!}{m^{k}} \big(e^{-1} - o(1)\big)
    \end{equation}
    where the final inequality uses \cref{lem:indecomp-chord-lower}. Applying the bounds $(m)_{k/2} \geq (m/e)^{k/2}$ and $(k-1)!! \geq (k/e)^{k/2}$ gives the final lower bound of $\Omega\left(\left(\frac{e^{-2}k}{m}\right)^{k/2}\right)$.
\end{proof}

We now complete our upper bound. The proof centers around upper-bounding the size of the following combinatorial set:
\begin{definition}
    Let $Z_{q,b}\subseteq \peven(2q)$ be the subset of even partitions of $2q$ elements with exactly $b$ blocks and satisfying the following conditions:
    \begin{itemize}
        \item Every block has size at least 4,
        \item Every block crosses every other block an even number of times, i.e., $\operatorname{cr}(V_i, V_j)$ is even for all $i\neq j$ (empty odd-intersection graph).
    \end{itemize}
\end{definition}
We will use the following lemma upper-bounding $|Z_{q,b}|$, whose proof is deferred to \cref{sec:combinatorial-proofs}:
\begin{lemma}\label{lem:even-block-count}
    There exists an absolute constant $C>0$ such that for all $q\geq 2$ and $1\leq b \leq \lfloor q/2\rfloor$,
    \begin{equation}\label{eq:even-block-count}
        |Z_{q,b}|\leq C^q b!\,q^{2q-4b}.
    \end{equation}
\end{lemma}
Additionally, we provide the conjectured improved bound on $|Z_{q,b}|$, whose proof would relax the condition on the number of terms $m$ in \cref{thm:cumulant-scaling}:
\begin{conjecture}\label{conjecture:combinatorial-linear}
    There exists an absolute constant $C>0$ such that for all $q\geq 2$ and $1\leq b \leq \lfloor q/2\rfloor$,
    \begin{equation}
        |Z_{q,b}| \leq C^q \frac{q!}{b!}.
    \end{equation}
\end{conjecture}
Using either \cref{lem:even-block-count} or \cref{conjecture:combinatorial-linear}, we now prove the upper bound on the dominating second moment contribution. 
\begin{lemma}[Dominating contribution upper bound]\label{lemma:second-moment-upper}
    There exist absolute constants $C, \gamma>0$ such that for all integers $m$ and even integers $k$ satisfying 
    \[
        (k/2)^3 \leq m \leq \poly(n),\quad 4\leq k \leq 2\gamma n/\log n,
    \]
    we have
    \begin{equation}
        \mu_{2}^*(k) \leq \left(\frac{Ck}{2m}\right)^{k/2}.
    \end{equation}        
    Additionally, if we further assume \cref{conjecture:combinatorial-linear}, then we only require $m\geq k/2$ instead of $m\geq (k/2)^3$ (which always holds since $k/2\leq n \leq m$).
\end{lemma}
\begin{proof}
    By \cref{eq:core-variance-contribution} and the counting argument in the proof of \cref{lem:graph-properties-covariance}, $\mu_2^*(k)$ is $m^{-k}$ times the sum of the weights $\chi(\pi)\chi(\sigma)$ over pairs of injectively labeled partitions $(\pi,\iota)$ and $(\sigma,\iota')$, with $\pi,\sigma\in\pevenconn(k)$, satisfying
    \[
        \iota(E(g(\pi)))=\iota'(E(g(\sigma))).
    \]
    Discard pairs with $\chi(\pi)\chi(\sigma)=0$. Taking absolute values and applying \cref{lem:growth_chi}, it suffices to count the remaining pairs, with an overall factor $36^{k}m^{-k}$.

    Concatenate the two labeled words to obtain
    \[
        w=(\iota\circ\pi)(\iota'\circ\sigma)\in[m]^{2k}.
    \]
    This map is injective, since each half determines its partition and its labeling. The positions of $w$ carrying the same label form a partition $\pi_w$ of $[2k]$ into $b$ blocks, where $b$ is the number of distinct labels. We claim that $\pi_w\in Z_{k,b}$. 
    
    Every label in the common edge set $\iota(E(g(\pi)))=\iota'(E(g(\sigma)))$ occurs in both halves, at least twice in each. Every other label occurs at least four times in any half containing it, by \cref{lemma:chi-vanish}. Thus every block of $\pi_w$ has even size at least four, and $b\leq k/2$. For distinct labels $x,y$, let $X=X_1\sqcup X_2$ and $Y=Y_1\sqcup Y_2$ be their (potentially empty) blocks of $\pi_w$ containing them, split between the two halves of $\pi_w$. All four sets have even size and the two inversion parities of sets within the same half agree by equality of the labeled edge sets and \cref{eq:inv-cr-parity}. Consequently,
    \[
        \operatorname{inv}(X,Y)=\operatorname{inv}(X_1,Y_1)+\operatorname{inv}(X_2,Y_2)+|X_2|\,|Y_1| \equiv0\pmod2.
    \]
    Applying \cref{eq:inv-cr-parity} again shows that $\Bar{g}(\pi_w)\cong\varnothing$, proving the claim. 
    
    Each partition in $Z_{k,b}$ has $(m)_b$ injective labelings, so
    \begin{equation}\label{eq:second-moment-glued-count}
        \mu_2^*(k)
        \leq\frac{36^{k}}{m^{k}}\sum_{b=1}^{k/2}(m)_b|Z_{k,b}|
        \leq36^{k}\sum_{b=1}^{k/2}m^{b-k}|Z_{k,b}|.
    \end{equation}

    Unconditionally, applying \cref{lem:even-block-count} at order $k$, using $b!\leq (k/2)^b$, and absorbing numerical exponential factors into $C^{k/2}$ gives
    \begin{equation}\label{eq:weighted-surplus-sum}
        \mu_2^*(k)\leq\frac{C^{k/2}}{m^{k}}\sum_{b=1}^{k/2}m^b b!\,(k/2)^{2k-4b}\leq\left(\frac{Ck}{2m}\right)^{k/2} \sum_{b=1}^{k/2}\left(\frac{(k/2)^3}{m}\right)^{k/2-b}.
    \end{equation}
    Under \cref{conjecture:combinatorial-linear}, instead use $k!/b!\leq k^{k-b}$ to obtain
    \begin{equation}\label{eq:conditional-linear-weighted-sum}
        \mu_2^*(k)\leq\frac{C^{k/2}}{m^{k}}\sum_{b=1}^{k/2}m^b\frac{k!}{b!}\leq\left(\frac{Ck}{2m}\right)^{k/2} \sum_{b=1}^{k/2}\left(\frac{k}{2m}\right)^{k/2-b}.
    \end{equation}
    The sums are at most $k/2$ when $m\geq (k/2)^3$ and $m\geq k/2$, respectively. Since $k/2\leq2^{k/2}$, increasing $C$ absorbs this factor and proves both claims.
\end{proof}

\subsection{Completing the proof and converting to free cumulants}
\label{subsec:hypercontractive-confidence}
Here we finalize the complete proof of our second moment asymptotic growth calculations for the loop-free expressions.

\begin{proof}[Proof of \cref{thm:cumulant-scaling}]
    Applying \cref{prop:dominating-variance-contribution}, $\mu_2^*(k)$ differs from the second moment of $\tilde{\kappa}_{k}(\Pm)$ by an additive $e^{-\Omega(n)}$ factor which does not change our final asymptotic expression when $k\leq2\gamma n/\log n$ for a small enough constant $\gamma$ depending on polynomial exponent in $m=\poly(n)$. Therefore combining \cref{lemma:second-moment-lower,lemma:second-moment-upper} gives the full proof.
\end{proof}

Lastly, as a corollary of \cref{thm:cumulant-scaling}, we give a high probability bound on the total absolute value of all free cumulants $|\kappa_k|$ within a specified range. We state the bound at any fixed inverse-polynomial failure probability, and include the improvement available when the lower order is fixed.
\begin{theorem}[Concentration of free cumulants]
    \label{thm:cumulant-concentration}
    Let $H$ be a sparse Pauli Hamiltonian instance with $n$ qubits, $m\leq\poly(n)$ terms, and free cumulants $(\kappa_k)_{k\in\N}$. There exist constants $A,C,\gamma>0$ such that, for all sufficiently large $n$, any failure probability $\delta=1/\poly(n)$, all even integers $k_0,k_1$ satisfying $4\leq k_0\leq k_1\leq L$, and
    \begin{equation}\label{eq:hypercontractive-window}
        Ak_1^3\leq m\leq\poly(n),\qquad L\leq\gamma n/\log n,
    \end{equation}
    we have with probability at least $1-\delta$ over the draw of $H$, 
    \begin{equation}\label{eq:cumulant-tail-sum}
        \sum_{\substack{k_0\leq k\leq k_1\\k\ \text{even}}}|\kappa_k|
        \leq 4\delta^{-1/2}\left(\frac{Ck_0}{m}\right)^{k_0/4}.
    \end{equation}
\end{theorem}
\begin{proof}
    Define $a_k\equiv(Ck/m)^{k/4}$. By \cref{thm:cumulant-scaling}, $\E_{\Pm}[\tilde{\kappa}_k^{\,2}]\leq a_k^2$ throughout the stated range. For consecutive even orders with $k+2\leq k_1$, since $m\geq Ak_1$,
    \begin{equation}
        \frac{a_{k+2}}{a_k}
        =\left(1+\frac2k\right)^{k/4}
          \left(\frac{C(k+2)}m\right)^{1/2}
        \leq e^{1/2}\left(\frac CA\right)^{1/2}
        \leq\frac12,
    \end{equation}
    for $A\geq4eC$. Hence
    \begin{equation}
        \sum_{\substack{k_0\leq k\leq k_1\\
            k\text{ even}}}a_k
        \leq a_{k_0}\sum_{j=0}^{\infty}2^{-j}
        =2a_{k_0}.
    \end{equation}
    Using Cauchy--Schwarz,
    \begin{equation}
        \E_\Pm\left[\left(\sum_{\substack{k_0\leq k\leq k_1\\k\ \text{even}}}|\tilde{\kappa}_k|\right)^2\right]
        \leq
        \left(\sum_{\substack{k_0\leq k\leq k_1\\k\ \text{even}}}\E_\Pm[\tilde{\kappa}_k^2]^{1/2}\right)^2
        \leq4a_{k_0}^2.
    \end{equation}
    By \cref{prop:loop-free-average,thm:cumulants-from-graph-statistics,lem:loop-freeness}, $\kappa_k=\tilde{\kappa}_k$ holds simultaneously for all $k\leq L$, except with probability $e^{-\Omega(n)}$. For any $0<\eta<1$, applying Markov's inequality to the square of the sum gives
    \begin{equation}\label{eq:cumulant-tail-probability}
        \Pr_H\left(\sum_{\substack{k_0\leq k\leq k_1\\k\ \text{even}}}|\kappa_k|>\frac{2a_{k_0}}{\sqrt{\eta}}
        \right)
        \leq e^{-\Omega(n)}+\frac{\eta}{4a_{k_0}^2}\E_\Pm\left[
            \left(\sum_{\substack{k_0\leq k\leq k_1\\k\text{ even}}}|\tilde{\kappa}_k|\right)^2
        \right]
        \leq e^{-\Omega(n)}+\eta.
    \end{equation}
    Taking $\eta=\delta/4$ and $n$ sufficiently large, so that the $e^{-\Omega(n)}$ term is at most $\delta/4$, proves \eqref{eq:cumulant-tail-sum} with failure probability at most $\delta/2$.
\end{proof}
\begin{corollary}[Stronger hypercontractivity bound]\label{corollary:cumulant-hypercontractivity}
    If $k_0$ is fixed independently of $n$, then with probability at least
    $1-\delta-e^{-\Omega(n)}$ over the draw of $H$,
    \begin{equation}\label{eq:hypercontractive-scalar-tail}
        \sum_{\substack{k_0\leq k\leq k_1\\k\ \text{even}}}|\kappa_k|
        \leq Cm^{-k_0/4}(\log n)^{\binom{k_0/2}{2}/2}.
    \end{equation}
\end{corollary}
The proof is deferred to \cref{sec:hypercontractivity}.
\begin{remark}
    If we further assume \cref{conjecture:combinatorial-linear}, then we can replace the $m\geq Ak_1^3$ assumption in \cref{thm:cumulant-concentration,corollary:cumulant-hypercontractivity} with $m\geq Ak_1$ (which always holds since $k_1\leq \gamma n/\log n\leq m/A$ for a small enough constant $\gamma$).
\end{remark}

Taking $k_0=4$ in \cref{corollary:cumulant-hypercontractivity} gives
\begin{equation}
    \sum_{\substack{4\leq k\leq k_1\\k\text{ even}}}|\kappa_k|\leq C\sqrt{\log n}/m
\end{equation}
at any fixed polynomial confidence; in the loop-free range, this sum is the quantity $\Ks(k_1)$ used in \cref{sec:classical-sfa-algorithm}.
\section{Fock space interpretation of the sparse Pauli model}\label{sec:fock-space}
Before we analyze our algorithm for the spectral amplitude, we first introduce a key framework which defines a mapping of non-commutative operators onto a corresponding single-particle quantum system. The moments of this operator correspond to steps in a one-dimensional quantum walk on a Fock space. We first apply this construction in \cref{subsec:fock-warmup} on the moments of an exact semicircular element as a warm-up. Then in \cref{subsec:fock-pauli}, we consider the moments of the sparse Pauli model and give a physical interpretation to our algorithm that will be essential for our analysis.

\begin{definition}[Full Fock space~{\cite[Definition~7.13]{nica2006lectures}}]\label{def:full-fock-space}
    Let $\Hs$ denote a Hilbert space. The \emph{full Fock space} over $\Hs$ is the Hilbert space given by
    \begin{equation}
        \Fs(\Hs) \equiv \bigoplus_{n=0}^{\infty} \Hs^{\otimes n}.
    \end{equation}
    The summand $\Hs^{\otimes 0}$ is a one-dimensional Hilbert space, which we denote $\C\Omega$, where $\Omega$ is a distinguished unit vector called the \emph{vacuum vector}.
    
    Given a Hilbert space $\Hs$, let $B(\Hs)$ denote the set of all bounded operators on $\Hs$. The linear functional $\psi$ on $B(\Fs(\Hs))$ defined by
    \begin{equation}
        \psi(T) \equiv \inprod{\Omega, T\Omega},\qquad T\in B(\Fs(\Hs)),
    \end{equation}
    is called the \emph{vacuum expectation state}. Then $\big(B(\Fs(\Hs)), \psi\big)$ forms a non-commutative probability space.
\end{definition}

This construction can be seen as a simplified version of the Fock space constructions for general elements with arbitrary (finite) sequence of non-zero free cumulants on a non-commutative probability space~\cite[Theorem~21.4]{nica2006lectures}. Interestingly, the mapping from the sparse Pauli model to the one-dimensional Fock space closely mirrors the boundary-to-bulk mapping in the double-scaled SYK model~\cite{berkooz2025cordial}, which reduces to the sparse Pauli model in the limit where its interaction terms are fully non-local and randomly sparsified.

\subsection{Warmup: Fock space for a semicircular element}\label{subsec:fock-warmup}
Suppose $H$ is a \emph{semicircular element}---it has exactly the semicircle spectral density of unit variance. One can calculate that the moments of $H$ are given by the Catalan numbers:
\begin{equation}
    \nTr(H^k) = C_{k/2} \equiv \frac{1}{k/2+1} \binom{k}{k/2},\qquad k\text{ even}
\end{equation}
The Catalan number $C_{k/2}$ counts the number of \emph{non-crossing chord diagrams} on $k$ sites for even $k$.

In this warmup section, we will see how to build a concrete Fock space for the semicircular element $H$. We consider the specialization of \cref{def:full-fock-space} where $\Hs=\C$. The Fock space $\Fs(\C)$ has an orthonormal basis of vectors $\ket{0}, \ket{1},\ket{2},\ldots$, where $\ket{0}\equiv \Omega$ denotes the vacuum vector.

\begin{figure}[ht]
\centering
\begin{subfigure}[t]{0.42\textwidth}
    \centering
    \includegraphics[height=3.5cm]{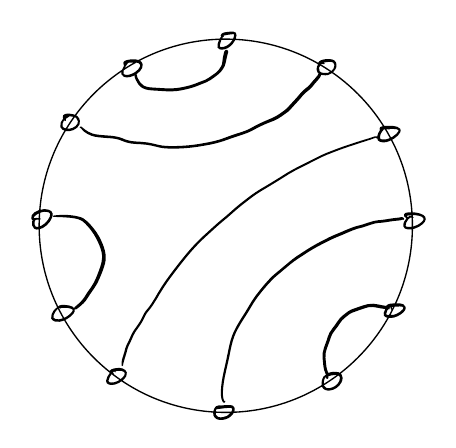}
    \caption{Non-crossing chord diagram on twelve sites.}
    \label{fig:chord_diagram}
\end{subfigure}
\hfill
\begin{subfigure}[t]{0.55\textwidth}
    \centering
    \includegraphics[height=2.5cm]{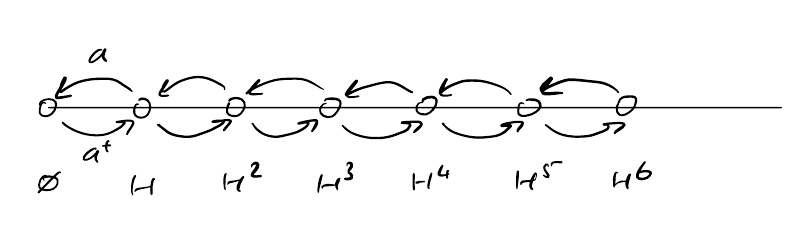}
    \caption{The Fock space representation.}
    \label{fig:fock_space_semicircle}
\end{subfigure}
\caption{Diagrammatic representation of the moments and free cumulants of a semicircular element.}
\end{figure}

Denote the creation and annihilation operators $\{a^\dag,a\}$, which act according to
\begin{equation} \label{eq:1d_creation_annihilation}
    \begin{aligned}
        a^\dag \ket{n}  &= \ket{n+1}\quad &&\forall n\geq 0,\\
        a \ket{n} &= \ket{n-1}\quad &&\forall n \geq 1,\\
        a \ket{0} &= 0.
    \end{aligned}
\end{equation}
We interpret the Fock space as a single quantum particle hopping on the semi-infinite line $\ell_2(\N)$; the creation and annihilation operators move this particle left or right (see~\cref{fig:fock_space_semicircle}). Now consider the hopping operator $\Ts \equiv a^\dag + a$. The moments of $\Ts$ give precisely the Catalan numbers:
\[
    \bra{0}\Ts^k\ket{0} = C_{k/2},\qquad k\text{ even},
\]
where we use $\psi(T)\equiv \bra{0}T\ket{0}$ to denote the vacuum expectation. Thus the expectations of the operator $\Ts$ on the vacuum state precisely match the moments of the semicircular distribution. Another interpretation of the vacuum expectations is a summation over all \emph{Dyck paths}, where a particle starts and ends at the vacuum state without ever moving to the left of $\ket{0}$. The number of length-$k$ Dyck paths for even $k$ is exactly the Catalan number $C_{k/2}$.

We can connect the particle hopping picture to the counting of non-crossing chord diagrams as follows. Imagine starting at the bottom of the circle in \cref{fig:chord_diagram} and proceeding counterclockwise. Every time the particle hops upwards we create a leg, and every time the particle hops downwards we annihilate the most recent open leg. Since we cannot hop downward from the vacuum state, there will always be a leg available to annihilate. Every valid path for the particle will result in a non-crossing chord diagram. This reproduces the well-known isomorphism between non-crossing chord diagrams and Dyck paths~\cite{Stanley_vol2}.

\subsection{Fock space for general non-commutative elements}\label{subsec:fock-pauli}
An actual instance of the sparse Pauli model $H$ will not have exactly a semicircular spectrum, but rather within the loop-free range will have vanishing odd free cumulants and even free cumulants $\kappa_k$ which decay as $k$ grows (\cref{thm:cumulant-concentration}). In this subsection, we develop a Fock space which allows us to analyze spectral properties from the free cumulants.

A first step towards a Fock space for free cumulants is to view the free cumulants as weightings for non-crossing partitions. A block of size four has weighting $\kappa_4$, a block of size six has weight $\kappa_6$, and so on.

\begin{figure}[ht]
\centering
\begin{subfigure}[t]{0.4\textwidth}
    \centering
  \includegraphics[height=4cm]{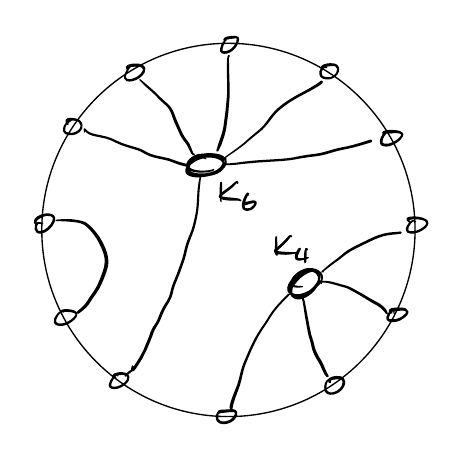}
  \caption{Non-crossing partition on twelve sites.}
  \label{fig:non_crossing_partition}
\end{subfigure}
\hfill
\begin{subfigure}[t]{0.55\textwidth}
    \centering
    \includegraphics[height=3cm]{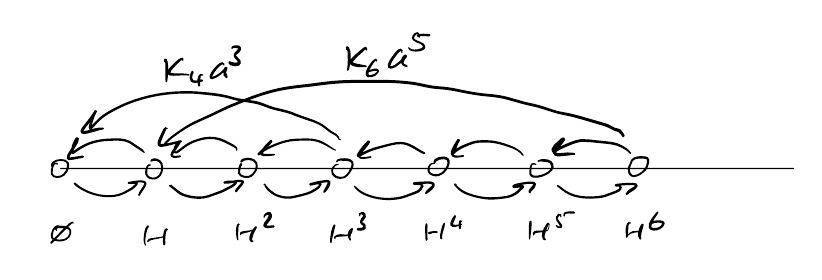}
    \caption{The Fock space representation.}
    \label{fig:fock_space_random_pauli_model}
\end{subfigure}
\caption{Diagrammatic representation of the moments and free cumulants of a sparse Pauli model instance.}
\end{figure}

We can now aim to build a Fock space which sums over these non-crossing partitions with the appropriate weightings. We will use the same creation and annihilation operators defined in \eqref{eq:1d_creation_annihilation}. Imagine beginning at the bottom of the circle in \cref{fig:non_crossing_partition} and proceeding counterclockwise. Any site which is not the final site of a non-crossing block can be associated to an upwards hop $a^\dag$. The annihilation of a chord, as in the previous section, is associated with a downwards hop $a$. We now also need to annihilate blocks of size larger than two. The final site of a block of size four effectively annihilates three legs at once, so we can associate to it a downwards hop of three steps $a^3$. This is accompanied by the corresponding weighting $\kappa_4$, giving altogether a term $\kappa_4 a^3$; likewise for larger blocks and higher cumulants. In this case our Fock space hopping operator models \emph{\L{}ukasiewicz paths}, which generalize Dyck paths by allowing downward steps of arbitrary size:

\begin{definition}[\L{}ukasiewicz paths]
    An ``almost-falling path\footnote{The more common definition uses ``almost-rising'' paths with step sizes in $\Z_{\geq 0} \cup \{-1\}$, however this is equivalent up to a horizontal reflection of the path.}'' is a path lying on $\Z^2$ which starts on $(0,0)$ and takes only steps of the vector $(1, i)$, $i\in \Z_{\leq 0} \cup \{+1\}$. A \emph{\L{}ukasiewicz path} of length $k$ is an almost-falling path which ends on $(k,0)$ and never goes below the $x$-axis.
\end{definition}
Every valid path of the Fock space follows a \L{}ukasiewicz path of length $k$ (see \cref{fig:fock_space_random_pauli_model}), reproducing the well-known isomorphism between non-crossing partitions and \L{}ukasiewicz paths~\cite{Stanley_vol2}. The conceptual relation between \L{}ukasiewicz paths and free probability was first presented by Voiculescu~\cite{voiculescu1986addition}. The hopping operator on the Fock space models the moments of the general non-commutative element. We define a specific hopping operator with a specified ``cutoff'' $\ell$ of its free cumulants, where the free cumulants above the cutoff are set to zero.
\begin{definition}[Fock space hopping operator]\label{def:hopping-operator}
    Let $H$ be a bounded Hermitian operator with free cumulants $(\kappa_k)_{k\in\N}$ such that $\kappa_2=1$ and all odd free cumulants through order $L$ vanish. For all even $2\leq\ell\leq L$, define the (not necessarily Hermitian) Fock space operator
    \begin{equation}
        \Ts_{\ell} \equiv a^\dag + a + \sum_{\substack{4\leq k\leq\ell\\k\text{ even}}} \kappa_k a^{k-1}.
    \end{equation}
\end{definition}
\begin{remark}[Semicircular case at large-error approximations]\label{rem:semicircular-reference}
    Throughout, free cumulant truncation will always keep the known second free cumulant $\kappa_2$. Thus $\ell=0$ corresponds to simply outputting the semicircle distribution, and in the scalar setting $\ell=2$ does the same. For color cumulants (see \cref{part:color-framework}), there is a mixed-color second free cumulant which is kept when $\ell=2$ and discarded when $\ell=0$.

    We say we are in the \emph{semicircular case} when the algorithm sets $\ell=0$ ($\ell=2$ in the scalar setting). In this case our algorithms compute the moments directly from the Catalan numbers.\footnote{For TOCs and OTOCs in \cref{sec:toc-otoc}, we compute products of Catalan series over the stretches of colors. For the spiked Pauli model in \cref{sec:spiked-pauli}, the moments are those of a free sum of a semicircle and the spike, which can be computed with a recurrence relation. Thermal expectation values in \cref{sec:thermal-expectations} use the same recurrence as well.} The error and runtime bounds still hold in all cases. Thus we assume we are not in the semicircular case without loss of generality.
\end{remark}

\begin{proposition}\label{prop:truncated_fock_space}
    Define $\hat{\mu}^{(\ell)}_k\equiv \bra{0}\Ts_{\ell}^k\ket{0}$, for $\Ts_{\ell}$ defined in \cref{def:hopping-operator}. Then
    \[
        \hat{\mu}^{(\ell)}_k=\nTr(H^k)
    \]
    for all even $k\leq\ell$.
\end{proposition}
\begin{proof}
    Let $\mathbf{1}_{E}$ denote the indicator random variable for event $E$. Expanding the vacuum expectation over \L{}ukasiewicz paths gives
    \begin{equation}
        \hat{\mu}^{(\ell)}_k=\sum_{\pi\in NC(k)}\prod_{V\in\pi}\kappa_{|V|}\mathbf{1}_{[|V|\leq \ell]}
    \end{equation}
    for every $k\geq 0$ (blocks of odd size contribute zero on both sides, since the odd free cumulants vanish through order $L\geq\ell$), i.e.\ $\hat{\mu}^{(\ell)}_k$ is the $k$-th moment of the free cumulant sequence $(\kappa_{k'}\cdot\mathbf{1}_{[k'\leq \ell]})_{k'\geq1}$ (we will omit the $(\ell)$ superscript when clear from context). When $k\leq\ell$ is even, the order-$k$ moments between the two free cumulant sequences are identical.
\end{proof}
\begin{remark}[Combinatorial interpretation of the free cumulant truncation]
    The hopping operator $\Ts_{\ell}$ satisfies the following moment-cumulant recurrence for even $k$:
    \begin{equation}
        \bra{0}\Ts_{\ell}^{k}\ket{0} = \sum_{\substack{\pi\in NC(k)\\\forall V\in\pi: |V|\leq \ell}}\prod_{V\in \pi} \kappa_{|V|}.
    \end{equation}
    In the combinatorial perspective of free probability, the moment order $k$ corresponds to the number of vertices in the non-crossing diagram, while the free cumulants correspond to the size of each block that appears in the non-crossing diagram. The truncated operator $\Ts_{\ell}$ corresponds to all of the non-crossing diagrams of $H$ which contain blocks of size $\ell$ or less. Since a non-crossing diagram on $\ell$ vertices contains blocks of size at most $\ell$, the moments $\nTr(H^{k})$ and $\bra{0}\Ts_{\ell}^{k}\ket{0}$ match exactly for all even $k\leq\ell$.
\end{remark}

We will use the hopping operator corresponding to a free cumulant truncation to analyze our algorithm in \cref{sec:classical-sfa-algorithm}. 
\section{A classical algorithm for the spectral amplitude}
\label{sec:classical-sfa-algorithm}
In this section, we establish our complete classical algorithm for the spectral amplitude. We start with an overview of the entire algorithm in \cref{subsec:algorithm-outline}. In \cref{subsec:truncation-approx-error}, we analyze the error of our free cumulant truncation utilizing the concentration of free cumulants shown in \cref{sec:free-cumulant-concentration} along with the Fock space interpretation of \cref{sec:fock-space}. Finally we put together all of the proofs in \cref{subsec:final-classical-proof}. We further generalize these results in \cref{sec:general-beta} to obtain our algorithm for estimating partition functions.

\subsection{Algorithm overview and runtime}\label{subsec:algorithm-outline}
We now give an outline of the algorithm.

\paragraph{Computing the low-order free cumulants.}
The goal of steps (1) and (2) is to compute the low-order free cumulants which will later be used to approximate higher-order moments. We compute the even moments of $H$ using standard sparse multiplication in the Pauli basis, as implemented, for example, in \cite[Algorithm~1]{Loizeau_2025}. Each operator is stored as a dictionary indexed by canonical Pauli strings, with all phases stored as coefficients. For even $k$, the value of $\nTr(H^k)$ is determined exactly by the expansion of the Pauli terms in $H^{k/2}$, due to orthogonality of Pauli strings in the Pauli basis.

The number of Pauli strings appearing in $H^j$, for $1\leq j\leq m/2$, is at most
\[
    \sum_{i=0}^j \binom{m}{i} \leq \left(\frac{em}{j}\right)^{j}.
\]
Each iteration multiplies each element currently stored by $m$ different Pauli terms from $H$, at a cost of $\Os(n)$ per product. Summing these costs over the powers $H^0,\ldots,H^{\frac{\ell}{2}-1}$, computing moments through order $\ell\geq 4$ can be done in time
\[
    \Os\left(n\ell\binom{m}{\ell/2}\right)
    \leq \Os\left(n\ell\left(\frac{2em}{\ell}\right)^{\ell/2}\right),
\]
provided $m\geq A\ell$ for a sufficiently large constant $A$.
\begin{remark}
    Applying this standard computation of the Pauli operators results in our $(1/\varepsilon)^{2+o(1)}$ scaling of our algorithm. An alternative computation of the low-order moments using fast matrix multiplication is given
    in \cref{sec:fmm-cumulants}.
\end{remark}

We use these moments to compute the free cumulants of $H$ to order $\ell$. Taking coefficients in \eqref{eqn:scalar-R-moment-relation} gives
\begin{equation}
    \nTr(H^k) = \sum_{s=1}^{k}\kappa_s[z^{k-s}]M_H(z)^s.
\end{equation}
For $k\leq\ell$, the $s=k$ term is $\kappa_k$, and every other term involves only lower-order cumulants and moments. We therefore compute the cumulants in increasing order by solving this recurrence for $\kappa_k$.

\paragraph{Approximating high-order moments.}
We compute the approximate moments $\hat{\mu}_k$, for all $k\leq L$, using the same recurrence, keeping the second free cumulant $\kappa_2=1$ and setting every other cumulant above order $\ell$ to zero. For either conversion, store the coefficient arrays of the powers $M^s$ of the moment series, for $1\leq s\leq\ell$, and fill them in increasing degree and then increasing $s$. The update
\[
    [z^d]M^s=\sum_{j=0}^d\hat{\mu}_j[z^{d-j}]M^{s-1}
\]
costs $\Os(d+1)$ per coefficient. Since each new
moment only uses earlier degrees of these arrays, the two conversions together cost $\Os(\ell L^2)$. The approximate moments agree with $\nTr(H^k)$ for $k\leq\ell$; the error at higher orders is bounded in \cref{subsec:truncation-approx-error}.

\paragraph{Approximating the spectral amplitude.}
Using the approximate moments, we compute
\[
    \widehat{Z}(t) \equiv \sum_{k=0}^{L}\frac{(-it)^k}{k!}\hat{\mu}_k.
\]
Evaluating this sum takes time $\Os(L)$. The conversion cost satisfies $\ell L^2\leq L^3=\Os(m)$ when $m\geq AL^3$, which is dominated by the Pauli propagation. For even $2\leq\ell<L$ and $m\geq AL^3$, the total running time is
\begin{equation}\label{eq:scalar-cutoff-runtime}
    \Os\left(n\ell\binom{m}{\ell/2}\right).
\end{equation}

\subsection{Approximation error of truncating moments and free cumulants}\label{subsec:truncation-approx-error}
In this subsection we analyze the error of the spectral amplitude which is incurred when we truncate the free cumulants at some order $\ell$ and the moments at some order $L\gg\ell$. Our analysis will crucially utilize the Fock space mapping from \cref{sec:fock-space}.

We will use the following lemmas in our proofs:
\begin{lemma}[Perturbed evolution]\label{lem:thermal-perturbed-evolution}
Let $S$ be a bounded self-adjoint operator and $N$ be a bounded operator on a Hilbert space. Then for all $z \in \C$,
\begin{equation}
    \|e^{i(S+N)z}\| \leq e^{|\operatorname{Im}(z)|\|S\| + |z|\|N\|}.
\end{equation}
\end{lemma}
\begin{proof}
    Fix $z\in\mathbb C$ and a vector $\psi_0$, and define $\psi(u)\equiv e^{i(S+N)zu}\psi_0$ for $u\in[0,1]$. Since $\langle\psi,S\psi\rangle$ is real,
    \begin{align*}
        \frac{d}{du}\|\psi(u)\|^{2}& =2\,\mathrm{Re}\,\langle\psi,\,iz(S+N)\psi\rangle
        \\&=-2\,\mathrm{Im}(z)\,\langle\psi,S\psi\rangle+2\,\mathrm{Re}\bigl(iz\langle\psi,N\psi\rangle\bigr)\\
        &\leq 2\bigl(|\mathrm{Im}\,z|\,\|S\|+|z|\,\|N\|\bigr)\|\psi(u)\|^{2}.
    \end{align*}
    Gr\"onwall's inequality gives
    \begin{equation}
        \|\psi(1)\|\le e^{|\mathrm{Im}\,z|\,\|S\|+|z|\,\|N\|}\|\psi_0\|,
    \end{equation}
completing the proof.
\end{proof}
\begin{corollary}[Perturbed unitary evolution]\label{lem:perturbed-unitary}
    Let $S$ be a bounded self-adjoint operator and $N$ be a bounded operator on a Hilbert space. Then for all $t\in\R$,
    \begin{equation}
        \|e^{i(S+N)t}\| \leq e^{|t|\|N\|}.
    \end{equation}
    \end{corollary}
\begin{proof}
    Proved as a corollary by setting $z=t\in\R$ in \cref{lem:thermal-perturbed-evolution}.
\end{proof}

\begin{lemma}[Taylor approximation theorem]
    \label{lemma:taylor-approximation}
    Let $X$ be a bounded operator on a Hilbert space, let $L\geq1$ be an integer, and let $t\geq 0$. Then we have the following bounds:
    \begin{equation}
        \begin{aligned}
            &\text{(Operator norm)}\qquad &&\left\|e^{-itX}-\sum_{k=0}^{L-1}\frac{(-it)^{k}}{k!}X^{k}\right\| \leq \frac{t^L\|X\|^L}{L!} \sup_{0\leq u\leq t}\left\|e^{-iuX}\right\|,\\
            &\text{(Trace)}\qquad &&\left| \nTr(e^{-itX})-\sum_{k=0}^{L-1}\frac{(-it)^{k}}{k!}\nTr(X^{k}) \right| \leq \frac{t^L}{L!}\nTr(|X|^L),\quad\ \text{if $X$ is Hermitian.}
        \end{aligned}
    \end{equation}
\end{lemma}
\begin{proof}
    From Taylor's formula \cite[Chapter XIII]{lang1993real} applied to $u\mapsto e^{-iutX}$,
    \begin{equation}
        e^{-itX}-\sum_{k=0}^{L-1}\frac{(-it)^{k}}{k!}X^{k}
        = \frac{(-itX)^L}{(L-1)!} \int_0^1(1-u)^{L-1}e^{-iutX}\,du.
    \end{equation}
    Taking the operator norms gives the first bound, and applying the scalar-version of the equation to each eigenvalue of $tX$ gives the trace bound.
\end{proof}

Define $\Ks(k')\equiv\sum_{4\leq k\leq k'}|\kappa_k|$. We will use the following intermediate bound in our analysis.

\begin{lemma}\label{lem:scalar-cutoff-comparison}
    Let $\Ts_{\ell}$ and $\Ts_L$ be the scalar hopping operators defined in \cref{def:hopping-operator}, for even integers $2\leq\ell<L$. Then, for every $t\geq0$,
    \begin{equation}\label{eqn:second-truncation}
        \left|\bra{0}e^{-i\Ts_{\ell}t}\ket{0}-\bra{0}e^{-i\Ts_Lt}\ket{0}\right|\leq\bigl(\Ks(L)-\Ks(\ell)\bigr)\,e^{t\Ks(L)}.
    \end{equation}
\end{lemma}
\begin{proof}
    The claim is immediate for $t=0$, so assume $t>0$. Define
    \[
        K_{\ell_0}(w)\equiv\frac1w+w+\sum_{4\leq k\leq \ell_0}\kappa_k w^{k-1}.
    \]
    The moment--cumulant relation \cite{nica2006lectures} and Lagrange inversion~\cite[Theorem~5.4.2]{Stanley_vol2} give $\hat{\mu}_k^{(\ell_0)}=(k+1)^{-1}[w^{-1}]K_{\ell_0}(w)^{k+1}$. Summing the exponential series, uniformly on $|w|=1$, therefore gives
    \[
        \bra{0}e^{-it\Ts_{\ell_0}}\ket{0}
        =-\frac{1}{it}\cdot\frac1{2\pi i}\oint_{|w|=1}e^{-itK_{\ell_0}(w)}\,dw.
    \]
    Write $\Delta(w)=K_L(w)-K_{\ell}(w)$. Subtracting and integrating along $K_{\ell}+s\Delta$, $0\leq s\leq1$, cancels the factor $-it$:
    \[
        \bra{0}e^{-it\Ts_L}\ket{0}-\bra{0}e^{-it\Ts_{\ell}}\ket{0}
        =\frac1{2\pi i}\int_0^1\oint_{|w|=1}
        e^{-it(K_{\ell}(w)+s\Delta(w))}\Delta(w)\,dw\,ds.
    \]
    On the unit circle, $w^{-1}+w$ is real, so
    \[
        \left|e^{-it(K_{\ell}(w)+s\Delta(w))}\right|\leq e^{t\Ks(L)},
        \qquad
        |\Delta(w)|\leq\Ks(L)-\Ks(\ell).
    \]
    The contour has length $2\pi$, proving \cref{eqn:second-truncation}.
\end{proof}

We now show the following bound which quantifies the error incurred when simultaneously truncating free cumulants and moments in the spectral amplitude. We generalize this truncation argument to general products of complex-time evolutions and the spiked model with \cref{lem:common-truncation} in \cref{sec:general-truncation}, for use in the subsequent sections.

\begin{lemma}[Approximation error of truncations]\label{lemma:fock_space_approx}
    Let $H$ be a Hermitian matrix with free cumulants $\{\kappa_k\}_{k\in\N}$ where $\kappa_2=1$ and odd free cumulants vanish through order $L$. Fix positive even integers $\ell,L$ where $2\leq\ell<L$, and define the truncated Fock space operator $\Ts_{\ell}$ according to \cref{def:hopping-operator}. Then, for every $t\geq0$,
    \begin{equation}
        \begin{aligned}
             &\left|\nTr(e^{-iHt})-\sum_{k=0}^L\frac{(-it)^k}{k!}\bra{0}\Ts_{\ell}^k\ket{0}\right|\\
             &\qquad\qquad\leq e^{t\Ks(L)}\left[\Ks(L)- \Ks(\ell) +\frac{\bigl(2t+t\Ks(L)\bigr)^L}{L!}\left(2+\frac{2t+t\Ks(L)}{L+1}\right)\right].
        \end{aligned}
    \end{equation}
\end{lemma}
\begin{proof}
    First, \cref{prop:truncated_fock_space} along with the triangle inequality gives, for every even $2\leq\ell_1\leq L$ and $0\leq k\leq\ell_1$,
    \begin{equation}\label{eqn:moment-trace-upper} 
        |\nTr(H^k)|= |\bra{0}\Ts_{\ell_1}^k\ket{0}|\leq \bigg\|a^\dagger+a+\sum_{\substack{4\leq k'\leq\ell_1\\ k'\text{ even}}}\kappa_{k'} a^{k'-1}\bigg\|^k
        \leq\bigg(2+\sum_{\substack{4\leq k'\leq\ell_1\\ k'\text{ even}}}|\kappa_{k'}|\bigg)^k =(2+\Ks(\ell_1))^k.
    \end{equation}
    Additionally, since $a^\dagger+a$ is self-adjoint, \cref{lem:perturbed-unitary} gives, with $t=-u$ for every $u\geq0$,
    \begin{equation}\label{eqn:cumulant-evolution-bound}
        \left\|e^{-i\Ts_{\ell_1}u}\right\|
        \leq e^{u\left\|\sum_{4\leq k\leq\ell_1}\kappa_k a^{k-1}\right\|} \leq e^{u\Ks(\ell_1)}.
    \end{equation}

    Our proof compares the Taylor series approximation of $\bra{0}e^{-i\Ts_{\ell}t}\ket{0}$ with the actual spectral amplitude $\nTr(e^{-iHt})$. To do this, we use an intermediate Fock space truncation given by
    \begin{equation}
        \Ts_L\equiv a^\dagger+a+\sum_{4\leq k\leq L}\kappa_k a^{k-1}.
    \end{equation}
    Intuitively, $L$ will be our moment truncation level and $\ell$ will be the free cumulant truncation level. Our goal is to show that the sequence of hybrids
    \begin{equation}\label{eqn:hybrid-method}
        \nTr(e^{-iHt})
        \xrightarrow{(1)}\bra{0}e^{-i\Ts_Lt}\ket{0}
        \xrightarrow{(2)}\bra{0}e^{-i\Ts_{\ell}t}\ket{0}
        \xrightarrow{(3)}\sum_{k=0}^L\frac{(-it)^k}{k!}\bra{0}\Ts_{\ell}^k\ket{0}
    \end{equation}
    incurs a small additive error at each step, and then apply the triangle inequality.

    We now prove each step of the hybrid method in \cref{eqn:hybrid-method}.
    \paragraph{Step (1): Initial cumulant truncation.}
    Starting from $\nTr(e^{-iHt})$, we first consider the Taylor series truncation of the Hermitian matrix $H$:
    \begin{equation}
        \left|\nTr(e^{-iHt})-\sum_{k=0}^{L-1}\frac{(-it)^k}{k!}\nTr(H^k)\right|
        \leq\frac{t^L}{L!}\nTr(H^L)
        \leq\frac{(2t+t\Ks(L))^L}{L!},
    \end{equation}
    where we have applied the trace bound from \cref{lemma:taylor-approximation} and \eqref{eqn:moment-trace-upper}. Also recall
    \begin{equation}
        \sum_{k=0}^{L-1}\frac{(-it)^k}{k!}\nTr(H^k)
        =\sum_{k=0}^{L-1}\frac{(-it)^k}{k!}\bra{0}\Ts_L^k\ket{0}.
    \end{equation}
    Now we wish to return to an exponential on the truncated-cumulant Fock space operator $\Ts_L$. To do this we once again utilize a Taylor series approximation, except this time in reverse:
    \begin{equation}
            \left|\sum_{k=0}^{L-1}\frac{(-it)^k}{k!}\bra{0}\Ts_L^k\ket{0}-\bra{0}e^{-i\Ts_Lt}\ket{0}\right|
            \leq\frac{t^L\|\Ts_L\|^L}{L!}
            \sup_{0\leq u\leq t}\left\|e^{-iu\Ts_L}\right\|
            \leq\frac{(2t+t\Ks(L))^L}{L!}e^{t\Ks(L)},
    \end{equation}
    where we have applied the operator norm bound from \cref{lemma:taylor-approximation} and \cref{eqn:moment-trace-upper,eqn:cumulant-evolution-bound}. Combining the two approximations with the triangle inequality gives
    \begin{equation}\label{eqn:first-truncation}
        \left|\nTr(e^{-iHt})-\bra{0}e^{-i\Ts_Lt}\ket{0}\right|
        \leq(1+e^{t\Ks(L)})\frac{(2t+t\Ks(L))^L}{L!}
        \leq2e^{t\Ks(L)}\frac{(2t+t\Ks(L))^L}{L!}.
    \end{equation}

    \paragraph{Step (2): Second cumulant truncation.}
    Applying \cref{lem:scalar-cutoff-comparison} gives
    \[
        \left|\bra{0}e^{-i\Ts_{\ell}t}\ket{0}-\bra{0}e^{-i\Ts_Lt}\ket{0}\right|
        \leq\bigl(\Ks(L)-\Ks(\ell)\bigr)e^{t\Ks(L)}.
    \]
    This is \cref{eqn:second-truncation}; the scalar comparison avoids the additional factor $t$ in the Duhamel bound.

    \paragraph{Step (3): Moment truncation.}
    Now that we have truncated the free cumulants to order $\ell$, we want to bound the error incurred from the Taylor series approximation for the spectral amplitude of this truncated operator. Applying the operator norm bound of \cref{lemma:taylor-approximation},
    \begin{equation}\label{eqn:third-truncation}
        \begin{aligned}
            \left|\bra{0}e^{-i\Ts_{\ell}t}\ket{0}-\sum_{k=0}^L\frac{(-it)^k}{k!}\bra{0}\Ts_{\ell}^k\ket{0}\right|
            &\leq\left\|e^{-i\Ts_{\ell}t}-\sum_{k=0}^L\frac{(-it)^k}{k!}\Ts_{\ell}^k\right\|\\
            &\leq\frac{\bigl(2t+t\Ks(\ell)\bigr)^{L+1}}{(L+1)!}e^{t\Ks(\ell)}.
        \end{aligned}
    \end{equation}

    \paragraph{Putting it all together.}
    Finally, we combine the contributions from \cref{eqn:first-truncation,eqn:second-truncation,eqn:third-truncation}. Using $\Ks(\ell)\leq\Ks(L)$ and applying the triangle inequality gives our final bound and completes the proof.
\end{proof}

To utilize \cref{lemma:fock_space_approx}, we need to upper bound $\Ks(L)$, and $\Ks(L)-\Ks(\ell)$ with high probability. We can do this in both the unconditionally proven regime and the conjectured regime:
\begin{enumerate}
    \item (Unconditional) We can apply \cref{thm:cumulant-scaling} whenever $m=\Omega(L^3)$ and can utilize \cref{thm:cumulant-concentration} to bound $\Ks(L)$ and $\Ks(L)-\Ks(\ell)$.

    \item (Conjectured) Assuming \cref{conjecture:combinatorial-linear} is true, \cref{thm:cumulant-scaling} additionally holds whenever $m=\Omega(L)$ so we can again apply \cref{thm:cumulant-concentration} to bound $\Ks(L)$, and $\Ks(L)-\Ks(\ell)$.
\end{enumerate}

\subsection{The full spectral amplitude algorithm}\label{subsec:final-classical-proof}
We now introduce the full algorithm in \cref{alg:sfa-algorithm} and prove that it satisfies \cref{thm:main-classical-algorithm}.

\begin{algorithm}[H]
\caption{Spectral amplitude algorithm}\label{alg:sfa-algorithm}
\begin{algorithmic}[1]
    \Require $H\sim$ sparse Pauli Hamiltonian, time $t\geq 0$, even $2\leq\ell<L\leq\gamma n/\log n$.
    \State Set $\hat{\mu}_0^{(\ell)}=\hat{\mu}_2^{(\ell)}=1$, and compute the exact moments $\hat{\mu}_k^{(\ell)}$ for even $4\leq k\leq\ell$ via direct Pauli expansion.
    \State Set $\kappa_2=1$, and compute the free cumulants $\kappa_k$ for even $4\leq k\leq\ell$ using the moment-cumulant recurrence. Set $\kappa_k=0$ for $k>\ell$
    \State Compute the approximate moments $\hat{\mu}_k^{(\ell)}$ for all even $k\leq L$ using \cref{eqn:scalar-R-moment-relation}.
    \State \Return
    \begin{equation}
        \hat{Z}(t)\equiv \sum_{\substack{0\leq k\leq L\\k\text{ even}}}\frac{(-it)^k}{k!}\hat{\mu}_k^{(\ell)}
    \end{equation}
\end{algorithmic}
\end{algorithm}

For a sparse Pauli Hamiltonian instance $H$, let $\hat{Z}(t)$ (which depends on $\ell$ and $L$) denote the output to \cref{alg:sfa-algorithm}. We now give our full theorem of the classical approximation algorithm for the spectral amplitude.

\begin{theorem}[Running time of the spectral amplitude approximation]\label{thm:scalar-runtime}
    Fix any $\delta = 1/\poly(n)$. There is a sufficiently small constant $c>0$ such that, for all sufficiently large $n$, $2n\leq m\leq \poly(n)$, $t\geq0$, and $0<\varepsilon < 1$ satisfying
    \begin{equation}\label{eq:scalar-runtime-window}
        1+t+\log(1/\varepsilon)\leq c\min\left\{m^{1/3},\frac n{\log n}\right\},
    \end{equation}
    there exists a classical algorithm which outputs an approximation $\hat{Z}(t)$ satisfying
    \[
        \left|\hat{Z}(t)-\nTr(e^{-iHt})\right|\leq\varepsilon
    \]
    with probability at least $1-\delta-e^{-\Omega(n)}$ over $H$ and total running time at most
    \begin{equation}\label{eq:scalar-accuracy-runtime}
        \Os(n/\delta)(1/\varepsilon)^{2+\Os(1/\log m)}.
    \end{equation}
\end{theorem}
\begin{proof}
    Choose a sufficiently large fixed constant $C_0$, and let $\ell\geq 2$ be the smallest even integer satisfying
    \begin{equation}\label{eq:optimized-cumulant-cutoff}
        \delta^{-1/2}\left(\frac{C_0(\ell+2)}m\right)^{(\ell+2)/4}\leq\frac{\varepsilon}{C_0}.
    \end{equation}
    Since $2n\leq m$ and $\log(1/\varepsilon)=\Os(m^{1/3})$, a sufficiently large constant multiple of $1+\log(1/\varepsilon)/\log m$ satisfies \eqref{eq:optimized-cumulant-cutoff}. Hence
    \begin{equation}\label{eq:optimized-cutoff-size}
    \ell=\Os\left(1+\frac{\log(1/\varepsilon)}{\log m}\right).
    \end{equation}
    Choose an even $L=\Theta(1+t+\log(1/\varepsilon))$ with a sufficiently large implied constant and $L\geq\ell+4$. Taking $c$ in \eqref{eq:scalar-runtime-window} sufficiently small ensures $m\geq AL^3$ and $L\leq\gamma n/\log n$, so \cref{thm:cumulant-concentration} applies.

    \paragraph{Approximation.}
    Apply the fixed lower-order estimate \eqref{eq:hypercontractive-scalar-tail} at order $4$ with failure probability $\delta/n^2$, and apply \eqref{eq:cumulant-tail-probability} at lower order $\ell+2$ with $\eta=\delta/2$. For sufficiently large $n$, these give simultaneously
    \begin{equation}\label{eq:scalar-runtime-cumulant-sums}
        \Ks(L)\leq C\frac{\sqrt{\log n}}m,\qquad 
        \Ks(L)-\Ks(\ell)\leq \delta^{-1/2}\left(\frac{C(\ell+2)}m\right)^{(\ell+2)/4},
    \end{equation}
    except with probability at most $\delta+e^{-\Omega(n)}$. Since $t\leq L$ and $L\leq(m/A)^{1/3}$, we have $t\Ks(L)\leq\Os(m^{-2/3}\sqrt{\log n})=o(1)$, using $m\geq 2n$. Thus the exponential prefactor in \cref{lemma:fock_space_approx} is bounded and the Taylor base is $\Os(1+t)$. The cumulant contribution is at most $\varepsilon/2$ by \eqref{eq:optimized-cumulant-cutoff}, and the Taylor contribution is at most $\varepsilon/2$ by $L!\geq(L/e)^L$ and the choice of $L$. This gives the claimed error and success probability. At $t=0$ the exact value $1$ may instead be returned directly.

    \paragraph{Running time.}
    Minimality at every chosen cutoff $\ell$ means that $\ell-2$ fails \eqref{eq:optimized-cumulant-cutoff}, giving
    \begin{equation}
        \left(\frac{m}{C_0\ell}\right)^{\ell/2}<C_0^2\delta^{-1}\varepsilon^{-2}.
    \end{equation}
    Using $\binom{m}{\ell/2}\leq(2em/\ell)^{\ell/2}$ in \eqref{eq:scalar-cutoff-runtime} yields running time upper bounded by
    \begin{equation}
        \Os\left(\frac{n}{\delta}\varepsilon^{-2}e^{\Os(\ell)}+L\right)\leq\Os(\frac{n}{\delta})(1/\varepsilon)^{2+\Os(1/\log m)},
    \end{equation}
    where the inequality uses \eqref{eq:optimized-cutoff-size}, $L\leq n$, and $\varepsilon<1$. This proves \eqref{eq:scalar-accuracy-runtime}.
\end{proof}

This proves the claim for the algorithm for approximating the spectral amplitude in \cref{thm:main-classical-algorithm}.

\newpage

\part{The color framework and correlation functions}\label{part:color-framework}
The preceding sections concern moments and free cumulants of a sparse Pauli Hamiltonian instance $H$ from \cref{eq:1}. We will now allow fixed Pauli operators to be inserted between copies of $H$ in trace computations; conjugating $H$ by different Paulis will be identified with different ``colors'' of $H$.

In \cref{sec:color-framework}, we formalize the concept of color and show how it applies to the sparse Pauli Hamiltonian. Joint moments and joint free cumulants between different colors generalize the single-variable moments and free cumulants of \cref{part:sfa-algorithm}. In the non-crossing partition diagrams and commutation graphs (\cref{sec:color-graphs}), each vertex will also be associated with a color. The Fock space construction is expanded upon in \cref{sec:multi-color-fock} to incorporate this new degree of freedom.

We will see how this simple framework allows us to compute time-ordered and out-of-time-order correlators (\cref{sec:toc-otoc}), the spectral amplitude of the spiked Pauli model (\cref{sec:spiked-pauli}), and thermal expectation values (\cref{sec:thermal-expectations}).

We provide a summary in \cref{table:color-framework} of the notation we define from \cref{sec:color-framework}. 

\begin{table}[ht]
    \centering
    \begin{tabular}{l l l}
        \toprule[.5mm]
        \textbf{} & \textbf{Name} & \textbf{Description} \\
        \midrule
        $\Qm$ & Generators & Sequence of Pauli operators independent modulo phase\\&&\qquad $\Qm\equiv (Q_1, \ldots, Q_s)$\\ \midrule
        $\Cs$ & Color group & Set of all possible Pauli insertions modulo phase\\
        &&\qquad $\Cs \equiv \langle [Q_1], \ldots, [Q_s]\rangle \cong \F_2^{s}$ \\ \midrule
        $\va{c}$ & Label & Binary string representing Pauli operator modulo phase\\&&\qquad$\va{c}\in\F_2^s$, identified with $[Q_{\va{c}}]\in\Cs$
        \\ \midrule
        $Q_{\va{c}}$ & Color representative & Pauli operator representing a color group element\\&&\qquad $Q_{\va{c}} \equiv Q_1^{c_1}\cdots Q_s^{c_s} \in M_{2^n}(\C)$\\ \midrule
        $H_{\va{c}}$ & Colored Hamiltonian & Pauli Hamiltonian instance conjugated by color representative\\&&\qquad $H_{\va{c}} \equiv Q_{\va{c}}HQ_{\va{c}}^\dagger$
        \\ \midrule
        $\Cm$ & Label sequence & Sequence of labels representing color representatives\\&&
        \qquad $\Cm \equiv (\va{c}_1, \ldots, \va{c}_k) \in (\F_2^s)^k \cong \Cs^k$\\ \midrule
        $\tau_k^{(\Cm)}$ & Color moment & Joint moment of $H$ with conjugated color representatives\\&&\qquad $\tau_k^{(\Cm)} \equiv \tau_k(H_{\va{c}_1},\ldots, H_{\va{c}_k})$
        \\ \midrule
        $\kappa_k^{(\Cm)}$ & Color cumulant & Joint free cumulant of $H$ with conjugated color representatives\\&&\qquad $\kappa_k^{(\Cm)} \equiv \kappa_k(H_{\va{c}_1},\ldots, H_{\va{c}_k})$
        \\
        \bottomrule[.5mm]
    \end{tabular}
    \caption{Summary of main color framework notation and definitions introduced in \cref{sec:color-framework}. We identify each label $\va{c}\in\F_2^s$ with the corresponding color $[Q_{\va{c}}]\in\Cs$, while $Q_{\va{c}}$ denotes its chosen matrix representative from the Pauli group.}
    \label{table:color-framework}
\end{table}

\section{The color framework of the sparse Pauli model}
\label{sec:color-framework}
In this section, we outline our color framework and provide extensions for several of the main tools and results that we developed in \cref{part:sfa-algorithm}. To build intuition, we first present the general construction for the two-color setting in \cref{subsec:two-color-warmup}, which suffices for the majority of the applications we consider. In \cref{subsec:general-color-framework}, we formalize the framework for an arbitrary $2^s$ different colors. 

\subsection{Warmup: two colors}\label{subsec:two-color-warmup}
We begin with a simple example to introduce the concept of colors. Let $Q$ be a fixed $n$-qubit Pauli operator. Write $H_0 \equiv H$ and $\mathcolor{red}{H_1}=QHQ$. The identity
\[
    QH_c=H_{c\oplus 1}Q,\quad c\in\{0,1\}
\]
shows that moving an insertion through a factor of $H$ swaps its subscript. For example shifting the left Pauli operator $Q$ all the way to the right gives
\[
    H_{0}^r Q H_{0}^s Q H_{0}^t Q H_{0}^u Q = H_0^r \mathcolor{red}{H_1^s} H_0^t \mathcolor{red}{H_1^u}.
\]
$H$ defines a sparse Pauli Hamiltonian instance, while its subscript identifies it with a labeling, which we call a ``color'', e.g., $H_0$ is colored black and $\mathcolor{red}{H_1}$ is colored red. This allows us to systematically deal with the joint moments of $(H_0, \mathcolor{red}{H_1})$ while respecting the Pauli operator relations.

\begin{figure}[ht]
\centering
\begin{subfigure}[t]{\textwidth}
\centering
    \includegraphics[height=5cm]{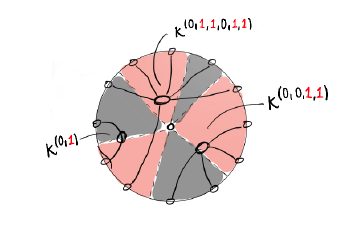}
    \caption{A colored non-crossing partition contributing the color moment. The colors distinguish copies of $H$ and $QHQ$ at each vertex, and each block contributes the color cumulant of its vertices.}
    \label{fig:color-wheel}
\end{subfigure}
\hfill
\begin{subfigure}[t]{\textwidth}
    \centering
    \includegraphics[height=6.2cm]{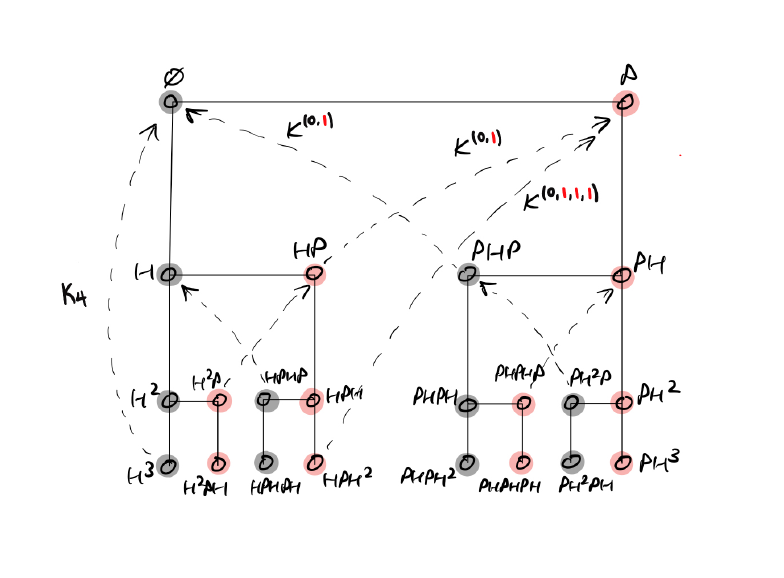}
    \caption{The two-color Fock space construction from \cref{sec:multi-color-fock}. Solid vertical edges represent creation and same-color annihilation, while horizontal edges represent color changes (Pauli insertions). Dashed edges illustrate annihilation weighted by higher or mixed color cumulants.}
    \label{fig:fock_space_spiked_pauli_model}
\end{subfigure}
\caption{Diagrammatic representation of the color moment $\nTr(H^2QH^3QHQH^2QH^2QH^2Q)$. Shifting the Pauli insertions $Q$ all to the right gives $H_0^2\mathcolor{red}{H_1^3}H_0\mathcolor{red}{H_1^2}H_0^2\mathcolor{red}{H_1^2}$.}
\end{figure}

For a two-color sequence $\Cm=(c_1,\ldots,c_k)\in\F_2^k$, we define the \emph{color moments} and \emph{color cumulants} as
\begin{equation}
    \tau_k^{(\Cm)}[H, \ldots, H]\equiv\tau(H_{c_1}\cdots H_{c_k}),\qquad 
    \kappa_k^{(\Cm)}[H, \ldots, H]\equiv\kappa_k(H_{c_1},\ldots,H_{c_k}).
\end{equation}
The all-0 sequence recovers the original moment and free cumulant definitions, i.e.,
\[
    \tau_k^{(0^k)}[H,\ldots, H]=\tau(H^k).
\]
These are joint moments and joint free cumulants from \cref{sec:preliminaries} with two variables. In particular, the moment--cumulant relation becomes
\begin{equation}\label{eq:color-moment-cumulant}
    \tau_k^{(\Cm)}[H, \ldots, H]=\sum_{\pi\in NC(k)}\prod_{V\in\pi} \kappa_{|V|}^{(\Cm|_V)}[H, \ldots, H],
\end{equation}
where $\Cm|_V$ is the sequence of colors at the positions contained in block $V$, in increasing order. The only difference from the single-variable expansion is that each block now carries a color sequence as well; see \cref{fig:color-wheel}.

Traciality gives invariance of color moments and cumulants under cyclic rotation of the color sequence. Conjugating every argument by a fixed Pauli operator $Q$ also leaves the trace unchanged, so swapping the two colors everywhere also does not change a moment. These symmetries also hold for the free cumulants due to \eqref{eq:color-moment-cumulant}. We may therefore normalize the first color to be $0$ without loss of generality. We also occasionally write the color moments and color cumulants as just $\tau_k^{(\Cm)}$ and $\kappa_k^{(\Cm)}$ when $H$ is given from context.

\paragraph{Fixed color sequence recurrence.}
For color sequence $\Cm\in\F_2^k$, we represent the contiguous-subsequence moments as
\begin{equation}
    \tau^{(\Cm)}_{p:q}[H, \ldots, H]\equiv
    \begin{cases}
        \tau(H_{c_p}\cdots H_{c_q}),&p\leq q,\\ 1,&p>q.
    \end{cases}
\end{equation}
\eqref{eqn:joint-R-recurrence} provides a recurrence relation which allows for moments to be computed using free cumulants. We now give an extension under our color framework. Decomposing a non-crossing partition according to the block containing element $1$ (as before) gives
\begin{equation}\label{eq:color-R-moment-recurrence}
    \tau^{(\Cm)}_{1:k}[H, \ldots, H]
    =\sum_{k'=1}^k\ \sum_{1=i_1<\cdots<i_{k'}\leq k}
       \kappa_{k'}^{(\Cm|_{I})}
       \prod_{j=1}^{k'}
       \tau^{(\Cm)}_{i_j+1:i_{j+1}-1},\qquad I\equiv (i_1,\ldots,i_{k'}),\quad  i_{k'+1}\equiv k+1.
\end{equation}
Every moment on the right side of the equation involves a strictly smaller subsequence of $\Cm$. Therefore \eqref{eq:color-R-moment-recurrence} gives a recursive computation for the color moments once the required color cumulants are supplied.

\subsection{General color framework setup}\label{subsec:general-color-framework}
We now extend the previous two-color construction to an arbitrary number of colors $2^s$ where $s\geq 0$. Consider a sequence of fixed $n$-qubit Pauli operators $\Qm \equiv (Q_1,\ldots,Q_s)$ which are independent modulo phase. We call these Pauli operators the \emph{color generators}; $s$ independent Paulis generate a total of $2^s$ colors. We call $\Cs\equiv\langle[Q_1],\ldots,[Q_s]\rangle$ the \emph{color group}, where $[Q]$ denotes the equivalence class of the Pauli operator $Q$ modulo phase. For each label $\va{c}=(c_1,\ldots,c_s)\in\F_2^s$, fix the ordered representative
\[
    Q_{\va{c}}\equiv Q_1^{c_1}\cdots Q_s^{c_s}.
\]
Independence modulo phase implies that the map $\va{c}\mapsto[Q_{\va{c}}]$ is a group isomorphism from $\F_2^s$ to $\Cs$. We use labels and their corresponding colors interchangeably, while $Q_{\va{c}}$ always denotes the chosen matrix representative in $M_{2^n}(\C)$. Given a color representative with label $\va{c}$, the associated conjugate Hamiltonian is
\[
    H_{\va{c}}\equiv Q_{\va{c}} H Q_{\va{c}}^\dagger.
\]
For each sampled Pauli term in $H$, the conjugation $H_{\va{c}}$ simply introduces its commutation sign with the color:
\begin{equation}\label{eq:multicolor-conjugation}
    Q_{\va{c}} P_jQ_{\va{c}}^\dagger=\langle Q_{\va{c}},P_j\rangle\,P_j,
    \qquad \langle Q_{\va{c}},P_j\rangle=\prod_{a=1}^s\langle Q_a,P_j\rangle^{c_a},
\end{equation}
where the second identity holds because the commutation sign is multiplicative in each argument.

The color moment and cumulant definitions, their symmetries, and \eqref{eq:color-R-moment-recurrence} remain mostly unchanged:
\begin{definition}[Color moments and color cumulants]\label{def:color-cumulants}
     Fix a Hamiltonian $H$ and generators $\Qm\equiv (Q_1, \ldots, Q_s)$. For a label sequence $\Cm=(\va{c}_1,\ldots,\va{c}_k)\in \Cs^k$, define the \emph{color moments} with respect to sequence $\Cm$ as
    \begin{equation}\label{eq:color-moment}
        \tau_k^{(\Cm)}[H, \ldots, H]
        \equiv\tau(H_{\va{c}_1}\cdots H_{\va{c}_k}),
    \end{equation}
    and the \emph{color cumulants} with respect to sequence $\Cm$ as
    \begin{equation} \label{eq:color-cumulant}
        \kappa_k^{(\Cm)}[H, \ldots, H]
    \equiv\kappa_k(H_{\va{c}_1},\ldots,H_{\va{c}_k}).
    \end{equation}
\end{definition}
The only difference from before is that now the color group is a general set $\Cs\cong \F_2^s$ instead of just a single bit representing a single Pauli operator. Let $\va{e}_a$ be the $a$-th standard basis vector of $\F_2^s$. The generalization of the identity from \cref{subsec:two-color-warmup} for general colors is
\[
    Q_a H_{\va{c}}=H_{\va{c}\oplus\va{e}_a} Q_a,
\]
where $\oplus$ denotes element-wise addition modulo 2 (XOR). For indices $g_1,\ldots,g_r\in[s]$ and nonnegative integers $k_0,\ldots,k_r$, set $\va{c}_0=\va{0}$ and $\va{c}_j=\va{e}_{g_1}\oplus\cdots\oplus\va{e}_{g_j}$. Shifting the Pauli operators to the right side of the product gives
  \begin{equation}\label{eq:color-insertion-identity}
    H^{k_0}Q_{g_1}H^{k_1}\cdots Q_{g_r}H^{k_r}
    =H_{\va{c}_0}^{k_0}H_{\va{c}_1}^{k_1}\cdots H_{\va{c}_r}^{k_r}
      Q_{g_1}\cdots Q_{g_r}.
    \end{equation}
The colors therefore record the commutation signs incurred by this process, while the product at the right side records the relationship between the inserted Pauli operators. We generally call this term the \emph{boundary term}. In \cref{part:color-framework}, we will only deal with the case where the boundary term $Q_{g_1}\cdots Q_{g_r}$ from \eqref{eq:color-insertion-identity} equals identity modulo phase ($[Q_{g_1}\cdots Q_{g_r}] = [\mathbb{I}]$). In this case we can ignore the boundary term except for its phase, and we only keep the colors incurred from the insertions. For example, in the two-color setting, any even number of Pauli insertions leaves $Q^{2k}=\mathbb{I}$ at the boundary, and the out-of-time-order correlators leave a boundary term $ABAB=\langle A,B\rangle \,\mathbb{I}$ (in \cref{part:operator-valued}, we consider the general case where $Q_{g_1}\cdots Q_{g_r}$ does not necessarily equal identity modulo phase). 

\section{Concentration of color cumulants of the sparse Pauli model}
\label{sec:color-graphs}
In this section, we show the color cumulants decay with respect to their order, extending \cref{sec:commutation-graphs,sec:free-cumulant-concentration}.

\subsection{Color cumulants from commutation graphs}
Let $H=m^{-1/2}\sum_{i=1}^m J_iP_i$ be a Pauli Hamiltonian as usual. By \eqref{eq:multicolor-conjugation}, a color does not change the Pauli terms of $H$, only their signs:
\begin{equation}
    H_{\va{c}}=\frac{1}{\sqrt{m}}\sum_{i=1}^m J_i\,\langle Q_{\va{c}},P_i\rangle\,P_i.
\end{equation}
In the graph representation, we record these signs by attaching a color to each vertex.

\begin{definition}[Color commutation graph statistics]\label{def:color-graph-statistic}
    Fix color generators $\Qm=(Q_1,\ldots,Q_s)$. For a simple graph $g$ with $r\equiv|V(g)|\leq m$ vertices and a vertex coloring $\Dm:V(g)\to\Cs$, define
    \begin{equation}
        C^{(\Dm)}[g] \equiv\frac{1}{(m)_r}\sum_{\iota:V(g)\hookrightarrow[m]}
        \left(\prod_{v\in V(g)}\langle Q_{\Dm(v)},P_{\iota(v)}\rangle\right)
        \left(\prod_{(u,v)\in E(g)}\langle P_{\iota(u)},P_{\iota(v)}\rangle\right).
    \end{equation}
\end{definition}
The color of a vertex therefore contributes one more commutation sign, with the fixed Pauli operator $Q_{\Dm(v)}$ in place of a sampled one. The coloring $\Dm\equiv\va{0}$ recovers $C[g]$ from \cref{def:graph-statistics}, every $C^{(\Dm)}[g]$ lies in $[-1,1]$, and the statistic of the empty graph is $1$. Unlike for $C[g]$, an isolated vertex can only be removed from $g$ if its color is $\va{0}$.

A label sequence $\Cm=(\va{c}_1,\ldots,\va{c}_k)$ of even length $k$ assigns a color to each position of a Pauli word. For $\pi\in\peven(k)$, we collect these colors block by block into the coloring of $g(\pi)$ given by
\begin{equation}
    \Cm_\pi(V)\equiv\bigoplus_{a\in V}\va{c}_a,\qquad V\in\pi.
    \label{eq:induced-block-color}
\end{equation}
Since the commutation sign is multiplicative in each argument, $\prod_{a\in V}\langle Q_{\va{c}_a},P\rangle=\langle Q_{\Cm_\pi(V)},P\rangle$ for every Pauli operator $P$, so the signs contributed by all occurrences of a Pauli operator combine into one color at its vertex. Expanding the product as in \eqref{eq:moment_distinct_indices}, averaging over the Rademacher signs again leaves the words whose partition $\pi$ of equal indices is even. 
Labeling the blocks of $\pi$ by an injection $\iota:\pi\hookrightarrow[m]$, the colors contribute the scalar $\prod_{V\in\pi}\langle Q_{\Cm_\pi(V)},P_{\iota(V)}\rangle$, and \cref{lem:even-word-trace} evaluates the remaining trace. Therefore
\begin{equation}
    \E_{\va{J}}\tau(H_{\va{c}_1}\cdots H_{\va{c}_{k}})
    =\frac{1}{m^{k/2}}\sum_{\pi\in\peven(k)}(m)_{|\pi|}\,C^{(\Cm_\pi)}[g(\pi)].
    \label{eq:color-even-partitions}
\end{equation}
This identity holds for all even $k\leq2m$ without assuming loop-freeness. Assuming loop-freeness through order $k$, every word containing some Pauli operator an odd number of times has vanishing trace, so the left-hand side equals the color moment of the fixed instance, exactly as in \eqref{eq:moments-graph-statistics}.

We next present the extension of loop-free expressions to color cumulants:
\begin{definition}[Loop-free expression of color cumulants]\label{def:color-loop-free-expression}
    For any (not necessarily loop-free) sparse Pauli Hamiltonian instance $H$, even integer $2\leq k\leq2m$, and label sequence $\Cm\in\Cs^{k}$, we define the \emph{loop-free expression} of its order-$k$ color cumulant as
    \begin{equation}
        \tilde{\kappa}_{k}^{(\Cm)}(\Pm) \equiv \frac{1}{m^{k/2}} \sum_{\pi \in \pevenconn(k)} (m)_{|\pi|}\,\chi(\pi)\, C^{(\Cm_\pi)}[g(\pi)].
    \end{equation}
\end{definition}
For the constant label sequence $\Cm=(\va{0},\ldots,\va{0})$, this is the loop-free expression $\tilde{\kappa}_{k}(\Pm)$ of \cref{def:loop-free-expression}.

\begin{theorem}[Color cumulants from commutation graph statistics]
    \label{thm:color-cumulants-from-graphs}
    Let $H$ be a sparse Pauli Hamiltonian instance with $n$ qubits, $m$ terms, and free cumulants $(\kappa_k)_{k\in\N}$ which is loop-free through order $k\leq 2m$. For a fixed label sequence $\Cm\in\Cs^{k}$, the sign-averaged color cumulant satisfies $\E_{\va{J}}\kappa_k^{(\Cm)}=0$ when $k$ is odd and equals its loop-free expression $\E_{\va{J}}\kappa_k^{(\Cm)}=\tilde{\kappa}_{k}^{(\Cm)}(\Pm)$ when $k$ is even. Moreover, $\kappa_k^{(\Cm)}=\E_{\va{J}}\kappa_k^{(\Cm)}$.
\end{theorem}
\begin{proof}
    The proof follows that of \cref{thm:cumulants-from-graph-statistics}, with one additional scalar per word. Expanding every $H_{\va{c}_a}$ into its Pauli terms, a word whose blocks of equal indices form $\pi$ and are labeled by $\iota:\pi\hookrightarrow[m]$ acquires from the colors the factor $\prod_{V\in\pi}\langle Q_{\Cm_\pi(V)},P_{\iota(V)}\rangle$. 
    This factor is a product over all positions of the word, so it is the same for every splitting of the word into the blocks of a non-crossing partition. The proof of \cref{lemma:loop-free-expression-partition} therefore applies verbatim and gives the coefficient $\chi(\pi)$ and the restriction to connected partitions $\pi$. 
    Summing the remaining traces over the labelings with \cref{lem:even-word-trace} gives $(m)_{|\pi|}\,C^{(\Cm_\pi)}[g(\pi)]$.

    Finally, by loop-freeness, every word of length at most $k$ containing some Pauli operator an odd number of times has vanishing trace, so the color moments of order at most $k$, and hence the color cumulants, do not depend on the Rademacher signs.
\end{proof}

\subsection{Concentration of color cumulants}
\label{subsec:color-cumulant-concentration}
We now utilize the loop-free expression from \cref{def:color-loop-free-expression} and extend the concentration argument of \cref{sec:free-cumulant-concentration} to color cumulants. Throughout this subsection, the generators $\Qm$ are fixed independently of the sampled Paulis $\Pm$, and $\overline{g}$ denotes the edge-induced subgraph of $g$ that ignores the coloring. The only new ingredient is a one-sided version of \cref{lem:graph-properties-covariance}.

\begin{lemma}[Mixed second moments of color graph statistics]
    \label{lem:color-graph-second-moment}
    Let $g$ and $g'$ be graphs on at most $m$ vertices with colorings $\Dm$ and $\Dm'$. Then
    \begin{equation}
        \left|\E_\Pm\left[
            C^{(\Dm)}[g]\, C^{(\Dm')}[g']
        \right]\right|
        \leq
        \begin{cases}
            \displaystyle
            \frac{|\Aut(\overline{g})|}{(m)_{|\overline{g}|}}+4^{-n},
                &\overline{g}\cong\overline{g}',\\[6pt]
            4^{-n},&\overline{g}\not\cong\overline{g}'.
        \end{cases}
        \label{eq:color-graph-second-moment}
    \end{equation}
\end{lemma}
\begin{proof}
    Expand both statistics over pairs of labelings $\iota,\iota'$ as in the proof of \cref{lem:graph-properties-covariance}. The pairs with equal labeled edge sets are counted as before, and each contributes at most $1$ in absolute value, and the $4^{-n}$ bound for a pair with $S_{\iota,\iota'}\neq\varnothing$ is due to \cref{lemma:k-wise-independence}. Applying the same count of labelings proves \eqref{eq:color-graph-second-moment}.
\end{proof}

\begin{theorem}[Concentration of color cumulants]
    \label{thm:color-cumulant-concentration}
    Let $k_0,k_1$ be even integers satisfying $4\leq k_0\leq k_1\leq L$, and suppose
    \begin{equation}
        A\max\{k_1^3,\,2^{4s}k_1\}\leq m\leq\poly(n),
        \qquad L\leq\gamma n/\log n,
    \end{equation}
    where $A>0$ is sufficiently large and $\gamma>0$ is sufficiently small. Then, for every $0<\eta<1$,
    \begin{equation}
        \sum_{\substack{k_0\leq k\leq k_1\\k\ \text{even}}}
        \sum_{\substack{\Cm\in \Cs^k\\\va{c}_1=\bm{0}}}
        |\kappa_k^{(\Cm)}|\leq\frac{2}{\sqrt{\eta}}
        \left(\frac{C\,2^{4s}k_0}{m}\right)^{k_0/4}
    \end{equation}
    with probability at least $1-\eta-e^{-\Omega(n)}$, for an absolute constant $C>0$.
\end{theorem}
\begin{proof}
    For each even $4\leq k\leq k_1$ and fixed color sequence $\Cm$, the proof of \cref{lemma:second-moment-upper} applies with \cref{lem:color-graph-second-moment} in place of its uncolored counterpart. Indeed, after taking absolute values, the gluing argument depends only on the underlying partitions and $\chi$, which are unchanged by the colors. Thus \cref{lemma:second-moment-upper} gives
    \begin{equation}\label{eq:color-cumulant-second-moment}
        \E_\Pm\left[|\tilde{\kappa}_k^{(\Cm)}|^2\right] \leq \left(\frac{Ck}{m}\right)^{k/2}.
    \end{equation}
    for all $\Cm$.

    There are $2^{s(k-1)}$ color sequences with first color $\bm{0}$. Expanding the square and applying Cauchy--Schwarz to each cross term, we obtain
    \begin{equation}
        \E_\Pm\left[\left(\sum_{\substack{k_0\leq k\leq k_1\\k\ \mathrm{even}}} \sum_{\substack{\Cm\in\Cs^k\\\va{c}_1=\bm{0}}} |\tilde{\kappa}_k^{(\Cm)}|\right)^2\right]
        \leq \left(\sum_{\substack{k_0\leq k\leq k_1\\k\ \mathrm{even}}} 2^{s(k-1)}\left(\frac{Ck}{m}\right)^{k/4}\right)^2
        \leq 4\left(\frac{C\,2^{4s}k_0}{m}\right)^{k_0/2}.
    \end{equation}
    The last inequality follows by the same geometric summation as in the proof of \cref{thm:cumulant-concentration}, since $m\geq A\,2^{4s}k_1$ for sufficiently large $A$.

    Markov's inequality applied to the square of this sum gives the desired bound for the loop-free expressions with failure probability at most $\eta$. Finally, \cref{lem:loop-freeness,thm:color-cumulants-from-graphs} identify these expressions with the actual color cumulants simultaneously, except with probability $e^{-\Omega(n)}$.
\end{proof}
\begin{corollary}[Stronger hypercontractivity bound]\label{corollary:color-cumulant-hypercontractivity}
    If additionally $k_0$ and $s$ are fixed and the generators are independent modulo phase, then, for every sufficiently large $n$, $\delta=1/\poly(n)$, and $m\leq\poly(n)$,
    \begin{equation}\label{eq:hypercontractive-color-tail}
        \sum_{\substack{k_0\leq k\leq k_1\\k\text{ even}}}\ \sum_{\substack{\Cm\in\Cs^k\\\va{c}_1=\va{0}}}\left|\kappa_k^{(\Cm)}\right|
        \leq C m^{-k_0/4}(\log n)^{D_{k_0,s}/2},\qquad D_{k,s}\equiv\binom{k/2}{2}+\frac{sk}{2},
    \end{equation}
    with probability at least $1-\delta-e^{-\Omega(n)}$.
\end{corollary}
The proof of \cref{corollary:color-cumulant-hypercontractivity} is deferred to \cref{sec:hypercontractivity}.
\begin{remark}
    Assuming \cref{conjecture:combinatorial-linear}, the lower bound assumption $m \geq A\max\{k_1^3,\,2^{4s}k_1\}$ in \cref{thm:color-cumulant-concentration,corollary:color-cumulant-hypercontractivity} may be replaced by $m\geq \max\{A\,2^{4s}k_1, n\}$.
\end{remark}

\section{Multi-color Fock space construction}\label{sec:multi-color-fock}
We can also interpret color moments and cumulants through a quantum walk, extending the scalar construction in \cref{sec:fock-space}. This construction will be used for the analysis of several subsequent algorithms, including correlation functions in \cref{sec:toc-otoc} and the spiked Pauli model in \cref{sec:spiked-pauli}. 

Define the Hilbert space
\begin{equation}
    \Hs_{\Cs} \equiv \ell^2(\Cs) \cong \operatorname{span}_{\C}\{\ket{c}:\va{c}\in\F_2^s\},
\end{equation}
where the basis vectors $\ket{c}$ are indexed by the labels of color representatives from the color group. We apply the full Fock space construction from \cref{def:full-fock-space}, together with a register denoting the current color of the walk:
\begin{equation}
    \Fs(\Hs_{\Cs})\otimes \Hs_{\Cs},
    \qquad\ket{\Wm;c}\equiv\ket{\Wm}\otimes\ket{c}.
\end{equation}
The $\Wm$ is a label sequence of arbitrary length which records the colors of the open legs, written in their order of creation; $\ket{c}$ represents a label $\va{c}$ recording the current color of the state. In this section, we only use the explicit specializations of $s=1$ (2 colors, used for TOCs, the spiked Pauli model, and thermal expectations), and $s=2$ (4 colors, used for OTOCs), but include a fully general construction which can be generalized to, e.g., arbitrary many-body correlations and arbitrarily many Pauli spikes.

\paragraph{Two-color Fock space construction for TOCs.}
In the two-color setting we have a single (non-identity modulo phase) Pauli generator. We represent this Pauli operator in the Fock space with an operator $b$, which we call the \emph{spike operator} (analogous to the spiked Pauli model). The label sequence $\Wm$ can be represented by a binary string of length $k$ where $k\geq 0$, and the current color $c$ can be represented by a single bit.

The creation, annihilation, and spike operators thus act by
\begin{equation}
    \begin{aligned}
        &a^\dagger\ket{\Wm;c}=\ket{\Wm c;c},&&
        a\ket{\Wm c';c}=\delta_{c,c'}\ket{\Wm ;c},\\
        &a\ket{\emptyset;c}=0,&&
        b\ket{\Wm;c}=\ket{\Wm;c\oplus1},
    \end{aligned}
    \label{eq:two-color-fock-operators}
\end{equation}
where $\Wm c$ denotes appending the bit $c$ to the end of the string $\Wm$. We have $a=(a^\dagger)^\dagger$, $b=b^\dagger$, $b^2=\mathbb{I}$, and all three operators have norm one. The spike operator changes the current color without affecting the legs already stored in the word, while the creation and annihilation operators act according to the current color. The vacuum state is $\ket{\emptyset; 0}$.

\paragraph{Four-color Fock space construction for OTOCs.}
In the four-color setting we have two independent Pauli generators, both not equal to identity modulo phase. We represent these Pauli operators in the Fock space with two spike operators $b_1, b_2$. The label sequence $\Wm$ can be represented by a binary string of length $2k$ where $k\geq 0$, and $\ket{c}$ represents the label of the current color $\va{c}=c_1c_2$ with two bits.

The creation, annihilation, and spike operators thus act by
\begin{equation} \label{eq:four-color-fock-operators}
    \begin{aligned}
        &a^\dagger\ket{\Wm;c}=\ket{\Wm c; c},\qquad
        &&a\ket{\Wm c';c}=\delta_{c,c'}\ket{\Wm; c},\qquad
        a\ket{\emptyset;c}=0,\\
        &b_1\ket{\Wm;c_1c_2}=\ket{\Wm;(c_1\oplus1)c_2},\qquad
    &&b_2\ket{\Wm;c_1c_2}=\langle Q_1,Q_2\rangle^{c_1}
    \ket{\Wm;c_1(c_2\oplus1)}.
    \end{aligned}
\end{equation}
where $\Wm c$ denotes appending the bits $c_1$ and $c_2$ to the end of the string $\Wm$. We similarly have $a=(a^\dagger)^\dagger$, $b_i=b_i^\dagger, b_i^2=\mathbb{I}$ for $i\in\{1,2\}$, and all four operators have norm one. Spike operators $b_1$ and $b_2$ are defined to follow the (anti)commutation of $Q_1,Q_2$ through $b_1b_2 = \langle Q_1,Q_2\rangle b_2 b_1$. The vacuum state is $\ket{\emptyset; 00}$.

\paragraph{Generalized color Fock space and hopping operator.}
For a general sequence of color generators $\Qm\equiv (Q_1, \ldots, Q_s)$, $\Wm$ is a sequence of labels in $\Cs \cong \F_2^{s}$, and $\ket{c}$ is indexed by a color with label $\va{c} \in \Cs$. We now have a collection of spike operators $b_1, \ldots b_s$ which are Hermitian, involutions, and follow the respective commutation relations of the Paulis $Q_1, \ldots Q_s$. We denote the vacuum state $\Omega_s \equiv \ket{\emptyset; 0^s}$.

It will be convenient to define \emph{color-indexed} creation and annihilation operators, which model conjugating the Hamiltonian $H$ with a Pauli operator: 
\begin{equation}
    a_{\va{c}} \equiv b_1^{c_1}\cdots b_s^{c_s} a \, b_s^{c_s}\cdots b_1^{c_1}, \qquad \va{c} \equiv (c_1, \ldots, c_{s})
\end{equation}
and $a^\dagger_{\va{c}} = (a_{\va{c}})^{\dagger}$. Using these color-indexed operators, we can define the analogue of \cref{def:hopping-operator} which generalizes to arbitrary color groups:

\begin{definition}[Fock space color hopping operator]\label{def:color-hopping-operator}
    Fix generators $\Qm\equiv (Q_1, \ldots, Q_s)$. Let $H$ be a bounded Hermitian operator with color cumulants $(\kappa_k^{(\Cm)})_{k\in\N, \Cm\in \Cs^k}$, such that $\kappa_2=1$ and all odd color cumulants through order $L$ of any color vanish. For all even cutoffs $2\leq \ell\leq L$, we define the Fock space \emph{color hopping operator}:
    \begin{equation}
        \Ts_{\ell}^{(\Qm)} \equiv a^\dag + a + \sum_{\substack{4\leq k\leq \ell\\k\ \mathrm{even}}} \kappa_k a^{k-1} + \sum_{\substack{2\leq k\leq \ell\\k\ \mathrm{even}}}\sum_{\substack{\Cm\in \Cs^{k},\ \va{c}_1=\va{0} \\ \Cm\neq(\va{0},\ldots,\va{0})}} \kappa_k^{(\Cm)} a_{\va{c}_{k}}\cdots a_{\va{c}_2},\quad \Cm \equiv (\va{c}_1, \ldots, \va{c}_{k}).
    \end{equation}
\end{definition}

In analogy with the truncated moments $\hat{\mu}^{(\ell)}_k$ of \cref{prop:truncated_fock_space}, for a label sequence $\Cm\in\Cs^k$ we define the \emph{truncated color moments}
\begin{equation}\label{eq:truncated-color-moments}
    \hat{\mu}^{(\ell;\Cm)}\equiv\sum_{\pi\in NC(k)}\prod_{V\in\pi}\kappa^{(\Cm|_V)}_{|V|}\,\mathbf{1}_{[|V|\leq\ell]}.
\end{equation}
The truncated color moments are the outputs of the recurrence \eqref{eq:color-R-moment-recurrence} when all color cumulants of order greater than $\ell$ are set to zero, and they equal the color moments $\tau^{(\Cm)}_k$ for $k\leq \ell$ by \eqref{eq:color-moment-cumulant}.

The following proposition shows that this Fock space hopping operator matches the color moments through order $\ell$.
\begin{proposition}\label{prop:hopping-operator-otoc}
    Fix a color hopping operator $\Ts_{\ell}^{(\Qm)}$ satisfying the conditions of \cref{def:color-hopping-operator}. Let $g_1,\ldots,g_r\in[s]$ be insertion indices in which every generator appears an even number of times, so that $\bigoplus_{i=1}^r\va{e}_{g_i}=\va{0}$ and $Q_{g_1}\cdots Q_{g_r}=\zeta\mathbb{I}$ for a phase $|\zeta|=1$. Then for all integers $k_1,\ldots,k_r\geq0$ with $\sum_{i=1}^r k_i\leq \ell$,
    \begin{equation}
        \bra{\Omega_s}\big(\Ts_{\ell}^{(\Qm)}\big)^{k_1} b_{g_1}\,\big(\Ts_{\ell}^{(\Qm)}\big)^{k_2} b_{g_2}\cdots b_{g_{r-1}}\,\big(\Ts_{\ell}^{(\Qm)}\big)^{k_r} b_{g_r} \ket{\Omega_s}=\nTr(H^{k_1} Q_{g_1} H^{k_2} Q_{g_2} \cdots Q_{g_{r-1}} H^{k_r} Q_{g_r}).
    \end{equation}
\end{proposition}
\begin{proof}
    Moving the probes to the right gives the boundary factor $\zeta$ and the label sequence $\Cm$ by \eqref{eq:color-insertion-identity}; the hypothesis $\bigoplus_i\va{e}_{g_i}=\va{0}$ implies a scalar-valued boundary term. Reading the Fock product in its order of application, from right to left, produces the same colors at the Hamiltonian sites, and the insertions $b_{g_i}$ return the current-color register to $\va{0}$ with the same phase $\zeta$.

    As in \cref{prop:truncated_fock_space}, each returning walk corresponds to a noncrossing partition of the $k\equiv \sum_{i}k_i$ Hamiltonian sites. The rightmost sites of a block of size $k'$ create legs, and its leftmost site removes the preceding open legs with their color
    cumulant. Because creation appends letters, the coefficient is ordered as the current color followed by the stored suffix in reverse creation order. This is exactly the action of $a_{\va{c}_{k'}}\cdots a_{\va{c}_2}$ in
    \cref{def:color-hopping-operator}, after translating the current color to $\va{0}$.

    Every block inherits a subsequence of $\Cm$. Every block has size at most $k\leq \ell$, and odd blocks contribute zero. Summing the weights therefore gives $\zeta\sum_{\pi\in NC(k)}\prod_{V\in\pi} \kappa_{|V|}^{(\Cm|_V)}$, which equals the required trace by \eqref{eq:color-moment-cumulant}.
\end{proof}

\section{Computing correlation functions}
\label{sec:toc-otoc}
In this section, we use our color cumulant and color Fock space constructions to give efficient classical algorithms for computing time-ordered and out-of-time-ordered correlators of the sparse Pauli model. \eqref{eq:color-even-partitions} gives the even-partition expansion of the sign-averaged coefficients, which under (standard) loop-freeness through order $L$ agree with the fixed-instance coefficients through order $L$, and \cref{thm:color-cumulants-from-graphs,thm:color-cumulant-concentration} give their concentration bound.

Time-ordered and out-of-time-ordered correlators prescribe a limited family of label sequences, so we can use the scalar recurrence from \eqref{eq:color-R-moment-recurrence} to compute moments directly without incurring a multiplicative $2^{L}$ factor in runtime. We will see in \cref{part:operator-valued} the case where this direct method does not work and requires an optimized approach.

Throughout this section, $A,B \in \{\mathbb{I}, X, Y, Z\}^{\otimes n}$ are two distinct non-identity $n$-qubit Pauli operators (equivalently, $A$ and $B$ are independent modulo phase), fixed independently of the sampled Pauli terms. Recall our definition of time-ordered correlators at $\beta=0$; we compute their Taylor expansions to get
\[
    G_0^A(t) \equiv \nTr(A(t)A(0)) = \sum_{h_0,h_1=0}^{\infty}\frac{(-1)^{h_1}(it)^{h_0+h_1}}{h_0!\,h_1!} \nTr(H^{h_0}A H^{h_1}A).
\]
Moving the first copy of $A$ past the $H^{h_1}$ term swaps its color and causes it to cancel with the second copy of $A$:
\begin{equation}\label{eq:toc-scalar-recurrence}
    G_0^A(t) = \sum_{h_0,h_1=0}^{\infty}\frac{(-1)^{h_1}(it)^{h_0+h_1}}{h_0!\,h_1!}
       \nTr(H_0^{h_0}H_1^{h_1}) = \sum_{h_0,h_1=0}^{\infty}\frac{(-1)^{h_1}(it)^{h_0+h_1}}{h_0!\,h_1!}  \tau_{h_0+h_1}^{(\Cm_{h_0,h_1})}[H,\ldots, H],
\end{equation}
where we define $H_0\equiv H$, $H_1\equiv AHA$, and $\Cm_{h_0,h_1}\equiv 0^{h_0}1^{h_1}$.

For out-of-time-ordered correlators, we index the multiplicities of the four stretches of $H$ by the color they acquire. Setting $\Qm=(A,B)$ gives the four conjugates
\[
    H_{00}=H,\qquad H_{10}=AHA,\qquad H_{01}=BHB,\qquad H_{11}=ABHBA,
\]
and for $\va{h}\equiv(h_{00},h_{10},h_{11},h_{01})\in\N^4$ we write $|\va{h}|\equiv h_{00}+h_{10}+h_{11}+h_{01}$, $\va{h}!\equiv h_{00}!\,h_{10}!\,h_{11}!\,h_{01}!$, and
\[
    \Cm_{\va{h}}\equiv (00)^{h_{00}}(10)^{h_{10}}(11)^{h_{11}}(01)^{h_{01}}.
\]
Expanding the exponentials and moving the four insertions to the right by \eqref{eq:color-insertion-identity} leaves $ABAB=\langle A,B\rangle \mathbb{I}$, so
\begin{equation}\label{eq:otoc-scalar-recurrence}
    \begin{aligned}
        F_0^{A,B}(t) \equiv \nTr(A(t) B(0) A(t) B(0))
        &= \sum_{\va{h}\in\N^4}\frac{(-1)^{h_{10}+h_{01}}(it)^{|\va{h}|}}{\va{h}!}\,\nTr(H^{h_{00}}AH^{h_{10}}BH^{h_{11}}AH^{h_{01}}B) \\
        &= \sum_{\va{h}\in\N^4}\frac{(-1)^{h_{10}+h_{01}}(it)^{|\va{h}|}}{\va{h}!}\,\nTr\big(H_{00}^{h_{00}}H_{10}^{h_{10}}H_{11}^{h_{11}}H_{01}^{h_{01}}\,ABAB\big)\\
        &= \sum_{\va{h}\in\N^4}\frac{(-1)^{h_{10}+h_{01}}(it)^{|\va{h}|}}{\va{h}!}\,\langle A,B\rangle\, \tau_{|\va{h}|}^{(\Cm_{\va{h}})}[H,\ldots, H].
    \end{aligned}
\end{equation}
This identity is exact for every instance and every order, even without loop-freeness.

\subsection{Approximation error of truncating color moments and cumulants}
\label{subsec:color-truncation-error}

We now show the analogue of \cref{lemma:fock_space_approx} for correlators: the error incurred by computing color cumulants only through order $\ell$ and color moments only through order $L$. Throughout, for an even integer $\ell\geq 2$ we abbreviate $\Ts_\ell\equiv\Ts^{(\Qm)}_{\ell}$, the color hopping operator of \cref{def:color-hopping-operator} built from all color cumulants of order at most $\ell$; it sums over every label sequence, and only the algorithms restrict to the label sequences they need. We take $\Qm=(A)$ for time-ordered correlators and $\Qm=(A,B)$ for out-of-time-ordered correlators, and write $b\equiv b_1$ when $s=1$.

In analogy with $\Ks(k')$ in \cref{lemma:fock_space_approx}, we define
\begin{equation}\label{eq:color-K-sums}
    \Ks^{\mathrm{col}}(k')\equiv\sum_{\substack{2\leq k\leq k'\\k\text{ even}}}\ \sum_{\substack{\Cm\in\Cs^k,\ \va{c}_1=\va{0}\\(k,\Cm)\neq(2,(\va{0},\va{0}))}}\big|\kappa^{(\Cm)}_{k}\big|.
\end{equation}
This sum runs over the absolute values of all coefficients of $\Ts_{k'}-(a^\dagger+a)$. Compared with $\Ks(k')$ it contains the label sequences of nonzero color, including the mixed-color quadratic cumulant $\kappa_2^{(\va{0},\va{c})}=\tau(HH_{\va{c}})$ for $\va{c}\neq\va{0}$, which is not covered by \cref{thm:color-cumulant-concentration} and is bounded separately in \cref{lem:mixed-quadratic}.

\begin{lemma}[Fock space matrix elements of truncated color moments]\label{lem:truncated-color-moments}
    Let $H$ satisfy the hypotheses of \cref{def:color-hopping-operator} for cutoff $\ell$. For all integers $h_0,h_1\geq0$ and all $\va{h}\in\N^4$,
    \begin{equation}\label{eq:toc-fock-element}
        \bra{\Omega_1}\Ts_\ell^{h_0}\,b\,\Ts_\ell^{h_1}\,b\ket{\Omega_1}=\hat{\mu}^{(\ell;\Cm_{h_0,h_1})},
    \end{equation}
    \begin{equation}\label{eq:otoc-fock-element}
        \bra{\Omega_2}\Ts_\ell^{h_{00}}\,b_1\,\Ts_\ell^{h_{10}}\,b_2\,\Ts_\ell^{h_{11}}\,b_1\,\Ts_\ell^{h_{01}}\,b_2 \ket{\Omega_2} =\langle A,B\rangle\,\hat{\mu}^{(\ell;\Cm_{\va{h}})}.
    \end{equation}
    In particular, for $h_0+h_1\leq\ell$ and $|\va{h}|\leq\ell$ the left-hand sides equal the coefficients $\nTr(H^{h_0}AH^{h_1}A)$ and $\nTr(H^{h_{00}}AH^{h_{10}}BH^{h_{11}}AH^{h_{01}}B)$ of \cref{eq:toc-scalar-recurrence,eq:otoc-scalar-recurrence}.
\end{lemma}
\begin{proof}
    The walk expansion in the proof of \cref{prop:hopping-operator-otoc} is valid for every order: a returning walk is a non-crossing partition of the Hamiltonian sites, each block carries the color cumulant of its label subsequence, and the insertions return the color register to $\va{0}$ with the boundary phase, which is $1$ for $AA$ and $\langle A,B\rangle$ for $ABAB$ by \eqref{eq:otoc-scalar-recurrence}. A block contributes only when its size is at most $\ell$, giving the indicator in \eqref{eq:truncated-color-moments}. The final statement is \cref{prop:hopping-operator-otoc} together with \cref{eq:toc-scalar-recurrence,eq:otoc-scalar-recurrence}.
\end{proof}

\begin{lemma}[Approximation error of truncations for correlators]\label{lemma:color-fock-space-approx}
    Let $H$ be a bounded Hermitian operator with color cumulants $(\kappa^{(\Cm)}_k)_{k\in \N,\Cm\in\Cs^k}$ with respect to $\Qm$, where $\kappa_2=1$ and all odd color cumulants vanish through order $L$. Fix even integers $\ell,L$ where $2\leq\ell<L$ and time $t\geq 0$. Then for $\Qm=(A)$,
    \begin{equation}\label{eq:toc-approx}
        \begin{aligned}
        &\left|G_0^A(t)-\sum_{\substack{h_0,h_1\geq0\\h_0+h_1\leq L}}\frac{(-1)^{h_1}(it)^{h_0+h_1}}{h_0!\,h_1!}\,\hat{\mu}^{(\ell;\Cm_{h_0,h_1})}\right|\\
        &\qquad\leq e^{2t\Ks^{\mathrm{col}}(L)}\left[2t\bigl(\Ks^{\mathrm{col}}(L)-\Ks^{\mathrm{col}}(\ell)\bigr)+\frac{\big(4t+2t\Ks^{\mathrm{col}}(L)\big)^L}{L!}\left(2+\frac{4t+2t\Ks^{\mathrm{col}}(L)}{L+1}\right)\right],
        \end{aligned}
    \end{equation}
    and for $\Qm=(A,B)$,
    \begin{equation}\label{eq:otoc-approx}
        \begin{aligned}
        &\left|F_0^{A,B}(t)-\langle A,B\rangle\sum_{\substack{\va{h}\in\N^4\\|\va{h}|\leq L}}\frac{(-1)^{h_{10}+h_{01}}(it)^{|\va{h}|}}{\va{h}!}\,\hat{\mu}^{(\ell;\Cm_{\va{h}})}\right|\\
        &\qquad\leq e^{4t\Ks^{\mathrm{col}}(L)}\left[4t\bigl(\Ks^{\mathrm{col}}(L)-\Ks^{\mathrm{col}}(\ell)\bigr)+\frac{\big(8t+4t\Ks^{\mathrm{col}}(L)\big)^L}{L!}\left(2+\frac{8t+4t\Ks^{\mathrm{col}}(L)}{L+1}\right)\right].
        \end{aligned}
    \end{equation}
\end{lemma}
\begin{proof}
    We apply \cref{lem:common-truncation}. Since $a$, $a^\dagger$ and $b_1,\ldots,b_s$ have norm one, every monomial $a^{k-1}$ and $a_{\va{c}_k}\cdots a_{\va{c}_2}$ in \cref{def:color-hopping-operator} has norm at most one, so the triangle inequality gives $\|\Ts_{\ell'}-(a^\dagger+a)\|\leq\Ks^{\mathrm{col}}(\ell')$ and, for all even $0\leq \ell' \leq L$,
    \begin{equation}\label{eqn:color-norm-upper}
        \|\Ts_{\ell'}\|\leq 2+\Ks^{\mathrm{col}}(\ell'),\qquad \|\Ts_L-\Ts_{\ell}\|\leq \Ks^{\mathrm{col}}(L)-\Ks^{\mathrm{col}}(\ell),
    \end{equation}
    and \cref{lem:perturbed-unitary} applied to the self-adjoint part $a^\dagger+a$ gives, for all real $u$,
    \begin{equation}\label{eqn:color-evolution-bound}
        \left\|e^{i\Ts_{\ell'}u}\right\|\leq e^{|u|\Ks^{\mathrm{col}}(\ell')}.
    \end{equation}
    Since $L$ is even, \eqref{eqn:moment-trace-upper} and $\Ks(L)\leq\Ks^{\mathrm{col}}(L)$ give
    \begin{equation}
        \nTr(|H|^L)=\nTr(H^L)\leq(2+\Ks(L))^L\leq(2+\Ks^{\mathrm{col}}(L))^L.
    \end{equation}
    Thus the operator and even-moment bounds in the common lemma hold with $S=a^\dagger+a$, $R=2$, and the color sums $\Ks^{\mathrm{col}}(\ell)$, $\Ks^{\mathrm{col}}(L)$ and $\Ks^{\mathrm{col}}(L) - \Ks^{\mathrm{col}}(\ell)$.

    For the TOC, take $X=H$, $(t_1,t_2)=(t,-t)$, $(U_1,U_2)=(A,A)$, $(V_1,V_2)=(b,b)$, and $\ket{\Omega}=\ket{\Omega_1}$. By \cref{lem:truncated-color-moments}, the moment-matching hypothesis holds and the Taylor polynomial through degree $L$ is the sum in \eqref{eq:toc-approx}. Applying \cref{lem:common-truncation} with $T=2t$ proves \eqref{eq:toc-approx}.

    For the OTOC, take $(t_1,t_2,t_3,t_4)=(t,-t,t,-t)$ and $(U_1,U_2,U_3,U_4)=(A,B,A,B)$. Set $(V_1,V_2,V_3,V_4)=(b_1,b_2,b_1,b_2)$ and $\ket{\Omega}=\ket{\Omega_2}$. The same moment-matching lemma identifies the Taylor polynomial with the sum in \eqref{eq:otoc-approx}, including the boundary phase $\langle A,B\rangle$. Applying \cref{lem:common-truncation} with $T=4t$ proves \eqref{eq:otoc-approx}.
\end{proof}

To apply \cref{lemma:color-fock-space-approx} we bound $\Ks^{\mathrm{col}}(L)$ and $\Ks^{\mathrm{col}}(L)-\Ks^{\mathrm{col}}(\ell)$ with high probability. \cref{thm:color-cumulant-concentration} covers all free cumulant orders $k\geq 4$, and we bound the mixed-color quadratic cumulant with the following lemma.

\begin{lemma}[Mixed quadratic color cumulants]\label{lem:mixed-quadratic}
Let the generators $Q_1,\ldots,Q_s$ be independent modulo phase and fixed
independently of the sampled Pauli terms, so that $[Q_{\va{c}}]\neq[\mathbb{I}]$ for
$\va{c}\neq\va{0}$. For every $\delta\in(0,1)$, with probability at least $1-\delta$ over the Pauli terms,
\begin{equation}\label{eq:mixed-quadratic-bound}
    \sum_{\va{c}\neq\va{0}}\Big|\frac1m\sum_{j=1}^m\langle Q_{\va{c}},P_j\rangle\Big|\leq2^{s}\sqrt{\frac{2\log(2^{s+1}/\delta)}{m}}.
\end{equation}
Moreover, whenever $H$ is loop-free through order $2$, we have
\[
    \kappa_2^{(\va{0},\va{c})}=\tau(HH_{\va{c}})=\frac1m\sum_{j=1}^m\langle Q_{\va{c}},P_j\rangle
\]
for every $\va{c}$, implying the left-hand side of \eqref{eq:mixed-quadratic-bound} equals $\sum_{\va{c}\neq\va{0}}\big|\kappa_2^{(\va{0},\va{c})}\big|$.
\end{lemma}
\begin{proof}
    By \eqref{eq:multicolor-conjugation},
    \[
        \tau(HH_{\va{c}})=\frac{1}{m}\sum_{i,k}J_iJ_k\langle Q_{\va{c}},P_k\rangle\tau(P_iP_k).
    \]
    Loop-freeness prevents $P_i\propto\mathbb{I}$ and $P_i\propto P_k$ for $i\neq k$, hence $\tau(P_iP_k)=\delta_{ik}$, $\tau(H)=0$, and 
    \[
        \kappa_2^{(\va{0},\va{c})}=\tau(HH_{\va{c}})-\tau(H)\tau(H_{\va{c}})=\frac{1}{m}\sum_j\langle Q_{\va{c}},P_j\rangle.
    \]
    For a fixed Pauli operator $Q_{\va{c}}\neq\mathbb{I}$, exactly half of all Pauli strings anticommute with it, so the signs $\langle Q_{\va{c}},P_j\rangle$, $j\in[m]$, are independent and uniform on $\{\pm1\}$. Hoeffding's inequality gives
    \[
        |\sum_j\langle Q_{\va{c}},P_j\rangle|\leq\sqrt{2m\log(2^{s+1}/\delta)}
    \]
    except with probability $\delta/2^s$, and a union bound over the $2^s-1$ labels $\va{c}\neq\va{0}$ gives \eqref{eq:mixed-quadratic-bound}.
\end{proof}

\begin{corollary}[Bounds on the color truncation sums]\label{cor:color-hypercontractive-sums}
    Fix any $s\geq 0$ generators independent modulo phase and independent of $H$. There exist constants $A,C,\gamma>0$ such that, for sufficiently large $n$, $n\leq m\leq \poly(n)$, failure probability $\delta=1/\poly(n)$, and even $2\leq\ell<L\leq\gamma n/\log n$ with $m\geq AL^3$, we have simultaneously
    \begin{equation}\label{eq:color-hypercontractive-sums}
        \Ks^{\mathrm{col}}(L)\leq C\sqrt{\frac{\log n}{m}},\qquad
        \Ks^{\mathrm{col}}(L) - \Ks^{\mathrm{col}}(\ell)\leq \delta^{-1/2}\left(\frac{C(\ell+2)}m\right)^{(\ell+2)/4}
    \end{equation}
    with probability at least $1-\delta-e^{-\Omega(n)}$.
\end{corollary}
\begin{proof}
    Apply \cref{corollary:color-cumulant-hypercontractivity} with failure probability $\delta/n^2$, and \cref{lem:mixed-quadratic} with failure probability $\delta/n^2$. The higher-order sum is at most $C(\log n)^{s+1/2}/m$, which is absorbed by the mixed quadratic bound $C\sqrt{\log n/m}$ because $s$ is fixed, $m\geq n$, and $\delta=1/\poly(n)$. We apply \cref{thm:color-cumulant-concentration} at lower order $\ell+2$ with $\eta=\delta/2$ for the omitted sum. A union bound proves both inequalities.
\end{proof}

\subsection{The full correlator algorithms}
\label{subsec:correlator-algorithms}

The algorithms compute only the color moments and cumulants of the label sequences $\Cm_{h_0,h_1}$, respectively $\Cm_{\va{h}}$, that appear in \cref{eq:toc-scalar-recurrence,eq:otoc-scalar-recurrence}; the hopping operator of the analysis contains all other label sequences, but they do not enter the output.

\begin{algorithm}[ht]
\caption{Time-ordered correlation algorithm}\label{alg:time-ordered-correlator}
\begin{algorithmic}[1]
\Require $H\sim$ sparse Pauli Hamiltonian, non-identity Pauli operator $A$, time $t\geq 0$, even $2\leq \ell < L \leq \gamma n/\log n$.
\State Compute the (exact) color moments $\hat{\mu}^{(\ell; \Cm_{h_0,h_1})}$ for all $1\leq h_0+h_1\leq\ell$ using direct Pauli expansion.
\State Compute the color cumulants $\kappa_{h_0+h_1}^{(\Cm_{h_0,h_1})}$ for all $1\leq h_0+h_1\leq\ell$ using the moment-cumulant recurrence.
\State Compute the (approximate) color moments $\hat{\mu}^{(\ell; \Cm_{h_0,h_1})}$ for all $\ell<h_0+h_1\leq L$ using \eqref{eq:color-R-moment-recurrence} and \cref{prop:correlator-cutoff-runtime}.
\State \Return
\begin{equation}
    \hat{G}_0^A(t)\equiv \sum_{\substack{h_0,h_1\geq 0\\h_0+h_1\leq L}}\frac{(-1)^{h_1}(it)^{h_0+h_1}}{h_0!\,h_1!}\,\hat{\mu}^{(\ell; \Cm_{h_0,h_1})}.
\end{equation}
\end{algorithmic}
\end{algorithm}

\begin{algorithm}[ht]
\caption{Out-of-time-order correlation algorithm}\label{alg:otoc}
\begin{algorithmic}[1]
\Require $H\sim$ sparse Pauli Hamiltonian, distinct non-identity Pauli operators $A,B$, time $t\geq 0$, even $2\leq\ell<L\leq\gamma n/\log n$.
\State Compute the (exact) color moments $\hat{\mu}^{(\ell;\Cm_{\va{h}})}$ for all $\va{h}\in\N^4$ with $1\leq |\va{h}|\leq\ell$ using direct Pauli expansion.
\State Compute the color cumulants $\kappa_{|\va{h}|}^{(\Cm_{\va{h}})}$ for all $1\leq |\va{h}|\leq\ell$ using the moment-cumulant recurrence.
\State Compute the (approximate) color moments $\hat{\mu}^{(\ell;\Cm_{\va{h}})}$ for all $\ell<|\va{h}|\leq L$ using \eqref{eq:color-R-moment-recurrence} and \cref{prop:correlator-cutoff-runtime}.
\State \Return
\begin{equation}
    \hat{F}_0^{A,B}(t)\equiv\langle A,B\rangle\sum_{\substack{\va{h}\in\N^4\\|\va{h}|\leq L}}\frac{(-1)^{h_{10}+h_{01}}(it)^{|\va{h}|}}{\va{h}!}\,\hat{\mu}^{(\ell;\Cm_{\va{h}})}.
\end{equation}
\end{algorithmic}
\end{algorithm}

Every subsequence and contiguous subword of these label sequences belongs to the same family, allowing zero exponents. Hence every moment and cumulant required by the recurrence is among the stored quantities.

\begin{proposition}[Running time at fixed cutoff]\label{prop:correlator-cutoff-runtime}
    Let $r=2$ for the TOC and $r=4$ for the OTOC. For even integers $2\leq\ell<L$ satisfying $m\geq CL^3$ and $L\leq\gamma n/\log n$, the corresponding algorithms \cref{alg:time-ordered-correlator,alg:otoc} can be implemented in time at most
    \begin{equation}\label{eq:correlator-cutoff-runtime}
        \Os\left(n\ell^{r+1}\left(\frac{2em}{\ell}\right)^{\ell/2}\right).
    \end{equation}
\end{proposition}
\begin{proof}
    There are $\Os(\ell^r)$ low-order color words. Expanding the two halves of each word in the Pauli basis and contracting their coefficients costs $\Os(n\ell)\binom m{\ell/2}$ per word, giving a total cost of $\Os\bigl(n\ell^{r+1}\binom m{\ell/2}\bigr)$. The commutation signs of the fixed probes cost $\Os(nm)$ and the OTOC boundary phase costs $\Os(n)$, which are absorbed.

    For $\ell=2$, only the second-order color cumulants are kept. Write $c_{ij}\equiv\kappa_2^{(\va{c}_i,\va{c}_j)}$ and let $F_{a:b}(z_a,\ldots,z_b)$ be the ordinary generating series of the truncated moments of $\va{c}_a^{h_a}\cdots\va{c}_b^{h_b}$ in the exponents $h_a,\ldots,h_b$. The one-stretch series are Catalan series, and pairing the first position gives, for $a<b$,
    \begin{equation}
        F_{a:b}=
    \frac{F_{a+1:b}+\sum_{j=a+1}^{b-1}c_{aj}z_az_jF_{a:j}F_{j:b}}
        {1-z_a^2F_{a:a}-c_{ab}z_az_bF_{b:b}}.
    \end{equation}
    Every series on the right involves a shorter interval, and the denominator has constant term one. Fast multivariate multiplication and power-series inversion therefore compute all coefficients through total degree $L$ in $\Os(L^r\log L)$ operations \cite[Chapters~8--9]{vonzurGathen2013}.

    For $\ell\geq4$, for each of the $\Os(L^r)$ stored words, the recurrence selects a block containing the first position of size at most $\ell$. There are at most $\sum_{j=1}^\ell\binom{L-1}{j-1}\leq\ell L^{\ell-1}$ choices, and each summand multiplies at most $\ell$ previously computed interval moments. Run-length indices identify the selected color subsequence and intervening intervals in $\Os(\ell)$ time. This costs $\Os(\ell^2L^{\ell+r-1})$ in total, including the low-order cumulant inversion and final Taylor sum.

    To absorb the recurrence, use $\binom m{\ell/2}\geq(m/\ell)^{\ell/2}$, $m\geq L^3$, and $4\leq\ell\leq L$. Then
    \begin{equation}\label{eq:correlator-recurrence-absorption}
        \frac{n\ell^{r+1}(m/\ell)^{\ell/2}}{\ell^2L^{\ell+r-1}}
        \geq\frac{n\ell^{r-1}}{L^{r-1}}\left(\frac L\ell\right)^{\ell/2}
        \geq n\ell^{r-3}L^{3-r}\geq1,
    \end{equation}
    where the last inequality uses $r\in\{2,4\}$ and $n\geq L$. The binomial upper bound $\binom{m}{\ell/2}\leq(2em/\ell)^{\ell/2}$ proves \eqref{eq:correlator-cutoff-runtime}.
\end{proof}

\begin{theorem}[Running time of the correlator approximations]\label{thm:correlator-algorithm}
    Fix the Pauli probes independently of the sampled Hamiltonian, with $A\neq\mathbb I$ for the TOC and $A,B$ independent modulo phase for the OTOC. There is a sufficiently small constant $c>0$ such that, for all sufficiently large $n$, $2n\leq m\leq \poly(n)$, $t\geq0$, failure probability $\delta=1/\poly(n)$ and approximation $0< \varepsilon < 1$ satisfying
    \begin{equation}\label{eq:correlator-runtime-window}
        1+t+\log(1/\varepsilon)\leq c\min\left\{m^{1/3},\frac n{\log n}\right\},
    \end{equation}
    \cref{alg:time-ordered-correlator} and \cref{alg:otoc} respectively output an approximation satisfying
    \begin{equation}
        |\hat G_0^A(t)-G_0^A(t)|\leq\varepsilon
        \qquad\text{and},\qquad |\hat F_0^{A,B}(t)-F_0^{A,B}(t)|\leq\varepsilon
    \end{equation}
    with probability at least $1-\delta-e^{-\Omega(n)}$ over $H$ and total running time at most
    \begin{equation}\label{eq:correlator-accuracy-runtime}
        \Os\left(\frac{n}{\delta}(1+t)^2(1/\varepsilon)^{2+\Os(1/\log m)}\right).
    \end{equation}
\end{theorem}
\begin{proof}
    Choose a sufficiently large fixed constant $C_0$. If the semicircular case holds, we use \cref{rem:semicircular-reference}. Otherwise, let $\ell\geq2$ be the smallest even integer satisfying
    \begin{equation}\label{eq:optimized-color-cutoff}
        (1+t)\delta^{-1/2}\left(\frac{C_0(\ell+2)}m\right)^{(\ell+2)/4}\leq\frac{\varepsilon}{C_0}.
    \end{equation}
    Since $t=\Os(m^{1/3})$, $m\geq2n$, and $\delta=1/\poly(n)$, we have $(1+t)\delta^{-1/2}\leq m^{\Os(1)}$. The same argument as in \eqref{eq:optimized-cutoff-size} gives $\ell=\Os(1+\log(1/\varepsilon)/\log m)$. Choose an even $L=\Theta(1+t+\log(1/\varepsilon))$ with a sufficiently large implied constant and $L\geq\ell+4$. Taking $c$ sufficiently small ensures $m\geq CL^3$ and $L\leq\gamma n/\log n$.

    \paragraph{Approximation.}
    Apply \cref{cor:color-hypercontractive-sums}. Since $t\leq L$ and $L\leq(m/A)^{1/3}$, its total-sum bound gives $t\Ks^{\mathrm{col}}(L)\leq\Os(m^{-1/6}\sqrt{\log n})=o(1)$, using $m\geq 2n$. Thus the exponential prefactor in \cref{lemma:color-fock-space-approx} is bounded and the Taylor base is $\Os(1+t)$. The general omitted-tail bound in \eqref{eq:color-hypercontractive-sums} and \eqref{eq:optimized-color-cutoff} make the cumulant contribution at most $\varepsilon/2$. The Taylor contribution is also at most $\varepsilon/2$ by $L!\geq(L/e)^L$ and the choice of $L$. This proves the claimed approximation with probability at least $1-\delta-e^{-\Omega(n)}$. At $t=0$, the exact values $G_0^A(0)=1$ and $F_0^{A,B}(0)=\langle A,B\rangle$ can instead be returned directly.

    \paragraph{Running time.}
    By minimality, the preceding cutoff $\ell-2$ fails \eqref{eq:optimized-color-cutoff}, so
    \begin{equation}
        \left(\frac{m}{C_0\ell}\right)^{\ell/2}<C_0^2(1+t)^2\delta^{-1}\varepsilon^{-2}.
    \end{equation}
    Applying \cref{prop:correlator-cutoff-runtime} with $r=4$ therefore gives runtime upper bounded by
    \[
        \Os\left(\frac{n}{\delta}(1+t)^2\varepsilon^{-2}e^{\Os(\ell)}\right).
    \]
    Using $\ell=\Os(1+\log(1/\varepsilon)/\log m)$ and $L\leq\gamma n/\log n$, this gives
    \[
        \Os\left(\frac{n}{\delta}(1+t)^2(1/\varepsilon)^{2+\Os(1/\log m)}\right),
    \]
    proving \eqref{eq:correlator-accuracy-runtime}.
\end{proof}

The analysis for general thermal correlators ($\beta>0$) is given in \cref{sec:general-beta}.

\newpage

\part{The spiked Pauli model}\label{part:operator-valued}
We begin with the \emph{spiked Pauli model}, which we recall is given as
\[
    H_{\text{spiked}} = H+\lambda Q,\quad \lambda\geq 0,
\]
where $H$ is the sparse Pauli Hamiltonian from \cref{eq:1} and $Q$ is a fixed, non-identity $n$-qubit Pauli operator called the ``spike.'' Our goal will be to compute the moments $\nTr\bigl((H+\lambda Q)^k\bigr)$ using the color framework of \cref{part:color-framework} and the two-color cumulants of $H_0=H$ and $H_1=QHQ$. This presents us with an issue: while correlation functions prescribed a limited set of label sequences (e.g., $0^{h_0}1^{h_1}$ for TOCs), if we expand out $\nTr\bigl((H+\lambda Q)^k\bigr)$, we end up with nearly all possible label sequences. Taking the same approach as we did for our correlation function algorithms will not work, as computing the moments of $H+\lambda Q$ to order $L$ via the color moments of $H$ requires a summation over roughly $2^L$ label sequences. 

Instead, our solution is to take a more direct approach to computing the moments and free cumulants of $H+\lambda Q$. To do this, we utilize the mathematical formalism of \emph{operator-valued free probability}, which we introduce in \cref{sec:operator-valued}. Our classical algorithm in \cref{sec:spiked-pauli} computes the spectral amplitude of the spiked Pauli model by computing its \emph{spiked cumulants}; these are operator-valued generalizations of the free cumulants utilized in \cref{part:sfa-algorithm}, specialized to the spiked Pauli model. Notably, we then are able to re-purpose the same techniques to compute thermal expectation values of the standard sparse Pauli model in \cref{sec:thermal-expectations}.

\section{Operator-valued free probability}\label{sec:operator-valued}
Operator-valued free probability is a generalization of free probability which replaces the scalar-valued expectation functional with an operator-valued \emph{conditional expectation}. The theory was (also) introduced by Voiculescu~\cite{voiculescu1995operations}, and the combinatorial perspective was (also) given by Speicher~\cite{speicherCombinatorialTheoryFree1998}. We give a brief introduction and highlight the tools that we will use. All results and definitions we introduce in this subsection are from the lecture notes of Speicher~\cite{speicher2020noncommutative} and the monograph by Mingo and Speicher~\cite{mingo2017free}.

\begin{definition}[Operator-valued non-commutative probability space]
    An \emph{operator-valued noncommutative probability space} $(\As, \Bs, E)$ consists of an algebra $\As$, a subalgebra $\Bs \subseteq \As$ containing $\mathds{1}_\As$, and a conditional expectation, i.e., a unital linear map $E:\As\rightarrow\Bs$ satisfying the \emph{bimodule property}:
    \begin{equation}
        E(b_1 A b_2) = b_1 E(A) b_2 \qquad \forall b_1, b_2 \in \Bs, A \in \As.
    \end{equation}
    The bimodule property also implies $E(b) = b$ for all $b\in\Bs$.

    Elements of $\As$ are called operator-valued, or $\Bs$-valued, non-commutative random variables. The \emph{operator-valued moments} of random variable $X\in \As$ are given by
    \begin{equation}\label{eqn:operator-moments}
        E(X),\quad E(Xb_1X), \quad E(Xb_1Xb_2X), \quad E(Xb_1Xb_2Xb_3X),\ldots
        \qquad \forall b_1,b_2,b_3,\ldots \in \Bs,
    \end{equation}
    and the collection of all operator-valued moments forms the \emph{operator-valued distribution} of $X$.
\end{definition}

The notion of free independence also extends to the operator-valued setting:
\begin{definition}[Operator-valued free independence]
    Let $(\As, \Bs, E)$ be an operator-valued non-commutative probability space. Subalgebras $(\As_i)_{i\in [s]}$ where $\Bs\subset \As_i$ are called \emph{freely independent with respect to $\Bs$} (or free with amalgamation over $\Bs$) if
    \begin{equation}
        E(a_1)=E(a_2)=\dots=E(a_k)=0\implies E(a_1a_2\dots a_k)=0
    \end{equation}
    for all $a_j\in \As_{i_j}$ when neighboring elements are from different subalgebras ($i_j \neq i_{j+1}$). $\Bs$-valued non-commutative random variables $a_1,\ldots,a_s\in\As$ are also called \emph{freely independent with respect to $\Bs$} if subalgebras $\big(\Bs\langle a_i\rangle\big)_{i\in [s]}$ are freely independent with respect to $\Bs$.
\end{definition}

\subsection{Operator-valued free cumulants} 
Free cumulants also have an operator-valued analogue. First, we introduce a definition which characterizes the operator-valued free cumulant maps:
\begin{definition}[$\Bs$-balanced maps]
    Given an algebra $\As$ and a subalgebra $\Bs \subset \As$, a $\Bs$-balanced map $T: \As^k \rightarrow \Bs$ is a $\C$-multilinear map, which satisfies for all $a_1, a_2,\ldots, a_k\in\As$, $b, b' \in \Bs$, and $j\in [k-1]$:
    \begin{enumerate}
        \item $T[ba_1, a_2, \ldots, a_kb'] = bT[a_1, a_2, \ldots, a_k]b'$
        \item $T[a_1, \ldots, a_{j}b, a_{j+1},\ldots, a_k] = T[a_1, \ldots, a_{j}, ba_{j+1}, \ldots, a_k]$
    \end{enumerate}
\end{definition}
Given a sequence of $\Bs$-balanced maps $(T_k)_{k\in \N}$, we can extend it to a multiplicative family $(T_\pi)_{k\in \N,\,\pi\in NC(k)}$. Set $T_{1_k}\equiv T_k$. For $\pi\neq 1_k$, choose an interval block $V=\{q+1,q+2,\ldots,q+r\}$ of $\pi$ that does not contain $1$, and let $\sigma\in NC(k-r)$ be the partition obtained by deleting $V$ and relabeling the remaining positions in order. Define $T_\pi$ as the unique extension of $T_k$ satisfying the \emph{insertion property}:
\begin{equation}
    T_\pi[a_1, \ldots, a_k] \equiv T_\sigma\bigl[a_1, \ldots, a_q \cdot T_r[a_{q+1}, \ldots, a_{q+r}], a_{q+r+1}, \ldots, a_k\bigr],
\end{equation}
where $q\geq 1$. The resulting family $(T_\pi)_{k\in \N,\,\pi\in NC(k)}$ is multiplicative and $\Bs$-balanced. Note that unlike scalar multiplicative functions (i.e., \cref{def:multiplicative-function}), operator-valued multiplicative functions are not simply products over their individual block values. The nesting rule is needed because an inner block supplies an operator-valued input to its surrounding block (this property extends to operator-valued free cumulants as well).

For any multiplicative family of $\Bs$-balanced maps, we can define the new multiplicative family
\begin{equation}
    S_\pi[a_1, \ldots, a_k] \equiv \sum_{\substack{\sigma \in NC(k)\\\sigma \leq \pi}} T_\sigma [a_1, \ldots, a_k].
\end{equation}
This relation can be inverted via M\"{o}bius inversion,
\begin{equation}
    T_\pi[a_1, \ldots, a_k] = \sum_{\substack{\sigma \in NC(k)\\\sigma \leq \pi}} \mu(\sigma, \pi)\, S_\sigma[a_1, \ldots, a_k].
\end{equation}

Now we can define the operator-valued free cumulants using $\Bs$-balanced maps:
\begin{definition}[Operator-valued free cumulants]
    Let $(\As, \Bs, E)$ be an operator-valued non-commutative probability space. Let $E_k^\Bs: \As^k \rightarrow \Bs$ denote the $\Bs$-balanced map 
    \begin{equation}
        (a_1, \ldots, a_k) \mapsto E_k^\Bs[a_1, \ldots, a_k] \equiv E(a_1 a_2\cdots a_k),
    \end{equation}
    and extend the sequence $(E^\Bs_k)_{k\in \N}$ to the $\Bs$-balanced multiplicative family $(E^\Bs_\pi)_{k\in \N, \pi\in NC(k)}$. The \emph{operator-valued free cumulants} $(\kappa_k^\Bs)_{k\in \N}$ are defined as
    \begin{equation}
        \kappa_k^\Bs[a_1,\ldots,a_k] \equiv \sum_{\pi \in NC(k)} \mu(\pi, 1_k)\, E^\Bs_\pi[a_1,\ldots,a_k],
    \end{equation}
    where $\mu$ is the $NC(k)$ M\"{o}bius function. 
\end{definition}

The general partition operator-valued moment-cumulant relations follow from the general $\Bs$-balanced construction:
\begin{equation}
    \begin{aligned}
        \kappa^\Bs_\pi[a_1, \ldots, a_k]
        &= \sum_{\substack{\sigma \in NC(k)\\\sigma \leq \pi}} \mu(\sigma, \pi)\, E^\Bs_\sigma[a_1, \ldots, a_k],\\
        E^\Bs_\pi[a_1, \ldots, a_k] &= \sum_{\substack{\sigma \in NC(k)\\\sigma \leq \pi}} \kappa^\Bs_\sigma[a_1, \ldots, a_k].
    \end{aligned}
\end{equation}
When considering the operator-valued free cumulants of a single non-commutative variable $a\in \As$, it suffices to consider
\[
    \kappa^\Bs_\pi[a, b_2 a, b_3 a, \ldots, b_{k}a],
\]
since the first $\Bs$ term can be factored out due to the bimodule property.

\subsection{The operator-valued \texorpdfstring{$R$}{R}-transform}
The operator-valued $R$-transform is the key tool we use to obtain our efficient algorithm for the spiked Pauli model.
\begin{definition}[Operator-valued $R$-transform]
    Given an element $a\in \As$, its operator-valued $R$-transform\footnote{We use the same convention as the scalar $R$-transform in \cite{nica2006lectures} for consistency, which differs from operator-valued free probability literature by a factor of $b$.} is the formal power series,
    \begin{equation}
        R_a^\Bs(b) \equiv \sum_{k=1}^\infty \kappa_{k}^{\Bs}[a,ba,\ldots,ba] \,b
    \end{equation}
\end{definition}

Let $M_{a}^\Bs$ be the $\Bs$-valued moment generating series defined by
\begin{equation}
    M_{a}^{\Bs}(z) \equiv \sum_{k=0}^\infty E(a^k)z^k
\end{equation}
for any $a\in \As$. The two series are related (e.g., using \cite[Proposition~9.6]{speicher2020noncommutative}) by
\begin{equation}\label{eqn:operator-R-moment-relation}
    M_{a}^{\Bs}(z)= \mathds{1}_\Bs + R_{a}^\Bs\bigl(zM_{a}^{\Bs}(z)\bigr).
\end{equation}
This expression holds either as a formal power series or as an analytic function near zero when the series converges. It is an operator-valued analogue of the relation in \cref{eqn:scalar-R-moment-relation}.

Once again, a key property of the operator-valued $R$-transform is that it is additive over freely independent variables~\cite[Chapter~9, Theorem~11]{mingo2017free}: for variables $a_1,a_2\in \As$ freely independent with respect to $\Bs$, the $R$-transform of $a_1+a_2$ is
\begin{equation}
    R^{\Bs}_{a_1+a_2}(b) = R^{\Bs}_{a_1}(b) + R^{\Bs}_{a_2}(b).
\end{equation}

\section{Spectral amplitude of the spiked Pauli model}\label{sec:spiked-pauli}
In this section we give our classical algorithm for computing the spectral amplitude of $H+\lambda Q$. We begin by introducing the color moments and cumulants of the spiked model in \cref{subsec:spiked-model-moments} using the same method as in \cref{part:color-framework}. Then in \cref{subsec:operator-valued-color}, we will give an extension to the color framework which incorporates operator-valued free probability and allows for more efficient algorithms. We end with the full construction of our algorithm and its analysis in \cref{subsec:spiked-algorithm-analysis}.

\subsection{Color moments and cumulants of the spiked Pauli model}\label{subsec:spiked-model-moments}
For every $k\geq1$, cyclicity of the trace gives
\begin{equation}\label{eq:spiked-moment-expansion}
    \nTr\bigl((H+\lambda Q)^k\bigr)=\nTr(H^k)+\sum_{r=1}^k\frac{k\lambda^r}{r}
      \sum_{\substack{h_1,\ldots,h_r\geq0\\h_1+\cdots+h_r=k-r}}
      \nTr\left(H^{h_1}Q\cdots H^{h_r}Q\right).
\end{equation}
For even $r$, moving the spikes to the right and cancelling them in pairs gives
\[
    \nTr\left(H^{h_1}QH^{h_2}Q\cdots H^{h_r}Q\right)
    =\nTr\left(H_0^{h_1}H_1^{h_2}\cdots H_0^{h_{r-1}}H_1^{h_r}\right),
\]
which gives the color moment
\begin{equation}
    \tau_{k_0}^{(\Cm)}[H,\ldots, H],\qquad k_0\equiv h_1 + \cdots + h_r,\quad \Cm \equiv 0^{h_1}1^{h_2}\cdots 0^{h_{r-1}}1^{h_r}. 
\end{equation}
We can again use the scalar recurrence relation \eqref{eq:color-R-moment-recurrence} to compute every even-spike contribution given the necessary color cumulants. For odd $r$, the cancellation leaves a boundary term of $Q$ rather than $\mathbb{I}$
\[
    \nTr\left(H^{h_1}QH^{h_2}Q\cdots H^{h_r}Q\right) = \nTr\left(H_0^{h_1}H_1^{h_2}\cdots H_1^{h_{r-1}}H_0^{h_r}Q\right).
\]
Here is a scenario where the boundary term in \eqref{eq:color-insertion-identity} does not necessarily cancel to identity. These terms vanish when averaging over the Rademacher signs in $H$: every surviving sampled Pauli word from $H$ has even multiplicity, leaving a scalar-valued word followed by the traceless operator $Q$. To show this same vanishing holds with high probability over sampled instances of $H$, we need to extend our definition of loop-freeness to include the spike operator. More generally, fix a set of color generators 
$\Qm = (Q_1,\ldots, Q_s)$:
\begin{definition}[Relative loop-freeness]\label{def:relative-loop-free}
    Given a sequence of $m$ Pauli operators $\Pm=(P_1,\ldots, P_{m})$, we say that $\Pm$ is \emph{loop-free relative to $\Qm$ through order $L$} (or $\Qm$-loop-free for short) if
    \begin{equation}
        [P_{i_1}P_{i_2}\cdots P_{i_k}] \notin \Cs,
    \end{equation}
    for any $1\leq k \leq L$, any sequence of distinct indices $i_1, i_2,\ldots, i_{k}\in [m]$. We say that a Pauli Hamiltonian instance $H$ is $\Qm$-loop-free through order $L$ if its support on Pauli strings is $\Qm$-loop-free through order $L$.
\end{definition}
We call the original definition from \cref{def:loop-free} ``standard loop-freeness'' when we wish to distinguish it from this new definition. The case where $s=0$ recovers the definition of standard loop-freeness, and the case where $s=1$ gives the setting we need for the spiked Pauli model. 

Additionally, the proof of \cref{lem:loop-freeness} extends directly to this case and implies $H$ is $\Qm$-loop-free through order $L\leq \gamma n/\log n$ with high probability for some constant $\gamma>0$.\footnote{See \cref{rem:relative-loop-free}; the only distinction is now we have a total of $m + s$ Pauli operators instead of $m$. Since we only consider the case where $s$ is a small, fixed constant, we ignore this and simply use \cref{lem:loop-freeness}.} In fact, for every nonempty set of distinct sampled indices, the product modulo phase is uniform over the $4^n$ Pauli operators. Hence
\begin{equation}\label{eq:loop-free-relative}
    \begin{aligned}
        &\Pr\bigl(H\text{ is not loop-free relative to }(Q)\text{ through order }L\bigr)\\
        &\qquad\leq 2\cdot4^{-n}\sum_{j=1}^L\binom{m}{j}
        \leq 2\cdot4^{-n}\left(\frac{em}{L}\right)^L
        =e^{-\Omega(n)},
    \end{aligned}
\end{equation}
where the second inequality uses $L\leq m/2$ and the final bound holds for a sufficiently small constant $\gamma$. We now see how relative loop-freeness simplifies color moment expressions similar to standard loop-freeness. 
\begin{proposition}[Loop-free color moments]\label{prop:loop-free-color-moments}
    If $H$ is $\Qm$-loop-free through order $L$, then, for $k\leq L$, Pauli operator $Q$ satisfying $[Q]\in\Cs$, and $\Cm=(\va{c}_1, \ldots, \va{c}_k)$,
    \begin{equation}
        \tau(H_{\va{c}_1}\cdots H_{\va{c}_k}Q) =\tau_k^{(\Cm)}[H,\ldots, H]\,\tau(Q).
    \end{equation}
\end{proposition}
\begin{proof}
    Expand the Hamiltonians into Pauli terms. Conjugation changes only scalar signs, so each monomial reduces, modulo phase, to the product of its distinct Pauli terms of odd multiplicity. (Standard) loop-freeness forces this set to be empty whenever the scalar trace is nonzero.

    From $\Qm$-loop-freeness, any non-empty set of at most $k$ Pauli terms cannot multiply to be $[Q]$. Therefore if $[Q]\neq [\mathbb{I}]$, then none of the Pauli strings equal identity so the trace is zero. If $[Q] = [\mathbb{I}]$, then $\tau(Q)$ is a non-zero scalar and we may extract it from the trace. Either case gives the desired identity.
\end{proof}

Returning to the spiked Pauli model, \cref{prop:loop-free-color-moments} implies that through order $L=\Omega(n/\log n)$, only words with an even number of spike insertions contribute. This gives the exact same ``evenness'' property that we saw in \cref{part:sfa-algorithm}: if $H$ is $\Qm$-loop-free to even order $k$, then
\begin{equation}
    \nTr\bigl((H+\lambda Q)^{k}\bigr)=\nTr(H^k)+\sum_{q=1}^{k/2}\frac{k\lambda^{2q}}{2q}
      \sum_{\substack{h_1,\ldots,h_{2q}\geq 0\\h_1+\cdots+h_{2q}=k-2q}}
      \nTr\left(H^{h_1}Q\cdots H^{h_{2q}}Q\right).
\end{equation}

Even now we can use the same algorithm as in \cref{part:color-framework} to compute the spectral amplitude of the spiked Pauli model: compute color cumulants for all sequences to order $\ell=\Os(1+\log(1/\varepsilon)/\log n)$, compute the color moments for all sequences to order $L=\Os(1+(2+\lambda)t+\log(1/\varepsilon))$ using \eqref{eq:color-R-moment-recurrence}, then compute the Taylor series approximation of $\nTr(e^{-i(H+\lambda Q)t})$. Unfortunately, enumerating all color sequences to order $L$ gives a $\Omega(2^L)$ overhead, which is exponential when $t=\poly(n)$. As an alternative, we instead design a specialized recurrence relation using operator-valued probability that will allow us to efficiently compute the moments of the spiked Pauli model directly.

\subsection{An operator-valued color framework and recurrence relation}\label{subsec:operator-valued-color}
Fix a valid set of color generators $\Qm = (Q_1,\ldots, Q_s)$. Starting from the non-commutative probability space 
\[
    (\As, \tau) = \bigl(M_{2^n}(\C), \nTr\bigr)
\]
considered throughout \cref{part:sfa-algorithm,part:color-framework}, we extend this to an operator-valued non-commutative probability space by considering the subalgebra
\[
    \Bs = \operatorname{span}_{\C}\{Q_{\va{c}}:\va{c}\in\F_2^s\}
\]
and the conditional expectation
\begin{equation}\label{eq:color-conditional-expectation}
    E: \As\rightarrow\Bs,\qquad E(X) \equiv \sum_{\va{c}\in\F_2^s} \tau\bigl(Q_{\va{c}}^\dagger \,X\bigr) Q_{\va{c}}.
\end{equation}
Since the color generators are independent modulo phase by definition, the collection $(Q_{\va{c}})_{\va{c}\in\F_2^s}$ forms an orthonormal basis of $\Bs$. Thus the conditional expectation $E$ can be seen as a trace-preserving ($\tau\bigl(E(X)\bigr)=\tau(X)$), orthogonal projection onto $\Bs$. Following the constructions from \cref{sec:operator-valued}, we can define operator-valued color moment and cumulant maps $(E^\Bs_{\pi})_{k\in\N, \pi\in NC(k)}$ and $(\kappa^\Bs_{\pi})_{k\in\N, \pi\in NC(k)}$. 

We can prove the following relation between operator-valued and color cumulants:
\begin{proposition}[Operator-color cumulant equivalence]\label{prop:color-operator-cumulant-equivalence}
    Suppose $H$ is loop-free relative to $\Qm=(Q_1,\ldots, Q_s)$ through order $L$. For all $k\leq L$ and $\Cm=(\va{c}_1,\ldots,\va{c}_k)\in\Cs^k$,
    \begin{equation}
        \kappa_k^{\Bs}(H_{\va{c}_1},\ldots,H_{\va{c}_k})
        =\kappa_k^{(\Cm)}[H,\ldots, H]\, \mathbb{I}.
    \end{equation}
\end{proposition}
\begin{proof}
    By \cref{prop:loop-free-color-moments} and the definition of the conditional expectation in \eqref{eq:color-conditional-expectation},
    \[
        E_k^\Bs[H_{c_1},\ldots, H_{c_k}] =\tau(H_{c_1}\cdots H_{c_k})\, \mathbb{I}.
    \]
    The same identity holds for the multiplicative extensions of the moment maps as well. Indeed, we can evaluate an inner interval block, pull out its scalar value, and recurse on the remaining positions. Hence $E_\pi^{\Bs}(H_{c_1},\ldots,H_{c_k})=\tau_\pi(H_{c_1},\ldots,H_{c_k})\,\mathbb{I}$. M\"obius inversion gives
    \[
        \kappa_k^{\Bs}(H_{c_1},\ldots,H_{c_k})=\left(\sum_{\pi\in NC(k)} \mu_{NC}(\pi,1_k)\tau_\pi(H_{c_1},\ldots,H_{c_k})\right) \mathbb{I},
    \]
    proving the desired equation.
\end{proof}
This allows us to give an expression for arbitrary operator-valued free cumulants for the Pauli Hamiltonian up to its loop-free order.
\begin{lemma}\label{lemma:general-color-operator-cumulant}
    Suppose $H$ is loop-free relative to $\Qm=(Q_1,\ldots, Q_s)$ through order $L$, and $(\va{c}_1, \ldots, \va{c}_k)\in\Cs^k$ is a label sequence. Then for all $k\leq L$,
    \begin{equation}
        \kappa^\Bs_{k}\left[Q_{\va{c}_1}H, Q_{\va{c}_2}H, \ldots, Q_{\va{c}_{k-1}}H, Q_{\va{c}_{k}}H\right] = \kappa_k^{(\Bm)}[H,\ldots, H]\prod_{i=1}^k Q_{\va{c}_i},
    \end{equation}
    where
    \[
        \Bm\equiv (\va{b}_1, \va{b}_2,\ldots, \va{b}_{k})\qquad \text{and}\qquad \va{b}_{j}\equiv \bigoplus_{i=1}^j \va{c}_{i}.
    \]
\end{lemma}
\begin{proof}
Since $Q_{\va{c}}^\dagger Q_{\va{c}} = \mathbb{I}$, we can insert adjacent pairs $Q_{\va{c}}^\dagger Q_{\va{c}}$ (denoted in bold) into the operator-valued cumulant map, as follows:
\[
    \kappa^\Bs_{k}\left[Q_{\va{c}_1}H\mathbf{Q}_{\va{c}_1}^\dagger\mathbf{Q}_{\va{c}_1}, Q_{\va{c}_2}H \mathbf{Q}_{\va{c}_2}^\dagger\mathbf{Q}_{\va{c}_2},\ldots, Q_{\va{c}_{k-1}} H \mathbf{Q}_{\va{c}_{k-1}}^\dagger\mathbf{Q}_{\va{c}_{k-1}}, Q_{\va{c}_k}H \mathbf{Q}_{\va{c}_k}^\dagger\mathbf{Q}_{\va{c}_k}\right],
\]
which does not change the value of the operator-valued free cumulant. Now using $H_{\va{c}} \equiv Q_{\va{c}}HQ_{\va{c}}^\dagger$, we get the expression
\[
    \kappa^\Bs_{k}\left[H_{\va{c}_1}Q_{\va{c}_1},  H_{\va{c}_2}Q_{\va{c}_2}, \ldots, H_{\va{c}_{k-1}} Q_{\va{c}_{k-1}}, H_{\va{c}_{k}}Q_{\va{c}_k}\right].
\]
Shifting all the colors to the right side using \eqref{eq:color-insertion-identity} and the bimodule property to move between arguments,
\begin{equation}
    \kappa^\Bs_{k}\left[H_{\va{c}_1},  H_{\va{c}_1\oplus\va{c}_2},\ldots, H_{\va{c}_{1}\oplus\cdots\oplus\va{c}_{k}}\right]\prod_{i=1}^k Q_{\va{c}_i}.
\end{equation}
Applying \cref{prop:color-operator-cumulant-equivalence} completes the proof.
\end{proof}

Now we finally show why we introduced all this background on operator-valued free probability: we utilize the key fact that the $R$-transform of the sum of freely independent variables is the sum of their individual $R$-transforms. As a preliminary example, consider the variable $H+\lambda\mathds{1}_{\As}$, where $\lambda\geq 0$. Since the set of scalars $\C\mathds{1}_{\As}$ is free from all other non-commutative random variables \cite[Chapter~1, Proposition~16]{mingo2017free}, the $R$-transform of $H+\lambda$ is simply
\[
    R_{H+\lambda}(z) = R_H(z) + \lambda R_{\mathds{1}_{\As}}(z) = R_H(z) + \lambda z.
\]
Thus computing the free cumulants of $H+\lambda\mathds{1}_{\As}$ reduces to computing the free cumulants of $H$ itself. 

Now we can consider our full spiked Pauli model $H+\lambda Q$. Unfortunately, $H$ and $\lambda Q$ are not freely independent for a fixed non-identity Pauli operator $Q$. However, when we instead consider an operator-valued free probability space, the subalgebra $\Bs$ is free with amalgamation over $\Bs$ from every other non-commutative random variable~\cite[Remark~5.4]{speicher2020noncommutative}; in a sense, $\Bs$ acts like a set of ``non-commutative scalars''! Using the additivity of the operator-valued $R$-transform, we get
\begin{equation}
    R^\Bs_{H+\lambda Q}(b) = R^\Bs_H(b) + \lambda R^\Bs_{Q}(b) = R_H^\Bs(b) + \lambda Qb.
\end{equation}
This implies that all operator-valued free cumulants of $H+\lambda Q$ are identical to those of $H$, with the single exception being $\kappa_{1}^\Bs(H+\lambda Q) = \kappa_{1}^\Bs(H) + \lambda Q$.

More generally, consider the operator-valued moments of the spiked Pauli model $E\bigl((H+\lambda Q_{\va{c}_*})^k\bigr) = \sum_{\va{c}\in\F_2^s} \tau\bigl(Q_{\va{c}}^\dagger \,(H+\lambda Q_{\va{c}_*})^k\bigr) Q_{\va{c}}$ for some $\va{c}_*\in\F_2^s$. Since $E$ is the orthogonal projection onto $\Bs$, the coefficient of $(H+\lambda Q_{\va{c}_*})^k$ along each color basis vector $Q_{\va{c}}$ is the inner product $\tau\bigl(Q_{\va{c}}^\dagger\,(H+\lambda Q_{\va{c}_*})^k\bigr)$. The following lemma provides a recurrence relation which allows us to directly compute these inner products given the color cumulants of $H$:

\begin{lemma}[Operator-valued recurrence relation]\label{lemma:operator-valued-recurrence}
    Fix a sequence of \emph{commuting} color generators $\Qm=(Q_1,\ldots, Q_s)$ independent modulo phase and a spike operator $\lambda Q_{\va{c}_*}$, where $\va{c}_*\in\F_2^s$ and $\lambda\geq 0$. Denote $\mu(\va{c};k)\equiv \tau\bigl(Q_{\va{c}}^\dagger \,(H+\lambda Q_{\va{c}_*})^k\bigr)\in\C$ and formal power series $\alpha_{\va{c}}(z)\equiv\sum_{k\geq 0}\mu(\va{c};k)z^k \in \C[[z]]$. 
    
    If $H$ is $\Qm$-loop-free through order $L$, then for all labels $\va{c}\in\F_2^s$,
    \begin{equation}\label{eqn:color-series-recurrence}
        \alpha_{\va{c}}(z) \equiv \delta_{\va{c},\va{0}} + \lambda z\,\alpha_{\va{c}_*\oplus\va{c}}(z) + \sum_{k=2}^L z^{k} \sum_{\va{c}_2,\ldots, \va{c}_{k}\in\F_2^s} \kappa_{k}^{(\va{b}_1,\ldots,\va{b}_k)}[H,\ldots,H] \prod_{i=2}^{k+1}\alpha_{\va{c}_{i}}(z) \pmod{z^{L+1}},
    \end{equation}
    where $\va{b}_1\equiv \va{0}$ and $\va{b}_j\equiv\bigoplus_{i=2}^{j}\va{c}_i$ are the colors generated by the color changes $\va{c}_2,\ldots,\va{c}_k$, and the last change $\va{c}_{k+1}\equiv\va{b}_k\oplus\va{c}$ brings the color to $\va{b}_{k+1}=\va{c}$.
\end{lemma}
\begin{proof}
    From the argument above, the operator-valued $R$-transform of $H + \lambda Q_{\va{c}_*}$ is given as
    \begin{equation}
        R^\Bs_{H+\lambda Q_{\va{c}_*}}(b) = R_H^{\Bs}(b) + \lambda Q_{\va{c}_*}b.
    \end{equation}
    Using the relation between the moment generating series and the $R$-transform \eqref{eqn:operator-R-moment-relation},
    \begin{align*}
        M_{H+\lambda Q_{\va{c}_*}}^\Bs(z)&=\mathbb{I}+ R^\Bs_{H+\lambda Q_{\va{c}_*}}\bigl(zM_{H+\lambda Q_{\va{c}_*}}^\Bs(z)\bigr) \\
        &= \mathbb{I} + \lambda Q_{\va{c}_*} zM_{H+\lambda Q_{\va{c}_*}}^\Bs(z) + R^\Bs_{H}\bigl(zM_{H+\lambda Q_{\va{c}_*}}^\Bs(z)\bigr)\\
        &=  \mathbb{I} + \lambda Q_{\va{c}_*} zM_{H+\lambda Q_{\va{c}_*}}^\Bs(z)\\&\qquad\qquad+\sum_{k=1}^\infty \kappa_{k}^{\Bs}\bigl[H, zM_{H+\lambda Q_{\va{c}_*}}^\Bs(z)H, \ldots, zM_{H+\lambda Q_{\va{c}_*}}^\Bs(z)H\bigr]  zM_{H+\lambda Q_{\va{c}_*}}^\Bs(z).
    \end{align*}
    By the multilinearity of the free cumulant maps
    \begin{equation}\label{eqn:expanded-moment-R-recurrence}
        \begin{aligned}
            M_{H+\lambda Q_{\va{c}_*}}^\Bs(z) &= \mathbb{I}+\lambda Q_{\va{c}_*} zM_{H+\lambda Q_{\va{c}_*}}^\Bs(z)\\
            &\qquad\qquad+ \sum_{k=1}^\infty z^{k}\kappa_{k}^{\Bs}\bigl[H,M_{H+\lambda Q_{\va{c}_*}}^\Bs(z)H,\ldots, M_{H+\lambda Q_{\va{c}_*}}^\Bs(z)H\bigr] M_{H+\lambda Q_{\va{c}_*}}^\Bs(z).
        \end{aligned}
    \end{equation}
    The operator-valued moment generating function is a formal power series with $\Bs$-valued coefficients, so we decompose it explicitly according to its color group basis:
    \begin{equation}\label{eqn:coefficient-color-decomposition}
        M_{H+\lambda Q_{\va{c}_*}}^\Bs(z) = \sum_{\va{c}\in\F_2^s} \alpha_{\va{c}}(z)\, Q_{\va{c}},
    \end{equation}
    by definition of the moment generating series and the fact that distinct color representatives are orthogonal; in particular, $\alpha_{\va{c}}(0)=\mu(\va{c};0)=\delta_{\va{c},\va{0}}$. Plugging \eqref{eqn:coefficient-color-decomposition} into \eqref{eqn:expanded-moment-R-recurrence} and expanding via multilinearity, with $\va{c}_i$ denoting the color label of the $(i-1)$-th copy of $M_{H+\lambda Q_{\va{c}_*}}^\Bs(z)$ in the $k$-th term, produces the cumulants $\kappa_{k}^{\Bs}[H,Q_{\va{c}_2}H,\ldots,Q_{\va{c}_k}H]$. The $k=1$ term vanishes, using the fact that the first order moment and free cumulant vanish under loop-freeness, and the terms with $k>L$ are of order $z^{L+1}$. Applying \cref{lemma:general-color-operator-cumulant} to the remaining terms, namely $\kappa_{k}^{\Bs}[H,Q_{\va{c}_2}H,\ldots,Q_{\va{c}_k}H]=\kappa_{k}^{(\Bm)}[H,\ldots,H]\,Q_{\va{c}_2}\cdots Q_{\va{c}_k}$ for $2\leq k\leq L$, and using that $\prod_{i=2}^{k+1} Q_{\va{c}_i} = Q_{\va{b}_{k+1}}$ since all the color generators commute and square to $\mathbb{I}$, we get
    \begin{equation}\label{eqn:c-summation}
        \begin{aligned}
            \sum_{\va{c}\in\F_2^s}Q_{\va{c}}\,\alpha_{\va{c}}(z) &\equiv \mathbb{I} + \lambda z\, Q_{\va{c}_*} \sum_{\va{c}\in\F_2^s}Q_{\va{c}}\,\alpha_{\va{c}}(z) \\&\qquad+ \sum_{k=2}^{L} z^{k} \sum_{\substack{\va{c}_2,\ldots, \va{c}_k\in\F_2^s\\\va{c}_{k+1}\in\F_2^s}} \kappa_{k}^{(\Bm)}[H,\ldots,H]\, Q_{\va{b}_{k+1}} \prod_{i=2}^{k+1} \alpha_{\va{c}_{i}}(z) \pmod{z^{L+1}},
        \end{aligned}
    \end{equation}
    where
    \[
        \Bm\equiv (\va{0}, \va{b}_2,\ldots, \va{b}_{k})\qquad \text{and}\qquad \va{b}_{j}\equiv \bigoplus_{i=2}^j \va{c}_{i}.
    \]
    Since the color representatives $Q_{\va{c}}$ form a basis of $\Bs$, we can equate the coefficients of each color representative in \eqref{eqn:c-summation}, after rewriting the second term on the right as $\lambda z\sum_{\va{c}\in\F_2^s}Q_{\va{c}}\,\alpha_{\va{c}_*\oplus\va{c}}(z)$ using $Q_{\va{c}_*}Q_{\va{c}}=Q_{\va{c}_*\oplus\va{c}}$ and the substitution $\va{c}\mapsto\va{c}_*\oplus\va{c}$. In the last term, $Q_{\va{b}_{k+1}}=Q_{\va{c}}$ exactly when $\va{c}_{k+1}=\va{b}_{k}\oplus\va{c}$, so only this value of $\va{c}_{k+1}$ survives, resulting in recurrence
    \begin{equation}
        \alpha_{\va{c}}(z) \equiv \delta_{\va{c},\va{0}} + \lambda z\,\alpha_{\va{c}_*\oplus\va{c}}(z) + \sum_{k=2}^L z^{k} \sum_{\va{c}_2,\ldots, \va{c}_{k}\in\F_2^s} \kappa_{k}^{(\Bm)}[H,\ldots,H] \prod_{i=2}^{k+1}\alpha_{\va{c}_{i}}(z) \pmod{z^{L+1}},
    \end{equation}
    where $\va{c}_{k+1}=\va{b}_k\oplus\va{c}$, which is \eqref{eqn:color-series-recurrence} since $\Bm=(\va{b}_1,\ldots,\va{b}_k)$. This completes the proof of \cref{lemma:operator-valued-recurrence}.
\end{proof}

Apart from the cumulant, the color changes enter \eqref{eqn:color-series-recurrence} only through the product $\prod_{i=2}^{k+1}\alpha_{\va{c}_i}(z)$, which depends only on how many of the changes $\va{c}_2,\ldots,\va{c}_{k+1}$ equal each label (for $s=1$, on the Hamming weight of $(\va{c}_2,\ldots,\va{c}_{k+1})$). Since $\va{c}_2\oplus\cdots\oplus\va{c}_{k+1}=\va{c}$, these counts also determine which equation a term enters (for $s=1$, through the parity of the weight). The inner sum can therefore be grouped by these counts, with the cumulants $\kappa_{k}^{(\va{b}_1,\ldots,\va{b}_k)}[H,\ldots,H]$, which depend only on $\va{c}_2,\ldots,\va{c}_k$, summed within each group.

In the following subsection we consider the specialization to the two-color setting for the spiked Pauli model and show how \cref{lemma:operator-valued-recurrence} can be used to produce an efficient algorithm for the spectral amplitude.

\subsection{Analysis of the spiked Pauli model algorithm}\label{subsec:spiked-algorithm-analysis}
In the two-color spiked Pauli model setting, $\Qm=(Q)$, where $Q$ is a fixed non-identity Hermitian Pauli operator and $\lambda\geq 0$. Suppose that $H$ is loop-free relative to $(Q)$ through order $L$. The recurrence relation from \cref{lemma:operator-valued-recurrence} gives
\begin{equation}\label{eqn:operator-recurrence-original}
    \begin{aligned}
        \alpha_0(z)&\equiv 1+\lambda z\,\alpha_1(z)+\sum_{k=2}^L z^k\sum_{c_2,\ldots,c_k\in\{0,1\}}\kappa_k^{(b_1,\ldots,b_k)}[H,\ldots,H]\prod_{i=2}^{k+1}\alpha_{c_i}(z),&&c_{k+1}\equiv b_k,\\
        \alpha_1(z)&\equiv \lambda z\,\alpha_0(z)+\sum_{k=2}^L z^k\sum_{c_2,\ldots,c_k\in\{0,1\}}\kappa_k^{(b_1,\ldots,b_k)}[H,\ldots,H]\prod_{i=2}^{k+1}\alpha_{c_i}(z),&&c_{k+1}\equiv b_k\oplus1,
    \end{aligned}
\end{equation}
modulo $z^{L+1}$, where $b_1\equiv0$ and $b_j\equiv\bigoplus_{i=2}^j c_i$. Trace orthogonality gives the readout
\begin{equation}\label{eq:spiked-moment-extraction}
    [z^k]\alpha_0(z)=\tau\bigl((H+\lambda Q)^k\bigr),\qquad [z^k]\alpha_1(z)=\tau\bigl(Q(H+\lambda Q)^k\bigr),
\end{equation}
for all $0\leq k\leq L$.

For $\Cm=(0,c_2,\ldots,c_k)$, the product $\prod_{i=2}^{k}\alpha_{c_i}(z)$ depends only on the Hamming weight $|\Cm|$. Its parity also determines $b_k$ and hence the final factor $\alpha_{c_{k+1}}(z)$. Thus, for $2\leq k\leq L$ and $0\leq w\leq k-1$, define
\begin{equation}
    G_k^{(w)}(z)\equiv\alpha_0(z)^{k-1-w}\alpha_1(z)^w\sum_{\substack{\Cm\in\{0,1\}^k\\c_1=0,\ |\Cm|=w}}\kappa_k^{(b_1,\ldots,b_k)}[H,\ldots,H],
\end{equation}
and
\begin{equation}
    G_k^{(\mathrm{odd})}(z)\equiv\sum_{\substack{0\leq w\leq k-1\\w\ \mathrm{odd}}}G_k^{(w)}(z),\qquad
    G_k^{(\mathrm{even})}(z)\equiv\sum_{\substack{0\leq w\leq k-1\\w\ \mathrm{even}}}G_k^{(w)}(z).
\end{equation}
Then
\begin{equation}\label{eqn:operator-recurrence-grouped}
    \begin{aligned}
        \alpha_0(z)&\equiv1+\lambda z\,\alpha_1(z)+\sum_{k=2}^L z^k\left(G_k^{(\mathrm{even})}(z)\alpha_0(z)+G_k^{(\mathrm{odd})}(z)\alpha_1(z)\right),\\
        \alpha_1(z)&\equiv\lambda z\,\alpha_0(z)+\sum_{k=2}^L z^k\left(G_k^{(\mathrm{odd})}(z)\alpha_0(z)+G_k^{(\mathrm{even})}(z)\alpha_1(z)\right)
    \end{aligned}
\end{equation}
modulo $z^{L+1}$.

For an even cutoff $2\leq \ell\leq L$, let $\alpha_c^{(\ell)}(z)$, $c\in\{0,1\}$, denote the formal power series obtained from \eqref{eqn:operator-recurrence-original} by replacing the upper summation limit $L$ by $\ell$ and imposing equality in $\C[[z]]$. Thus all color cumulants above order $\ell$ are set to zero, while all other color cumulants are unchanged. 
Every non-constant term on the right has a positive power of $z$, so the constant terms are $\alpha_0^{(\ell)}(0)=1$, $\alpha_1^{(\ell)}(0)=0$, and coefficient $k$ depends only on coefficients of degree less than $k$. This defines the two series uniquely at every order. They can also be computed using \eqref{eqn:operator-recurrence-grouped}, with the sums truncated at $\ell$ and each $G_k^{(w)}$ evaluated using $\alpha_0^{(\ell)},\alpha_1^{(\ell)}$. In particular, computing their coefficients through degree $L$ does not require enumerating color sequences of length $L$.

\begin{algorithm}[ht]
\caption{Spectral amplitude algorithm for the spiked Pauli model.}\label{alg:spiked}
\begin{algorithmic}[1]
\Require $H\sim$ sparse Pauli Hamiltonian, fixed non-identity Hermitian Pauli operator $Q$, spike strength $\lambda\geq 0$, time $t\geq0$, even $2\leq\ell<L$.
\State Compute the exact color moments $\tau_k^{(\Cm)}[H,\ldots,H]$ for $1\leq k\leq\ell$ and all $\Cm\in\{0,1\}^k$ with $c_1=0$, using direct Pauli expansion.
\State Compute the color cumulants $\kappa_k^{(\Cm)}[H,\ldots,H]$ for even $2\leq k\leq\ell$ using the color moment-cumulant recurrence. Set odd color cumulants and color cumulants above order $\ell$ to zero.
\State Compute $[z^k]\alpha_c^{(\ell)}(z)$ for $c\in\{0,1\}$ and $0\leq k\leq L$, using \eqref{eqn:operator-recurrence-grouped} with the cumulant sums truncated at $\ell$.
\State \Return
\begin{equation}\label{eq:spiked-algorithm-output}
    \hat{Z}^{(\lambda)}(t)\equiv\sum_{k=0}^L\frac{(-it)^k}{k!}[z^k]\alpha_0^{(\ell)}(z).
\end{equation}
\end{algorithmic}
\end{algorithm}

\paragraph{Fock space representation.}
We use the two-color Fock space, vacuum $\Omega_1$, and spike operator $b$ from \eqref{eq:two-color-fock-operators}. For every even $2\leq \ell\leq L$, define
\begin{equation}\label{eq:spiked-hopping-operator}
    \Ts_\ell^{(\lambda)}\equiv\Ts_\ell^{(\Qm)}+\lambda b,\qquad \Qm=(Q),
\end{equation}
where $\Ts_\ell^{(\Qm)}$ is the color hopping operator of \cref{def:color-hopping-operator}. Relative loop-freeness implies standard loop-freeness, so $\kappa_2=1$ and all odd color cumulants through order $L$ vanish, as required by that definition.

\begin{lemma}[Fock space matrix elements of truncated spiked moments]\label{lem:truncated-spiked-moments}
    Suppose that $H$ is loop-free relative to $(Q)$ through order $L$, and let $2\leq \ell\leq L$ be even. Then, for every integer $k\geq0$ and $c\in\{0,1\}$,
    \begin{equation}\label{eq:spiked-fock-moments}
        [z^k]\alpha_c^{(\ell)}(z)=\bra{\Omega_1}b^c\bigl(\Ts_\ell^{(\lambda)}\bigr)^k\ket{\Omega_1}.
    \end{equation}
    For $0\leq k\leq \ell$, these matrix elements also satisfy
    \begin{equation}\label{eq:spiked-fock-moment-matching}
        \bra{\Omega_1}b^c\bigl(\Ts_\ell^{(\lambda)}\bigr)^k\ket{\Omega_1}
        =\tau\bigl(Q^c(H+\lambda Q)^k\bigr).
    \end{equation}
\end{lemma}
\begin{proof}
    We use the same walk expansion as in \cref{prop:hopping-operator-otoc,lem:truncated-color-moments}, now allowing a spike step $\lambda b$ at any site. Expand $\bigl(\Ts_\ell^{(\lambda)}\bigr)^k$ into words in $\Ts_\ell^{(\Qm)}$ and $\lambda b$. Each spike changes the current color and has weight $\lambda$. For a fixed spike word, the returning walks at the Hamiltonian sites give non-crossing partitions, each block carrying the color cumulant of its label subsequence. A block of size greater than $\ell$ has no corresponding down-step and contributes zero. The final factor $b^c$ selects walks whose current color ends at $c$.

    To identify the generating series, decompose a walk by the block containing its leftmost site, as in the proof of \cref{lemma:operator-valued-recurrence}. A spike at that site contributes $\lambda z$ and interchanges the two endpoint colors. Otherwise, the site belongs to a Hamiltonian block of even size $k'\leq \ell$. If the $k'$ intervening intervals have color changes $c_2,\ldots,c_{k'+1}$, then the colors of the block, relative to its first site, are $b_1=0$ and $b_j=\bigoplus_{i=2}^j c_i$. The endpoint condition is $c_{k'+1}=b_{k'}\oplus c$, and the block contributes its color cumulant times the product of the $k'$ interval series. Thus the two generating series of the matrix elements satisfy exactly \eqref{eqn:operator-recurrence-original} with cutoff $\ell$, now as formal identities at every order. The coefficient-by-coefficient uniqueness above proves \eqref{eq:spiked-fock-moments}.

    By \cref{lemma:operator-valued-recurrence}, the spiked Pauli Hamiltonian moment series satisfy the same recurrence modulo $z^{\ell+1}$, since $H$ is loop-free relative to $(Q)$ through order $\ell$. Their coefficients through degree $\ell$ therefore agree with $\alpha_c^{(\ell)}$, and \eqref{eq:spiked-moment-extraction} proves \eqref{eq:spiked-fock-moment-matching}.
\end{proof}

\paragraph{Approximation error of truncations.}
We use the quantities $\Ks^{\mathrm{col}}(\ell')$ from \eqref{eq:color-K-sums}, specialized to $\Qm=(Q)$.

\begin{lemma}[Approximation error of truncations for the spiked model]\label{lemma:spiked-fock-space-approx}
    Let $H$ be a sparse Pauli Hamiltonian instance which is loop-free relative to $(Q)$ through order $L$, where $Q$ is a non-identity Hermitian Pauli operator. Fix even integers $\ell,L$ with $2\leq\ell<L$. Then, for every $\lambda\geq 0$ and $t\geq0$,
    \begin{equation}\label{eq:spiked-approx}
        \begin{aligned}
            &\left|\nTr\bigl(e^{-it(H+\lambda Q)}\bigr)-\sum_{k=0}^L\frac{(-it)^k}{k!}\bra{\Omega_1}\bigl(\Ts_{\ell}^{(\lambda)}\bigr)^k\ket{\Omega_1}\right|\\
            &\quad\leq e^{t\Ks^{\mathrm{col}}(L)}
            \left[t\bigl(\Ks^{\mathrm{col}}(L)-\Ks^{\mathrm{col}}( \ell)\bigr)
            +\frac{\bigl(t(2+\lambda+\Ks^{\mathrm{col}}(L))\bigr)^L}{L!}
            \left(2+\frac{t(2+\lambda+\Ks^{\mathrm{col}}(L))}{L+1}\right)\right].
        \end{aligned}
    \end{equation}
    By \cref{lem:truncated-spiked-moments}, the finite sum on the left is precisely the output $\hat{Z}^{(\lambda)}(t)$ of \cref{alg:spiked}.
\end{lemma}
\begin{proof}
    First, \eqref{eqn:color-norm-upper} and $\|b\|=1$ give, for every even $2\leq \ell'\leq L$,
    \begin{equation}\label{eq:spiked-norm-upper}
        \|\Ts_{\ell'}^{(\lambda)}\|\leq2+\lambda+\Ks^{\mathrm{col}}(\ell'),\qquad
        \|\Ts_L^{(\lambda)}-\Ts_{\ell}^{(\lambda)}\|\leq\Ks^{\mathrm{col}}(L)-\Ks^{\mathrm{col}}( \ell).
    \end{equation}
    The spike is identical at both cutoffs, so it cancels in their difference. Moreover,
    \begin{equation}
        \Ts_{\ell'}^{(\lambda)}
        =\bigl(a^\dagger+a+\lambda b\bigr)
        +\bigl(\Ts_{\ell'}^{(\Qm)}-(a^\dagger+a)\bigr).
    \end{equation}
    The first summand is self-adjoint because $\lambda\geq 0$ and $b=b^\dagger$, and the second has norm at most $\Ks^{\mathrm{col}}(\ell')$. Thus \cref{lem:perturbed-unitary} gives, for every real $u$,
    \begin{equation}\label{eq:spiked-evolution-bound}
        \left\|e^{iu\Ts_{\ell'}^{(\lambda)}}\right\|\leq e^{|u|\Ks^{\mathrm{col}}(\ell')}.
    \end{equation}
    The spike is included in the self-adjoint summand, so it does not contribute to this exponential prefactor.

    Since $H+\lambda Q$ is Hermitian and $L$ is even, \cref{eq:spiked-fock-moment-matching,eq:spiked-norm-upper} imply
    \begin{equation}\label{eq:spiked-moment-upper}
        0\leq\nTr\bigl(|H+\lambda Q|^L\bigr)
        =\nTr\bigl((H+\lambda Q)^L\bigr)
        =\bra{\Omega_1}\bigl(\Ts_L^{(\lambda)}\bigr)^L\ket{\Omega_1}
        \leq\bigl(2+\lambda+\Ks^{\mathrm{col}}(L)\bigr)^L.
    \end{equation}
    Apply \cref{lem:common-truncation} with $X=H+\lambda Q$, $r=1$, $t_1=-t$, $T=t$, $U_1=V_1=\mathbb{I}$, and vacuum $\ket{\Omega_1}$. The preceding bounds give its operator and even-moment hypotheses with $S=a^\dagger+a+\lambda b$, $R=2+\lambda$, and the sums $\Ks^{\mathrm{col}}(\ell')$ and $\Ks^{\mathrm{col}}(L)-\Ks^{\mathrm{col}}( \ell)$. By \cref{lem:truncated-spiked-moments}, the required moment matching holds and the auxiliary Taylor polynomial is the stated output. Hence \cref{lem:common-truncation} gives \eqref{eq:spiked-approx}.
\end{proof}

The concentration estimates needed to apply \cref{lemma:spiked-fock-space-approx} are the same as those in \cref{cor:color-hypercontractive-sums}, with one color generator.

\subsection{Running time of the spiked model algorithm}\label{subsec:spiked-runtime}
We analyze \cref{alg:spiked} using the same sparse Pauli multiplication procedure as in \cref{sec:classical-sfa-algorithm}.

\begin{proposition}[Running time at fixed cutoffs]\label{prop:spiked-cutoff-runtime}
    Fix even integers $2\leq\ell<L$, and assume $m\geq CL^3$ for a sufficiently large constant $C$. Then \cref{alg:spiked} runs in time
    \begin{equation}\label{eq:spiked-cubic-runtime}
        \Os\left(n\ell2^\ell\binom m{\ell/2}\right).
    \end{equation}
\end{proposition}
\begin{proof}
    We bound the cost of each step of \cref{alg:spiked}.

    \paragraph{Computing the low-order color moments.}
    There are $\sum_{k=1}^{\ell}2^{k-1}=2^{\ell}-1$ binary color sequences of length at most $\ell$ whose first color is $0$. For each sequence, split the corresponding product of colored Hamiltonians into two halves, expand each half in the Pauli basis, and contract matching Pauli coefficients to obtain its normalized trace. Conjugating $H$ by $Q$ changes only the signs of its Pauli coefficients, so the support and multiplication bounds from \cref{sec:classical-sfa-algorithm} apply to both halves. Each half has length at most $\ell/2$, giving cost $\Os(n\ell)\binom{m}{\ell/2}$ per color moment and total cost
    \begin{equation}
        \Os\left(n\ell2^{\ell}\binom{m}{\ell/2}\right).
    \end{equation}
    The commutation signs between $Q$ and the sampled terms can be computed once in $\Os(nm)$ time, which is absorbed in this bound.

    \paragraph{Computing and grouping the color cumulants.}
    We recover the cumulants in increasing order by isolating the full-block term in \eqref{eq:color-R-moment-recurrence}. For a fixed word of length $k$, the recurrence sums over at most $2^{k-1}$ subsets containing its first position. Each proper subset uses a previously computed cumulant and the moments of the intervening intervals. A single scan of the word extracts these subsequences and intervals in $\Os(k)$ operations; color translation normalizes each required word to begin with $0$. Thus computing all cumulants through order $\ell$ costs
    \begin{equation}
        \sum_{k=1}^{\ell}\Os(k)\,2^{k-1}2^{k-1}
        =\Os\left(\ell4^{\ell}\right).
    \end{equation}
    The scalar coefficient sums in $G_k^{(w)}$ can then be precomputed in $\Os(\ell2^{\ell})$ time. Specifically, a cumulant indexed by $(b_1,\ldots,b_k)$ with $b_1=0$ is assigned to the group
    \begin{equation}\label{eq:spiked-group-index}
        w=\sum_{i=2}^{k}(b_{i-1}\oplus b_i).
    \end{equation}
    This is the Hamming weight of its change sequence $(c_2,\ldots,c_k)$ in \eqref{eqn:operator-recurrence-original}. There are only $\Os(\ell^2)$ resulting groups.

    \paragraph{Computing the higher-order moments.}
    To evaluate \eqref{eqn:operator-recurrence-grouped}, store the coefficient arrays of all monomials
    \begin{equation}
        \bigl(\alpha_0^{(\ell)}(z)\bigr)^a
        \bigl(\alpha_1^{(\ell)}(z)\bigr)^b,
        \qquad a,b\geq0,\quad a+b\leq\ell.
    \end{equation}
    There are $\Os(\ell^2)$ such arrays. Process the coefficient degrees $k=0,\ldots,L$ in increasing order. At degree $k$, first compute $[z^k]\alpha_0^{(\ell)}$ and $[z^k]\alpha_1^{(\ell)}$ from the grouped recurrence. The spike term uses degree $k-1$, and every cumulant term carries $z^{k'}$ with $k'\geq2$, so this step only reads previously computed monomial coefficients.

    Next update the degree-$k$ coefficients of the stored monomials in increasing order of $a+b$. For $a\geq1$, use
    \[
        \bigl(\alpha_0^{(\ell)}\bigr)^a\bigl(\alpha_1^{(\ell)}\bigr)^b
        =
        \alpha_0^{(\ell)}
        \bigl(\alpha_0^{(\ell)}\bigr)^{a-1}\bigl(\alpha_1^{(\ell)}\bigr)^b,
    \]
    and for $a=0$, multiply by $\alpha_1^{(\ell)}$ instead. Each degree-$k$ coefficient requires one convolution of length at most $k+1$. Its only possible same-degree reference is to a monomial of smaller total degree, which has already been updated. The cost is therefore $\Os(\ell^2(k+1))$ at degree $k$, and
    \begin{equation}\label{eq:spiked-series-runtime}
        \Os\left(\ell^2L^2\right)
    \end{equation}
    through degree $L$. Finally, evaluating the Taylor sum costs $\Os(L)$ operations. 
    
    For $m\geq2\ell$,
    \[
        \binom{m}{\ell/2}
        \geq\left(\frac{2m}{\ell}\right)^{\ell/2}
        \geq2^{\ell},
    \]
    so the cumulant-conversion cost is absorbed by the low-order moment computation. Moreover, $\binom m{\ell/2}\geq m$, $L^2\leq m$, and $\ell\leq2^\ell$ give $\ell^2L^2\leq\ell2^\ell\binom{m}{\ell/2}$, so the higher-order moment computation is absorbed as well. Together this proves \eqref{eq:spiked-cubic-runtime}.
\end{proof}

The color moments, cumulants, and grouped coefficients depend on $H$ and $Q$, but are independent of $\lambda$ and $t$. After computing them once, each additional spike strength requires only $\Os(\ell^2L^2)$ operations to obtain its two moment series. At a fixed spike strength, each additional time requires only $\Os(L)$ operations for the Taylor readout.

Combining this computational bound with \cref{cor:color-hypercontractive-sums} and \cref{lemma:spiked-fock-space-approx} gives the accuracy-dependent running time.

\begin{theorem}[Running time of the spiked model approximation]\label{thm:spiked-runtime}
    Fix any non-identity Hermitian Pauli operator $Q$. There is a sufficiently small constant $c>0$ such that, for all sufficiently large $n$, $2n\leq m\leq\poly(n)$, approximation $0<\varepsilon < 1$, spike $\lambda\geq 0$, evolution time $t\geq0$, and failure probability $\delta=1/\poly(n)$, satisfying
    \begin{equation}\label{eq:spiked-runtime-window}
        1+(2+\lambda)t+\log(1/\varepsilon)
        \leq c\min\left\{m^{1/3},\frac n{\log n}\right\},
    \end{equation}
    the cutoffs in \cref{alg:spiked} can be chosen so that
    \begin{equation}
        \left|\hat{Z}^{(\lambda)}(t)-\nTr\bigl(e^{-it(H+\lambda Q)}\bigr)\right|
        \leq\varepsilon
    \end{equation}
    with probability at least $1-\delta-e^{-\Omega(n)}$ and running time at most
    \begin{equation}\label{eq:spiked-accuracy-runtime}
        \Os\left(\frac{n}{\delta}(1+t)^2(1/\varepsilon)^{2+\Os(1/\log m)}\right).
    \end{equation}
\end{theorem}
\begin{proof}
    Choose a sufficiently large fixed constant $C_0$. Assuming we are not in the semicircular output case, let $\ell\geq 2$ be the smallest even integer satisfying
    \begin{equation}\label{eq:optimized-spiked-cutoff}
        (1+t)\delta^{-1/2}
        \left(\frac{C_0(\ell+2)}{m}\right)^{(\ell+2)/4}
        \leq\frac{\varepsilon}{C_0}.
    \end{equation}
    As in the cutoff choice for \eqref{eq:optimized-color-cutoff}, the bounds $(1+t)\delta^{-1/2}\leq m^{\Os(1)}$ and $\log(1/\varepsilon)=\Os(m^{1/3})$ imply
    \begin{equation}\label{eq:spiked-cutoff-size}
        \ell=\Os\left(1+\frac{\log(1/\varepsilon)}{\log m}\right).
    \end{equation}
    The first bound follows from $m\geq 2n$ and $t=\Os(m^{1/3})$. A sufficiently large constant multiple of $1+\frac{\log(1/\varepsilon)}{\log m}$ satisfies \eqref{eq:optimized-spiked-cutoff} since throughout this range $\log(m/(C_0(\ell+2)))$ is bounded below by a positive constant times $\log m$. 
    
    Choose an even
    \begin{equation}\label{eq:spiked-moment-cutoff}
        L=\Theta\bigl(1+(2+\lambda)t+\log(1/\varepsilon)\bigr), \qquad L\geq\ell+4.
    \end{equation}
    with a sufficiently large implied constant. Choosing $c$ in \eqref{eq:spiked-runtime-window} sufficiently small ensures $m\geq CL^3$ and $L\leq\gamma n/\log n$.

    \paragraph{Approximation.}
    Apply \cref{cor:color-hypercontractive-sums} with $s=1$ and intersect with the relative loop-free event of \cref{eq:loop-free-relative}. Since $m\geq2n$ and $t=\Os(m^{1/3})$, we have $t\Ks^{\mathrm{col}}(L)\leq\Os(m^{-1/6}\sqrt{\log n})=o(1)$. Thus the stability exponential is bounded. The omitted-tail estimate is the same, so \eqref{eq:optimized-spiked-cutoff} bounds the cumulant term in \eqref{eq:spiked-approx} by $\varepsilon/2$. The Taylor term is also at most $\varepsilon/2$, using $L!\geq(L/e)^L$ and \eqref{eq:spiked-moment-cutoff}. This gives the stated accuracy and success probability.

    \paragraph{Running time.}
    Minimality of the cutoff $\ell$ implies $\ell-2$ fails \eqref{eq:optimized-spiked-cutoff}, so
    \begin{equation}
        \binom{m}{\ell/2} \leq C_0^2(2eC_0)^{\ell/2}(1+t)^2\delta^{-1}\varepsilon^{-2}.
    \end{equation}
    Substituting this inequality into \eqref{eq:spiked-cubic-runtime} yields
    \begin{equation}
        \Os\left(\frac{n}{\delta}(1+t)^2\varepsilon^{-2}e^{\Os(\ell)}\right) \leq\Os\left(\frac{n}{\delta}(1+t)^2(1/\varepsilon)^{2+\Os(1/\log m)}\right).
    \end{equation}
    where the last inequality uses \eqref{eq:spiked-cutoff-size} and $L\leq\gamma n/\log n$, completing the proof.
\end{proof}

\section{Computing thermal expectation values}\label{sec:thermal-expectations}
We now apply the spiked model recurrence to the thermal expectation values from item~\ref{item:thermal-expectation} in the main text. Let
\[
    H=\frac{1}{\sqrt m}\sum_{k=1}^m J_kP_k,
    \qquad
    \rho_\beta=\frac{e^{-\beta H}}{\Tr(e^{-\beta H})},
\]
and fix a non-identity sampled Pauli $P_i$. Our target is
\begin{equation}\label{eq:thermal-ratio}
    \langle P_i\rangle_\beta
    =\frac{\nTr(P_ie^{-\beta H})}{\nTr(e^{-\beta H})}.
\end{equation}
We compute the numerator using the coupled series of the spiked algorithm and use the common scalar partition-function approximation $\hat{z}_\beta$ from \eqref{eq:thermal-partition-output} for the denominator.

\paragraph{Removing the observed Pauli term.}
Define the residual Hamiltonian
\begin{equation}\label{eq:thermal-residual}
    H^{(i)}\equiv H-\frac{J_i}{\sqrt m}P_i,
    \qquad
    H=H^{(i)}+\frac{J_i}{\sqrt m}P_i.
\end{equation}
Thus we use the spiked recurrence with background $H^{(i)}$, generator $\Qm=(P_i)$, and spike strength $\lambda=J_i/\sqrt m$.

If the full sampled support of $H$ is loop-free through order $L+1$, then the support of $H^{(i)}$ is loop-free relative to $(P_i)$ through order $L$, simultaneously for every $i$. A nonempty subset of at most $L$ residual terms cannot multiply projectively to identity by standard loop-freeness. If it multiplied projectively to $P_i$, adjoining the index $i$ would produce a loop of size at most $L+1$, again a contradiction.

The residual is kept in its original normalization. Its scalar second cumulant is
\begin{equation}\label{eq:thermal-residual-variance}
    \kappa_2[H^{(i)},H^{(i)}]=\frac{m-1}{m}
\end{equation}
on the loop-free event. The operator-valued recurrence in \cref{lemma:operator-valued-recurrence} does not assume that this cumulant equals $1$, so we simply retain its actual value along with the other residual color cumulants.

\paragraph{The two moment series.}
For an even cutoff $2\leq \ell\leq L$, let $\alpha_{c,i}^{(\ell)}(z)$, $c\in\{0,1\}$, be the truncated series obtained from \eqref{eqn:operator-recurrence-grouped} using the color cumulants of $H^{(i)}$ through order $\ell$ (including its variance $(m-1)/m$), with all higher cumulants set to zero, and spike strength $J_i/\sqrt m$. The resulting formal recurrence uniquely determines the series at every order.

\begin{proposition}[Thermal moment readout]\label{prop:thermal-moment-readout}
    If the sampled support of $H$ is loop-free through order $\ell+1$, then
    \begin{equation}\label{eq:thermal-moment-readout}
        [z^k]\alpha_{0,i}^{(\ell)}(z)=\nTr(H^k),
        \qquad
        [z^k]\alpha_{1,i}^{(\ell)}(z)=\nTr(P_iH^k),
        \qquad 0\leq k\leq \ell.
    \end{equation}
\end{proposition}
\begin{proof}
    By \eqref{eq:thermal-residual} and the relative loop-freeness of the residual, \cref{lemma:operator-valued-recurrence} gives the moment recurrence through degree $\ell$ with background $H^{(i)}$ and spike strength $J_i/\sqrt m$. Applying the cumulant cutoff does not change any coefficient through degree $\ell$. The trace readout in \eqref{eq:spiked-moment-extraction}, applied to this background, therefore gives \eqref{eq:thermal-moment-readout}.
\end{proof}

\paragraph{Fock space representation.}
Assume $m\geq2$ and that the full sampled support is loop-free through order $L+1$, and let $2\leq \ell\leq L$ be even. To apply the variance-one spiked construction, normalize the residual as
\begin{equation}\label{eq:thermal-normalized-residual}
    \widetilde H^{(i)}\equiv\sqrt{\frac{m}{m-1}}H^{(i)},
    \qquad
    H=\sqrt{\frac{m-1}{m}}\left(\widetilde H^{(i)}+\frac{J_i}{\sqrt{m-1}}P_i\right).
\end{equation}
Let $\Ts_{\ell,i}$ be the scaled spiked hopping operator
\begin{equation}\label{eq:thermal-expectation-hopping}
    \Ts_{\ell,i}\equiv\sqrt{\frac{m-1}{m}}\,\Ts_\ell^{(J_i/\sqrt{m-1})},
\end{equation}
where the operator on the right is constructed from $\widetilde H^{(i)}$. Homogeneity in \eqref{eqn:operator-recurrence-grouped} and \cref{lem:truncated-spiked-moments} give, for $c\in\{0,1\}$ and every $k\geq0$,
\begin{equation}\label{eq:thermal-expectation-fock-readout}
    [z^k]\alpha_{c,i}^{(\ell)}(z)=\bra{\Omega_1}b^c\Ts_{\ell,i}^{\,k}\ket{\Omega_1}.
\end{equation}
Indeed, multiplying the normalized background and spike by $\sqrt{(m-1)/m}$ multiplies every degree-$k$ coefficient by $((m-1)/m)^{k/2}$. By \cref{prop:thermal-moment-readout}, the cutoff-$\ell$ matrix elements in \eqref{eq:thermal-expectation-fock-readout} consequently equal $\nTr(P_i^cH^k)$ for $0\leq k\leq \ell$.

The two series are computed together, since their recurrences are coupled. We use $\alpha_{1,i}^{(\ell)}$ for the numerator and the shared approximation $\hat{z}_\beta$ for the denominator:
\begin{equation}\label{eq:thermal-algorithm-output}
    \widehat{\langle P_i\rangle}_\beta
    \equiv
    \frac{\displaystyle\sum_{k=0}^{L}\frac{(-\beta)^k}{k!}[z^k]\alpha_{1,i}^{(\ell)}(z)}
         {\hat{z}_\beta},
\end{equation}
whenever $\hat{z}_\beta>0$, and return zero otherwise. The series $\alpha_0$ remains part of the numerator computation, while the final normalization is reused across all observables at the same inverse temperature. The factor $2^n$ cancels in the ratio.

\begin{algorithm}[ht]
\caption{Thermal expectation algorithm}\label{alg:thermal-expectation}
\begin{algorithmic}[1]
\Require $H=m^{-1/2}\sum_{k=1}^m J_kP_k$, a sampled non-identity Pauli $P_i$, inverse temperature $\beta\geq0$, even cutoffs $2\leq\ell<L$.
\State Compute the color moments and color cumulants through order $\ell$ of $H^{(i)}$ from \eqref{eq:thermal-residual} for all normalized binary color sequences, using generator $P_i$, as well as $\kappa_2=(m-1)/m$.
\State Compute $\alpha_{0,i}^{(\ell)}$ and $\alpha_{1,i}^{(\ell)}$ through degree $L$ using \eqref{eqn:operator-recurrence-grouped} with spike strength $\lambda=J_i/\sqrt m$ and cumulant cutoff $\ell$.
\State Compute $\hat{z}_\beta$ using \eqref{eq:thermal-partition-output} with the same cutoffs.
\State \Return $\widehat{\langle P_i\rangle}_\beta$ in \eqref{eq:thermal-algorithm-output} if $\hat{z}_\beta>0$ and zero otherwise.
\end{algorithmic}
\end{algorithm}

\paragraph{Running time.}
The thermal algorithm uses the same type of low-order color data and the same grouped series computation as the spiked algorithm, now with $m-1$ background terms. Obtaining both moment series is already included in \cref{prop:spiked-cutoff-runtime}.

\begin{proposition}[Running time of the thermal computation]\label{prop:thermal-runtime}
    Fix even cutoffs $2\leq\ell<L$, and suppose $m-1 \geq CL^3$ for a sufficiently large constant $C$. Then \cref{alg:thermal-expectation} computes \cref{eq:thermal-algorithm-output} with total running time
    \begin{equation}\label{eq:thermal-cubic-runtime}
        \Os\left(n\ell 2^\ell\binom{m-1}{\ell/2}\right).
    \end{equation}
\end{proposition}
\begin{proof}
    Apply the proof of \cref{prop:spiked-cutoff-runtime} with $m-1$ background terms. The coefficients of these terms are $J_k/\sqrt{m}$ for all $k\neq i$; changing their normalization affects their scalar values but leaves all support sizes and runtimes unchanged.

    Evaluating the numerator in \eqref{eq:thermal-algorithm-output} costs $\Os(L)$ operations, computing the weights $(-\beta)^k/k!$ successively, which is absorbed in the series cost. The denominator satisfies \eqref{eq:thermal-scalar-cutoff-runtime}; since $\binom m{\ell/2}\leq2\binom{m-1}{\ell/2}$ under $m-1\geq CL^3$, its cost is also absorbed, proving \eqref{eq:thermal-cubic-runtime}.
\end{proof}

\newpage

\part{Numerical studies}\label{part:numerical-results}
Here, we provide additional details on our numerical studies in Figs.~\ref{fig:3} and~\ref{fig:2} of the main text, as well as our small heuristic study of the remaining quantum-classical gap at small precisions, in which $\varepsilon$ is exponentially small in $m$ and $m = \Theta(n)$.

\section{Details of numerical studies in the main text}
In Figs.~\ref{fig:3} and~\ref{fig:2}, we simulate the exact observables by replacing the infinite-temperature trace with an empirical average over a large number ($R=200$--$20{,}000$) of Haar-random states $\ket{\psi}$. This is done for numerical convenience, as it allows one to simulate state vectors in a size $2^n$ Hilbert space, rather than operators in a size $4^n$ space.
The empirical average is unbiased but leads to a small random numerical error of order $2^{-n/2}/\sqrt{R}$.
This is the ``Num. Error Floor'' marked in the inset of the top plot of (d); for all other figures, the floor is sufficiently small that it makes no visual difference.

The error statistics in the bottom plot of Fig.~\ref{fig:2}(c) use 10 random instances of the Hamiltonian, while the cumulant statistics in the top plot use 60 random instances to verify the precise match with theory predictions.
The theory predictions are obtained under the assumption that the Hamiltonian moments are loop-free and the commutation graph of the Pauli terms can be treated as Erd\H{o}s--R\'enyi [Eq.~\eqref{eq:core-variance-contribution}].
The $y$-axis values in Fig.~\ref{fig:2}(d) are obtained by taking the median over 100 random instances of the root-mean-square difference between $Z_\ell(t)$ and $Z_{\ell-2}(t)$ within each width-$5$ time window.
The time window is chosen to smooth over Bessel-function-like oscillations in the spectral amplitude.
The inset of the top plot of Fig.~\ref{fig:2}(d) shows the pointwise median absolute error over 5 random instances of the Hamiltonian, with $R=500$ and the same width-$5$ time average.

\section{Heuristic study of remaining advantage at small precisions}\label{sec:heuristic-precision}

Our classical algorithm is rigorous in the regimes $t = \mathcal{O}(n/\log n)$, $\varepsilon = 1/2^{\mathcal{O}(n/\log n)}$, and either $m = \Omega(n^3)$ (fully rigorous) or $m = \Omega(n)$ (rigorous under an additional combinatorial conjecture).
Beyond these regimes, it is not always clear whether our algorithm breaks down or merely our proof techniques.

For example, we are no longer able to rigorously prove that the loop-free approximation applies after near-linear time using our existing techniques from coding theory.
However, the anti-commutation of the Pauli terms tends to favor loop-free contributions with small crossings, and our numerical results suggest our algorithm might apply at late times as well.

In this section, we address a particular regime where our algorithm appears fundamentally unable to match the quantum runtime (i.e.~where the breakdown appears to not be of proof techniques).
This regime sets $m = 2n$, $\varepsilon = 2^{-\Omega(n)}$, and $t=\Omega(n)$.
It is hence relevant to moderate-size systems, where the number of qubits $n$ and number of terms $m$ grow only logarithmically in the runtime $\poly(1/\varepsilon)$.

Two important events occur for such small $\varepsilon$.
First, as mentioned in the introduction and discussions of the main text, the free cumulant expansion begins to diverge at order $\ell = \Theta(m)$, beyond which the large number of diagrams overwhelms the exponential decay in  each diagram's contribution.
This divergence can be seen for $m = 20$ in Fig.~\ref{fig:2}(c), top.
This breakdown is similar to the breakdowns of other diagrammatic expansions in physics and mathematics, where it is common to assume that the optimum order of the expansion is obtained at its minimum value, just before the divergence begins~\cite{zinnjustin2002quantum}.
In our context, recalling that the typical free cumulant magnitude scales as $\mathcal{O}((C\ell/m)^{\ell/4})$, the minimum  magnitude is at $\ell = a m$ and has value $b 2^{-cm}$, for constants $a = 1/(eC),b,c$.
This treatment is non-rigorous, as it uses our loop-free calculation of the free cumulants, which is only rigorously shown for $\ell = \mathcal{O}(n/\log n)$.
Nonetheless, if trusted, it suggests that our free cumulant expansion has a finite error floor, $b 2^{-cm}$, that it cannot go beyond.
If we then assume that this floor leads to a similar error floor in our algorithm prediction, $\mathcal{O}(2^{-cm})$, then it implies that, for accuracies below this value and at the late times, the only known classical algorithm is state-vector simulation.
This requires classical runtime $\mathcal{O}(2^{m/2})$ (see below), whereas the quantum runtime is $1/\varepsilon^2 = 2^{2cm}$.
Hence, if $c < 1/4$, a polynomial quantum advantage $2^{m/2} = (2^{2cm})^{1/(4c)}$, remains.

Second, even before the accuracy falls below this error floor, the runtime of our free cumulant algorithm becomes exponentially large.
In the regime $\ell = a m$, the propagation cost given below is exponential in $m$, as is state-vector simulation.
For sufficiently large $a$, the cost of our algorithm can thus exceed state-vector simulation and one should use state-vector simulation instead.
This again opens the door to a potential polynomial quantum advantage identically to the above.

\begin{figure}[t]
    \centering
    \includegraphics[width=\textwidth]{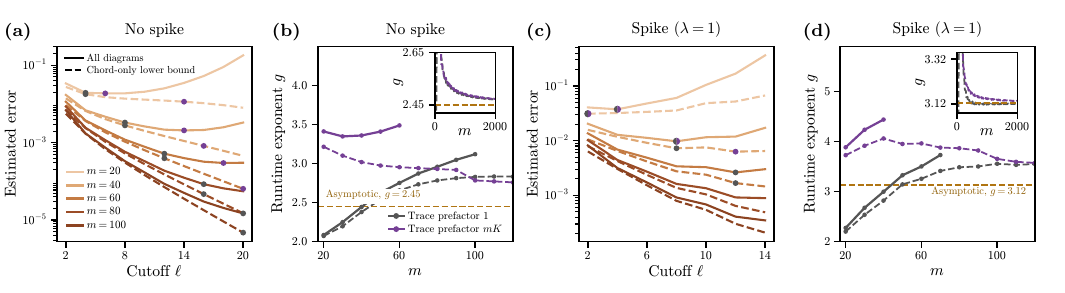}
    \caption{Heuristic study of the classical runtime at small precisions.
    \textbf{(a)} The estimated algorithm error from Eq.~\eqref{eq:numerical-error-proxy} as a function of the free cumulant cutoff $\ell$. 
    Solid curves include all diagram sectors; dashed curves denote a rigorous lower bound from including diagrams that contain only chords.
    Gray and purple dots mark the crossover values of $\ell$ at which state-vector simulation becomes more efficient than our algorithm.
    \textbf{(b)} Runtime exponent $g$ at the crossover, using trace prefactors $1$ (gray) or $mK$ (purple), with $K=m/2$. 
    Insets extend to larger $m$, to show convergence of the chord-only lower bound results (dashed) to their asymptotic value (gold dashed) $2.45$.
    \textbf{(c,d)} Analogous computations for the spiked model.
    We refer to the text for further details.}
    \label{fig:numerical-cumulant-error}
\end{figure}

We perform a modest numerical study of the free cumulants, described below and shown in Fig.~\ref{fig:numerical-cumulant-error}, which suggests that the constant factors above indeed work out in a manner supportive of super-quadratic polynomial quantum advantages.
Our study uses our exact expressions [Eqs.~\eqref{eq:cumulants_graph_properties} and~\eqref{eq:core-variance-contribution}] for the free cumulants and their second moments within the loop-free approximation.
This involves a summation over diagrams which can be evaluated numerically for any $m,n$ up to cutoff $\ell = 20$.
(For larger orders, the number of diagrams becomes too large for us to evaluate in reasonable time.)
Our study makes some heuristic approximations, most notably, that the loop-free expression for the free cumulants continues to apply at the linear orders of interest, and that the error of our algorithm's prediction is proportional to the magnitude of the leading truncated free cumulant term. 
Determining whether these approximations are appropriate, and developing a fuller understanding of this small-precision regime, are beyond the reach of our methods and remain an interesting challenge for future work.

Our study proceeds as follows.
First, we estimate the error in our classical algorithm for the spectral amplitude as a function of the cutoff $\ell$ for various $m$ [Fig.~\ref{fig:numerical-cumulant-error}(a)].
To do so, we estimate the root-mean-square cumulants by assuming the loop-free approximation holds, treating the commutation graph as Erd\H{o}s--R\'enyi, and using Eq.~\eqref{eq:core-variance-contribution}.
We also obtain a rigorous lower bound on the root-mean-square cumulants within the loop-free Erd\H{o}s--R\'enyi approximation, obtained by keeping only diagrams composed entirely of chords; this is useful for lower-bounding behavior at larger $\ell$, which cannot be reached by our direct enumeration of diagrams.
We then translate this magnitude into an estimated algorithm error by Taylor expanding the semicircle value of the spectral amplitude, $Z(t)$, about the value of the first truncated free cumulant, at order $\ell+2$, to first order.
The first derivative is a Bessel function, $(-i)^{\ell+2}\Js_{\ell+2}(2t)$, which gives an estimated maximum error
\begin{equation}\label{eq:numerical-error-proxy}
    \varepsilon_\ell(m)^2
    = \max_{t\geq0}|\Js_{\ell+2}(2t)|^2\,
      \mathbb E[\widetilde\kappa_{\ell+2}^{\,2}],
\end{equation}
which is plotted in Fig.~\ref{fig:numerical-cumulant-error}(a) for both the full numerical estimates and lower bounds.
The maximum occurs at $t=j'_{\ell+2,1}/2$, where $j'_{k,1}$ is the first positive zero of $\Js_k'$, and its magnitude scales as $|\Js_{\ell+2}(2t)| \sim \ell^{-1/3}$.
We also repeat this procedure for the spiked model at $\lambda = 1$ [Fig.~\ref{fig:numerical-cumulant-error}(c)]; to handle the many color cumulants in the spiked model, we compute the cross-correlations between all color cumulants of order $\ell+2$ and the derivative of the spectral amplitude, and maximize the resulting mean-square error over time.
To obtain an analytically tractable rigorous lower bound within the loop-free Erd\H{o}s--R\'enyi model, we retain only the contributions from chord-only diagrams in which all $q=(\ell+2)/2$ distinct Hamiltonian terms appearing in a diagram commute with the spike. This gives a lower bound on the estimated mean-square error.

Second, we estimate the costs of direct classical simulation, for future comparison.
A straightforward mapping shows that any $m$-term Pauli Hamiltonian can be compressed to act on at most $m$ qubits. 
Hence, any $m$-term Hamiltonian can be simulated with state-vector simulation on a Hilbert space of size $2^m$.
A further reduction, to size $\mathcal{O}(2^{m/2})$, is possible by combining the ideas of Ref.~\cite{gunderman2024minimal} with the randomness of our specific Hamiltonian of interest.
This follows because the rank of the commutation matrix of the sparse random Pauli Hamiltonian is $m-d$ where $d$ is small with high probability.
The spectral amplitude of a Hamiltonian on $m/2$ qubits can be estimated by replacing the trace with an empirical average over $R$ Haar-random states, with standard deviation $\mathcal{O}(2^{-m/4}/R^{1/2})$ in runtime $T_{\text{tr}} \sim A (2^{m/2} + \varepsilon^{-2})$. The parameter $A$ accounts for the overhead of each Haar-state simulation compared to the Hilbert space dimension, e.g.~a reasonable estimate is $A = m K$ where $m$ is the number of Hamiltonian terms and $K$ is the maximum moment required for the time-evolution. We evaluate this classical cost for two choices, $A = 1$ and $A = mK = m^2/2$, in Fig.~\ref{fig:numerical-cumulant-error}, to gain an understanding of the spread of our numerical estimates depending on this non-exponential prefactor. Here, we set $K = m/2$ as a rough estimate.

Third, we compare these two costs to determine the threshold algorithm error and order $\ell$ at which state-vector simulation becomes more favorable than our algorithm.
We assume that the classical runtime of our algorithm is $C(m,\ell)=m\sum_{j=0}^{\ell/2-1}\sum_{0\leq k\leq j, k\equiv j\ {\rm mod}\ 2}\binom{m}{k}$, which precisely counts the number of Pauli operators that can appear in $H^{\ell/2}$.
For each demanded accuracy, we select the smallest cutoff $\ell$ with estimated error less than the desired accuracy.
We then compare the classical runtime of our algorithm at this $\ell$ with $T_{\text{tr}}$. 
The location of the crossover from our algorithm to state-vector simulation is denoted with a gray ($A=1$) and purple ($A = m^2/2$) dot on each curve of Fig.~\ref{fig:numerical-cumulant-error}(a) and (c).

Finally, we use this crossover to estimate the effective degree of the runtime of the classical algorithm with the accuracy.
We do so extremely coarsely, by assuming that the classical cost is equal to $(1/\varepsilon)^g$, where $\varepsilon$ is the desired accuracy which we have estimated as described.
For each $\varepsilon$, we take the logarithm with base $1/\varepsilon$ of the classical runtime at the threshold obtained above to obtain the effective degree $g$.
The results are shown in Fig.~\ref{fig:numerical-cumulant-error}(b) and (d) for the unspiked and spiked model, respectively.
The obtained degrees are between $2$ and $3.5$ for the unspiked model up to $\ell = 20$, and between $2$ and $4.5$ for the spiked model up to $\ell = 14$.
These are obtained for numbers of terms between $m = 20$ and $m=100$.
Asymptotically, our chord-only lower bound predicts that the exponent $g$ approaches $g \approx 2.45$ for the unspiked model, and $g \approx 3.12$ for the spiked model.
Nonetheless, we observe an increasing difference between our chord-only lower bound and the estimated $g$ from all diagrams as $m$ increases, suggesting that the true asymptotic $g$ accounting for all diagrams approaches a larger value.

\newpage

\part{Select technical proofs}\label{part:technical-proofs}
In this part we complete select technical proofs deferred from the previous sections. Most of these are not strictly necessary for the correctness of our algorithms or their main ideas; they instead provide slight optimizations for certain parameters.

In \cref{sec:general-truncation}, we extend the moment and free cumulant truncation analysis from \cref{sec:classical-sfa-algorithm} to a general form which can be utilized for many of our proofs. In \cref{sec:general-beta}, we extend the infinite temperature results of \cref{sec:toc-otoc} and \cref{sec:spiked-pauli} to finite-temperature thermal states, including those in \cref{sec:thermal-expectations}. In \cref{sec:combinatorial-proofs} we provide the proof of a combinatorial lemma utilized in \cref{sec:free-cumulant-concentration}. Finally, we provide an alternative algorithm with improved runtime in certain parameter regimes in \cref{sec:fmm-cumulants}, and a tighter analysis utilizing hypercontractivity of the success probability for all our algorithms in \cref{sec:hypercontractivity}.

\section{General Fock space truncation bounds}\label{sec:general-truncation}

In this section, we provide general results for approximation guarantees when truncating moments and free cumulants. We first give a common truncation error bound for products of real-time evolutions, then extend it to general complex values. \cref{lem:common-truncation} is stated in a general way so that it can be used for both TOC/OTOC algorithms in \cref{part:color-framework} and the spiked Pauli algorithm in \cref{part:operator-valued}.

In the following setting, function $\Ks$ will be used to represent the free cumulant summations. However, we just treat it as a generic upper-bound function for now. We consider the following common setting between the real- and complex-time cases:
\begin{setting}\label{setting:truncation}
    Fix even integers $2\leq\ell<L$. Let $X$ be a Hermitian operator and let $U_1,\ldots,U_r$ be operators acting on the same space with $\|U_j\|\leq1$. Let $\Ts_{\ell},\Ts_L,V_1,\ldots,V_r, S$ be bounded operators on a Hilbert space with unit vectors $\ket{\xi},\ket{\zeta}$, where $\|V_j\|\leq1$ and $S$ is self-adjoint. Suppose for some non-decreasing function $\Ks:\mathbb{Z}_{\geq0}\rightarrow\R_{\geq0}$,
    \begin{equation}\label{eq:thermal-common-operator-bounds}
        \|S\|\leq R,\qquad \|\Ts_{\ell}-S\|\leq\Ks(\ell),\qquad \|\Ts_{L}-S\|\leq\Ks(L),\qquad
        \|\Ts_L-\Ts_{\ell}\|\leq\Ks(L)-\Ks(\ell),
    \end{equation}
    where $R\geq2$. Assume also that, for all nonnegative integers $h_1,\ldots,h_r$ with $\sum_jh_j<L$,
    \begin{equation}\label{eq:thermal-common-moment-matching}
        \nTr(X^{h_1}U_1\cdots X^{h_r}U_r)
        =\bra{\xi}\Ts_L^{h_1}V_1\cdots\Ts_L^{h_r}V_r\ket{\zeta}.
    \end{equation}
    For $z_1,\ldots,z_r\in\C$, define
    \begin{equation}
        N\equiv\nTr\left(\prod_{j=1}^r e^{iz_jX}U_j\right),\qquad
        N_{\ell}\equiv\bra{\xi}\prod_{j=1}^r e^{iz_j\Ts_{\ell}}V_j\ket{\zeta},\qquad N_{L}\equiv\bra{\xi}\prod_{j=1}^r e^{iz_j\Ts_{L}}V_j\ket{\zeta},
    \end{equation}
    and $\widehat{N}$ to be the Taylor polynomial of $N_{\ell}$ of degree $L$.
\end{setting}

We will also make use of the following standard result:
\begin{proposition}[Duhamel formula]\label{lem:duhamel}
    Let $A$ and $B$ be bounded operators on a Hilbert space and let $t\geq0$. Then
    \begin{equation}
        e^{iAt}-e^{iBt}=i\int_0^t e^{iA(t-s)}\,(A-B)\,e^{iBs}\,ds.
    \end{equation}
\end{proposition}

Our generic truncation lemma is the following:
\begin{lemma}[Approximation error of truncations (real-time)]\label{lem:common-truncation}
    Consider \cref{setting:truncation} where $\ket{\xi}=\ket{\zeta}=\ket{\Omega}$ and $z_j=t_j\in \R$. Define $T\equiv \sum_{j}|t_j|$, and assume $\nTr(|X|^L)\leq (R+\Ks(L))^L$. Then
    \begin{equation}\label{eq:common-truncation}
        |N-\widehat{N}| \leq e^{T\Ks(L)}\left[T\bigl(\Ks(L)-\Ks(\ell)\bigr)+\frac{\bigl(T(R+\Ks(L))\bigr)^L}{L!}\left(2+\frac{T(R+\Ks(L))}{L+1}\right)\right].
    \end{equation}
\end{lemma}
\begin{proof}
    First, \cref{eq:thermal-common-operator-bounds} and the triangle inequality bound $\|\Ts_{\ell'}\|\leq\|S\|+\|\Ts_{\ell'}-S\|$ for $\ell'\in\{\ell,L\}$. Additionally, since $S$ is self-adjoint, \cref{lem:perturbed-unitary} applies to $\Ts_{\ell'}=S+(\Ts_{\ell'}-S)$ (and to $-S$ and $-(\Ts_{\ell'}-S)$ when $u<0$). Together these give, for $\ell'\in\{\ell,L\}$ and every $u\in\R$,
    \begin{equation}\label{eq:common-evolution-bound}
        \|\Ts_{\ell'}\|\leq R+\Ks(L),\qquad
        \bigl\|e^{iu\Ts_{\ell'}}\bigr\|\leq e^{|u|\,\|\Ts_{\ell'}-S\|}\leq e^{|u|\Ks(L)}.
    \end{equation}

    To compare products of evolutions, we rescale all times by a common parameter $u\in[0,1]$. Define
    \begin{equation}
        f(u)\equiv\nTr\left(\prod_{j=1}^r e^{iut_jX}U_j\right),\qquad
        f_{\ell'}(u)\equiv\bra{\Omega}\prod_{j=1}^r e^{iut_j\Ts_{\ell'}}V_j\ket{\Omega},
    \end{equation}
    so that $N=f(1)$ and $N_{\ell'}=f_{\ell'}(1)$. A monomial of total degree $k$ in $(t_1,\ldots,t_r)$ contributes $u^k$ to $f_{\ell}(u)$, so the Taylor polynomial $\widehat{N}$ of $N_{\ell}$ of degree $L$ is
    \begin{equation}\label{eq:common-output-polynomial}
        \widehat{N}=\sum_{k=0}^{L}\frac{f_{\ell}^{(k)}(0)}{k!}.
    \end{equation}
    By Leibniz's rule, for every $k\geq0$,
    \begin{equation}\label{eq:common-leibniz}
        f^{(k)}(u)=\sum_{\substack{h_1,\ldots,h_r\geq0\\h_1+\cdots+h_r=k}}\frac{k!}{h_1!\cdots h_r!}\left(\prod_{j=1}^r(it_j)^{h_j}\right)\nTr\left(\prod_{j=1}^r X^{h_j}e^{iut_jX}U_j\right),
    \end{equation}
    and the same formula holds for $f_{\ell'}$ with $\Ts_{\ell'}$, $V_j$, and $\bra{\Omega}\cdot\ket{\Omega}$ in place of $X$, $U_j$, and $\nTr$. For $k=L$, H\"older's inequality for the normalized Schatten norms with exponents $L/h_j$ bounds every trace in \cref{eq:common-leibniz} by $\nTr(|X|^L)$, since each $e^{iut_jX}U_j$ is a contraction. For $f_{\ell'}$, every matrix element is at most $\prod_j\|\Ts_{\ell'}\|^{h_j}\|e^{iut_j\Ts_{\ell'}}\|$. The multinomial theorem, the assumption on $\nTr(|X|^L)$, and \cref{eq:common-evolution-bound} therefore give, for all $0\leq u\leq1$ and $k\geq0$,
    \begin{equation}\label{eq:common-derivative-bound}
        |f^{(L)}(u)|\leq T^L\,\nTr(|X|^L)\leq\bigl(T(R+\Ks(L))\bigr)^L,\qquad
        |f_{\ell'}^{(k)}(u)|\leq\bigl(T(R+\Ks(L))\bigr)^k e^{uT\Ks(L)}.
    \end{equation}

    Our goal is to show that the sequence of hybrids
    \begin{equation}\label{eq:common-hybrid-method}
        N\xrightarrow{(1)}N_L\xrightarrow{(2)}N_{\ell}\xrightarrow{(3)}\widehat{N}
    \end{equation}
    incurs a small additive error at each step, and then apply the triangle inequality.

    \paragraph{Step (1): Initial cumulant truncation.}
    Setting $u=0$ in \cref{eq:common-leibniz}, the moment-matching assumption \cref{eq:thermal-common-moment-matching} gives $f^{(k)}(0)=f_L^{(k)}(0)$ for all $k<L$, so $f$ and $f_L$ have the same Taylor polynomial $P_{L-1}$ of degree $L-1$ at $u=0$. We first truncate the Taylor series of $f$. By Taylor's formula with integral remainder and \cref{eq:common-derivative-bound},
    \begin{equation}
        \left|N-P_{L-1}(1)\right|
        \leq\frac{1}{L!}\sup_{0\leq u\leq1}|f^{(L)}(u)|
        \leq\frac{\bigl(T(R+\Ks(L))\bigr)^L}{L!}.
    \end{equation}
    We then return to the product of exponentials of $\Ts_L$ by the same Taylor approximation in reverse:
    \begin{equation}
        \left|P_{L-1}(1)-N_L\right|
        \leq\frac{1}{L!}\sup_{0\leq u\leq1}|f_L^{(L)}(u)|
        \leq\frac{\bigl(T(R+\Ks(L))\bigr)^L}{L!}e^{T\Ks(L)}.
    \end{equation}
    Combining the two approximations with the triangle inequality gives
    \begin{equation}\label{eq:common-first-truncation}
        |N-N_L|
        \leq\bigl(1+e^{T\Ks(L)}\bigr)\frac{\bigl(T(R+\Ks(L))\bigr)^L}{L!}
        \leq2e^{T\Ks(L)}\frac{\bigl(T(R+\Ks(L))\bigr)^L}{L!}.
    \end{equation}

    \paragraph{Step (2): Second cumulant truncation.}
    Telescoping the difference of the two products defining $N_L$ and $N_{\ell}$ gives
    \begin{equation}
        N_L-N_{\ell}
        =\sum_{j=1}^r\bra{\Omega}\Bigl(\prod_{i<j}e^{it_i\Ts_L}V_i\Bigr)\bigl(e^{it_j\Ts_L}-e^{it_j\Ts_{\ell}}\bigr)V_j\Bigl(\prod_{i>j}e^{it_i\Ts_{\ell}}V_i\Bigr)\ket{\Omega}.
    \end{equation}
    By \cref{lem:duhamel} (applied to $-\Ts_L$ and $-\Ts_{\ell}$ when $t_j<0$) together with \cref{eq:thermal-common-operator-bounds,eq:common-evolution-bound},
    \begin{equation}
        \bigl\|e^{it_j\Ts_L}-e^{it_j\Ts_{\ell}}\bigr\|
        \leq\int_0^{|t_j|}e^{(|t_j|-s)\Ks(L)}\,\|\Ts_L-\Ts_{\ell}\|e^{s\Ks(L)}\,ds
        \leq|t_j|\,\bigl(\Ks(L)-\Ks(\ell)\bigr) e^{|t_j|\Ks(L)}.
    \end{equation}
    Bounding the remaining factors by \cref{eq:common-evolution-bound} and $\|V_i\|\leq1$, and summing over $j$, gives
    \begin{equation}\label{eq:common-second-truncation}
        |N_L-N_{\ell}|
        \leq T\bigl(\Ks(L)-\Ks(\ell)\bigr)e^{T\Ks(L)}.
    \end{equation}

    \paragraph{Step (3): Moment truncation.}
    Now that we have truncated the cumulants to order $\ell$, we bound the error incurred from the Taylor series approximation of this truncated quantity. By \cref{eq:common-output-polynomial}, $\widehat{N}$ is the Taylor polynomial of $f_{\ell}$ of degree $L$, evaluated at $u=1$. Taylor's formula with integral remainder and \cref{eq:common-derivative-bound} with $k=L+1$ give
    \begin{equation}\label{eq:common-third-truncation}
        |N_{\ell}-\widehat{N}|
        \leq\frac{1}{(L+1)!}\sup_{0\leq u\leq1}\bigl|f_{\ell}^{(L+1)}(u)\bigr|
        \leq\frac{\bigl(T(R+\Ks(L))\bigr)^{L+1}}{(L+1)!}e^{T\Ks(L)}.
    \end{equation}

    \paragraph{Putting it all together.}
    Finally, we combine the contributions from \cref{eq:common-first-truncation,eq:common-second-truncation,eq:common-third-truncation}. Applying the triangle inequality gives
    \begin{equation}
        |N-\widehat{N}|
        \leq e^{T\Ks(L)}\left[2\frac{\bigl(T(R+\Ks(L))\bigr)^L}{L!}+T\bigl(\Ks(L)-\Ks(\ell)\bigr)+\frac{\bigl(T(R+\Ks(L))\bigr)^{L+1}}{(L+1)!}\right],
    \end{equation}
    which is \cref{eq:common-truncation}.
\end{proof}

\cref{tab:real-time-truncation-parameters} shows the parameters which \cref{lem:common-truncation} is instantiated with for each algorithm. 

\begin{table}[htbp]
    \centering
    \renewcommand{\arraystretch}{1.2} \setlength{\tabcolsep}{1em}
    \begin{tabular}{@{}ccccc@{}}
        \toprule
        Parameter & Spectral amplitude & TOC & OTOC & Spiked spectral amplitude \\
        \midrule
        $X$ & $H$ & $H$ & $H$ & $H+\lambda Q$ \\
        $(t_j)$ & $(-t)$ & $(t,-t)$ & $(t,-t,t,-t)$ & $(-t)$ \\
        $(U_j)$ & $(\mathbb I)$ & $(A,A)$ & $(A,B,A,B)$ & $(\mathbb I)$ \\
        $(V_j)$ & $(\mathbb I)$ & $(b,b)$ & $(b_1,b_2,b_1,b_2)$ & $(\mathbb I)$ \\
        $\ket{\Omega}$ & $\ket{0}$ & $\ket{\Omega_1}$ & $\ket{\Omega_2}$ & $\ket{\Omega_1}$ \\
        $S$ & $a^\dagger+a$ & $a^\dagger+a$ & $a^\dagger+a$ & $a^\dagger+a+\lambda b$ \\
        $R$ & $2$ & $2$ & $2$ & $2+\lambda$ \\
        $T$ & $t$ & $2t$ & $4t$ & $t$ \\
        \bottomrule
    \end{tabular}
    \caption{Parameters used in \cref{lem:common-truncation}. The spectral amplitude proof uses the free cumulant sums and \cref{lem:scalar-cutoff-comparison}; the other observables use the color cumulant sums and the Duhamel bound.}
    \label{tab:real-time-truncation-parameters}
\end{table}

\begin{lemma}[Approximation error of truncations (complex-time)]\label{lem:thermal-common-truncation}
    In \cref{setting:truncation}, suppose $X$ satisfies \eqref{eq:thermal-absolute-tail} with parameter $0<\eta<1$. Set $\beta=\sum_j|\operatorname{Im}z_j|$ and choose $T\geq\sum_j|z_j|$ with $T\leq L/C$, where $C$ is the constant in \cref{lem:thermal-physical-taylor}. Then we have
    \begin{equation}\label{eq:thermal-common-truncation}
        |N-\widehat{N}|\leq\eta^{-1}2^{1-L}+e^{R\beta+T\Ks(L)}\Bigg[T\bigl(\Ks(L)-\Ks(\ell)\bigr)+\frac{\big(T(R+\Ks(L))\big)^L}{L!}\left(1+\frac{T(R+\Ks(L))}{L+1}\right)\Bigg].
    \end{equation}
\end{lemma}
\begin{proof}
    By \cref{lem:thermal-perturbed-evolution,eq:thermal-common-operator-bounds}, for $\ell'\in\{\ell,L\}$,
    \begin{equation}\label{eq:thermal-common-evolution}
        \|\Ts_{\ell'}\|\leq R+\Ks(\ell'),\qquad
        \|e^{iuz\Ts_{\ell'}}\|\leq e^{uR|\operatorname{Im}z|+u|z|\Ks(\ell')},\qquad 0\leq u\leq1.
    \end{equation}
    We use the same three hybrids as in the real-time proofs:
    \begin{equation}\label{eq:thermal-common-hybrids}
        N\xrightarrow{(1)}N_L\xrightarrow{(2)}N_{\ell}\xrightarrow{(3)}\widehat{N}.
    \end{equation}

    \paragraph{Step (1): Initial cumulant truncation.}
    Insert the common Taylor polynomial of total degree less than $L$, using \eqref{eq:thermal-common-moment-matching}. The remainder for $X$ is at most $\eta^{-1}2^{1-L}$ by \eqref{eq:thermal-product-taylor}. For the auxiliary remainder, set $f_{\ell'}(u)=\bra{\xi}\prod_j e^{iuz_j\Ts_{\ell'}}V_j\ket{\zeta}$. Leibniz's rule, \eqref{eq:thermal-common-evolution}, and the multinomial theorem give
    \begin{equation}\label{eq:thermal-common-derivative-bound}
        |f_{\ell'}^{(k)}(u)|\leq\big(T(R+\Ks(\ell'))\big)^k e^{R\beta+T\Ks(\ell')},\qquad 0\leq u\leq1.
    \end{equation}
    Taylor's formula with integral remainder therefore bounds this step by
    \begin{equation}
        \eta^{-1}2^{1-L}+e^{R\beta+T\Ks(L)}\frac{\big(T(R+\Ks(L))\big)^L}{L!}.
    \end{equation}

    \paragraph{Step (2): Second cumulant truncation.}
    Telescope the difference of the two products defining $N_L$ and $N_{\ell}$. Apply \cref{lem:duhamel} to each pair $z_j\Ts_L,z_j\Ts_{\ell}$ at time $1$, and bound the remaining factors by \eqref{eq:thermal-common-evolution}. Since $\sum_j|z_j|\leq T$, this gives
    \begin{equation}
        |N_L-N_{\ell}|\leq T\bigl(\Ks(L)-\Ks(\ell)\bigr)e^{R\beta+T\Ks(L)}.
    \end{equation}

    \paragraph{Step (3): Moment truncation.}
    The output is the Taylor polynomial of $f_{\ell}$ through degree $L$, evaluated at $u=1$. Apply \eqref{eq:thermal-common-derivative-bound} with $k=L+1$ to bound the remainder by
    \begin{equation}
        |N_{\ell}-\widehat{N}|\leq e^{R\beta+T\Ks(\ell)}\frac{\big(T(R+\Ks(\ell))\big)^{L+1}}{(L+1)!}.
    \end{equation}
    Adding the three errors and using $\Ks(\ell)\leq\Ks(L)$ proves the claim.
\end{proof}
\section{Extension for thermal state observables}\label{sec:general-beta}
We extend the algorithms of Parts~\ref{part:sfa-algorithm}, \ref{part:color-framework}, and \ref{part:operator-valued} to Gibbs states at inverse temperature $\beta\geq0$. We first establish general moment bounds for the sparse Pauli Hamiltonian needed for the approximation analysis in \cref{subsec:thermal-physical-tails}. We then give the partition function approximation and apply the same truncation arguments to correlation functions and thermal expectation values. 

\subsection{Moment bounds and Taylor series truncation}\label{subsec:thermal-physical-tails}
We first bound the Taylor remainders for the Pauli Hamiltonian needed for the thermal algorithms. We use the moment bound of Bandeira et al.~\cite[Theorem~4.2]{bandeira2026matrix}, specialized to the sparse Pauli model, together with the commutation-sign estimates from \cref{subsec:commutation-graph}.

\begin{lemma}[Sign-averaged moments]\label{lem:thermal-sign-averaged-moments}
    Fix the Pauli support $\Pm=(P_1,\ldots,P_m)$, and let $H=m^{-1/2}\sum_{i=1}^m J_i P_i$ with independent uniform signs $J_i\in\{\pm1\}$. Let $W$ be its commutation sign matrix (\cref{def:commutation-sign-matrices}). Then, for every even integer $k\geq2$,
    \begin{equation}\label{eq:thermal-sign-averaged-moments}
        \left(\E_{\va{J}}\nTr(H^k)\right)^{1/k}
        \leq2\sqrt{1+\frac{k}{2m}\lambda_{\max}^{+}(W-2\mathbb{I}_m)},
    \end{equation}
    where $\lambda_{\max}^{+}(M)\equiv\max\{0,\lambda_{\max}(M)\}$ for a Hermitian matrix $M$.
\end{lemma}
\begin{proof}
    We specialize \cite[Theorem~4.2]{bandeira2026matrix} to independent Rademacher variables. On the orthonormal basis $\{\ket{S}:S\subseteq[m]\}$, let $\ell_i\ket{S}=\ket{S\cup\{i\}}$ when $i\notin S$, and let $\ell_i\ket{S}=0$ otherwise. The operators $\ell_i+\ell_i^*$ have the joint law of independent Rademacher variables in the vacuum state $\ket{\varnothing}$: a product returns to the vacuum exactly when every index appears an even number of times. Their creation and annihilation operators satisfy
    \begin{equation}
        \ell_i^*\ell_j-(1-2\delta_{ij})\ell_j\ell_i^*=\delta_{ij}\mathbb{I}.
    \end{equation}
    The number operator $\sum_i\ell_i\ell_i^*$ acts on $\ket{S}$ by multiplication by $|S|$, so its norm on the subspace with $|S|\leq k/2$ is at most $k/2$.

    Apply the theorem with $q=0$ and matrix coefficients $P_i/\sqrt m$. Their squared sum is $\mathbb{I}$, and the partial commutator in that theorem is
    \begin{equation}
        C_0=\left[\frac{1-2\delta_{ij}}m P_jP_i\right]_{i,j=1}^m.
    \end{equation}
    The block-diagonal matrix $D=\operatorname{diag}(P_1,\ldots,P_m)$ is unitary and satisfies
    \begin{equation}
        DC_0D=\frac{W-2\mathbb{I}_m}{m}\otimes\mathbb{I},
    \end{equation}
    because $P_iP_jP_iP_j=\langle P_i,P_j\rangle\mathbb{I}$. Thus $\lambda_{\max}^{+}(C_0)=m^{-1}\lambda_{\max}^{+}(W-2\mathbb{I}_m)$, which proves \eqref{eq:thermal-sign-averaged-moments}.
\end{proof}

This estimate holds at every moment order without a loop-freeness assumption. We next bound the commutation sign matrix using the same approximate independence of commutation signs as in \cref{part:sfa-algorithm}.

\begin{lemma}[Norm of the commutation sign matrix]\label{lem:thermal-commutation-matrix-norm}
    There exists a constant $C>0$ such that, whenever $n\geq2$, $C\log n\leq m\leq\poly(n)$ and $\delta=1/\poly(n)$, the commutation sign matrix of the random Pauli support satisfies
    \begin{equation}\label{eq:thermal-commutation-matrix-norm}
        \|W\|\leq C\sqrt m
    \end{equation}
    with probability at least $1-\delta-e^{-\Omega(n)}$ over $\Pm$.
\end{lemma}
\begin{proof}
    Let $S$ be a symmetric $m\times m$ matrix with zero diagonal and independent uniform signs above the diagonal. For even $k$, expand $\Tr((W-\mathbb{I}_m)^{k})$ as a sum over closed walks of length $k$. Only walks with no diagonal steps contribute. If every unordered edge of such a walk appears an even number of times, its contribution is $1$, both for $W-\mathbb{I}_m$ and in expectation for $S$. Every other nonzero term reduces to a nonempty product of distinct commutation signs, whose expectation is at most $4^{-n}$ by \cref{lemma:k-wise-independence}; its expectation for $S$ is zero. There are at most $m^{k}$ walks, so
    \begin{equation}\label{eq:thermal-commutation-moment-comparison}
        \E_{\Pm}\Tr\bigl((W-\mathbb{I}_m)^{k}\bigr)
        \leq\E\Tr(S^{k})+4^{-n}m^{k}.
    \end{equation}

    For every real unit vector $x$, Hoeffding's inequality applied to $x^\top Sx=2\sum_{a<b}S_{ab}x_ax_b$ gives $\Pr(|x^\top Sx|>u)\leq2e^{-u^2/4}$. A $1/4$-net of the unit sphere with at most $9^m$ points therefore gives $\Pr(\|S\|>u)\leq2\cdot9^m e^{-u^2/16}$. Integrating this tail yields
    \begin{equation}
        \E\Tr(S^{k})\leq m\bigl(C(\sqrt m+\sqrt{k/2})\bigr)^{k}.
    \end{equation}
    Choose an even integer $k=\Theta(\log n)$ with a sufficiently large implied constant. The assumption $m\geq C\log n$ ensures $k/2\leq m$. Markov's inequality applied to \eqref{eq:thermal-commutation-moment-comparison}, after increasing the constant in the norm threshold, gives
    \begin{equation}
        \Pr_{\Pm}\bigl(\|W-\mathbb{I}_m\|>C\sqrt m\bigr)
        \leq m\,4^{-k/2}+4^{-n}m^{k/2}
        \leq \delta+e^{-\Omega(n)}.
    \end{equation}
    The last inequality uses $\frac{k}{2}\log m=\Os(\log^2 n)=o(n)$. Finally, $\|W\|\leq1+\|W-\mathbb{I}_m\|$, which is absorbed by increasing $C$.
\end{proof}

\begin{lemma}[Taylor series truncation for sparse Pauli Hamiltonians]\label{lem:thermal-physical-taylor}
    There exists a constant $C>0$ such that, for every $n\geq 2$, failure probability $\delta=1/\poly(n)$, $0<\eta<1$, integer $L\geq1$ and $m$ satisfying
    \begin{equation}\label{eq:thermal-physical-tail-range}
        \max\{L^2,C\log n\}\leq m\leq\poly(n),
    \end{equation}
    a sparse Pauli Hamiltonian $H$ satisfies
    \begin{equation}\label{eq:thermal-absolute-tail}
        \sum_{k=L}^{\infty}\frac{(L/C)^k}{k!}\nTr(|H|^k)
        \leq\eta^{-1}2^{1-L}
    \end{equation}
    with probability at least $1-\delta-\eta-e^{-\Omega(n)}$.

    On the same event, for every integer $r\geq1$, all $n$-qubit operators $U_1,\ldots,U_r$ with $\|U_j\|\leq1$, and all $z_1,\ldots,z_r\in\C$ satisfying $\sum_j|z_j|\leq L/C$,
    \begin{equation}\label{eq:thermal-product-taylor}
        \left|\nTr\left(e^{iz_1H}U_1\cdots e^{iz_rH}U_r\right)-\sum_{\substack{h_1,\ldots,h_r\geq0\\h_1+\cdots+h_r<L}}\left(\prod_{j=1}^r\frac{(iz_j)^{h_j}}{h_j!}\right)\nTr\left(H^{h_1}U_1\cdots H^{h_r}U_r\right)\right|\leq\eta^{-1}2^{1-L}.
    \end{equation}
\end{lemma}
\begin{proof}
    Condition on a support satisfying \eqref{eq:thermal-commutation-matrix-norm}. By \cref{lem:thermal-sign-averaged-moments}, for every even integer $k'\geq2$,
    \begin{equation}
        \left(\E_{\va{J}}\nTr(H^{k'})\right)^{1/k'}
        \leq2\sqrt{1+\frac{C_0k'}{2\sqrt m}}.
    \end{equation}
    For an arbitrary integer $k\geq1$, apply H\"older's inequality to the normalized trace and the sign expectation, with $k'=2\lceil k/2\rceil$, to obtain
    \begin{equation}\label{eq:thermal-all-absolute-moments}
        \left(\E_{\va{J}}\nTr(|H|^k)\right)^{1/k}
        \leq C_1\left(1+\frac{\sqrt k}{m^{1/4}}\right).
    \end{equation}
    For $k\geq L$, the bound $k!\geq(k/e)^k$ and the assumption $m\geq L^2$ imply
    \begin{equation}
        \left(\frac{(L/C)^k}{k!}\E_{\va{J}}\nTr(|H|^k)\right)^{1/k}
        \leq\frac{eC_1L}{C}\left(\frac1k+\frac1{m^{1/4}\sqrt k}\right)
        \leq\frac{2eC_1}{C}\leq\frac12,
    \end{equation}
    for a sufficiently large $C$. Monotone convergence and summation of the geometric series give
    \begin{equation}
        \E_{\va{J}}\left[\sum_{k=L}^{\infty}\frac{(L/C)^k}{k!}\nTr(|H|^k)\right]
        \leq\sum_{k=L}^{\infty}2^{-k}=2^{1-L}.
    \end{equation}
    Markov's inequality proves \eqref{eq:thermal-absolute-tail} with conditional probability at least $1-\eta$ over $\va{J}$. Combining this with \cref{lem:thermal-commutation-matrix-norm} gives the stated probability over $H$.

    To prove \eqref{eq:thermal-product-taylor}, expand each exponential in its Taylor series and group terms by the total degree $k=h_1+\cdots+h_r$. The normalized Schatten H\"older inequality used in the proof of \cref{lem:common-truncation} gives
    \begin{equation}
        \left|\nTr\left(H^{h_1}U_1\cdots H^{h_r}U_r\right)\right|
        \leq\prod_{j:h_j>0}\left(\nTr(|H|^k)\right)^{h_j/k}
        =\nTr(|H|^k),\qquad k\geq1,
    \end{equation}
    where factors with $h_j=0$ are bounded in operator norm. By the multinomial theorem, the sum of the absolute values of all degree-$k$ terms is at most
    \begin{equation}
        \frac{\bigl(\sum_j|z_j|\bigr)^k}{k!}\nTr(|H|^k)
        \leq\frac{(L/C)^k}{k!}\nTr(|H|^k).
    \end{equation}
    Summing over $k\geq L$ and applying \eqref{eq:thermal-absolute-tail} proves the result simultaneously for all the stated choices.
\end{proof}

Taking $r=1$ and $U_1=\mathbb{I}$ gives the Taylor remainders for the partition function and spectral amplitude; taking $U_1=P_i$ gives the inserted trace needed for thermal expectations. For the TOC and OTOC numerators, combining the thermal factor with an adjacent evolution gives total absolute complex time at most $\beta+2|t|$ and $\beta+4|t|$, respectively. Thus the same event controls the Taylor remainders for $H$ in all four applications. The bound concerns the actual sampled Hamiltonian and uses no loop-free cutoff; the finite-order cumulant assumptions enter when we compare these Taylor coefficients with the auxiliary moments.

\subsection{Partition functions}\label{subsec:thermal-partition}
We begin by computing the partition function, which is the common normalization for the thermal spectral amplitude, correlation functions, and expectation values. Recall that $\Zs(\beta)\equiv \Tr(e^{-\beta H})$, and write
\begin{equation}\label{eq:thermal-normalized-partition}
    z_\beta\equiv\nTr(e^{-\beta H})=2^{-n}\Zs(\beta).
\end{equation}
We approximate $z_\beta$ to relative accuracy; multiplication by $2^n$ gives the same relative accuracy for $\Zs(\beta)$.

The computation uses the same moments $\hat{\mu}_k^{(\ell)}=\bra{0}\Ts_{\ell}^k\ket{0}$ as \cref{alg:sfa-algorithm}. For even $2\leq\ell<L$, define the scalar Taylor polynomial
\begin{equation}\label{eq:thermal-scalar-polynomial}
    \widehat{Z}_L(z)\equiv\sum_{k=0}^{L}\frac{(iz)^k}{k!}\hat{\mu}_k^{(\ell)},\qquad z\in\C.
\end{equation}
Our partition-function algorithm performs steps (1)--(3) of \cref{alg:sfa-algorithm} and replaces its final readout by
\begin{equation}\label{eq:thermal-partition-output}
    \hat{z}_\beta\equiv\widehat{Z}_L(i\beta),\qquad
    \widehat{\Zs}_\beta\equiv2^n\hat{z}_\beta.
\end{equation}
For $m\geq CL^3$, the running time at fixed cutoffs is
\begin{equation}\label{eq:thermal-scalar-cutoff-runtime}
    \Os\left(n\ell\binom{m}{\ell/2}\right).
\end{equation}
The coefficient recurrence costs $\Os(\ell L^2)\leq\Os(L^3)\leq\Os(m)$ and is absorbed by the low-order moment computation. We next bound the error at the scale of the true partition function.

\begin{lemma}[Partition function lower bound]\label{lem:thermal-denominator}
    Let $H$ be Hermitian with $\kappa_2=1$ and all odd free cumulants vanishing through an even order $L\geq4$ and let $\Ks(L)$ be as in \cref{lemma:fock_space_approx}. If
    \begin{equation}\label{eq:thermal-denominator-conditions}
        (L+1)\Ks(L)\leq\log(3/2),\qquad L\geq2\beta+2,
    \end{equation}
    then, for an absolute constant $c>0$,
    \begin{equation}\label{eq:thermal-lower-partition}
        z_\beta\geq\max\left\{1,\frac{c e^{2\beta}}{(1+\beta)^{3/2}}\right\}.
    \end{equation}
\end{lemma}
\begin{proof}
    Lagrange inversion applied to \eqref{eqn:scalar-R-moment-relation}, with the odd cumulants set to zero through even order $k$, gives
    \begin{equation}
        \nTr(H^k)=\frac1{k+1}[u^{k/2}]
        \left(1+u+\sum_{\substack{4\leq k'\leq k\\k'\text{ even}}}\kappa_{k'}u^{k'/2}\right)^{k+1}.
    \end{equation}
    The contribution with no higher-cumulant factor is $C_{k/2}$. If $h\geq1$ higher-cumulant factors are chosen, their product has absolute coefficient sum at most $\Ks(L)^h$, and every coefficient of the remaining $(1+u)^{k+1-h}$ is at most $\binom{k+1}{k/2}$. Consequently,
    \begin{equation}
        |\nTr(H^k)-C_{k/2}|
        \leq C_{k/2}\sum_{h=1}^{k+1}\binom{k+1}{h}\Ks(L)^h
        =C_{k/2}\bigl((1+\Ks(L))^{k+1}-1\bigr).
    \end{equation}
    Under \eqref{eq:thermal-denominator-conditions}, this is at most $C_{k/2}/2$, so $\nTr(H^k)\geq C_{k/2}/2$ throughout the controlled range.

    Since $\nTr(H)=0$, Jensen's inequality gives $z_\beta\geq1$. For $\beta\geq4$, put $k=2\lfloor\beta\rfloor$ and $h=\lfloor\sqrt\beta\rfloor$. Taylor's integral remainder gives $e^x\geq\sum_{k'=0}^{k+1}x^{k'}/k'!$ for every real $x$. Apply this inequality to the eigenvalues of $-\beta H$. All odd moments in the polynomial vanish because $k+1\leq L-1$, and every even moment is nonnegative. Keeping the even degrees $k'$ with $k-2h\leq k'\leq k$ therefore gives
    \begin{equation}
        z_\beta\geq\frac12\sum_{\substack{k-2h\leq k'\leq k\\k'\text{ even}}}\frac{\beta^{k'}}{(k'/2)!(k'/2+1)!}
        \geq\frac{c(h+1)e^{2\beta}}{(1+\beta)^2}
        \geq\frac{c e^{2\beta}}{(1+\beta)^{3/2}},
    \end{equation}
    where Stirling's bound gives the middle inequality uniformly over this range, since $0\leq\beta-k'/2\leq\sqrt\beta+1$ and $k'/2\asymp\beta$. For $0\leq\beta<4$, the same bound follows from $z_\beta\geq1$ after reducing $c$.
\end{proof}

The lower bound converts the common truncation estimate into an error at the scale of the partition function.
\begin{corollary}[Normalization of the truncation error]\label{cor:thermal-common-normalized}
    Under the hypotheses of \cref{lem:thermal-common-truncation}, suppose that \eqref{eq:thermal-lower-partition} holds and that
    \begin{equation}\label{eq:thermal-common-stability}
        2\leq R\leq3,\qquad \beta(R-2)+T\Ks(L)\leq1,\qquad L\geq C(1+T),
    \end{equation}
    for a sufficiently large constant $C$. Then
    \begin{equation}\label{eq:thermal-common-normalized-error}
        \frac{|N-\widehat{N}|}{z_\beta}
        \leq C(1+\beta)^{3/2}\left[T\bigl(\Ks(L)-\Ks(\ell)\bigr)+\eta^{-1}2^{-L}\right].
    \end{equation}
\end{corollary}
\begin{proof}
    Divide the first term of \eqref{eq:thermal-common-truncation} by $z_\beta\geq1$. For the remaining terms, \cref{eq:thermal-lower-partition,eq:thermal-common-stability} give
    \begin{equation}
        \frac{e^{R\beta+T\Ks(L)}}{z_\beta}
        \leq C(1+\beta)^{3/2}e^{\beta(R-2)+T\Ks(L)}
        \leq C(1+\beta)^{3/2}.
    \end{equation}
    Also, $T(R+\Ks(L))\leq3T+1$. Thus $L!\geq(L/e)^L$ and the choice of $L$ bound the auxiliary Taylor terms by $C2^{-L}$. Since $\eta<1$, these terms are absorbed by the displayed bound.
\end{proof}

\begin{lemma}\label{lemma:thermal-scalar-approx}
    Let $H$ be Hermitian with $\kappa_2=1$ and all odd free cumulants vanishing through an even order $L$. Fix even $2\leq\ell<L$ and $t,\beta\geq0$, and put $z=-t+i\beta$. Suppose the Taylor-tail event \eqref{eq:thermal-absolute-tail} holds with parameter $0<\eta<1$, and that
    \begin{equation}\label{eq:thermal-scalar-approx-conditions}
        (L+1)\Ks(L)\leq\log(3/2),\qquad
        L\geq C(1+\beta+t),
    \end{equation}
    for a sufficiently large constant $C$, also large enough for the complex-time domain in \cref{lem:thermal-physical-taylor}. Then, simultaneously for $w=z$ and $w=i\beta$,
    \begin{equation}\label{eq:thermal-scalar-approx}
        \frac{|\widehat{Z}_L(w)-\nTr(e^{iwH})|}{z_\beta}
        \leq C(1+\beta)\bigl(\Ks(L)-\Ks(\ell)\bigr)+C(1+\beta)^{3/2}\eta^{-1}2^{-L}.
    \end{equation}
\end{lemma}
\begin{proof}
    Apply \cref{lem:thermal-common-truncation} with $r=1$, $z_1=w$, $U_1=V_1=\mathbb I$, $\ket{\xi}=\ket{\zeta}=\ket{0}$, and $T=\beta+t$. The moment-matching hypothesis follows from \cref{prop:truncated_fock_space}. The norm bounds in \cref{lemma:fock_space_approx} give the perturbation bounds in the common lemma with $S=a^\dagger+a$, $R=2$, and the scalar sums $\Ks(\ell')$ and $\Ks(L)-\Ks(\ell)$. The assumptions imply $T\Ks(L)\leq1$ and the hypotheses of \cref{lem:thermal-denominator}.

    We sharpen only Step~(2) of the common proof, using the scalar contour identity from \cref{lem:scalar-cutoff-comparison}. Write
    \begin{equation}
        K_{\ell'}(u)\equiv\frac1u+u+\sum_{4\leq k\leq\ell'}\kappa_k u^{k-1},\qquad
        \Delta(u)\equiv K_L(u)-K_{\ell}(u).
    \end{equation}
    The residue identity in that proof extends to every complex time by the same absolutely convergent series. For $w\neq0$, it gives
    \begin{equation}
        \bra{0}e^{iw\Ts_L}\ket{0}-\bra{0}e^{iw\Ts_{\ell}}\ket{0}
        =\frac1{iw}\cdot\frac1{2\pi i}\oint_{|u|=1}e^{iwK_{\ell}(u)}\bigl(e^{iw\Delta(u)}-1\bigr)\,du.
    \end{equation}
    For $u=e^{i\theta}$ and $w\in\{-t+i\beta,i\beta\}$, we have
    \begin{equation}
        |e^{iwK_{\ell}(u)}|\leq e^{-2\beta\cos\theta+|w|\Ks(\ell)},\qquad
        |e^{iw\Delta(u)}-1|\leq |w|\bigl(\Ks(L)-\Ks(\ell)\bigr)e^{|w|\bigl(\Ks(L)-\Ks(\ell)\bigr)}.
    \end{equation}
    Since $1+\cos\theta\geq2(\theta-\pi)^2/\pi^2$ for $0\leq\theta\leq2\pi$, a Gaussian integral bounds the remaining angular integral by
    \begin{equation}
        \frac1{2\pi}\int_0^{2\pi}e^{-2\beta\cos\theta}\,d\theta
        \leq\frac{C e^{2\beta}}{\sqrt{1+\beta}}.
    \end{equation}
    Consequently, the factor $|w|$ cancels and
    \begin{equation}\label{eq:thermal-scalar-cutoff-comparison}
        \left|\bra{0}e^{iw\Ts_L}\ket{0}-\bra{0}e^{iw\Ts_{\ell}}\ket{0}\right|
        \leq\frac{C e^{2\beta+T\Ks(L)}}{\sqrt{1+\beta}}\bigl(\Ks(L)-\Ks(\ell)\bigr).
    \end{equation}
    The same estimate holds at $w=0$, when the difference vanishes. Dividing by \eqref{eq:thermal-lower-partition} and using $T\Ks(L)\leq1$ bounds the normalized middle hybrid by $C(1+\beta)\bigl(\Ks(L)-\Ks(\ell)\bigr)$. Steps~(1) and~(3) are unchanged; as in the proof of \cref{cor:thermal-common-normalized}, their normalized Taylor errors are at most $C(1+\beta)^{3/2}\eta^{-1}2^{-L}$. Adding the three errors proves \eqref{eq:thermal-scalar-approx} for both values of $w$.
\end{proof}

We next choose the cutoffs and perform the running-time optimization used in all four applications.
\begin{definition}[Free cumulant cutoffs]\label{def:truncation-cutoffs}
    Suppose $n\leq m\leq\poly(n)$, $0<\varepsilon<1$, and $\delta=1/\poly(n)$. Let $C_0$ be a sufficiently large constant and $T\geq\beta\geq0$ satisfy 
    \begin{equation}
        T+\log(n/\varepsilon)\leq c\min\left\{m^{1/3},\frac n{\log n}\right\},
    \end{equation}
    for sufficiently small constant $c>0$.
    Then we define the \emph{colored cutoff} as the smallest positive even integer $\ell$ satisfying
    \begin{equation}\label{eq:thermal-common-cumulant-cutoff}
        (1+\beta)^{3/2}(1+T)\delta^{-1/2}
        \left(\frac{C_0(\ell+2)}m\right)^{(\ell+2)/4}
        \leq\frac{\varepsilon}{C_0},
    \end{equation}
    and the \emph{scalar cutoff} as the smallest positive even integer $\ell$ satisfying
    \begin{equation}\label{eq:thermal-scalar-cumulant-cutoff}
        (1+\beta)\delta^{-1/2}\left(\frac{C_0(\ell+2)}m\right)^{(\ell+2)/4}\leq\frac{\varepsilon}{C_0}.
    \end{equation}
    In either case, we choose an even integer
    \begin{equation}\label{eq:thermal-common-moment-cutoff}
        L=\Theta\bigl(T+\log(n/\varepsilon)\bigr),\qquad L\geq\ell+4,
    \end{equation}
    for a sufficiently large constant.
\end{definition}

\begin{proposition}[Free cumulant cutoff properties]\label{prop:thermal-cutoff-choice}
    We have the following properties of $\ell$ and $L$ defined in \cref{def:truncation-cutoffs}:
    \begin{enumerate}
        \item $\ell=\Os\left(1+\frac{\log(1/\varepsilon)}{\log m}\right)$, $m\geq CL^3$, $L+1\leq\gamma n/\log n$, $L\sqrt{\log n/m}=o(1)$, and $(L+1)\sqrt{\log n}/m=o(1)$.
        \item If
        \begin{equation}\label{eq:thermal-common-cumulant-tail}
            \Ks(L)-\Ks(\ell) \leq \delta^{-1/2}\left(\frac{\Os(\ell+2)}m\right)^{(\ell+2)/4},
        \end{equation}
        then, with $\eta=\delta$ and after choosing these constants sufficiently large, the right-hand side of \eqref{eq:thermal-common-normalized-error} (colored cutoff) and \eqref{eq:thermal-scalar-approx} (scalar cutoff), respectively, is at most $\varepsilon/4$.
    
        \item Moreover, beyond outputting the semicircular background, a computation using 
        \begin{equation}
            \Os\left(n\poly(\ell) C_1^{\ell}\binom{m}{\ell/2}\right)
        \end{equation}
        operations, for fixed constant $C_1>0$, has running time at most
        \begin{equation}\label{eq:thermal-common-runtime}
            \Os\left(\frac{n}{\delta}(1+\beta)^3\right)(1+T)^2(1/\varepsilon)^{2+\Os(1/\log m)}
        \end{equation}
        for the colored cutoff and 
        \begin{equation}\label{eq:thermal-scalar-optimized-runtime}
            \Os\left(\frac{n}{\delta}(1+\beta)^2\right)(1/\varepsilon)^{2+\Os(1/\log m)}
        \end{equation}
        for the scalar cutoff.
    \end{enumerate}
\end{proposition}
\begin{proof}
    Since $n\leq m$ and $\delta=1/\poly(n)$, $(1+\beta)^{3/2}(1+T)\delta^{-1/2}=m^{\Os(1)}$ and $\log(1/\varepsilon)=\Os(m^{1/3})$, so the logarithm of the prefactor in \eqref{eq:thermal-common-cumulant-cutoff} is $\Os(\log m)$. As in \eqref{eq:optimized-cutoff-size}, a sufficiently large constant multiple of $1+\log(1/\varepsilon)/\log m$ satisfies the cutoff inequality: for such a candidate, $C_0(\ell+2)\leq m^{1/2}$, so its power of $m^{-1/2}$ absorbs the prefactor and the required accuracy. This proves $\ell=\Os\left(1+\frac{\log(1/\varepsilon)}{\log m}\right)$. The choice of $L$ and a sufficiently small $c$ give the claimed bounds on $m$ and $L+1$. Since $L=\Os(m^{1/3})$ and $m\geq n$, both stated stability quantities tend to zero.

    \paragraph{Approximation.}
    The cumulant term in \eqref{eq:thermal-common-normalized-error} is at most $\varepsilon/8$ by \cref{eq:thermal-common-cumulant-cutoff,eq:thermal-common-cumulant-tail}, after increasing $C_0$ to absorb the fixed constants. The Taylor term is at most $\varepsilon/8$ because
    \begin{equation}
        C(1+\beta)^{3/2}\delta^{-1}2^{-L}\leq\varepsilon/8
    \end{equation}
    for the choice \eqref{eq:thermal-common-moment-cutoff}. Thus their sum is at most $\varepsilon/4$. The scalar
    estimate follows in the same way from
    \eqref{eq:thermal-scalar-cumulant-cutoff} and
    \eqref{eq:thermal-scalar-approx}.

    \paragraph{Running time.}
   Minimality of the colored cutoff gives
    \begin{equation}
        \binom{m}{\ell/2}
    \leq C_0^2(2eC_0)^{\ell/2}(1+\beta)^3(1+T)^2\delta^{-1}\varepsilon^{-2}.
    \end{equation}
    Here we use the failure of the preceding cutoff $\ell-2$ and $\binom m{\ell/2}\leq(2em/\ell)^{\ell/2}$.
    Substituting this into the operation count and absorbing the fixed polynomial in $\ell$ into $e^{\Os(\ell)}$ gives
    \begin{equation}
        \Os\left(\frac{n}{\delta}(1+\beta)^3(1+T)^2\varepsilon^{-2}e^{\Os(\ell)}\right)
        \leq\Os\left(\frac{n}{\delta}(1+\beta)^3(1+T)^2(1/\varepsilon)^{2+\Os(1/\log m)}\right),
    \end{equation}
    since $\ell=\Os\left(1+\frac{\log(1/\varepsilon)}{\log m}\right)$. This proves \eqref{eq:thermal-common-runtime}.

    The same cutoff-size argument applies to \eqref{eq:thermal-scalar-cumulant-cutoff}. For the remaining scalar cutoffs, failure of the preceding cutoff gives
    \begin{equation}
        \binom m{\ell/2}\leq C_0^2(2eC_0)^{\ell/2}(1+\beta)^2\delta^{-1}\varepsilon^{-2}.
    \end{equation}
    The same substitution in the operation count proves \eqref{eq:thermal-scalar-optimized-runtime}.
\end{proof}

\begin{theorem}[Partition function algorithm]\label{thm:thermal-partition}
    There exists a constant $c>0$ such that, for sufficiently large $n$, $2n\leq m\leq \poly(n)$, $0<\varepsilon < 1$, $\delta=1/\poly(n)$, and $\beta\geq0$ satisfying
    \begin{equation}\label{eq:thermal-partition-window}
        \beta+\log(n/\varepsilon)
        \leq c\min\left\{m^{1/3},\frac{n}{\log n}\right\},
    \end{equation}
    the cutoffs in \eqref{eq:thermal-partition-output} can be chosen so that
    \begin{equation}\label{eq:thermal-partition-relative-error}
        |\hat{z}_\beta-z_\beta|\leq\varepsilon z_\beta,
        \qquad
        |\widehat{\Zs}_\beta-\Zs(\beta)|\leq\varepsilon\Zs(\beta),
    \end{equation}
    with probability at least $1-\Os(\delta)-e^{-\Omega(n)}$ and running time at most
    \begin{equation}\label{eq:thermal-partition-runtime}
        \Os(\frac{n}{\delta}(1+\beta)^2)(1/\varepsilon)^{2+\Os(1/\log m)}.
    \end{equation}
\end{theorem}
\begin{proof}
    Choose the scalar cutoffs \cref{eq:thermal-scalar-cumulant-cutoff,eq:thermal-common-moment-cutoff} from \cref{prop:thermal-cutoff-choice} with $T=\beta$. Apply \eqref{eq:hypercontractive-scalar-tail} at order $4$ and \eqref{eq:cumulant-tail-probability} at lower order $\ell+2$, with failure probabilities of order $\delta$. Together with loop-freeness through order $L$, these give
    \begin{equation}\label{eq:thermal-scalar-cumulant-event}
        \Ks(L)\leq\Os\left(\frac{\sqrt{\log n}}m\right),\qquad
        \bigl(\Ks(L)-\Ks(\ell)\bigr)
        \leq \delta^{-1/2}\left(\frac{\Os(\ell+2)}m\right)^{(\ell+2)/4}
    \end{equation}
    with probability at least $1-\Os(\delta)-e^{-\Omega(n)}$. Intersect this event with \cref{lem:thermal-physical-taylor}, taking $\eta=\delta$. The cutoff proposition ensures its hypotheses and gives $(L+1)\Ks(L)=o(1)$, so \eqref{eq:thermal-denominator-conditions} holds for sufficiently large $n$. Thus \cref{lemma:thermal-scalar-approx} and the same proposition bound the relative error by $\varepsilon$. Multiplication by $2^n$ gives the guarantee for $\Zs(\beta)$.

    For the running time, apply \eqref{eq:thermal-scalar-optimized-runtime} to \eqref{eq:thermal-scalar-cutoff-runtime}. This gives \eqref{eq:thermal-partition-runtime}.
\end{proof}

The same approximation $\hat{z}_\beta$ can be reused for all observables at the chosen inverse temperature. We record the normalization estimate used in each application.
\begin{lemma}[Normalization of approximate numerators]\label{lem:thermal-ratio}
    Suppose $D>0$, $|N|\leq D$, and
    \begin{equation}
        |\widehat{D}-D|\leq\eta D,\qquad
        |\widehat{N}-N|\leq\xi D,\qquad 0\leq\eta<1,\quad \xi\geq0.
    \end{equation}
    Then $|\widehat{D}|\geq(1-\eta)D$ and
    \begin{equation}\label{eq:thermal-quotient-error}
        \left|\frac{\widehat{N}}{\widehat{D}}-\frac ND\right|
        \leq\frac{\xi+\eta}{1-\eta}.
    \end{equation}
    If $\widehat{D}$ is real, it is positive. In particular, $\xi,\eta\leq\varepsilon/4$ imply additive error at most $\varepsilon$ for $0<\varepsilon\leq1$.
\end{lemma}
\begin{proof}
    The reverse triangle inequality gives $|\widehat{D}|\geq(1-\eta)D$. Using $|N|\leq D$,
    \begin{equation}
        \left|\frac{\widehat{N}}{\widehat{D}}-\frac ND\right|
        \leq\frac{|\widehat{N}-N|}{|\widehat{D}|}
        +\frac{|N|\,|D-\widehat{D}|}{D|\widehat{D}|}
        \leq\frac{\xi+\eta}{1-\eta}.
    \end{equation}
\end{proof}
For $D=z_\beta$ and $N=\nTr(Ue^{-\beta H})$ with $\|U\|\leq1$, the inequality $|N|\leq D$ follows from the trace norm bound. Thus \cref{lem:thermal-ratio} applies to all four observables. We set each normalized algorithm's output to zero if the computed partition function is nonpositive; the approximation event ensures that this case does not occur.

\subsection{Correlation functions}\label{subsec:thermal-correlators}
We now extend \cref{sec:toc-otoc} to thermal time-ordered and out-of-time-ordered correlators. We retain the convention $A(t)=e^{iHt}Ae^{-iHt}$, the color words $\Cm_{h_0,h_1}$ and $\Cm_{\va{h}}$, and the multiplicity notation $|\va{h}|$ and $\va{h}!$ from \cref{eq:toc-scalar-recurrence,eq:otoc-scalar-recurrence}. With the common normalization $z_\beta=\nTr(e^{-\beta H})$, define
\begin{equation}\label{eq:thermal-correlator-numerators}
    \begin{aligned}
        N_\beta^A(t)&\equiv z_\beta G_\beta^A(t)=\nTr\big(e^{(it-\beta)H}Ae^{-itH}A\big),\\
        N_\beta^{A,B}(t)&\equiv z_\beta F_\beta^{A,B}(t)=\nTr\big(e^{(it-\beta)H}Ae^{-itH}Be^{itH}Ae^{-itH}B\big).
    \end{aligned}
\end{equation}
Thus the thermal factor changes only the coefficient of the first stretch of $H$. Moving the insertions to the right gives the same color words as at $\beta=0$, including the boundary phase $ABAB=\langle A,B\rangle\mathbb I$ for the OTOC.

For even integers $2\leq\ell<L$, compute the truncated color moments $\hat{\mu}^{(\ell;\Cm)}$ by the same recurrences as in \cref{alg:time-ordered-correlator,alg:otoc}, and set
\begin{equation}\label{eq:thermal-toc-polynomial}
    \hat N_\beta^A(t)\equiv\sum_{\substack{h_0,h_1\geq0\\h_0+h_1\leq L}}
    \frac{(it-\beta)^{h_0}(-it)^{h_1}}{h_0!\,h_1!}\,\hat{\mu}^{(\ell;\Cm_{h_0,h_1})},
\end{equation}
\begin{equation}\label{eq:thermal-otoc-polynomial}
    \hat N_\beta^{A,B}(t)\equiv\langle A,B\rangle\sum_{\substack{\va{h}\in\N^4\\|\va{h}|\leq L}}
    \frac{(it-\beta)^{h_{00}}(-it)^{h_{10}}(it)^{h_{11}}(-it)^{h_{01}}}{\va{h}!}\,\hat{\mu}^{(\ell;\Cm_{\va{h}})}.
\end{equation}
Using the partition-function approximation \eqref{eq:thermal-partition-output}, the outputs are
\begin{equation}\label{eq:thermal-correlator-outputs}
    \hat G_\beta^A(t)\equiv\frac{\hat N_\beta^A(t)}{\hat z_\beta},\qquad
    \hat F_\beta^{A,B}(t)\equiv\frac{\hat N_\beta^{A,B}(t)}{\hat z_\beta}.
\end{equation}
As above, we return zero if $\hat z_\beta\leq0$; the success event below guarantees $\hat z_\beta>0$. The same value of $\hat z_\beta$ is used for both correlators.

\paragraph{Approximation error of truncations.}
We use the color hopping operators $\Ts_\ell=\Ts_\ell^{(\Qm)}$ and the sums $\Ks^{\mathrm{col}}(L)$ and $\Ks^{\mathrm{col}}(L)-\Ks^{\mathrm{col}}( \ell)$ from \eqref{eq:color-K-sums}. As in \cref{lemma:color-fock-space-approx}, the hopping operator in the analysis contains every color word, whereas the algorithms compute only the words in \cref{eq:thermal-toc-polynomial,eq:thermal-otoc-polynomial}.

\begin{lemma}[Approximation error of thermal correlator truncations]\label{lemma:thermal-color-fock-space-approx}
    Let $H$ be Hermitian with $\kappa_2=1$ and all odd color cumulants vanishing through order $L$, where $2\leq\ell<L$ are even. Take $\Qm=(A)$ and $r=2$ for the TOC, or $\Qm=(A,B)$ and $r=4$ for the OTOC, and write $N_\beta(t)$ and $\hat N_\beta(t)$ for the corresponding numerator and polynomial. Set $T\equiv\beta+r|t|$. Suppose that \eqref{eq:thermal-absolute-tail} holds with parameter $\eta$, that the scalar assumptions \eqref{eq:thermal-denominator-conditions} hold, and that
    \begin{equation}\label{eq:thermal-color-stability}
        T\Ks^{\mathrm{col}}(L)\leq1,\qquad L\geq C(1+T),
    \end{equation}
    for a sufficiently large constant $C$. Then
    \begin{equation}\label{eq:thermal-color-normalized-error}
        \frac{|N_\beta(t)-\hat N_\beta(t)|}{z_\beta}
        \leq C(1+\beta)^{3/2}\left[T\bigl(\Ks^{\mathrm{col}}(L)-\Ks^{\mathrm{col}}( \ell)\bigr)+\eta^{-1}2^{-L}\right].
    \end{equation}
\end{lemma}
\begin{proof}
    For the TOC, apply \cref{lem:thermal-common-truncation} with $s=1$, $(z_1,z_2)=(t+i\beta,-t)$, $(U_1,U_2)=(A,A)$, and $(V_1,V_2)=(b,b)$. For the OTOC, take $s=2$, $(z_1,z_2,z_3,z_4)=(t+i\beta,-t,t,-t)$, $(U_1,U_2,U_3,U_4)=(A,B,A,B)$, and $(V_1,V_2,V_3,V_4)=(b_1,b_2,b_1,b_2)$. In both cases, use $\ket{\xi}=\ket{\zeta}=\ket{\Omega_s}$. The total absolute complex time is at most $T$, and the sum of the absolute imaginary parts is $\beta$.

    By \cref{lem:truncated-color-moments}, the moment-matching hypothesis holds, and the auxiliary degree-$L$ polynomials are exactly \cref{eq:thermal-toc-polynomial,eq:thermal-otoc-polynomial}, including the OTOC boundary phase. The bounds \eqref{eqn:color-norm-upper} supply the common lemma's perturbation estimates with $S=a^\dagger+a$, $R=2$, and $\Ks^{\mathrm{col}}(\ell'), \Ks^{\mathrm{col}}(L)-\Ks^{\mathrm{col}}( \ell)$ in place of the scalar sums. Finally, \cref{lem:thermal-denominator} gives \eqref{eq:thermal-lower-partition}, so \cref{cor:thermal-common-normalized} proves the claim.
\end{proof}

For real $t$, the products $A(t)A$ and $A(t)BA(t)B$ are unitary, so $|N_\beta^A(t)|,|N_\beta^{A,B}(t)|\leq z_\beta$ by H\"older's inequality. Consequently, \cref{lem:thermal-ratio} converts \eqref{eq:thermal-color-normalized-error} and the relative error of $\hat z_\beta$ directly into additive error for \eqref{eq:thermal-correlator-outputs}.

\paragraph{The thermal correlator algorithms.}
The first three steps of \cref{alg:time-ordered-correlator,alg:otoc} are unchanged. We replace their return expressions by \eqref{eq:thermal-correlator-outputs} and compute the common denominator from \eqref{eq:thermal-partition-output} with the same cutoffs. The closure of the two color-word families under subsequences, translation, and reversal is exactly the one established in \cref{subsec:correlator-algorithms}; changing the Taylor coefficients introduces no additional color moments or cumulants.

\begin{theorem}[Thermal correlator algorithms]\label{thm:thermal-correlator-algorithm}
    Fix any non-identity Pauli insertions $A\neq\mathbb I$ for TOC or $A,B$  independent modulo phase for the OTOC. There is a constant $c>0$ such that, for sufficiently large $n$, $2n\leq m\leq\poly(n)$, $0<\varepsilon<1$, $\delta=1/\poly(n)$, $\beta\geq0$, and $t\geq 0$ satisfying
    \begin{equation}\label{eq:thermal-correlator-window}
        \beta+t+\log(n/\varepsilon)
        \leq c\min\left\{m^{1/3},\frac{n}{\log n}\right\},
    \end{equation}
    the thermal versions of \cref{alg:time-ordered-correlator,alg:otoc} give
    \begin{equation}\label{eq:thermal-correlator-accuracy}
        |\hat G_\beta^A(t)-G_\beta^A(t)|\leq\varepsilon,\qquad
        |\hat F_\beta^{A,B}(t)-F_\beta^{A,B}(t)|\leq\varepsilon
    \end{equation}
    with probability at least $1-\Os(\delta)-e^{-\Omega(n)}$ over $H$. Computing either correlator, or both with the same denominator, has total running time
    \begin{equation}\label{eq:thermal-correlator-runtime}
        \Os\left(\frac{n}{\delta}(1+\beta)^3(1+\beta+t)^2\right)(1/\varepsilon)^{2+\Os(1/\log m)},
    \end{equation}
    including the cost of the partition-function approximation.
\end{theorem}
\begin{proof}
    Choose the cutoffs from \cref{prop:thermal-cutoff-choice} with $T=\beta+4t$. This also covers the TOC, whose total absolute complex time is at most $\beta+2t$.

    \paragraph{Approximation.}
    Apply \cref{cor:color-hypercontractive-sums}, which is valid at $\ell$. On its event and the loop-free event through order $L$,
    \begin{equation}\label{eq:thermal-color-concentration}
        \begin{aligned}
            \Ks^{\mathrm{col}}(L)-\Ks^{\mathrm{col}}( \ell)
            &\leq \delta^{-1/2}\left(\frac{\Os(\ell+2)}m\right)^{(\ell+2)/4},\\
            \Ks^{\mathrm{col}}(L)
            &\leq\Os\left(\sqrt{\frac{\log n}m}\right).
        \end{aligned}
    \end{equation}
    By \cref{prop:thermal-cutoff-choice}, $T\leq L$ and $L\sqrt{\log n/m}=o(1)$, so $T\Ks^{\mathrm{col}}(L)\leq1$ for sufficiently large $n$. Separately, the scalar concentration bound \eqref{eq:thermal-scalar-cumulant-event} and $(L+1)\sqrt{\log n}/m=o(1)$ give the hypotheses of \cref{lem:thermal-denominator}.

    Intersect these events with \cref{lem:thermal-physical-taylor}, taking $\eta=\delta$. By \cref{lemma:thermal-color-fock-space-approx,prop:thermal-cutoff-choice}, each numerator error is at most $\varepsilon z_\beta/4$. Since
    \[
        \Ks(L)-\Ks(\ell)\leq\Ks^{\mathrm{col}}(L) -\Ks^{\mathrm{col}}(\ell),
    \]
    the same cutoffs give $|\hat z_\beta-z_\beta|\leq\varepsilon z_\beta/4$ by \cref{lemma:thermal-scalar-approx}. Applying \cref{lem:thermal-ratio} proves \eqref{eq:thermal-correlator-accuracy}. A union bound gives the stated probability for the fixed probes in the theorem.

    \paragraph{Running time.}
    By the proof of \cref{prop:correlator-cutoff-runtime}, the low-order computations, continuation, and thermal readout cost $\Os\bigl(n\ell^5\binom {m}{\ell/2}\bigr)$, with $r=4$ covering both correlators. The all-zero color words supply the scalar data for the shared denominator, whose cost in \eqref{eq:thermal-scalar-cutoff-runtime} is also absorbed. Applying \cref{prop:thermal-cutoff-choice} gives \eqref{eq:thermal-correlator-runtime}.
\end{proof}

\subsection{Thermal expectation values}\label{subsec:thermal-expectation-analysis}
We now analyze the thermal expectation construction from \cref{sec:thermal-expectations}. Fix a sampled non-identity Pauli $P_i$, and recall the residual $H^{(i)}$ and the two moment series $\alpha_{0,i}^{(\ell)},\alpha_{1,i}^{(\ell)}$ from \cref{eq:thermal-residual,prop:thermal-moment-readout}. We use the shared approximation $\hat{z}_\beta$ from \eqref{eq:thermal-partition-output} with the same cutoffs for the final denominator in \cref{alg:thermal-expectation}. Thus the readout is
\begin{equation}\label{eq:thermal-expectation-shared-output}
    \widehat{N}_{\beta,i}\equiv\sum_{k=0}^L\frac{(-\beta)^k}{k!}[z^k]\alpha_{1,i}^{(\ell)}(z),
    \qquad
    \widehat{\langle P_i\rangle}_\beta\equiv\frac{\widehat{N}_{\beta,i}}{\hat{z}_\beta}.
\end{equation}
The two series are still computed together, since their recurrences are coupled. The residual decomposition reconstructs the original Hamiltonian $H$, so its normalizing partition function is the same $z_\beta$ as for the other observables.

For $m\geq2$, recall the normalized residual $\widetilde H^{(i)}$ and scaled spiked hopping operator $\Ts_{\ell',i}$ from \cref{eq:thermal-normalized-residual,eq:thermal-expectation-hopping}. Let $\Ks_i^{\mathrm{col}}(\ell')$ and $\Ks_i^{\mathrm{col}}(L)-\Ks_i^{\mathrm{col}}(\ell)$ denote the sums from \eqref{eq:color-K-sums} for $\widetilde H^{(i)}$ and generator $\Qm=(P_i)$, including the mixed quadratic cumulant. If the full support of $H$ is loop-free through order $L+1$, then \eqref{eq:thermal-expectation-fock-readout} with $\ell'=L$ gives the physical moments $\nTr(P_i^cH^k)$ for $k\leq L$.

\begin{lemma}[Approximation error for thermal expectations]\label{lemma:thermal-expectation-approx}
    Suppose that $H$ is loop-free through order $L+1$, where $2\leq\ell<L$ are even, and that \eqref{eq:thermal-absolute-tail} holds with parameter $\eta$. Suppose also that
    \begin{equation}\label{eq:thermal-expectation-stability}
        \begin{gathered}
            (L+1)\Ks(L)\leq\log(3/2),\qquad L\geq C(1+\beta),\\
            \beta\bigl(m^{-1/2}+\Ks_i^{\mathrm{col}}(L)\bigr)\leq1,
        \end{gathered}
    \end{equation}
    for a sufficiently large constant $C$. Then
    \begin{equation}\label{eq:thermal-expectation-gibbs-error}
        \frac{|\nTr(P_ie^{-\beta H})-\widehat{N}_{\beta,i}|}{z_\beta}
        \leq C(1+\beta)^{3/2}\left[\beta\bigl(\Ks_i^{\mathrm{col}}(L)-\Ks_i^{\mathrm{col}}(\ell)\bigr)+\eta^{-1}2^{-L}\right].
    \end{equation}
\end{lemma}
\begin{proof}
    Apply \cref{lem:thermal-common-truncation} with $r=1$, $z_1=i\beta$, $T=\beta$, $U_1=P_i$, $V_1=\mathbb I$, $\ket{\xi}=b^\dagger\ket{\Omega_1}$, and $\ket{\zeta}=\ket{\Omega_1}$. Thus the auxiliary matrix element is $\bra{\Omega_1}b e^{-\beta\Ts_{\ell',i}}\ket{\Omega_1}$, and \eqref{eq:thermal-expectation-fock-readout} gives the required moment matching and the output polynomial.

    The scaled semicircular part and the spike $J_i b/\sqrt m$ form a common self-adjoint operator $S$ of norm at most $R=2+m^{-1/2}$. Scaling \eqref{eq:spiked-norm-upper} gives
    \begin{equation}
        \|\Ts_{\ell',i}-S\|\leq\Ks_i^{\mathrm{col}}(\ell'),\qquad
        \|\Ts_{L,i}-\Ts_{\ell,i}\|\leq\Ks_i^{\mathrm{col}}(L)-\Ks_i^{\mathrm{col}}(\ell).
    \end{equation}
    The common lemma therefore applies with these residual color sums. The scalar assumptions give \eqref{eq:thermal-lower-partition} by \cref{lem:thermal-denominator}, while the remaining assumption is exactly $\beta(R-2)+T\Ks_i^{\mathrm{col}}(L)\leq1$. Hence \cref{cor:thermal-common-normalized} proves the result.
\end{proof}

Since $|\nTr(P_ie^{-\beta H})|\leq z_\beta$, \cref{lem:thermal-ratio} converts this bound and the relative accuracy of $\hat{z}_\beta$ into additive accuracy for the expectation.

\begin{theorem}[Thermal expectation approximation]\label{thm:thermal-expectation}
    Fix any index $i\in[m]$. There is a sufficiently small constant $c>0$ such that, for sufficiently large $n$, $2n\leq m\leq\poly(n)$, $0<\varepsilon<1$, $\delta=1/\poly(n)$, $\beta\geq0$, and
    \begin{equation}\label{eq:thermal-expectation-window}
        \beta+\log(n/\varepsilon)
        \leq c\min\left\{m^{1/3},\frac{n}{\log n}\right\},
    \end{equation}
    the cutoffs in \eqref{eq:thermal-expectation-shared-output} can be chosen so that
    \begin{equation}
        \left|\widehat{\langle P_i\rangle}_\beta-\langle P_i\rangle_\beta\right|\leq\varepsilon
    \end{equation}
    with probability at least $1-\Os(\delta)-e^{-\Omega(n)}$. Including the partition-function computation, the running time is at most
    \begin{equation}\label{eq:thermal-expectation-accuracy-runtime}
        \Os(\frac{n}{\delta}(1+\beta)^5)(1/\varepsilon)^{2+\Os(1/\log m)}.
    \end{equation}
\end{theorem}
\begin{proof}
    Since $m-1\geq n$, apply \cref{prop:thermal-cutoff-choice} with $T=\beta$ and $m$ replaced by $m-1$. The parameter range \eqref{eq:thermal-expectation-window} implies the proposition's window after reducing its constant, since $m-1\asymp m$. In particular, $m-1\geq CL^3$ and $L+1\leq\gamma n/\log n$.

    \paragraph{Approximation.}
    The full support is loop-free through $L+1$ except with probability $e^{-\Omega(n)}$. Conditional on a non-identity value of $P_i$, the remaining Pauli terms are independent uniform strings. \cref{cor:color-hypercontractive-sums}, applied to $\widetilde H^{(i)}$, gives
    \begin{equation}\label{eq:thermal-expectation-cumulant-event}
        \begin{aligned}
            \Ks_i^{\mathrm{col}}(L)-\Ks_i^{\mathrm{col}}(\ell)
            &\leq \delta^{-1/2}\left(\frac{\Os(\ell+2)}{m-1}\right)^{(\ell+2)/4},\\
            \Ks_i^{\mathrm{col}}(L)
            &\leq\Os\left(\sqrt{\frac{\log n}{m-1}}\right)
        \end{aligned}
    \end{equation}
    except with probability $\Os(\delta)+e^{-\Omega(n)}$, uniformly over the conditioned non-identity probe. The scalar estimate for the original $H$ gives $\Ks(L)\leq\Os(\sqrt{\log n}/m)$. These estimates imply \eqref{eq:thermal-expectation-stability}: the cutoff proposition gives $(L+1)\sqrt{\log n}/m=o(1)$, while $\beta\leq L$, $m-1\geq CL^3$, and the polynomial lower bound on $m-1$ give $\beta(m^{-1/2}+\sqrt{\log n/(m-1)})=o(1)$.

    Apply \cref{lem:thermal-physical-taylor} to the original Hamiltonian with $\eta=\delta$. On the common event, \cref{lemma:thermal-expectation-approx,prop:thermal-cutoff-choice} give numerator error at most $\varepsilon z_\beta/4$. The same cutoffs suffice for the shared scalar denominator by \cref{lemma:thermal-scalar-approx}, since its cumulant-tail upper bound has $m$ in place of $m-1$. The ratio lemma proves the additive-error guarantee, and a union bound gives the stated probability.

    \paragraph{Running time.}
    By \cref{prop:thermal-runtime}, the numerator, including both coupled series, costs
    \begin{equation}
        \Os\left(n\ell2^{\ell}\binom{m-1}{\ell/2}\right),
    \end{equation}
    using $m-1\geq CL^3$. The scalar denominator cost is absorbed: its low-order moment computation omits the factor $2^{\ell}$, and
    \begin{equation}
        \frac{\binom m{\ell/2}}{\binom{m-1}{\ell/2}}=\frac{m}{m-\ell/2}\leq2.
    \end{equation}
    Its moment continuation and evaluation are also smaller than the displayed cost. Applying \cref{prop:thermal-cutoff-choice} with $m-1$ in place of $m$ gives \eqref{eq:thermal-expectation-accuracy-runtime}.
\end{proof}

\section{Even-parity crossing partition counts}\label{sec:combinatorial-proofs}
We prove \cref{lem:even-block-count}, using the sets $Z_{q,b}$ defined in \cref{sec:free-cumulant-concentration}. The proof has three steps: we bound the number of pairs of chord diagrams with the same intersection graph (\cref{lem:chord-word-second-moment}), then the number of partitions with all blocks of size four (\cref{lem:quartic-even-crossing-count}), and finally compare larger even blocks with blocks of size four (\cref{lem:even-label-insertion}). Throughout, $C>0$ denotes an absolute constant which may increase between inequalities.

\paragraph{Words and inversion parities.}
We represent a block-labeled partition by the word $w$ obtained by reading its block labels in order. For a label $a$ let $V_a(w)$ be the set of positions carrying $a$. For distinct labels $a,b$ define
\begin{equation}\label{eq:word-inversion-parity}
    \operatorname{inv}_{ab}(w)\equiv\operatorname{inv}\big(V_a(w),V_b(w)\big)\bmod 2,
    \qquad\text{i.e.\ the parity of }\big|\{(i,j):i<j,\ w_i=b,\ w_j=a\}\big|.
\end{equation}
When every label has even multiplicity, \cref{lem:even-word-trace} shows that $\operatorname{inv}_{ab}(w)$ is the parity of the crossing number of $V_a(w)$ and $V_b(w)$, so the odd-intersection graph $g(w)$ is empty precisely when every $\operatorname{inv}_{ab}(w)$ vanishes. We call such a word \emph{balanced}. Being balanced is invariant under permuting the labels. It is also invariant under reversing a word all of whose multiplicities are even, since reversal replaces $\operatorname{inv}(V_a,V_b)$ by $|V_a|\,|V_b|-\operatorname{inv}(V_a,V_b)$. In this language, $b!\,|Z_{q,b}|$ is the number of balanced words on $b$ labels in which every label occurs an even number $\geq4$ of times, with total length $2q$.

\subsection{Blocks of size two}
For a graph $G$ on the labeled vertex set $[b]$, let $N_b(G)$ denote the number of words with two copies of each label and intersection graph exactly $G$. Such words are block-labeled chord diagrams, so the labeled/unlabeled identities \cref{eq:labeled-automorphism-factor,eq:labeled-factorial-factor} apply to them.

\begin{lemma}[Chord diagrams with identical intersection graphs]
    \label{lem:chord-word-second-moment}
    There exists an absolute constant $C>0$ such that
    \begin{equation}\label{eq:chord-word-second-moment}
        \sum_{G\text{ on }[b]}N_b(G)^2\leq C^b(b!)^2
        \qquad(b\geq1).
    \end{equation}
\end{lemma}
\begin{proof}
    \emph{Connected graphs.} We use the canonical split-tree representation of circle graphs from \cite[Section~2.2 and Theorem~12]{bruckner2024extending}. For a connected circle graph, its canonical split tree has one leaf for each vertex and every internal node has degree at least three; the graph on the markers incident to an internal node is prime, a clique, or a star, and the tree together with its marker graphs determines $G$. Once a reference chord is fixed, each prime node admits at most four chord representations, and a clique or star node of degree $d$ admits $(d-1)!$. Summing over the $b$ choices of reference label,
    \begin{equation}\label{eq:split-representation-bound}
        N_b(G)\leq b\,4^{\#\text{ prime nodes}}
        \prod_{\nu\text{ a clique or star node}}(d_\nu-1)!\,,
    \end{equation}
    which counts linear words with a fixed initial position, as required.

    On $d$ distinguishable markers there are at most $(2d)!/2^d$ prime circle graphs, the total number of labeled chord diagrams, and at most $d+1$ cliques or stars. After squaring \eqref{eq:split-representation-bound} and summing over the marker graphs, a node of degree $d\geq3$ therefore contributes at most
    \begin{equation}\label{eq:split-node-bound}
        16\frac{(2d)!}{2^d}+(d+1)((d-1)!)^2
        \leq C^d((d-1)!)^2,
    \end{equation}
    using $(2d)!/(2^d((d-1)!)^2)=d^2\binom{2d}{d}/2^d\leq d^22^d$.

    It remains to sum over split trees. Root each tree at the leaf labeled $1$ and delete that leaf; the result is a rooted tree with $b-1$ labeled leaves in which an internal vertex of original degree $d_\nu$ has $d_\nu-1\geq2$ children. There are at most $b-2$ internal vertices, so $\sum_\nu d_\nu\leq3b$ and the exponential factors in \eqref{eq:split-node-bound} are absorbed into $C^b$. For a fixed leaf-labeled tree there are exactly $\prod_\nu(d_\nu-1)!$ ways to order the children at all internal vertices, and these orderings are distinct because the child subtrees carry disjoint nonempty sets of leaf labels. Ordering the children thus absorbs one of the two factorials at every node:
    \begin{equation}\label{eq:connected-chord-collision}
        \sum_{G\text{ connected}}N_b(G)^2
        \leq C^b b^2\sum_{T\text{ ordered}}\prod_{\nu\text{ internal}}(d_\nu-1)!\,,
    \end{equation}
    where the sum now runs over all rooted ordered trees with $b-1$ labeled leaves and at least two children at every internal vertex. Such a tree has at most $2b-3$ vertices, so there are at most $C^b$ ordered tree shapes and $(b-1)!$ leaf labelings, and
    \begin{equation}\label{eq:ordered-tree-factorial}
        \prod_\nu(d_\nu-1)!\leq(b-1)!\,,
    \end{equation}
    because for a fixed unordered tree, different child orders give different left-to-right permutations of the labeled leaves. Hence, for $b\geq3$,
    \begin{equation}\label{eq:connected-chord-second-moment}
        \sum_{G\text{ connected}}N_b(G)^2
        \leq C^b b^2((b-1)!)^2
        =C^b(b!)^2,
    \end{equation}
    and the cases $b=1,2$ are included by increasing $C$.

    \emph{Disconnected graphs.} By \eqref{eq:labeled-factorial-factor}, the left-hand side of \eqref{eq:chord-word-second-moment} divided by $b!$ counts triples consisting of two unlabeled chord diagrams and an isomorphism of their intersection graphs. The connected components of a chord diagram have non-crossing endpoint supports (\cref{sec:preliminaries}), so each support partition can be chosen in at most $|NC(2b)|\leq16^b$ ways; a correspondence between $r$ components can be chosen in $r!$ ways; and by \eqref{eq:connected-chord-second-moment} and the same labeling identity, a matched pair of components of size $b_i$ contributes at most $C^{b_i}b_i!$. Since $b_i\geq1$ and $\sum_ib_i=b$,
    \begin{equation}
        r!\prod_{i=1}^r b_i!\leq b!\,,
    \end{equation}
    as one sees by placing one copy of each component label in the first $r$ positions in an arbitrary order and appending the remaining copies in a fixed order. The component sizes are determined by the support partitions and the correspondence, so the number of triples is at most $C^bb!$, and multiplying by $b!$ completes the proof.
\end{proof}

\paragraph{A midpoint counting estimate.}
The next lemma controls the interleavings that appear when a word is cut into two halves; it is used with multiplicities $p=3$ and $p=4$.
\begin{lemma}[Midpoint profiles]\label{lem:word-midpoint-profiles}
    Fix $p\in\{3,4\}$ and $B\geq1$. There exists $K\geq B$ such that the following holds for every $b\geq2$. Put $r=\lfloor b/2\rfloor$ and sum over nonnegative integers $b_0,\ldots,b_p$ with
    \begin{equation}\label{eq:word-profile-constraints}
        \sum_{i=0}^p b_i=b,
        \qquad \sum_{i=0}^p i b_i=pr.
    \end{equation}
    For $s=\sum_{i=1}^{p-1}b_i$, define
    \begin{equation}\label{eq:word-profile-factor}
        R_p(b_0,\ldots,b_p)\equiv
        \frac{\prod_{i=0}^p b_i!}{b!}
        \frac{(pr)!}{(p b_p)!\prod_{i=1}^{p-1}(i b_i)!}
        \frac{(p(b-r))!}{(p b_0)!\prod_{i=1}^{p-1}((p-i)b_i)!}.
    \end{equation}
    Then
    \begin{equation}\label{eq:word-profile-sum}
        \sum_{(b_0,\ldots,b_p)}
        R_p(b_0,\ldots,b_p)(B/K)^s\leq1.
    \end{equation}
\end{lemma}
\begin{proof}
    We first show that every profile satisfies
    \begin{equation}\label{eq:uniform-profile-bound}
        R_p(b_0,\ldots,b_p)
        \leq\frac{C^s}{\binom br}\binom bs^{p+1}.
    \end{equation}
    If $s=0$ the unique profile is $b_p=r$, $b_0=b-r$, and \eqref{eq:uniform-profile-bound} holds with equality. Suppose $s>0$. Each half contains between $s$ and $(p-1)s$ positions occupied by the mixed classes $V_1,\ldots,V_{p-1}$, and $ps$ such positions in total. Choosing these positions and then their classes bounds the product of the two multinomials by
    \begin{equation}\label{eq:profile-interleavings}
        \frac{(pr)!}{(p b_p)!\prod_{i=1}^{p-1}(i b_i)!}
        \frac{(p(b-r))!}{(p b_0)!\prod_{i=1}^{p-1}((p-i)b_i)!}
        \leq\left(\frac{ep(p-1)b}{s}\right)^{ps}.
    \end{equation}
    Put $d=b-r-b_0=p^{-1}\sum_{i=1}^{p-1}(p-i)b_i$, so that $1\leq d\leq s$. Since the product of the factorials of all classes other than $V_0$ is at most $(b-b_0)!$,
    \begin{align}
        \frac{\prod_i b_i!}{b!}
        &\leq\frac1{\binom b{b_0}}
        =\frac1{\binom br}
          \frac{\binom{r+d}{d}}{\binom{b-r}{d}}\notag\\
        &\leq\frac1{\binom br}\left(\frac{eb}{d}\right)^d
        \leq\frac1{\binom br}\left(\frac{eb}{s}\right)^s,
        \label{eq:profile-factorial-bound}
    \end{align}
    where the last inequality holds because $x\log(eb/x)$ is increasing on $0<x\leq b$. Combining \cref{eq:profile-interleavings,eq:profile-factorial-bound} with $\binom bs\geq(b/s)^s$ proves \eqref{eq:uniform-profile-bound}.

    For fixed $s$ there are at most $(p-1)^s$ profiles, since a profile is encoded by a word of length $s$ over the $p-1$ mixed classes and its pure-class sizes are then determined by \eqref{eq:word-profile-constraints}. Absorbing this factor into $C^s$,
    \begin{align}
        \sum_{(b_0,\ldots,b_p)}R_p(B/K)^s
        &\leq\frac1{\binom br}
            \sum_{s=0}^b\binom bs^{p+1}(CB/K)^s
        \leq\frac1{\binom br}
            \left(\sum_{s=0}^b\binom bs(CB/K)^{s/(p+1)}\right)^{p+1}\notag\\
        &=\frac{\bigl(1+(CB/K)^{1/(p+1)}\bigr)^{(p+1)b}}{\binom br}.
        \label{eq:profile-binomial-sum}
    \end{align}
    Since $r=\lfloor b/2\rfloor$ and $b\geq2$, $\binom br\geq(b/r)^r\geq2^r\geq2^{b/3}$. Choosing $K\geq B$ so large that $\bigl(1+(CB/K)^{1/(p+1)}\bigr)^{p+1}\leq2^{1/3}$ makes \eqref{eq:profile-binomial-sum} at most one for every $b\geq2$.
\end{proof}

\subsection{Blocks of size four}
\begin{lemma}[Partitions with blocks of size four]
    \label{lem:quartic-even-crossing-count}
    For an absolute constant $C>0$,
    \begin{equation}\label{eq:quartic-even-crossing-count}
        |Z_{2b,b}|\leq C^b b!\qquad(b\geq1).
    \end{equation}
\end{lemma}
\begin{proof}
    Both parts of the proof cut a word into two halves, classify each label by how many of its copies fall in the left half, bound the number of possibilities within each class, and sum over the class sizes with \cref{lem:word-midpoint-profiles}.

    \emph{An auxiliary count.} Let $T_b$ be the number of pairs $(u,w)$ on a fixed alphabet of $b$ labels such that $u$ is an ordering of the labels, $w$ has three copies of each label, and the concatenation $uw$ is balanced. We claim
    \begin{equation}\label{eq:triple-word-bound}
        T_b\leq C^b(b!)^2.
    \end{equation}
    For fixed $w$ there is at most one admissible $u$, since
    \begin{equation}\label{eq:triple-order-condition}
        \operatorname{inv}_{ab}(uw)
        =\mathbf{1}_{\{b\text{ precedes }a\text{ in }u\}}
        +\operatorname{inv}_{ab}(w)+1\pmod2
    \end{equation}
    determines the relative order in $u$ of every pair of labels.

    We prove \eqref{eq:triple-word-bound} by induction on $b$, with $T_0=T_1=1$. For $b\geq2$, cut $w$ after $3r$ positions, $r=\lfloor b/2\rfloor$, and let $V_i$ be the set of labels with exactly $i$ copies in the left half, $b_i=|V_i|$. The profile $(b_0,\ldots,b_3)$ satisfies \eqref{eq:word-profile-constraints} with $p=3$. Restricted to a pure class $V_0$ or $V_3$, the pair $(u,w)$ is a smaller pair of the same kind, counted by $T_{b_i}$. Restricted to a mixed class $V_1$ or $V_2$, the word is $(u|_{V_i})\,v\,D$ or $(u|_{V_i})\,D\,v$, where $v$ is the chunk with one copy of each label and $D$ the chunk with two. Moving $D$ past $v$ changes the order of exactly $2\cdot1+2\cdot1=4$ position pairs for every pair of labels, so it preserves balance; afterwards $(u|_{V_i})v$ is a chord diagram, its cross-inversions with $D$ are even, and the balance condition becomes
    \begin{equation}\label{eq:mixed-class-chord-pair}
        g\bigl((u|_{V_i})v\bigr)=g(D).
    \end{equation}
    The map $(u|_{V_i},v,D)\mapsto((u|_{V_i})v,D)$ is injective (split the first word after $b_i$ positions), so by \cref{lem:chord-word-second-moment} a mixed class admits at most $B^{b_i}(b_i!)^2$ configurations, for an absolute constant $B$.

    There are $b!/\prod_ib_i!$ assignments of labels to classes, the two halves of $w$ are interleaved in at most the two multinomials of \eqref{eq:word-profile-factor} ways, and once these choices and the within-class data are fixed, $w$ is determined and $u$ is determined by \eqref{eq:triple-order-condition}. Under the induction hypothesis $T_j\leq K^j(j!)^2$ for $j<b$, which applies because every pure class is smaller than $b$, a profile therefore contributes at most
    \begin{equation}
        K^b(b!)^2 R_3(b_0,b_1,b_2,b_3)(B/K)^{b_1+b_2}.
    \end{equation}
    Summing over profiles with \cref{lem:word-midpoint-profiles} ($p=3$) closes the induction and proves \eqref{eq:triple-word-bound}.

    \emph{Blocks of size four.} We count the $b!\,|Z_{2b,b}|$ balanced words with four copies of each of $b$ labels, by induction on $b$ with the base cases $b=0,1$ (the empty word counts once). Cut such a word after $4r$ positions, $r=\lfloor b/2\rfloor$; the classes $V_0,\ldots,V_4$ satisfy \eqref{eq:word-profile-constraints} with $p=4$. The pure classes $V_0,V_4$ are smaller words of the same kind. The class $V_1$ is a pair counted by $T_{b_1}$, and so is $V_3$ after reversing the restricted word, which preserves balance because all multiplicities are even. In $V_2$ the two halves are chord diagrams whose cross-inversions are even, so they must have the same intersection graph, and \cref{lem:chord-word-second-moment} applies. Choosing $B$ to dominate the constants of all three mixed classes, the label assignment and interleaving counts are as before, so under the induction hypothesis $j!\,|Z_{2j,j}|\leq K^j(j!)^2$ for $j<b$ a profile contributes at most
    \begin{equation}
        K^b(b!)^2 R_4(b_0,b_1,b_2,b_3,b_4)
        (B/K)^{b_1+b_2+b_3};
    \end{equation}
    the balance conditions between different classes can only decrease this count. \cref{lem:word-midpoint-profiles} ($p=4$) gives $b!\,|Z_{2b,b}|\leq K^b(b!)^2$, and dividing by $b!$ proves the claim.
\end{proof}

\subsection{Blocks of arbitrary even size}
We compare inserting an even number of copies of a new label with inserting four copies. The existing labels may have arbitrary multiplicities.
\begin{lemma}[Insertion of one label]\label{lem:even-label-insertion}
    Fix a word $w$ of length $h$. For even $r\geq4$, let $I_r(w)$ be the number of words obtained from $w$ by inserting $r$ indistinguishable copies of a new label so that the new label has inversion parity zero with every existing label. Then
    \begin{equation}\label{eq:even-label-insertion}
        I_r(w)\leq\bigl(2(h+r+1)\bigr)^{r-4}I_4(w).
    \end{equation}
\end{lemma}
\begin{proof}
    Suppose $w$ uses $b$ labels, and for $0\leq j\leq h$ let $v_j\in\F_2^b$ record the parities of the multiplicities in the prefix ending at the $j$th gap. Inserting $c_j$ copies of the new label into that gap, with the new label taken smaller than the existing ones, contributes $c_jv_j$ to the vector of inversion parities; the opposite orientation gives the same condition because $r$ is even. Fourier inversion over $\F_2^b$ therefore gives
    \begin{equation}\label{eq:insertion-fourier}
        I_r(w)=2^{-b}\sum_{\xi\in\F_2^b}
        [z^r]\prod_{j=0}^{h}
        \left(1-(-1)^{\langle\xi,v_j\rangle}z\right)^{-1},
    \end{equation}
    where the factors correspond to the $h+1$ gaps, whose insertion multiplicities are chosen independently.

    Fix $\xi$. Replacing every sign by its negative leaves the even coefficients unchanged, so we may assume there are at least as many positive as negative signs. Pairing opposite signs writes the product as
    \begin{equation}\label{eq:positive-insertion-product}
        (1-z^2)^{-u}(1-z)^{-v},
        \qquad u,v\geq0,\quad 2u+v=h+1,
    \end{equation}
    whose coefficients are nonnegative. Set $a_j=[z^j](1-z)^{-u}$, $b_k=[z^k](1-z)^{-v}$ and $H_k=[z^k](1-z^2)^{-u}(1-z)^{-v}$, with the convention that a zero exponent gives the sequence $1,0,0,\ldots$. The binomial formulas give, for $k\leq r$ and $j\leq r/2$,
    \begin{equation}
        b_k\leq(h+r+1)^2 b_{k-2}\quad(k\geq2),
        \qquad a_{j+1}\leq(h+r+1)a_j .
    \end{equation}
    For even $4\leq k\leq r-2$, expand
    \begin{equation}
        H_{k+2}=\sum_{j=0}^{k/2}a_jb_{k+2-2j}+a_{k/2+1}.
    \end{equation}
    The sum is at most $(h+r+1)^2H_k$ and the last term is at most $(h+r+1)H_k$, because $a_{k/2}$ is one of the terms of $H_k$. Thus
    \begin{equation}
        H_{k+2}\leq2(h+r+1)^2H_k,
        \qquad
        H_r\leq\bigl(2(h+r+1)\bigr)^{r-4}H_4,
    \end{equation}
    with equality for $r=4$. Averaging this coefficientwise comparison over $\xi$ in \eqref{eq:insertion-fourier} proves the lemma.
\end{proof}

\begin{proof}[Proof of \cref{lem:even-block-count}]
    Fix even multiplicities $r_1,\ldots,r_b\geq4$ with $\sum_ir_i=2q$, and count the balanced words with these multiplicities. Deleting one label leaves a balanced word $w$, and the number of ways to reinsert that label is $I_{r_i}(w)$. By \cref{lem:even-label-insertion}, and since every word in the process has length at most $2q$, replacing $r_i$ by four changes the count by a factor of at most $(6q)^{r_i-4}$; summing over $w$ preserves this inequality. Repeating for all $b$ labels and applying \cref{lem:quartic-even-crossing-count}, the count for this multiplicity vector is at most
    \begin{equation}
        (6q)^{2q-4b}\,b!\,|Z_{2b,b}|
        \leq (6q)^{2q-4b}C^b(b!)^2 .
    \end{equation}
    Writing $r_i=4+2s_i$, the number of multiplicity vectors is $\binom{q-b-1}{b-1}\leq2^q$. Summing over them and dividing by $b!$, since every unlabeled partition has exactly $b!$ block labelings,
    \begin{equation}
        |Z_{q,b}|\leq2^qC^b b!\,(6q)^{2q-4b}
        \leq C^q b!\,q^{2q-4b}
    \end{equation}
    after increasing $C$.
\end{proof}
\section{Moment computation with matrix multiplication}
\label{sec:fmm-cumulants}
Step~(1) of \cref{alg:sfa-algorithm} computes the moments $\nTr(H^k)$, for even $k\leq\ell$, by sparse multiplication in the Pauli basis at a cost of $\Os(n\ell\binom{m}{\ell/2})$. In this section we provide an alternative algorithm which sums over subgraphs of the commutation graph using matrix multiplication. This is based on the method introduced by Ne\v{s}et\v{r}il and Poljak for the $k$-clique problem~\cite{NesetrilPoljak1985}.

By \eqref{eq:moments-graph-statistics}, in the loop-free regime each moment is a sum over even partitions of injectively labeled products of commutation signs. Replacing injective labelings by unrestricted ones turns each term into a homomorphism sum of the odd-intersection graph, which fast matrix multiplication evaluates in $\Os(m^{\omega\lceil r/3\rceil})$ for a pattern on $r$ vertices instead of $(m)_r$; here $2\leq\omega\leq 9/4$ is the exponent of square matrix multiplication~\cite{pass2026matrix}. The change of normalization is the M\"obius inversion between injective and unrestricted homomorphism counts \cite[\S5.2.3]{Lovasz2012}, \cite[eq.~(2)]{CurticapeanDellMarx2017}, whose coefficients here factor into tangent numbers. The cumulants are then obtained from the moments by the recurrence of step~(2) of \cref{alg:sfa-algorithm}.

\begin{definition}[Signed homomorphism sum]\label{def:fmm-hom}
    For a simple graph $h$ on a vertex set $V(h)$ and a commutation sign matrix $W$ from \cref{def:commutation-sign-matrices}, define
    \begin{equation}\label{eq:fmm-hom}
        \hom_W(h)\triangleq\sum_{\phi:V(h)\to[m]}\ \prod_{\{u,v\}\in E(h)}W_{\phi(u)\phi(v)},
    \end{equation}
    with the summation running over all maps $\phi$.
\end{definition}

Each summand factors over the connected components of $h$, so $\hom_W$ is multiplicative,
\[
    \hom_W(h_1\sqcup h_2)=\hom_W(h_1)\,\hom_W(h_2),\qquad \hom_W(K_1)=m .
\]
For an even partition $\pi\in\peven(k)$ and $H$ loop-free through order $k$, \eqref{eq:moments-graph-statistics} gives
\begin{equation}\label{eq:fmm-moment-M}
    \nTr(H^k)=\frac{1}{m^{k/2}}\sum_{\pi\in\peven(k)}(m)_{|\pi|}\,C[g(\pi)].
\end{equation}

\begin{lemma}[Coarsening identity]\label{lem:fmm-coarsening}
    For every $\pi\in\peven(k)$,
    \begin{equation}\label{eq:fmm-coarsening}
        \hom_W\big(g(\pi)\big)=\sum_{\pi'\geq\pi}(m)_{|\pi'|}\,C[g(\pi')],
    \end{equation}
    the sum running over all coarsenings $\pi'$ of $\pi$ in the partition lattice of $[k]$, all of which are even.
\end{lemma}
\begin{proof}
    By \cref{lem:even-word-trace}, applied to an arbitrary map $\phi$ from the blocks of $\pi$ to $[m]$, the summand of \eqref{eq:fmm-hom} for $\phi$ is the normalized trace of the Pauli word with kernel $\pi'\geq\pi$ obtained by merging the blocks of $\pi$ with equal label. Grouping the maps $\phi$ by $\pi'$, they become the injective labelings of $\pi'$, and \cref{lem:even-word-trace} applied to $\pi'$ identifies their sum with $(m)_{|\pi'|}\,C[g(\pi')]$.
\end{proof}

\subsection{Evaluating moments with homomorphism sums}

Let $T_s$ denote the tangent numbers,
\[
    \tan x=\sum_{s\geq1}T_s\,\frac{x^{2s-1}}{(2s-1)!},\qquad (T_1,T_2,T_3,T_4,T_5)=(1,2,16,272,7936),
\]
and for an even partition $\pi$ set
\begin{equation}\label{eq:fmm-weight}
    \epsilon(\pi)\triangleq\prod_{U\in\pi}(-1)^{|U|/2-1}\,T_{|U|/2}.
\end{equation}

\begin{proposition}[Loop-free moments from homomorphism sums]\label{prop:fmm-moments}
    If $k$ is even and $H$ is loop-free through order $k$, then
    \begin{equation}\label{eq:fmm-moment-expansion}
        \nTr(H^k)=\frac{1}{m^{k/2}}\sum_{\pi\in\peven(k)}\epsilon(\pi)\,\hom_W\big(g(\pi)\big).
    \end{equation}
\end{proposition}
\begin{proof}
    M\"obius inversion of \eqref{eq:fmm-coarsening} on the interval $[\pi,1_{k}]$ of the partition lattice (\cref{lem:mobius_inversion} for the dual order) gives
    \[
        (m)_{|\pi|}\,C[g(\pi)]=\sum_{\pi'\geq\pi}\mu(\pi,\pi')\,\hom_W\big(g(\pi')\big),
    \]
    the identity between injective and unrestricted homomorphism counts of \cite[eq.~(5.18)]{Lovasz2012}, \cite[eq.~(2)]{CurticapeanDellMarx2017} for the weighted host $W$, with the quotients indexed by coarser partitions through \cref{lem:fmm-coarsening}. Summing over $\pi\in\peven(k)$ and exchanging the order of summation, \eqref{eq:fmm-moment-M} gives
    \[
        \nTr(H^k)=\frac{1}{m^{k/2}}\sum_{\pi'\in\peven(k)}\hom_W\big(g(\pi')\big)\sum_{\substack{\pi\in\peven(k)\\\pi\leq\pi'}}\mu(\pi,\pi').
    \]
    The even refinements of $\pi'$ are the partitions whose restriction to each block $U\in\pi'$ is an even partition of $U$, and the M\"obius function of the partition lattice factors over the blocks, so the inner sum equals $\prod_{U\in\pi'}e_{|U|/2}$ with
    \[
        e_s\triangleq\sum_{\tau\in\peven(2s)}\mu(\tau,1_{2s})=\sum_{\tau\in\peven(2s)}(-1)^{|\tau|-1}(|\tau|-1)!\,.
    \]
    By the exponential formula, $\sum_{s\geq1}e_s\,x^{2s}/(2s)!=\log\big(1+\sum_{j\geq1}x^{2j}/(2j)!\big)=\log\cosh x$, whose derivative is $\tanh x=\sum_{s\geq1}(-1)^{s-1}T_s\,x^{2s-1}/(2s-1)!$. Hence $e_s=(-1)^{s-1}T_s$ and the inner sum is $\epsilon(\pi')$.
\end{proof}

\subsection{The alternative moment computation}
We first give the general algorithmic result of Ne\v{s}et\v{r}il--Poljak~\cite{NesetrilPoljak1985}:
\begin{lemma}[Ne\v{s}et\v{r}il--Poljak~\cite{NesetrilPoljak1985}]\label{lem:fmm-three-way-split}
    Let $h$ be a simple graph on $r$ vertices. Then $\hom_W(h)$ can be computed in time $\Os\big(|E(h)|\,m^{2\lceil r/3\rceil}+m^{\omega\lceil r/3\rceil}\big)$.
\end{lemma}
\begin{proof}
    Partition $V(h)=A\sqcup B\sqcup C$ with $|A|,|B|,|C|\leq\lceil r/3\rceil$. Every edge of $h$ lies inside one part or joins two distinct parts, so each edge belongs to exactly one of the six classes $AA$, $BB$, $CC$, $AB$, $BC$, $CA$. A map $\phi:V(h)\to[m]$ is a triple $(\alpha,\beta,\gamma)$ of its restrictions to the three parts, where $\alpha\in[m]^{A}$, $\beta\in[m]^{B}$ and $\gamma\in[m]^{C}$. Define the matrices
    \begin{equation}\label{eq:fmm-xyz}
        X[\alpha,\beta]\triangleq\!\!\prod_{e\in E_{AA}\cup E_{AB}}\!\!W_e(\phi),\qquad
        Y[\beta,\gamma]\triangleq\!\!\prod_{e\in E_{BB}\cup E_{BC}}\!\!W_e(\phi),\qquad
        Z[\gamma,\alpha]\triangleq\!\!\prod_{e\in E_{CC}\cup E_{CA}}\!\!W_e(\phi),
    \end{equation}
    where $W_e(\phi)\triangleq W_{\phi(u)\phi(v)}$ for $e=\{u,v\}$; each matrix depends only on the two displayed tuples. Since every edge appears in exactly one of the three products,
    \begin{equation}\label{eq:fmm-trace}
        \hom_W(h)=\sum_{\alpha,\beta,\gamma}X[\alpha,\beta]\,Y[\beta,\gamma]\,Z[\gamma,\alpha]
        =\sum_{\alpha,\gamma}(XY)[\alpha,\gamma]\,Z[\gamma,\alpha]=\Tr(XYZ).
    \end{equation}
    The matrices have at most $m^{2\lceil r/3\rceil}$ entries, each a product of at most $|E(h)|$ signs. The product $XY$ is a multiplication of matrices of side at most $m^{\lceil r/3\rceil}$, which costs $\Os(m^{\omega\lceil r/3\rceil})$, and the final trace is an entrywise product summed over $m^{|A|+|C|}\leq m^{2\lceil r/3\rceil}$ pairs.
\end{proof}

For the case of the sparse Pauli model, grouping the partitions in \eqref{eq:fmm-moment-expansion} by the isomorphism class of their odd-intersection graph and using the multiplicativity of $\hom_W$,
\begin{equation}\label{eq:fmm-grouped}
    \nTr(H^k)=\frac{1}{m^{k/2}}\sum_{\gamma}a^{(k)}_{\gamma}\;m^{i(\gamma)}\prod_{h\in\mathrm{comp}(\gamma)}\hom_W(h),\qquad
    a^{(k)}_{\gamma}\triangleq\sum_{\substack{\pi\in\peven(k)\\ g(\pi)\cong\gamma}}\epsilon(\pi),
\end{equation}
where $\gamma$ ranges over isomorphism classes of simple graphs on at most $k/2$ vertices, $i(\gamma)$ is the number of isolated vertices of $\gamma$, and $\mathrm{comp}(\gamma)$ is the multiset of connected components of $\gamma$ with at least two vertices. The integers $a^{(k)}_\gamma$ and the list of connected patterns $h$ depend only on $k$.

\begin{algorithm}[ht]
\caption{Low-order moments from commutation graph statistics}\label{alg:fmm-cumulants}
\begin{algorithmic}[1]
\Require Pauli operators $P_1,\ldots,P_m$ on $n$ qubits, even $\ell\geq 4$.
\State \textbf{Precomputation}: for each even $k\leq\ell$, enumerate $\pi\in\peven(k)$, compute $\epsilon(\pi)$ and $g(\pi)$, and assemble the coefficients $a^{(k)}_\gamma$ of \eqref{eq:fmm-grouped} and the list of connected patterns $h$ that occur as components.
\For{each connected pattern $h$ in the list}
    \State Compute $\hom_W(h)=\Tr(XYZ)$ by \cref{lem:fmm-three-way-split}.
\EndFor
\For{each even $k\leq\ell$}
    \State $\hat{\mu}_{k}\gets m^{-k/2}\sum_\gamma a^{(k)}_\gamma\,m^{i(\gamma)}\prod_{h\in\mathrm{comp}(\gamma)}\hom_W(h)$.
\EndFor
\State \Return $\hat{\mu}_2,\hat{\mu}_4,\ldots,\hat{\mu}_{\ell}$.
\end{algorithmic}
\end{algorithm}

\begin{proposition}[Correctness and cost]\label{prop:fmm-cost}
    On the loop-free event of order $\ell$ (\cref{lem:loop-freeness}), \cref{alg:fmm-cumulants} outputs the exact moments $\nTr(H^k)$ of $H$ for even $k\leq\ell$. Its running time is
    \begin{equation}\label{eq:fmm-cost}
        \Os(m^2n)+\sum_{\substack{2\leq k\leq\ell\\k\text{ even}}}|\peven(k)| \cdot \Os\big((k/2)^2m^{\omega\lceil k/6\rceil}\big)
        \leq \Os\big(m^2n\big)+ \ell^{\ell+\Os(1)} m^{\omega\lceil\ell/6\rceil}.
    \end{equation}
    The precomputation costs $\ell^{\Os(\ell)}$ and is independent of $n$ and $m$.
\end{proposition}
\begin{proof}
    Correctness of the moments is \cref{prop:fmm-moments} with \eqref{eq:fmm-grouped}. We have $|\peven(k)|\leq k^k$, and forming $W$ takes $\Os(m^2n)$ bit operations. Each pattern has at most $k/2$ vertices and $\binom{k/2}{2}$ edges, and \cref{lem:fmm-three-way-split} bounds the cost of each homomorphism sum.
\end{proof}
\begin{corollary}[Running time dependence on $\varepsilon$]\label{cor:fmm-runtime}
    In the setting of \cref{thm:scalar-runtime}, suppose in addition that $\log(1/\varepsilon)\leq m^{o(1)}$. If Step~(1) of \cref{alg:sfa-algorithm} is replaced by \cref{alg:fmm-cumulants}, then the result of \cref{thm:scalar-runtime} holds with the same success probability and with total running time
    \begin{equation}\label{eq:fmm-accuracy-runtime}
        \Os(m^2n)+\delta^{-1}\,m^{2\omega/3+o(1)}\,(1/\varepsilon)^{2\omega/3+o(1)}.
    \end{equation}
\end{corollary}
\begin{proof}
     \cref{prop:fmm-cost} shows that \cref{alg:fmm-cumulants} returns the same moments as Step~(1) of \cref{alg:sfa-algorithm}, so the output and error bounds are unchanged. For the running time we may assume $\ell\geq4$, since otherwise the semicircular output is returned directly. Minimality of the cutoff $\ell$ in \eqref{eq:optimized-cumulant-cutoff} gives
    \[
        m^{\ell/2}<C_0^2(C_0\ell)^{\ell/2}\delta^{-1}\varepsilon^{-2}.
    \]
    Since $\ell$ is even, $\lceil\ell/6\rceil\leq(\ell+4)/6$, and therefore
    \begin{equation}
        m^{\omega\lceil\ell/6\rceil}\leq m^{2\omega/3}\bigl(m^{\ell/2}\bigr)^{\omega/3} \leq m^{2\omega/3}\,\ell^{\Os(\ell)}\,\delta^{-\omega/3}\,\varepsilon^{-2\omega/3}.
    \end{equation}
    By \eqref{eq:optimized-cutoff-size} and the assumption $\log(1/\varepsilon)\leq m^{o(1)}$ we have $\ell\leq m^{o(1)}$, hence $\ell\log\ell=o(\log m)+o(\log(1/\varepsilon))$ and $\ell^{\Os(\ell)}=m^{o(1)}(1/\varepsilon)^{o(1)}$. Substituting into \eqref{eq:fmm-cost}, and using $\omega\leq3$ together with the fact that the moment--cumulant conversions cost $\Os(\ell L^2)\leq\Os(m)$, proves \eqref{eq:fmm-accuracy-runtime}.
\end{proof}

\section{Sharper concentration via hypercontractivity}
\label{sec:hypercontractivity}
In this section we prove the hypercontractivity bounds in \cref{corollary:cumulant-hypercontractivity,corollary:color-cumulant-hypercontractivity}. Hypercontractivity bounds higher-order moments of low-degree polynomials on the Boolean hypercube utilizing their second moments, which we have already bounded. We first consider low-degree polynomials over independent signs on the Boolean hypercube, then convert these estimates to the actual Pauli model commutation signs.

Throughout this appendix, $s\geq0$ is fixed. When $s\geq1$, the Pauli generators $Q_1,\ldots,Q_s$ are fixed independently of the sampled support and independent modulo phase. The case $s=0$ denotes the scalar problem, with a single empty color sequence and no root variables. All Pauli supports are sampled as in \cref{eq:1}. Define $X_{ij}$ and $Y_{ai}$ to be matrices with the same index sets as $W, V$ from \cref{def:commutation-sign-matrices}, but with mutually independent uniform signs instead of Pauli commutations. We use $\E_{\mathrm{ind}}$ for expectation over $(X,Y)$, and $\E_\Pm$ for expectation over the sampled Pauli support. No expectation below is conditioned on loop-freeness; we utilize it only after proving the concentration bounds.

\subsection{Low-degree polynomials over independent signs}
For a real multilinear polynomial $F$ in finitely many sign variables, let $\|F\|_{1}$ be the sum of the absolute values of its coefficients. Reducing powers by $x^2=1$ does not increase this norm, and multiplication followed by reduction is submultiplicative for this norm.

\begin{lemma}\label{lem:hypercontractive-monomial-comparison}
    Let $W,V$ be the commutation matrices from \cref{def:commutation-sign-matrices}. For every real polynomial $F$ reduced to multilinear form and every even integer $p\geq2$,
    \begin{equation}\label{eq:hypercontractive-moment-transfer}
        \left|\E_\Pm|F(W,V)|^p-\E_{\mathrm{ind}}|F(X,Y)|^p\right|
        \leq4^{-n}\|F\|_{1}^{p}.
    \end{equation}
\end{lemma}
\begin{proof}
    Under the independent signs in $X,Y$, every nonempty square-free monomial has expectation zero, while the empty monomial has expectation one in both models.

    Expand $F^p$ and reduce all powers modulo two. The sum of the absolute coefficients of the resulting multilinear polynomial is at most $\|F\|_{1}^p$. The constant term cancels when subtracting the expectations in the two models, and each remaining monomial contributes a difference of absolute value at most $4^{-n}$. Therefore
    \[
        \left|\E_\Pm F(W,V)^p-\E_{\mathrm{ind}}F(X,Y)^p\right|
        \leq4^{-n}\|F\|_{1}^{p}.
    \]
    Since $p$ is even and $F$ is real, $F^p=|F|^p$, proving \eqref{eq:hypercontractive-moment-transfer}.
\end{proof}

For even $k\geq4$ and a label sequence $\Cm\in\Cs^k$, define the polynomial over independent signs by the same expansion as \cref{def:color-loop-free-expression}:
\begin{equation}\label{eq:hypercontractive-graph-polynomial}
    F_{k,\Cm}(X,Y)\equiv\frac1{m^{k/2}}
    \sum_{\pi\in\pevenconn(k)}\chi(\pi)
    \sum_{\iota:\pi\hookrightarrow[m]}
    \left(\prod_{\{U,V\}\in E(g(\pi))}X_{\iota(U),\iota(V)}\right)
    \left(\prod_{U\in\pi}\prod_{a=1}^s Y_{a,\iota(U)}^{(\Cm_\pi(U))_a}\right),
\end{equation}
where $X_{ij}=X_{ji}$ for distinct labels. The graph has one vertex per block before any isolated vertices are removed. At the actual signs, this polynomial is $F_{k,\Cm}(W,V)=\tilde\kappa_k^{(\Cm)}$ for every sampled support; on the loop-free event it is the true cumulant by \cref{thm:color-cumulants-from-graphs}. At $s=0$, the same statement uses \cref{def:loop-free-expression} and the scalar setting.

\begin{lemma}\label{lem:hypercontractive-polynomial-inputs}
    For any color sequence $\Cm\in\Cs^k$, the polynomial in \eqref{eq:hypercontractive-graph-polynomial}, after multilinear reduction, satisfies
    \begin{equation}\label{eq:hypercontractive-polynomial-inputs}
        \deg F_{k,\Cm}\leq D_{k,s}\equiv\binom{k/2}{2}+\frac{sk}{2},
        \qquad
        \|F_{k,\Cm}\|_{1}\leq(6k)^k.
    \end{equation}
    For sufficiently large constants $A,C$, if $m\geq Ak^3$, then
    \begin{equation}\label{eq:hypercontractive-independent-second-moment}
        \E_{\mathrm{ind}}|F_{k,\Cm}(X,Y)|^2
        \leq\left(\frac{Ck}{m}\right)^{k/2}.
    \end{equation}
    Assuming \cref{conjecture:combinatorial-linear}, it suffices to assume $m\geq Ak$ instead of $m\geq Ak^3$.
\end{lemma}
\begin{proof}
    An even partition of $k$ positions has at most $k/2$ blocks. The factors between Pauli Hamiltonian terms have degree at most $\binom{k/2}{2}$, and there are at most $sk/2$ factors between Pauli Hamiltonian terms and colors. This proves the degree bound. Since $(m)_{|\pi|}/m^{k/2}\leq1$, \cref{lem:growth_chi} gives
    \[
        \|F_{k,\Cm}\|_{1}
        \leq\frac1{m^{k/2}}\sum_{\pi\in\pevenconn(k)}(m)_{|\pi|}|\chi(\pi)|
        \leq6^k|\pevenconn(k)|\leq(6k)^k.
    \]
    For the second moment, expand the square as a sum over two partitions and their injective labelings. A product survives the independent-sign expectation only if its labeled Pauli Hamiltonian edge sets agree and each sign from a color has even multiplicity. Taking absolute values of the coefficients and ignoring the even multiplicity condition, this leaves exactly the absolute-value count of matching labeled edge sets used in the proof of \cref{lemma:second-moment-upper}. By \cref{lemma:chi-vanish}, every isolated block with nonzero coefficient has size at least four, and the gluing map from that proof therefore gives
    \begin{equation}\label{eq:hypercontractive-independent-gluing}
        \E_{\mathrm{ind}}|F_{k,\Cm}(X,Y)|^2
        \leq\frac{36^k}{m^k}\sum_{b'=1}^{k/2} m^{b'}|Z_{k,b'}|.
    \end{equation}
    Applying \cref{lem:even-block-count} gives the geometric sum \eqref{eq:weighted-surplus-sum}, and applying \cref{conjecture:combinatorial-linear} gives \eqref{eq:conditional-linear-weighted-sum}. These imply \eqref{eq:hypercontractive-independent-second-moment} in the respective ranges.
\end{proof}

\subsection{Transferring to Pauli commutation signs}
We recall the Bonami--Gross inequality~\cite{odonnell2014aobf}:
\begin{lemma}[Bonami--Gross inequality]\label{lem:bonami-even-moments}
    Let $f:\{\pm 1\}^n\rightarrow\R$ be a multilinear polynomial of degree at most $D$. Then for all $q\geq 2$,
    \begin{equation}\label{eq:bonami-even-moments}
        \|f\|_{q}\leq(q-1)^{D/2}\|f\|_2,
        \qquad \|f\|_q\equiv\E_{x\sim\{\pm 1\}^{n}}\bigl[|f(x)|^q\bigr]^{1/q}.
    \end{equation}
\end{lemma}

Combining \cref{lem:hypercontractive-monomial-comparison,lem:hypercontractive-polynomial-inputs,lem:bonami-even-moments} gives the following estimate at arbitrary even moment order. This is the only use of hypercontractivity in the proof.

\begin{lemma}[Higher moments of a fixed-order loop-free expression]\label{lem:hypercontractive-pauli-moments}
    In the range of \cref{lem:hypercontractive-polynomial-inputs}, for every even integer $p\geq2$ and color sequence $\Cm\in\Cs^k$,
    \begin{equation}\label{eq:hypercontractive-pauli-moments}
        \E_\Pm\left|\tilde\kappa_k^{(\Cm)}\right|^p
        \leq\left[(p-1)^{D_{k,s}/2}\left(\frac{Ck}{m}\right)^{k/4}\right]^p
        +4^{-n}(6k)^{k p}.
    \end{equation}
    In particular, fix an even integer $k\geq4$ and integer $s\geq0$. For sufficiently large $n$, $n\leq m\leq \poly(n)$, and $\delta=1/\poly(n)$,
    \begin{equation}\label{eq:hypercontractive-fixed-order}
        \Pr_\Pm\left(\bigl|\tilde\kappa_k^{(\Cm)}\bigr|>
        Cm^{-k/4}(\log n)^{D_{k,s}/2}\right)
        \leq \delta+e^{-c n},
    \end{equation}
    uniformly over the color sequence and fixed generators, for some $c,C>0$.
\end{lemma}
\begin{proof}
    The first bound follows by applying \cref{eq:bonami-even-moments,eq:hypercontractive-independent-second-moment} to $F_{k,\Cm}(X,Y)$ and then \eqref{eq:hypercontractive-moment-transfer}. For the second, choose $p=2\lceil\log(n/\delta)\rceil$ and put
    \[
        u=e(p-1)^{D_{k,s}/2}\left(\frac{Ck}{m}\right)^{k/4}.
    \]
    Markov's inequality gives
    \begin{equation}\label{eq:hypercontractive-markov}
        \Pr_\Pm\left(\bigl|\tilde\kappa_k^{(\Cm)}\bigr|>u\right)
        \leq e^{-p}+4^{-n}\left(\frac{(6k)^k}{u}\right)^p.
    \end{equation}
    The first term is at most $\delta$. Since $k,s$ are fixed, $m\leq \poly(n)$, and $p=\Os(\log n)$, the logarithm of the second term is at most $-n\log4+\Os(\log^2 n)\leq-cn$ for sufficiently large $n$. Finally, $u\leq Cm^{-k/4}(\log n)^{D_{k,s}/2}$. This proves \eqref{eq:hypercontractive-fixed-order}.
\end{proof}

We now prove the two concentration statements.

\begin{proof}[Proofs of \cref{corollary:cumulant-hypercontractivity,corollary:color-cumulant-hypercontractivity}]
    Fix an even integer $k_0\geq4$. Since $\delta=1/\poly(n)$, there is a constant $c_{\delta}>0$ such that $\delta\geq n^{-c_\delta}$ for all sufficiently large $n$. Choose an even integer $k_*\geq k_0$ such that
    \begin{equation}\label{eq:hypercontractive-split-choice}
        k_*+2-k_0>2c_{\delta}+4.
    \end{equation}
    Split the loop-free expression sum into orders $k_0\leq k\leq\min\{k_*,k_1\}$ and $k_*+2\leq k\leq k_1$.

    \paragraph{The fixed low orders.}
    Apply \cref{lem:hypercontractive-pauli-moments} with failure probability $\delta/n^2$ to every even $k\in[k_0,k_*]$ and every normalized label sequence $\Cm\in\Cs^k$ with $\va{c}_1=\va{0}$. There are at most $\sum_{k=k_0}^{k_*}2^{s(k-1)}$ such pairs, a fixed number. A union bound gives all these estimates simultaneously, with failure at most $\delta/4+e^{-c n}$ for sufficiently large $n$. For each fixed $k>k_0$ in this interval,
    \begin{equation}\label{eq:hypercontractive-low-order-domination}
        \frac{m^{-k/4}(\log n)^{D_{k,s}/2}}{m^{-k_0/4}(\log n)^{D_{k_0,s}/2}}
        \leq n^{-(k-k_0)/4}(\log n)^{(D_{k,s}-D_{k_0,s})/2}=o(1).
    \end{equation}
    Thus, including the fixed number of color words, the low-order sum is at most
    \begin{equation}\label{eq:hypercontractive-low-sum}
        Cm^{-k_0/4}(\log n)^{D_{k_0,s}/2}
    \end{equation}
    for some constant $C>0$.

    \paragraph{The remaining orders.}
    If $k_1\leq k_*$, this part is empty. Otherwise, the second-moment proof of \cref{thm:color-cumulant-concentration}, with failure parameter $\delta/4$ and lower order $k_*+2$, gives for the loop-free expressions
    \begin{equation}\label{eq:hypercontractive-high-sum}
        \sum_{\substack{k_*+2\leq k\leq k_1\\k\text{ even}}}\ \sum_{\substack{\Cm\in\Cs^k\\\va{c}_1=\va{0}}}
        |\tilde\kappa_k^{(\Cm)}|
        \leq C \delta^{-1/2}m^{-(k_*+2)/4}
    \end{equation}
    except with probability at most $\delta/4+e^{-\Omega(n)}$. For $s=0$, use the scalar proof; its same second-moment estimate also applies if the sum contains just its first order. The factors $2^{4s}$ and $k_*+2$ in the geometric sum are constants here. By $m\geq n$, $\delta\geq n^{-c_{\delta}}$, and \eqref{eq:hypercontractive-split-choice}, the right-hand side of \eqref{eq:hypercontractive-high-sum} is at most
    \[
        C m^{-k_0/4}n^{c_\delta/2-(k_*+2-k_0)/4}
        \leq C m^{-k_0/4}n^{-1}.
    \]
    It is therefore absorbed by \eqref{eq:hypercontractive-low-sum}.

    \paragraph{Converting to actual free cumulants.}
    Take the intersection of these events with the loop-free event through order $k_1$ from \cref{lem:loop-freeness}. \cref{thm:cumulants-from-graph-statistics,thm:color-cumulants-from-graphs} then give $\tilde\kappa_k^{(\Cm)}=\kappa_k^{(\Cm)}$ simultaneously for every order and color sequence used above, with an additional probability failure of $e^{-\Omega(n)}$. Combining the two sums and enlarging $C$ proves \cref{corollary:color-cumulant-hypercontractivity}, with failure at most $\delta+e^{-\Omega(n)}$; setting $s=0$ gives \cref{corollary:cumulant-hypercontractivity}.
\end{proof}

\clearpage
\bibliographystyle{unsrt}
\bibliography{refs}

\end{document}